\documentclass[12pt]{article}
\usepackage{color}
\usepackage{a4}
\usepackage{latexsym}
\usepackage{cite}

\usepackage{array}
\newcolumntype{T}{>{\ttfamily} c}
\newcolumntype{M}{>{$\displaystyle} c <{$}}

\usepackage{epsfig}
\usepackage{graphicx}

\usepackage{pslatex}
\usepackage[latin1]{inputenc}
\usepackage[T1]{fontenc}

\usepackage{color,colordvi}

\renewcommand{\theequation}{\thesection.\arabic{equation}}
\newcommand{\gsim}{\raisebox{-0.07cm}{$\:\:\stackrel{>}{{\scriptstyle
 \sim}}\:\: $} }

\def\Slash#1{\rlap{\hbox{$\mskip 3 mu /$}}#1}      

\newcommand{\beq}{\begin{equation}}
\newcommand{\eeq}{\end{equation}}
\newcommand{\bea}{\begin{eqnarray}}
\newcommand{\eea}{\end{eqnarray}}
\newcommand{\nn}{\nonumber}
\newcommand{\MSb}{$\overline{\mbox{MS}}$}
\newcommand{\ra}{\rightarrow}

\newcommand{\als}{\alpha_{\rm s}}
\newcommand{\ars}{a_{\rm s}}
\newcommand{\ep}{\varepsilon}

\newcommand{\hspn}{{\hspace{-5mm}}}
\newcommand{\hspp}{{\hspace{5mm}}}

\def\frct#1#2{\mbox{\large{$\frac{#1}{#2}$}}}

\def\as(#1){{\alpha_{\rm s}^{\,#1}}}
\def\ar(#1){{a_{\rm s}^{\,#1}}}

\def\mus{{\mu^{\,2}}}

\def\b#1{{\beta_{#1}}}
\def\B(#1,#2){{\beta_{#1}^{\,#2}}}

\def\nc{{n_c}}
\def\ncs{{n_{c}^{\,2}}}
\def\nct{{n_{c}^{\,3}}}
\def\ncf{{n_{c}^{\,4}}}
\def\ca{{C^{}_A}}
\def\cas{{C^{\,2}_A}}
\def\cat{{C^{\,3}_A}}

\def\cf{{C^{}_F}}
\def\cfs{{C^{\, 2}_F}}
\def\cft{{C^{\, 3}_F}}
\def\cff{{C^{\, 4}_F}}
\def\nf{{n^{}_{\! f}}}
\def\nfz{{n^{\,0}_{\! f}}}
\def\nfo{{n^{\,1}_{\! f}}}
\def\nfs{{n^{\,2}_{\! f}}}
\def\nft{{n^{\,3}_{\! f}}}
\def\tf{{\:\!T^{}_{F}}}

\def\dabcnc{{\frac{d^{abc}d_{abc}}{N_R }}}
\def\dfFAnc{{\frac{d_F^{\,abcd}d_A^{\,abcd}}{N_R }}}
\def\dfFFnc{{\frac{d_F^{\,abcd}d_F^{\,abcd}}{N_R }}}

\def\xm1{{(1 \! - \! x)}}
\def\xp1{{(1 \! + \! x)}}

\def\Lnt(#1){\ln^{\,#1}(1\!-\!x)}
\def\pqq(#1){p_{\rm{qq}}(#1)}

\def\zts{{\zeta_{3}^{\,2}}}

\def\S(#1){{{S}_{#1}}}
\def\Ss(#1,#2){{{S}_{#1,#2}}}
\def\Sss(#1,#2,#3){{{S}_{#1,#2,#3}}}
\def\Ssss(#1,#2,#3,#4){{{S}_{#1,#2,#3,#4}}}
\def\Sssss(#1,#2,#3,#4,#5){{{S}_{#1,#2,#3,#4,#5}}}
\def\Ssssss(#1,#2,#3,#4,#5,#6){{{S}_{#1,#2,#3,#4,#5,#6}}}
\def\Sssssss(#1,#2,#3,#4,#5,#6,#7){{{S}_{#1,#2,#3,#4,#5,#6,#7}}}

\def\H(#1){{\rm{H}}_{#1}}
\def\Hh(#1,#2){{\rm{H}}_{#1,#2}}
\def\Hhh(#1,#2,#3){{\rm{H}}_{#1,#2,#3}}
\def\Hhhh(#1,#2,#3,#4){{\rm{H}}_{#1,#2,#3,#4}}
\def\Hhhhh(#1,#2,#3,#4,#5){{\rm{H}}_{#1,#2,#3,#4,#5}}
\def\Hhhhhh(#1,#2,#3,#4,#5,#6){{\rm{H}}_{#1,#2,#3,#4,#5,#6}}

\begin{document}
\setlength{\parskip}{0.2cm}
\setlength{\baselineskip}{0.55cm}

\begin{titlepage}
\noindent
DESY 17--106 \hfill July 2017\\
Nikhef 2017-034 \\
LTH 1139 \\
\vspace{0.6cm}
\begin{center}
{\LARGE \bf Four-Loop Non-Singlet Splitting Functions\\[1ex]
  in the Planar Limit and Beyond}\\ 
\vspace{1.4cm}
\large
S. Moch$^{\, a}$, B. Ruijl$^{\, b,c}$, T. Ueda$^{\, b}$, 
J.A.M. Vermaseren$^{\, b}$ and A. Vogt$^{\, d}$\\
\vspace{1.4cm}
\normalsize
{\it $^a$II.~Institute for Theoretical Physics, Hamburg University\\
\vspace{0.1cm}
D-22761 Hamburg, Germany}\\
\vspace{0.5cm}
{\it $^b$Nikhef Theory Group \\
\vspace{0.1cm}
Science Park 105, 1098 XG Amsterdam, The Netherlands} \\
\vspace{0.5cm}
{\it $^c$Leiden Centre of Data Science, Leiden University \\
\vspace{0.1cm}
Niels Bohrweg 1, 2333 CA Leiden, The Netherlands}\\
\vspace{0.5cm}
{\it $^d$Department of Mathematical Sciences, University of Liverpool\\
\vspace{0.1cm}
Liverpool L69 3BX, United Kingdom}\\
\vspace{1.4cm}
{\large \bf Abstract}
\vspace{-0.2cm}
\end{center}
We present the next-to-next-to-next-to-leading order (N$^3$LO) contributions 
to the non-singlet splitting functions for both parton distribution and 
fragmentation functions in perturbative QCD.  The exact expressions are derived
for the terms contributing in the limit of a large number of colours.  For the 
remaining contributions, approximations are provided that are sufficient for 
all collider-physics applications.  From their threshold limits we derive  
analytical and high-accuracy numerical results, respectively, for all
contributions to the four-loop cusp anomalous dimension for quarks, including 
the terms proportional to quartic Casimir operators.  We briefly illustrate the
numerical size of the four-loop corrections, and the remarkable 
renormalization-scale stability of the N$^3$LO results, for the evolution of 
the non-singlet parton distribution and the fragmentation functions.  
Our~results appear to provide a first point of contact of four-loop QCD
calculations and the so-called wrapping corrections to anomalous dimensions in
$\,{\cal N}\!=4$ super Yang-Mills theory.
\vspace*{0.3cm}
\end{titlepage}
%
%
\section{Introduction}
\label{sec:intro}

Within the gauge theory of the strong interaction, Quantum Chromodynamics 
(QCD), the precision of theory predictions for hard reactions at colliders 
crucially depends on our knowledge of hadronic matrix elements for the 
description of the long-distance hadronic degrees of freedom, 
once the hard-interaction part due to short-distance physics has been 
separated by means of QCD factorization.
For scattering reactions with initial-state protons the relevant matrix 
elements are given by the well-known parton distribution functions (PDFs) 
of the proton, which provide information about the fractions of the proton's 
longitudinal momentum carried by the partons.

The dependence of these PDFs on the scale $Q^2$ is generated by evolution 
equations for the corresponding local operator matrix elements (OMEs). 
The relevant anomalous dimensions as functions of the Mellin moment $N$, 
or splitting functions as functions of the momentum fraction~$x$, can be 
computed order by order in perturbative QCD.
The corresponding one- and two-loop results have been known since
long~\cite{Gross:1973ju,Georgi:1951sr,Altarelli:1977zs,Kim:1977hp,%
Floratos:1977au,Floratos:1978ny,GonzalezArroyo:1979df,GonzalezArroyo:1979he,%
Curci:1980uw,Furmanski:1980cm,Floratos:1981hs,Hamberg:1991qt,Ellis:1996nn}.
The current precision is at the three-loop level~\cite{Moch:2004pa,Vogt:2004mw}
-- see refs.~\cite{Ablinger:2010ty,Ablinger:2014vwa,Ablinger:2014nga,%
Ablinger:2017tan} for partial recalculations of these results --
i.e., at the next-to-next-to-leading order (NNLO), 
which is nowadays the accepted standard for analyses of 
PDFs~\cite{Accardi:2016ndt} and forms the backbone of precision predictions 
at the Large Hadron Collider (LHC).  

However, computations for a number key observables at hadron colliders have 
been performed even at next-to-next-to-next-to-leading order (N$^3$LO), 
including the cross section for Higgs-boson production in gluon-gluon 
fusion~\cite{Anastasiou:2015ema} and structure functions in deep-inelastic 
scattering (DIS) \cite{Vermaseren:2005qc,Moch:2008fj,Davies:2016ruz,DMVVprep}. 
The latter results have also found an application in predicting Higgs-boson 
production in vector-boson fusion at the LHC~\cite{Dreyer:2016oyx}.
Due to QCD factorization, the resulting predictions carry a residual 
uncertainty and dependence on the factorization scheme due to the missing 
N$^3$LO (i.e., four-loop) splitting functions.
This situation motivates the computation of the QCD splitting functions at
four loops. First steps in this direction have already been taken in 
refs.~\cite{Baikov:2006ai,Velizhanin:2011es,Velizhanin:2014fua,Baikov:2015tea,%
Ruijl:2016pkm} at low $N$, and in ref.~\cite{Davies:2016jie} where large-$n_f$ 
contributions have been derived at all $N$.

In the present article, we address the splitting functions for the non-singlet 
quark evolution equations at four loops in QCD.  We use {\sc Forcer} 
\cite{Ruijl:2017cxj}, 
a {\sc Form}~\cite{Vermaseren:2000nd,Kuipers:2012rf,Tentyukov:2007mu} 
program for four-loop massless propagators, to compute the anomalous dimensions
at fixed integer values of the Mellin variable $N$.
In the planar limit, i.e., for large $\nc$ for a general colour $SU(\nc)$ gauge 
group, the exact four-loop results for moments up to $N = 20$ turn out to be
 sufficient to find and validate the analytic expressions as functions of $N$ 
in terms of harmonic sums~\cite{Vermaseren:1998uu,Blumlein:1998if} by LLL-based 
techniques~\cite{Lenstra1982,axbAlg,DBLP:journals/dcc/Silverman00,Calc} 
for solving systems of Diophantine equations.
Such an approach has been used for anomalous dimensions at the three-and 
four-loop level before, 
cf.~refs.~\cite{Velizhanin:2012nm,Moch:2014sna,Davies:2016jie}.
Our analytic results in the threshold limit $x \to 1$ ($\,N \to \infty$)
include the (light-like) four-loop cusp anomalous dimension, 
see ref.~\cite{Korchemsky:1988si}, which has also been obtained in 
refs.~\cite{Henn:2016men,Lee:2016ixa} by different means.

Beyond the large-$\nc$ limit, we have computed the moments up to $N = 16$ 
for a general gauge group. These results are insufficient for a reconstruction 
of the analytic all-$N$ results. They can be used, though, to obtain 
approximations for the four-loop splitting functions including $x$-dependent 
estimates of their residual uncertainties, see, e.g., earlier work at the 
three-loop level \cite{vanNeerven:1999ca,vanNeerven:2000uj,vanNeerven:2000wp}. 
The~approximations presented below are sufficiently accurate for the evolution
of non-singlet PDFs down to small $x$, and include numerical results for the 
non-planar contributions to the four-loop cusp anomalous dimension that are 
sufficiently precise for phenomenological applications.

For processes with identified hadrons in the final state, QCD factorization 
requires fragmentation functions (FFs) that account for the physics of 
hadronization at long distances.  Completely analogous to PDFs, the scale 
dependence of FFs can be computed within perturbative QCD.
However, in contrast to the case of initial state hadrons, where the 
evolution equations for the scale-dependence of the PDFs are controlled by 
space-like kinematics, $Q^2 \le 0$, the scale evolution of the FFs with 
$Q^2 \ge 0$ requires the so-called time-like splitting functions.
These functions are known completely at two loops
\cite{Curci:1980uw,Furmanski:1980cm,Floratos:1981hs,Kalinowski:1980wea,%
Kalinowski:1980ju,Munehisa:1981ke}, see also 
refs.~\cite{Mitov:2006wy,Gituliar:2015iyq}.
The three-loop corrections have been obtained in refs.~\cite{Mitov:2006ic,%
Moch:2007tx,Almasy:2011eq} up to a phenomenologically irrelevant small 
uncertainty in the result for the time-like NNLO quark-gluon splitting 
function. 
First NNLO analyses of FFs have been performed 
recently~\cite{Anderle:2015lqa,Bertone:2017tyb}.

The three-loop results in refs.~\cite{Mitov:2006ic,Moch:2007tx,Almasy:2011eq} 
have been derived using well-known relations between space- and time-like
kinematics, i.e., the Drell-Yan-Levy relation for the analytic continuation 
in energy $q^2 \to -q^2$ and the Gribov-Lipatov relation in $x$-space 
\cite{Gribov:1972ri,Gribov:1972rt}, 
see also refs.~\cite{Stratmann:1996hn,Blumlein:2000wh},
and generalizations based on conformal symmetry yielding a universal 
reciprocity-respecting evolution kernel 
\cite{Dokshitzer:2005bf,Dokshitzer:2006nm,Basso:2006nk}.
Exploiting these relations, it is possible to use (space-like) DIS results to
predict (time-like) cross sections for single-particle inclusive
electron-positron annihilation. 
Thus, we are able to present here also the flavour non-singlet evolution
equations for FFs at four loops in QCD.

This article is organized as follows. In section~\ref{sec:calc} we specify
our notations and present the theoretical framework for obtaining our results.
In particular we address the basis of non-singlet operators, their 
renormalization and the respective anomalous dimensions.
We sketch the work-flow of the perturbative computation up to four loops, list 
all colour factors to this order and discuss general and end-point properties 
of the anomalous dimensions and splitting functions.

\vspace*{-0.5mm}
In section~\ref{sec:Nres} we present the results of our fixed-$N$ diagram 
calculations of the four-loop non-singlet anomalous dimensions and their 
all-$N$ generalization in the large-$\nc$ limit. We discuss the large-$N$ 
behaviour of the latter which includes the four-loop cusp anomalous dimension.
The \mbox{$x$-space} counterparts of these anomalous dimensions, i.e., the 
splitting functions, are addressed in section~\ref{sec:Xres}. 
We present the exact formulae and compact parametrizations for 
the large-$\nc$ splitting functions, and approximate expressions for 
all cases that cannot be obtained exactly for now.
 
\vspace*{-0.5mm}
Two important applications of these results are presented in section~\ref
{sec:numerics}: we present high-accuracy numerical results for large-$x$
coefficients, in particular the four-loop cusp anomalous dimension in QCD,
and illustrate the N$^3$LO evolution of all three types of non-singlet
quark distributions. 
The N$^3$LO non-singlet evolution is extended to the `time-like' case 
of final-state fragmentation functions in section~\ref{sec:timelike}. 
We summarize our main results and provide a brief outlook in 
section~\ref{sec:summ}.
 
\vspace*{-0.5mm}
The appendices contain the Feynman rules in appendix~\ref{sec:appA}, the 
exact results for the anomalous dimensions at $1 \le N \le 16$ at four loops 
in appendix~\ref{sec:appB}, and the analytic expression for the difference of 
the time-like and space-like four-loop splitting functions in appendix~
\ref{sec:appC}. Finally appendix~\ref{sec:appD} provides the complete all-$N$ 
result for the terms with $\zeta_5$, which may be of theoretical interest.

%
\section{Theoretical framework and calculations}
\label{sec:calc}
\setcounter{equation}{0}
 
The standard set of spin-$N$ twist-two irreducible flavour non-singlet 
quark operators is given by
\bea
\label{eq:loc-ops}
 O^{\,\rm ns}_{\{\,\mu^{\,}_1, \ldots ,\,\mu^{\,}_N\}} &\! =\! & 
   \overline{\psi}\;\lambda^{\!\alpha}\,\gamma_{\,\{\mu^{}_1}
   D_{\,\mu^{}_2} \ldots D_{\,\mu^{}_N\}}\,\psi 
   \:\: ,
   \qquad \alpha \,=\,3,8, \ldots ,(\nfs-1) \, , 
\eea
where $\psi$ represents the quark field, 
$D_{\mu}=\partial_{\mu}-{\rm i}g\:\!A_{\mu}$ the covariant derivative, and 
$\lambda^{\!\alpha}$ the diagonal generators of the flavour group $SU(\nf)$.
It is understood in eq.~(\ref{eq:loc-ops}) that the symmetric and traceless 
part is taken with respect to the Lorentz indices $\mu_i$ in the curly 
brackets.

\begin{figure}[hbt]
\begin{center}
\includegraphics[width=0.475\textwidth, angle=0]{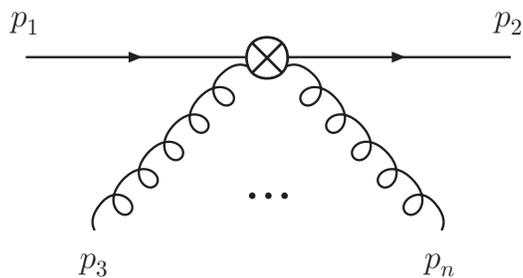}
\caption{\small 
\label{fig:OMEng} 
Vertices with additional gluons arising from the operators 
$O^{\,\rm ns}_{\{\,\mu^{\,}_1,\ldots ,\,\mu^{\,}_N\}}$ in perturbative QCD.
\mbox{Vertices} with up to $L$ gluons need to be considered at $L$ loops
at Mellin moments $N > L$.
}
\end{center}
\end{figure}

We consider (spin-averaged) matrix elements of these operators (OMEs), 
specifically
\bea
\label{eq:OME}
  \langle p_1^{} \vert\, O^{\,\rm ns}_{\{\mu^{\,}_1, \ldots,\,\mu^{\,}_N\}} 
  \,\vert p_2^{} \rangle 
\:\: ,
\eea
for external quark (or anti-quark) fields with momenta $p_{1}^{}$ and $p_2^{}$.
The operators $O^{\,\rm ns}$ in eq.~(\ref{eq:OME}) are contracted with tensors 
of rank $N$, 
\begin{eqnarray}
\label{eq:Delta}
  \Delta^{\mu_1^{}}\, \dots \,\Delta^{\mu_N^{}}
\:\: ,
\end{eqnarray}
where $\Delta$ is a light-like vector, $\Delta^2 = 0$. 
In the present context we need to compute OMEs of renormalized operators with 
zero momentum flow through the operator vertex, thus $p_1^{}=p_2^{}=p$ in 
eq.~(\ref{eq:OME}) for the (off-shell, $p^2 \neq 0$) momenta of the external 
(anti-)$\,$quarks, 
\bea
\label{eq:renOME}
  [ \,A^{\rm ns}](N) \;=\;
  \Delta^{\mu_1^{}}\, \dots \,\Delta^{\mu_N^{}} \, \langle p \vert \,  
  [O^{\,\rm ns\,}]_{\{\,\mu^{\,}_1, \ldots ,\,\mu^{\,}_N\}} \,\vert p \rangle 
\:\: .
\eea
Here and below we use square brackets $[\dots]$ to denote renormalized 
operators (in a minimal subtraction scheme \cite{tHooft:1973mfk,Bardeen:1978yd}
of dimensional regularization \cite{Bollini:1972ui,tHooft:1972tcz}).
This reduces the vertex diagrams for the OMEs to quark two-point functions 
and, therefore, the computational complexity to propagator-type diagrams.
The perturbative expansion of the operator in eq.~(\ref{eq:loc-ops}) 
contracted with eq.~(\ref{eq:Delta}) 
generates vertices with additional gluons as depicted in fig.~\ref{fig:OMEng}.
The current four-loop calculation requires up to four additional gluons. 
The corresponding Feynman rules are presented in appendix~\ref{sec:appA}, 
see refs.~\cite{Floratos:1977au,Bierenbaum:2009mv} 
for earlier calculations at two- and three-loop accuracy.

In order to derive the anomalous dimensions for the scale dependence of the 
non-singlet PDFs we need to perform the renormalization of those operators 
$O^{\,\rm ns}$, which proceeds multiplicatively as 
\beq
\label{eq:renO}
  \left[\,O^{\,\rm ns\,}\right] \;=\;  Z_{\,\rm ns}\, O^{\,\rm ns}
\:\: .
\eeq
The anomalous dimensions $\gamma_{\rm ns}$ governing the scale dependence
of these operators,
\beq
\label{eq:gamma}
  \frac{d}{d \ln \mus }\, \left[O^{\,\rm ns\,}\right]  \;=\; 
  - \,\gamma_{\rm ns}^{}\, \left[O^{\,\rm ns\,}\right] 
\eeq
are connected to the factors $Z_{\rm ns}$ in eq.~(\ref{eq:renO}) by
\beq
\label{eq:gamZ}
 \gamma_{\rm ns}^{} \: = \: 
 -\,\bigg( \,\frac{d }{d\ln\mus }\; Z_{\rm ns} \bigg)\, Z^{\,-1}_{\rm ns} 
\:\: .
\eeq

All flavour differences of quark--anti-quark sums (+) and differences ($-$)
evolve in with the same anomalous dimensions $\gamma_{\rm ns}^{\,+}(N)$ 
and $\gamma_{\rm ns}^{\,-}(N)$, and the total valence distribution with
$  \gamma_{\rm ns}^{\,\rm v}(N) 
 = \gamma_{\rm ns}^{\,-}(N) + \gamma_{\rm ns}^{\;\rm s}(N) $, 
see, e.g., ref.~\cite{Moch:2004pa}.
These quantities are related to the corresponding splitting functions 
$P_{\rm ns}^{\,\pm}(x)$ and $P_{\rm ns}^{\,\rm s}(x) $ by a Mellin transform, 
\beq
\label{eq:Mtrf}
  \gamma_{\rm ns}^{}(N) \;=\; - \int_0^1 \!dx\; x^{\:N-1}\, P_{\rm ns}^{}(x)
\:\: ,
\eeq
where the relative sign is a standard convention.
In perturbation theory these quantities can be expanded in powers of the 
strong coupling constant $\als$.  
Here and below we normalize $\ars\,\equiv\, \als /(4\pi)$, so that up to four 
loops 
\beq
\label{eq:asexp}
  \gamma_{\rm ns}^{}(N) \; =\; 
        \ars\,       \gamma_{\rm ns}^{\,(0)}(N) 
  \,+\, \ars^{\,2}\, \gamma_{\rm ns}^{\,(1)}(N) 
  \,+\, \ars^{\,3}\, \gamma_{\rm ns}^{\,(2)}(N) 
  \,+\, \ars^{\,4}\, \gamma_{\rm ns}^{\,(3)}(N) 
\; ,
\eeq
and similarly for the splitting functions $P_{\rm ns}(x)$ and other quantities.
The first-order quantity $\gamma_{\rm ns}^{\,(0)}$ is the same for all three 
cases given above. 
$\gamma_{\rm ns}^{\,+}$ and $\gamma_{\rm ns}^{\,-}$ differ at order 
$\ars^{\,2}$, and  a non-vanishing flavour-independent (`sea') contribution 
$\gamma_{\rm ns}^{\,\rm s}$ occurs at order $\ars^3$ for the first 
time~\cite{Moch:2004pa}. 
The fourth-order contributions $\gamma_{\rm ns}^{\,(3)}(N)$ to all three 
quantities are addressed in the present article.

The actual computation follows a well-established production chain.
The Feynman diagrams for the OMEs in eq.~(\ref{eq:OME}) are generated up to
four loops using {\sc Qgraf} \cite{Nogueira:1991ex}.
The latest version \cite{Ruijl:2017dtg} of the symbolic manipulation program
{\sc Form}~\cite{Vermaseren:2000nd,Kuipers:2012rf} and its multi-threaded
version {\sc TForm}~\cite{Tentyukov:2007mu} are used for all further steps.
The {\sc Qgraf} output is processed by a program that assigns the topology and
computes the colour factor using the code of ref.~\cite{vanRitbergen:1998pn};
the group invariants occurring in the present case are listed in
table~\ref{tab:color}.
Diagrams of the same topology and colour factor are combined to meta diagrams
for computational efficiency, where lower-order self-energy insertions are
treated as described in ref.~\cite{Herzog:2016qas}.
Considering all color factors, this procedure 
leads to 1 one-loop, 7 two-loop, 53 three-loop and 650 four-loop
meta diagrams for $\gamma_{\rm ns}^{\,\pm\,}$; and 1 three-loop and 29
four-loop meta diagrams for $\gamma_{\rm ns}^{\:\rm s}$.
For comparison: the output of {\sc Qgraf} consists of 15901 four-loop diagrams.
The running of the meta diagrams is managed using the database program
{\sc Minos} \cite{MINOS}.

\begin{table}[t!]
  \centering
  \renewcommand{\arraystretch}{1.2}
  \begin{tabular}{MTMM}
    \hline\\[-16pt]
                               &               & SU(\nc)                                  & \mbox{QCD}    \\[1pt]
    \hline \rule{0pt}{4ex}
    C_A                        & ca            & \nc                                      & 3             \\[3pt]
    C_F                        & cf            & \frac{\ncs-1}{2\nc}                      & \frac{4}{3}   \\[10pt]
    \dabcnc                    & [dabc\^{}2/nr]& \frac{(\ncs-1)(\ncs-4)}{16\,\ncs}          & \frac{5}{18}  \\[10pt]
    \dfFAnc                    & [d4RA/nr]     & \frac{(\ncs+6)(\ncs-1)}{48}              & \frac{5}{2}   \\[10pt]
    \dfFFnc                    & [d4RR/nr]     & \frac{(\ncf-6\ncs+18)(\ncs-1)}{96\,\nct}   & \frac{5}{36}  \\[12pt]
    \hline
  \end{tabular}
  \vspace*{2mm}
  \caption{\small 
  \label{tab:color}
  The colour factors for non-singlet OMEs up to four loops with their numerical 
  values in $SU(\nc)$ with $N_R = \nc$ and QCD, see also 
  ref.~\cite{Moch:2015usa} for a discussion on the normalization of 
  $d^{abc}d_{abc}$.      
  The second column gives the notations of the result files, which are 
  distributed with this article on \texttt{https://arxiv.org}. 
  As in many other articles, we suppress the colour factor $\tf$ ( = 1/2 in 
  $SU(\nc)$) which can be readily re-instated.
  }
  \vspace*{-1mm}
\end{table}

\bigskip

The diagram calculations are done in dimensional regularization
\cite{tHooft:1972tcz,Bollini:1972ui} with the {\sc Forcer} program 
\cite{Ruijl:2017cxj} which was already used for the $N \leq 6$ and 
high-$\nf$ computations in refs.~\cite{Ruijl:2016pkm,Davies:2016jie}. 
Our agreement (after renormalization, see below) with those results, which 
were obtained in a different theoretical framework, provides a strong check 
of our present setup.
The {\sc Forcer} program itself has been validated in calculations of the 
four-loop renormalization of Yang-Mills theories to all powers of the gauge 
parameter, see ref.~\cite{Ruijl:2017eht}, and has recently been applied 
-- together with the algorithms for the $R^\ast$ operation 
\cite{Chetyrkin:1982nn,Chetyrkin:1984xa} developed in 
ref.~\cite{Herzog:2017bjx} -- in five-loop computations of the beta function,
Higgs-boson decays to hadrons and the $R$-ratio in $e^+e^-$-annihilation in 
refs.~\cite{Herzog:2017ohr,Herzog:2017dtz}.

\bigskip

The bare results $A^{\rm ns}$ for the OMEs in eq.~(\ref{eq:renOME}) obtained
in this way are then subject to renormalization which we perform in the
standard modified minimal subtraction scheme 
\MSb~\cite{tHooft:1973mfk,Bardeen:1978yd}.
In this scheme and in $D=4-2\epsilon$ dimensions the strong coupling $\als$ 
evolves according to
\beq
\label{eq:arun}
  \frac{d}{d \ln \mus}\: \frac{\als}{4\pi} \:\: \equiv \:\: 
  \frac{d\,\ars}{d \ln \mus} \:\: = \:\: \beta(\ars) \:\: = \:\:
  - \ep\, \ars - \beta_0\, \ar(2) - \beta_1\, \ar(3) 
  - \beta_2\, \ar(4) - \ldots \:\: ,
\eeq
where $\beta(\ars)$ denotes the usual four-dimensional beta function in QCD,
with coefficients $\,\beta_0 = 11/3\,\ca - 2/3\,\nf\,$ etc, and $\nf$ 
represents the number of active quark flavours.

Using eq.~(\ref{eq:renO}) the renormalized OMEs $[A^{\rm ns}]$ are obtained by 
\beq
\label{eq:ZqZns-OME}
  [A^{\rm ns}](N) \;=\; Z_{\psi}\,\, Z_{\rm ns}(N)\, A^{\rm ns}(N)
\:\: ,
\eeq
where we have made all dependences on $N$ explicit. 
The factor $Z_{\psi}$ denotes the quark wave function renormalization constant 
accounting for the external quarks field with off-shell momenta in 
eq.~(\ref{eq:renOME}), see, for instance ref.~\cite{Ruijl:2017eht}.
Unlike $Z_{\rm ns}$, the quantities $Z_{\psi}$ and $A^{\rm ns}$ are 
gauge-dependent, hence also the renormalization constant $Z_{\,\xi}$ of the 
gauge parameter is required.

The resulting operator renormalization factors $Z_{\rm ns}$ in 
eq.~(\ref{eq:renO}) can be expressed as a Laurent series in $\ep$ as
\bea
\label{eq:Zns}
 \nonumber
{\lefteqn{
  Z_{\rm ns} \;=\; 
  1 
  \:+\:
  \ars \,\biggl\{ 
  \frct{1}{\epsilon}\, 
  \gamma_{\rm ns}^{\, (0)}
  \biggr\}
  \:+\:
  \ars^{\,2} \,\biggl\{
  \frct{1}{\epsilon^2} 
  \biggl( 
    \frct{1}{2}\,\biggl({{\gamma_{\rm ns}^{\,(0)}}}\biggr)^{2}
    -\frct{1}{2}\,{\gamma_{\rm ns}^{\, (0)}}\,\beta_0 
  \biggr)
  +
  \frct{1}{2 \epsilon}\,
  {\gamma_{\rm ns}^{\, (1)}} 
  \biggr\}
}}
  \\
  && \nonumber \mbox{}
  \:+\:
  \ars^{\,3} \, \biggl\{
  \frac{1}{\epsilon^3} 
  \biggl( 
    \frac{1}{6}\,\biggl({{\gamma_{\rm ns}^{\, (0)}}}\biggr)^{3}
    -\frac{1}{2}\,\beta_0\,\biggl({{\gamma_{\rm ns}^{\, (0)}}}\biggr)^{2}
    +\frac{1}{3}\,\beta_0^{\, 2}{\gamma_{\rm ns}^{\, (0)}}
  \biggr) 
  +
  \frac{1}{\epsilon^2} 
  \biggl( 
    \frct{1}{2}\,{\gamma_{\rm ns}^{\, (1)}}\,{\gamma_{\rm ns}^{\, (0)}}
    -\frct{1}{3}\,\beta_0\,{\gamma_{\rm ns}^{\, (1)}}
    -\frct{1}{3}\,\beta_1\,{\gamma_{\rm ns}^{\, (0)}}
  \biggr) 
  \\
  && \nonumber \mbox{}
  +
  \frct{1}{3 \epsilon}\, 
  {\gamma_{\rm ns}^{\, (2)}} 
  \biggr\}
  \:+\:
  \ars^{\,4} \,\biggl\{
  \frct{1}{\epsilon^4} 
  \biggl( 
    \frct{1}{24}\,\biggl({{\gamma_{\rm ns}^{\, (0)}}}\biggr)^{4}
    -\frct{1}{4}\,\beta_0\,\biggl({{\gamma_{\rm ns}^{\, (0)}}}\biggr)^{3}
    +\frct{11}{24}\,\beta_0^{\, 2}\biggl({{\gamma_{\rm ns}^{\,(0)}}}\biggr)^{2}
    -\frct{1}{4}\,\beta_0^{\, 3}{\gamma_{\rm ns}^{\, (0)}}
  \biggr) 
  \\
  && \nonumber \mbox{}
  +
  \frct{1}{\epsilon^3} 
  \biggl(
    \frct{1}{4}\,{\gamma_{\rm ns}^{\, (1)}}\,\biggl({{\gamma_{\rm ns}^{\, (0)}}}
     \biggr)^{2}
    -\frct {7}{12}\,\beta_0\,{\gamma_{\rm ns}^{\, (1)}}\,
     {\gamma_{\rm ns}^{\, (0)}}
    +\frct{1}{4}\,\beta_0^{\, 2}{\gamma_{\rm ns}^{\, (1)}}
    +\frct{1}{2}\,\beta_0\,\beta_1\,{\gamma_{\rm ns}^{\, (0)}}
    -\frct{1}{3}\,\beta_1\,\biggl({{\gamma_{\rm ns}^{\, (0)}}}\biggr)^{2}\, 
  \biggr) 
  \\
  &&  \mbox{}
  +
  \frct{1}{\epsilon^2}
  \biggl(
    \frct{1}{3}\,{\gamma_{\rm ns}^{\, (2)}}\,{\gamma_{\rm ns}^{\, (0)}}
    +\frct{1}{8}\,\biggl({\gamma_{\rm ns}^{\, (1)}}\biggr)^{2}
    -\frct{1}{4}\,{\gamma_{\rm ns}^{\, (2)}}\,\beta_0
    -\frct{1}{4}\,\beta_1\,{\gamma_{\rm ns}^{\, (1)}}
    -\frct{1}{4}\,\beta_2\,{\gamma_{\rm ns}^{\, (0)}}
  \biggr) 
  +
  \frct{1}{4 \epsilon}\,
  {\gamma_{\rm ns}^{\, (3)}} 
  \biggr\}
  \:\: .
\eea
In this manner, the anomalous dimensions $\gamma_{\rm ns}$ have been computed
for a general gauge group at $1 \le N \le 16$, i.e, $\gamma_{\rm ns}^{\,+}$ 
at even $N$ and $\gamma_{\rm ns}^{\,-,\rm s}$ at odd $N$. 
The exact results are listed in appendix~\ref{sec:appB}; numerical values for 
QCD can be found in section~\ref{sec:Nres}. 
The hardest (non-planar) diagrams do not contribute in the limit of a large 
number of colours $\nc$, where the functions $\gamma_{\rm ns}^{\,+}(N)$ and 
$\gamma_{\rm ns}^{\,-}(N)$ are identical,
as it is evident from diagrammatical analyses and
the known $x$-space expressions for $P_{\rm ns}^{\,\pm}(x)$, 
see refs.~\cite{Broadhurst:2004jx,Moch:2004pa,Dokshitzer:2006nm,Davies:2016jie}.
Consequently we were able to obtain the even-$N$ and odd-$N$ values of the
\mbox{large-$\nc$} anomalous dimension, which is structurally simpler than
full QCD results, even up to $N=20$.

\bigskip

So far, fixed-$N$ values of anomalous dimensions have been found to be 
fractions of (large) integer numbers, multiplied at most by values 
$\zeta_3 \ldots \zeta_{2L-3}$ of the Riemann zeta-function at $L$ loops. 
The denominator structure of the fractions suggests analytic all-$N$ 
expressions in terms of harmonic sums~\cite{Vermaseren:1998uu,Blumlein:1998if}
up to weight $2L-1$.
Assuming no numerator-$N$ terms, 
cf.~refs.~\cite{Vermaseren:2005qc,Moch:2008fj,Davies:2016ruz},
the most complicated parts (without a factor $\zeta_n$) of the non-singlet 
anomalous dimensions at $n$ loops read
\beq
\label{eq:ansatz}
  \gamma_{\rm ns}^{\,(n)}(N) \:\: = \:\: 
  \sum_{w=0}^{2n+1} \, c_{00w}^{}\, S_w(N) \,+\, 
  \sum_{\;a^{\phantom{a}}} \: \sum_{k=1}^{2n+1} \: \sum_{w=0}^{2n+1-k}
  c_{akw}^{}\, D_a^{\:k}\, S_w(N)
\:\: ,
\eeq
where $D_a^{\:k}$ are simple denominators, 
\beq
\label{eq:simple-den}
  D_a^{\:k} \:\:=\:\: (N+a)^{-k}
\:\: , 
\eeq
and $S_w(N)$ is a shorthand for all harmonic sums of a given weight $w$ 
with $S_0(N) \equiv 1$. The calculated moments suggest $a = 0,\,1$ for 
$\gamma_{\rm ns}^{\,(3)\pm}(N)$, as at three loops \cite{Moch:2004pa} and 
for the $\nfs$ and $\nft$ four-loop contributions \cite{Davies:2016jie}.
The function $\gamma_{\rm ns}^{\:\rm s}(N)$, on the other hand, includes 
terms with $a=-1$ and $a=2$.

The functions $\gamma_{\rm ns}^{\,(3)}(N)$ contain harmonic sums up to weight 
$w=7$, hence the ansatz~(\ref{eq:ansatz}) includes far too many unknown 
coefficients $c_{akw}$ for a direct determination from the (small) number 
of calculated moments.
However, these coefficients are integer modulo some predictable powers of 
2 and 3.
Therefore the systems of equations derived from eq.~(\ref{eq:ansatz}) can be 
turned into Diophantine systems which require far fewer equations than 
unknowns and which can be solved by LLL-based 
techniques~\cite{Lenstra1982,axbAlg,DBLP:journals/dcc/Silverman00,Calc}. 
This approach has been successfully applied before in 
refs.~\cite{Velizhanin:2012nm,Moch:2014sna,Davies:2016jie}.

In this context it is crucial to constrain eq.~(\ref{eq:ansatz}) as far as 
possible based on general properties of the anomalous dimensions 
$\gamma_{\rm ns}^{}(N)$. Here three issues are worth pointing out.
First, the functional forms of the $\gamma_{\rm ns}^{}(N)$ are (conjectured 
to be) constrained by `self-tuning' \cite{Dokshitzer:2006nm,Basso:2006nk},
\bea
\label{eq:selftune}
  \gamma_{\rm ns}^{}(N) &\!=\!& \gamma_{\rm u}^{} 
  \left( N + \sigma \,\gamma_{\rm ns}^{}(N)-\beta(\ars)/\ars ) \right)
\:\: ,
\eea
where $\,\sigma = -1\, (+1)$ for the space-like (time-like) anomalous 
dimensions, and the non-singlet universal evolution kernel $\gamma_{\rm u}^{}$ 
is reciprocity-respecting (RR), i.e., invariant under the replacement 
$N \to (1-N)$.
By expanding the r.h.s.~of eq.~(\ref{eq:selftune}) about $N$ and inserting
the perturbation series of all quantities involved, $\gamma_{{\rm u}}^{}$ 
can be expressed in terms of the \MSb\ anomalous dimensions, see also 
ref.~\cite{Braun:2017cih}.
Expressing the latter in terms of 
$\gamma_0^{} \,=\, \gamma_{\rm ns}^{\:S(n)} \,=\, \gamma_{\rm ns}^{\:T(n)}$ 
and the average of the space-like and time-like expansion coefficients 
$\overline{\gamma}_n^{}= \frac{1}{2} \big( \:\!\gamma_{\rm ns}^{\:S(n)} 
+ \gamma_{\rm ns}^{\:T(n)\,} \big)$, one arrives at
\bea
  \gamma_{\rm u}^{} &\!=\!&
         \ars \,\* \gamma_0^{} 
   \:+\: \ar(2) \* \left( \,
         \overline{\gamma}_1^{} 
      \,-\, \b0 \,\* d_N^{} \* \gamma_0^{} 
       \right)
   \:+\: \ar(3) \* \Big( 
         \overline{\gamma}_2^{} 
      \,-\, \frct{1}{6}\, \* d_N^{\,2\,} \* \gamma_0^{\:3}
      \,-\, \b1 \,\* d_N^{} \* \gamma_0^{} 
      \,-\, \b0 \,\* d_N^{} \* \overline{\gamma}_1^{} 
      \,+\, \frct{1}{2}\, \* \B(0,2) \,\* d_N^{\,2\,} \* \gamma_0^{}  
      \Big)
\nn \\ & & \mbox{\hspn} 
   \:+\: \ar(4) \* \Big(
         \overline{\gamma}_3^{} 
      \,-\, \frct{1}{2}\, \* d_N^{\,2\,} \* 
         ( \gamma_0^{\:2}\,\overline{\gamma}_1^{} ) 
      \,-\, \b2\,\* d_N^{} \* \gamma_0^{} 
      \,-\, \b1\,\* d_N^{} \* \overline{\gamma}_1^{} 
      \,-\, \b0\,\* d_N^{} \* \overline{\gamma}_2^{} 
      \,+\, \frct{1}{6}\, \* \b0 \,\* d_N^{\,3\,} \* \gamma_0^{\:3}
      \,+\, \b0\,\* \b1\,\* d_N^{\,2\,} \* \gamma_0^{}
\nn \\ & & \mbox{\hspp} 
      \,+\, \frct{1}{2}\, \* \B(0,2) \,\* d_N^{\,2\:} \* \overline{\gamma}_1^{}
      \,-\, \frct{1}{6}\, \* \B(0,3) \,\* d_N^{\,3\:} \* \gamma_0^{}
      \Big)
  \:+\: {\cal O}(\ar(5))
\:\: ,
\eea
where we have used the abbreviation $d_N^{} = d/dN$ and suppressed the
$N$-dependences for brevity. 
A~convenient way to take these derivatives is via inverse Mellin transforms 
to $x$-space, where the multiplication with $\ln^{\,n\!}x$ corresponds to 
the $N$-space operator $d^{\,n\!}/dN^{n}$, and Mellin transforms of the
result. 
The required manipulations can be readily performed using algorithms for
harmonic sums, harmonic polylogarithms and their (inverse) Mellin 
transformations \cite{Vermaseren:1998uu,Remiddi:1999ew,Moch:1999eb} 
which have been implemented in publicly available {\sc Form} packages 
described in ref.~\cite{Vermaseren:2000nd}.

Since the difference between the time-like and space-like anomalous 
dimensions is known to four loops, eq.~(\ref{eq:ansatz}) can be applied to 
the RR quantity $\gamma_{\rm u}^{\,(3)}$ instead of $\gamma_{\rm ns}^{\,(3)}$.
This implies that the denominators $1/N$ and $1/(N+1)$ can only enter in the 
combination $1/(N(N+1))$, and that only RR (combinations of) harmonic sums
occur, see refs.~\cite{Lukowski:2009ce,wwwVelizh}, which reduces the number 
of sums at weight $w$ from $2\cdot 3^{w-1}$ to $2^{w-1}$. 
Assuming that only powers of $1/(N+1)$ enter in addition, the total number of 
basis functions in eq.~(\ref{eq:ansatz}) up to weight $w$ is $2^{w+1}-1$, e.g.,
255 for $w=7$.
Even taking account end-point constraints, see below, this is a prohibitively
large number for now.

Second, the identical leading-$\nc$ terms of $\gamma_{\rm ns}^{\:\pm}(N)$
contain only non-alternating harmonic sums,
i.e., only positive indices in eq.~(\ref{eq:Hsum1}). This reduces the number 
of RR sums of weight $w$ to the Fibonacci number $F(w)$, i.e., 1, 1, 2, 3, 5,
8, 13 for $w=1$ to $w=7$, 
as can be seen by counting the number of binomial harmonic sums 
at weight $w$~\cite{Lukowski:2009ce}.
Considering all combinations with additional powers of the weight-1 object $1/(N(N+1)$,
the total number of functions up to weight $w$ in 
eq.~(\ref{eq:ansatz}) amounts to $F(w+4) -2$, e.g., 87 for $w=7$.


The third and final point is that the $\,N\to \infty\,$ (large-$x$) and 
$\,N\to 0\,$ (small-$x$) limits of the anomalous dimensions (splitting 
functions) provide a substantial number of constraints. If one disregards
terms of order ${\cal O}(1/N^2)$ for $\,N\to \infty\,$, then all three
non-singlet anomalous dimensions $\gamma_{\rm ns}^{\;a}(N)$, $a = +,-,\rm v,$
are identical and given by \cite{Dokshitzer:2005bf}~
($\gamma_{\:\!e}$ is the Euler-Mascheroni constant)
\beq
\label{eq:ntoinf}
  \gamma_{\,\rm ns}^{\:(n-1)}(N) \; = \:\;
    A_n \, ( \ln N +\gamma_{\:\!e} ) - B_n
  + C_n \; ( \ln N + \gamma_{\:\!e} ) \, N^{-1}
  - ( \widetilde{D}_n - \frct{1}{2}\: A_n ) \, N^{-1}
\:\: .
\eeq
Here the coefficients $A_n$ -- the $n$-loop (light-like) cusp anomalous 
dimension -- and $B_n$ provide genuine $n$-loop information. 
The coefficients $C_n$ and $\widetilde{D}_n$, on the other hand, can be 
expressed in terms of lower-order information (see eqs.~(\ref{eq:CDofAB}) and
(\ref{eq:CD4ofAB}) below). 
This and the absence of second and higher powers of $\ln\, N$ in 
eq.~(\ref{eq:ntoinf}), and similar if less stringent constraints on 
$N^{-k}\,\ln N$ terms with $k>1$, provide a substantial number of constraints 
on the coefficients in eq.~(\ref{eq:ansatz}).

The small-$x$ expansion of the splitting function $P_{\rm ns}^{\:(n)}(x)$ 
shows a double-logarithmic enhancement, i.e., there are contributions of the 
form $x^{\,a} \ln^{\,b\!}x$ with $a>0$ and $b \leq 2n$. 
The leading-logarithmic (LL) contributions to $P_{\rm ns}^{\:\pm}$ have been 
known to all orders for a long time \cite{Kirschner:1983di,Blumlein:1995jp}. 
This resummation has been extended to next-to-next-to-logarithmic 
accuracy for the $x^{\,2k} \ln^{\,b\!}x$ contributions to $P_{\rm ns}^{\:+}$ 
and the $x^{\,2k+1} \ln^{\,b\!}x$ contributions to $P_{\rm ns}^{\:-}$ at all 
$k\geq 0$ \cite{Vogt:2012gb,KDVprep}. 
The formal structure of these results is analogous to their time-like 
counterparts \cite{Vogt:2011jv,Kom:2012hd}, but the numerical pattern is 
completely different such that the space-like resummation is of no direct 
phenomenological use. 
The functions $P_{\rm ns}^{\:+}(x)$ and $P_{\rm ns}^{\:-}(x)$ are the same in 
the large-$\nc$ limit, hence in this case the small-$x$ resummation constrains 
the coefficients contributing to 
\beq
  x^{\,a} \ln^{\,b\!}x \qquad \mbox{for~~~} a \,\geq\, 0 
  \mbox{~~~and~~~} 4 \,\leq\, b \,\leq\, 6 
\:\: .
\eeq
 
An alternative approach to the limit $N \ra 0$, i.e., the small-$x$ logarithms
for $a=0$, has been pursued in ref.~\cite{Velizhanin:2014dia}. 
In the large-$\nc$ limit, the generalization
\beq
\label{gnsto0V}
  \gamma_{\rm ns}^{}(N,\ars) \cdot \left( \,\gamma_{\rm ns}^{}(N,\ars)
  + N - \beta(\ars) / \ars \right) \;=\; O(1)
\eeq
of the LL relation in refs.~\cite{Kirschner:1983di,Blumlein:1995jp}
correctly (re-)$\,$produces all $a=0$ small-$x$ logarithms obtained in 
refs.~\cite{Moch:2004pa,Vogt:2012gb,Davies:2016jie}, after correcting  
typos in eqs.~(25) and (26) of ref.~\cite{Velizhanin:2014dia}. 
Hence we can assume that eq.~(\ref{gnsto0V}) is also correct for the 
$\nfz$ and $\nfo$ four-loop contributions at large $\nc$.

\bigskip

Together these relations comprise 18 large-$N$ and 28 small-$x$ constraints 
for the $\nfz$-terms at four loops eliminating 
more than half of the 87 free parameters of the $w\!=\!7$ large-$\nc$ ansatz, 
after which it is possible to solve the remaining system of Diophantine equation using 
the moments $N=1,\, \ldots, 18$ with the program {\tt axb()} of the 
{\sc Calc} package \cite{Calc}.
The resulting analytic expressions for $\gamma_{\,\rm ns}^{\:(3)}(N)$ 
agree with the result of the diagram calculations at $N=19$ and $N=20$. 
This agreement renders it extremely likely -- although, of course, not
mathematically certain -- that these results (and, therefore, the above 
structural conjectures and features used in their derivation) are correct.

As mentioned above, present information and understanding appears not to 
be sufficient for extending these analytic results beyond the large-$\nc$ 
contributions for the $\nfz$ and $\nfo$ parts of $\gamma_{\,\rm ns}^{\:(3)a}(N)$
for any $a = +, -, \rm s$. For the remaining functions we resort to $x$-space
approximations based on the first eight even-$N$ or odd-$N$ moments 
supplemented by the large-$x$ and small-$x$ constraints discussed above.
These approximations and their error estimates can be constructed in the same 
manner as those for the three-loop splitting functions in 
refs.~\cite{vanNeerven:1999ca,vanNeerven:2000uj,vanNeerven:2000wp}. 
The present results are more accurate, though, due to the higher number of 
available moments and the improved understanding of the end-point limits.
The fact that the large-$N$ limit (\ref{eq:ntoinf}) includes only the two 
free parameters $A_4$ and $B_4$, in particular, results in a high accuracy 
of these coefficients which are relevant also in the context of the soft-gluon 
exponentiation, 
see refs.~\cite{Moch:2005ba,Moch:2005ky,Ravindran:2006cg,Ahmed:2014cla} 
and references therein, and beyond.

\newpage
 
%
\setcounter{equation}{0}
\section{Results in $N$-space}
\label{sec:Nres}

We start presenting our results by writing down the moments to $N=16$ of 
the non-singlet four-loop anomalous dimensions for QCD in a numerical form. 
The exact results for a general gauge group with one set of fermions can be 
found in appendix~\ref{sec:appB}.
For $\gamma_{\,\rm ns}^{\:(3)\pm}(N)$ we separately display the leading 
(subscript $L$) and non-leading (subscript $N$) contributions in the 
large-$\nc$ limit of $SU(\nc)$ at $\nc=3$. 
The former correspond to the colour factors $\cf n_c^{\,3-k} n_{\!f}^{\:k}$. 
The latter collect all other terms, which are suppressed by two or more powers 
of $\nc$, cf.~table \ref{tab:color} above.

The first eight even-$N$ values of $\gamma_{\,\rm ns}^{\:(3)+}(N)$,
normalized as in eq.~(\ref{eq:asexp}) above -- division by $2.5 \cdot 10^4$ 
provides an approximate conversion to an expansion in $\als$ -- are given by
\bea
\label{gam+to16}
  \gamma_{\,\rm ns}^{\:(3)+}(2) &\!=\!&
          15079.979904_L^{} + 1583.245584_N^{} 
  \,-\, ( 3610.4544273_L^{} + 137.0956130_N^{} ) \,\nf
\nn \\[-0.5mm] & & \mbox{}
  \,+\, ( 114.04980426_L^{} +  3.73169344_N^{} ) \,\nfs 
  \,+\,    1.908433871_L^{} \, \nft
\:\: , \nn \\[0.5mm] 
  \gamma_{\,\rm ns}^{\:(3)+}(4) &\!=\!&
          26818.645638_L^{} + 996.7898428_N^{} 
  \,-\, ( 6549.4629888_L^{} + 154.7093878_N^{} ) \,\nf
\nn \\[-0.5mm] & & \mbox{}
  \,+\, ( 231.33355779_L^{} +  7.35324122_N^{} ) \,\nfs 
  \,+\,    3.884444957_L^{} \, \nft
\:\: , \nn \\[0.5mm] 
  \gamma_{\,\rm ns}^{\:(3)+}(6) &\!=\!&
          33391.136191_L^{} + 801.6538203_N^{} 
  \,-\, ( 8219.2929199_L^{} + 167.0140214_N^{} ) \,\nf
\nn \\[-0.5mm] & & \mbox{}
  \,+\, ( 299.12524225_L^{} +  9.35853808_N^{} ) \,\nfs 
  \,+\,    5.073482406_L^{} \, \nft
\:\: , \nn \\[0.5mm]
  \gamma_{\,\rm ns}^{\:(3)+}(8) &\!=\!&
          38122.913329_L^{} + 685.2274355_N^{}
  \,-\, ( 9424.0745879_L^{} + 175.4883891_N^{} ) \,\nf
\nn \\[-0.5mm] & & \mbox{}
  \,+\, ( 34.752553808_L^{} + 10.77524037_N^{} ) \,\nfs
  \,+\,    5.937491139_L^{} \, \nft
\:\: , \nn \\[0.5mm]
  \gamma_{\,\rm ns}^{\:(3)+}(10) &\!=\!&
          41872.765979_L^{} + 600.8875674_N^{}
  \,-\, ( 10377.826722_L^{} + 181.7070215_N^{} ) \,\nf
\nn \\[-0.5mm] & & \mbox{}
  \,+\, ( 385.45469324_L^{} + 11.87446948_N^{} ) \,\nfs
  \,+\,    6.618723172_L^{} \, \nft
\:\: , \nn \\[0.5mm]
  \gamma_{\,\rm ns}^{\:(3)+}(12) &\!=\!&
          44999.013143_L^{} + 533.8372124_N^{}
  \,-\, ( 11171.437026_L^{} + 186.4798090_N^{} ) \,\nf
\nn \\[-0.5mm] & & \mbox{}
  \,+\, ( 416.75989357_L^{} + 12.77304489_N^{} ) \,\nfs
  \,+\,    7.181732766_L^{} \, \nft
\:\: , \nn \\[0.5mm]
  \gamma_{\,\rm ns}^{\:(3)+}(14) &\!=\!&
          47688.997330_L^{} + 477.6377749_N^{}
  \,-\, ( 11852.885145_L^{} + 190.2641681_N^{} ) \,\nf
\nn \\[-0.5mm] & & \mbox{}
  \,+\, ( 443.46765303_L^{} + 13.53286843_N^{} ) \,\nfs
  \,+\,    7.661731916_L^{} \, \nft
\:\: , \nn \\[0.5mm]
  \gamma_{\,\rm ns}^{\:(3)+}(16) &\!=\!&
          50054.446557_L^{} + 428.9331378_N^{}
  \,-\, ( 12450.928224_L^{} + 193.3399276_N^{} ) \,\nf
\nn \\[-0.5mm] & & \mbox{}
  \,+\, ( 466.78457532_L^{} + 14.19090917_N^{} ) \,\nfs
  \,+\,    8.080152509_L^{} \, \nft
\:\: ,
\eea
and the first eight odd-$N$ values of $\gamma_{\,\rm ns}^{\:(3)-}(N)$ read
\bea
\label{gam-to15}
  \gamma_{\,\rm ns}^{\:(3)-}(1) &\!=\!& 0
\:\: , \nn \\[1.5mm]
  \gamma_{\,\rm ns}^{\:(3)-}(3) &\!=\!&
          22101.772707_L^{} + 999.3940009_N^{}
  \,-\, ( 5359.5335080_L^{} + 139.7317364_N^{} ) \,\nf
\nn \\[-0.5mm] & & \mbox{}
  \,+\, ( 183.02229066_L^{} +  5.91723185_N^{} ) \,\nfs
  \,+\,    3.058631170_L^{} \, \nft
\:\: , \nn \\[0.5mm]
  \gamma_{\,\rm ns}^{\:(3)-}(5) &\!=\!&
          30431.615620_L^{} + 841.7002595_N^{}
  \,-\, ( 7466.3287550_L^{} + 160.0856189_N^{} ) \,\nf
\nn \\[-0.5mm] & & \mbox{}
  \,+\, ( 268.62851396_L^{} +  8.47774637_N^{} ) \,\nfs
  \,+\,    4.534727508_L^{} \, \nft
\:\: , \nn \\[0.5mm]
  \gamma_{\,\rm ns}^{\:(3)-}(7) &\!=\!&
          35914.342449_L^{} + 722.2583518_N^{}
  \,-\, ( 8861.7720221_L^{} + 171.1596475_N^{} ) \,\nf
\nn \\[-0.5mm] & & \mbox{}
  \,+\, ( 325.00219825_L^{} + 10.13425314_N^{} ) \,\nfs
  \,+\,    5.534366480_L^{} \, \nft
\:\: , \nn \\[0.5mm]
  \gamma_{\,\rm ns}^{\:(3)-}(9) &\!=\!&
          40092.293568_L^{} + 632.8933507_N^{}
  \,-\, ( 9925.1782168_L^{} + 178.6640945_N^{} ) \,\nf
\nn \\[-0.5mm] & & \mbox{}
  \,+\, ( 367.49639662_L^{} + 11.36914948_N^{} ) \,\nfs
  \,+\,    6.295934539_L^{} \, \nft
\:\: , \nn \\[0.5mm]
  \gamma_{\,\rm ns}^{\:(3)-}(11) &\!=\!&
          43499.696829_L^{} + 561.6966621_N^{}
  \,-\, ( 10791.034298_L^{} + 184.1995847_N^{} ) \,\nf
\nn \\[-0.5mm] & & \mbox{}
  \,+\, ( 401.78232680_L^{} + 12.35557815_N^{} ) \,\nfs
  \,+\,    6.912368819_L^{} \, \nft
\:\: , \nn \\[0.5mm]
  \gamma_{\,\rm ns}^{\:(3)-}(13) &\!=\!&
          46390.354670_L^{} + 502.2596221_N^{}
  \,-\, ( 11524.077148_L^{} + 188.4837700_N^{} ) \,\nf
\nn \\[-0.5mm] & & \mbox{}
  \,+\, ( 430.59996163_L^{} + 13.17711657_N^{} ) \,\nfs
  \,+\,    7.430544877_L^{} \, \nft
\:\: , \nn \\[0.5mm]
  \gamma_{\,\rm ns}^{\:(3)-}(15) &\!=\!&
          48907.084426_L^{} + 451.0029039_N^{}
  \,-\, ( 12160.990233_L^{} + 191.9086558_N^{} ) \,\nf
\nn \\[-0.5mm] & & \mbox{}
  \,+\, ( 455.49394328_L^{} + 13.88095204_N^{} ) \,\nfs
  \,+\,    7.877634036_L^{} \, \nft
\:\: .
\eea
It is clear from these results, that the large-$\nc$ limit alone provides
an excellent approximation to the individual $n_{\!f}^{\:a}$ coefficients
except for the lowest values of $N$. 
The non-large-$\nc$ `correction' amounts to 10\% and 4\% for the $\nfz$ and 
$\nfo$ terms, respectively, but 2\% or less at $N \geq 7$ in both cases.

We have computed the first nine odd-$N$ values of the `sea' contribution
$\gamma_{\,\rm ns}^{\:(3)\rm s}(N)$ to the four-loop anomalous dimension
for the overall valence distribution, and find 
\bea
\label{gamsto17}
  \gamma_{\,\rm ns}^{\:(3)\rm s}(1) &\!=\!& \phantom{1} 0
\:\: , \nn \\[1.5mm]
  \gamma_{\,\rm ns}^{\:(3)\rm s}(3) &\!=\!& 
          18.9700898832 \:\nf \:-\: 1.6109396433 \:\nfs
\:\: , \nn \\[0.5mm]
  \gamma_{\,\rm ns}^{\:(3)\rm s}(5) &\!=\!& \phantom{1}
           9.1402406178 \:\nf \:+\: 0.1525610933 \:\nfs
\:\: , \nn \\[0.5mm]
  \gamma_{\,\rm ns}^{\:(3)\rm s}(7) &\!=\!& \phantom{1}
           4.0470106556 \:\nf \:+\: 0.3095914493\:\nfs
\:\: , \nn \\[0.5mm]
  \gamma_{\,\rm ns}^{\:(3)\rm s}(9) &\!=\!& \phantom{1}
           1.9658456985 \:\nf \:+\: 0.2887522942 \:\nfs
\:\: , \nn \\[0.5mm]
  \gamma_{\,\rm ns}^{\:(3)\rm s}(11) &\!=\!& \phantom{1}
           1.0102117327 \:\nf \:+\: 0.2474854100 \:\nfs
\:\: , \nn \\[0.5mm]
  \gamma_{\,\rm ns}^{\:(3)\rm s}(13) &\!=\!& \phantom{1}
           0.5255291656 \:\nf \:+\: 0.2099342070 \:\nfs
\:\: , \nn \\[0.5mm]
  \gamma_{\,\rm ns}^{\:(3)\rm s}(15) &\!=\!& \phantom{1}
           0.2612077170 \:\nf \:+\: 0.1791623892 \:\nfs
\:\: , \nn \\[0.5mm]
  \gamma_{\,\rm ns}^{\:(3)\rm s}(17) &\!=\!& \phantom{1}
           0.1094470531 \:\nf \:+\: 0.1544243611 \:\nfs
\:\: .
\eea

\bigskip

We now turn to the analytic all-$N$ expressions for the $\nfz$ and $\nfo$ 
parts of the four-loop non-singlet anomalous dimensions 
$\gamma_{\rm ns}^{\,(3)\pm}$ in the large-$\nc$ limit. 
The complete lower-order contributions can be found, in a different 
notation but the same normalization, in eqs.~(3.4)$\,$--$\,$(3.8) of 
ref.~\cite{Moch:2004pa}. 
The anomalous dimensions can be expressed in terms of the denominators 
$D_a^{\,k}$ in eq.~(\ref{eq:simple-den}) and harmonic sums 
\cite{Vermaseren:1998uu,Blumlein:1998if} at argument $N$, which are 
recursively defined by
\bea
\label{eq:Hsum1}
  S_{\pm m}(N) &\!=\!& \sum_{n=1}^{N}\; (\pm 1)^n \, n^{\, -m}
\:\: , \nn \\
  S_{\pm m_1^{},\,m_2^{},\,\ldots,\,m_d}(N) &\!=\!& \sum_{n=1}^{N}\:
  (\pm 1)^{n} \; n^{\, -m_1^{}}\; S_{m_2^{},\,\ldots,\,m_d}(n)
\:\: .
\eea
The weight $w$ of the harmonic sums is defined by the sum of the absolute
values of the indices~$m_d$.  Sums up to $w = 2\:\!n-1$ occur in the
$n$-loop anomalous dimensions. 
The argument $N$ of the sums is suppressed for brevity below, and we use
the shorthand $\eta \,=\, 1/(N(N+1)) = D_0\, D_1$.

The identical large-$\nc$ parts of the functions $\gamma_{\rm ns}^{\,(3)+}(N)$ 
and $\gamma_{\rm ns}^{\,(3)-}(N)$ are given by 
\beq
\label{eq:g3Lnc}
\gamma_{\rm ns,\,L}^{\,(3)}(N) \;\;=\;\;  \cf \left( 
  \nct\,     \gamma_{\,\rm L,\:0}^{\,(3)}(N) 
  \:+\: \ncs \nf\, \gamma_{\,\rm L,\:1}^{\,(3)}(N)
  \:+\: \nc \nfs\, \gamma_{\,\rm L,\:2}^{\,(3)}(N)
  \:+\: \nft\,     \gamma_{\,\rm L,\:3}^{\,(3)}(N) \right)
\:\: ,
\eeq
where the $\nfs$ and $\nft$ contributions to eq.~(\ref{eq:g3Lnc}) have been
given in eqs.~(3.1) and (3.6) of ref.~\cite{Davies:2016jie}; the latter has
first been derived in ref.~\cite{Gracey:1994nn}.
Our new results are
\bea
\label{eq:g3LncNF0}
{\lefteqn{
   \;\;\;\gamma_{\,\rm L,\,0}^{\:(3)}(N) \; = \;
}}
\nn
\\
&& \nonumber
         16 \,\* \biggl(
            1379569/82944
          - 7453/24\,\*D_1^2
          - 102641/648\,\*D_1^3
          - 9883/864\,\*D_1^4
          - 209/12\,\*D_1^5
\\[-1mm]
&& \nonumber
          - 95/12\,\*D_1^6
          - 5/4\,\*D_1^7
          + 534767/5184\,\*\eta
          - 231341/2592\,\*\eta^2
          - 15469/1296\,\*\eta^3
          - 12997/864\,\*\eta^4
\\
&& \nonumber
          - 83/12\,\*\eta^5
          - 25/24\,\*\eta^6
          - 5/8\,\*\eta^7
          - 55/4\,\*\zeta_5
          + 40\,\*\zeta_5\*\eta
          - 25\,\*\zeta_5\*\eta^2
          - 1517/144\,\*\zeta_3
          + 839/108\,\*\zeta_3\*\eta
\\
&& \nonumber
          + 13/24\,\*\zeta_3\*\eta^2
          - 2/3\,\*\zeta_3\*\eta^3
          - 1/2\,\*\zeta_3\*\eta^4
          + 42139/648\,\*\S(1)
          - 130795/648\,\*\S(1)\*D_1^2
          - 298/9\,\*\S(1)\*D_1^3
\\
&& \nonumber
          - 995/24\,\*\S(1)\*D_1^4
          - 92/3\,\*\S(1)\*D_1^5
          - 15/2\,\*\S(1)\*D_1^6
          + 278627/1296\,\*\S(1)\*\eta
          + 19757/2592\,\*\S(1)\*\eta^2
\\
&& \nonumber
          + 3625/48\,\*\S(1)\*\eta^3
          + 1789/72\,\*\S(1)\*\eta^4
          + 9\*\S(1)\*\eta^5
          + 7/2\,\*\S(1)\*\eta^6
          + 422/27\,\*\S(1)\*\zeta_3
          - 62/3\,\*\S(1)\*\zeta_3\*\eta
\\
&& \nonumber
          - 7/3\,\*\S(1)\*\zeta_3\*\eta^2
          + 10\*\S(1)\*\zeta_3\*\eta^3
          - 24211/432\,\*\S(2)
          + 23153/216\,\*\S(2)\*D_1^2
          + 143/3\,\*\S(2)\*D_1^3
          + 16\*\S(2)\*D_1^4
\\
&& \nonumber
          + 10\*\S(2)\*D_1^5
          - 18725/1296\,\*\S(2)\*\eta
          + 23689/432\,\*\S(2)\*\eta^2
          + 3187/144\,\*\S(2)\*\eta^3
          + 229/12\,\*\S(2)\*\eta^4
          + 29/4\,\*\S(2)\*\eta^5
\\
&& \nonumber
          - 20\*\S(2)\*\zeta_3\*\eta
          + 8\*\S(2)\*\zeta_3\*\eta^2
          + 71591/864\,\*\S(3)
          - 5099/72\,\*\S(3)\*D_1^2
          - 5/3\,\*\S(3)\*D_1^3
          - 2\*\S(3)\*D_1^4
          + 4373/108\,\*\S(3)\*\eta
\\
&& \nonumber
          + 1051/72\,\*\S(3)\*\eta^2
          + 40/3\,\*\S(3)\*\eta^3
          + 27/4\,\*\S(3)\*\eta^4
          - 13/3\,\*\S(3)\*\zeta_3
          + 2\*\S(3)\*\zeta_3\*\eta
          - 55291/864\,\*\S(4)
\\
&& \nonumber
          + 155/6\,\*\S(4)\*D_1^2
          + 7\*\S(4)\*D_1^3
          + 3233/144\,\*\S(4)\*\eta
          + 103/12\,\*\S(4)\*\eta^2
          + 11/2\,\*\S(4)\*\eta^3
          + 4\*\S(4)\*\zeta_3
          + 227/4\,\*\S(5)
\\
&& \nonumber
          - 12\*\S(5)\*D_1^2
          + 1/3\,\*\S(5)\*\eta
          + 7\*\S(5)\*\eta^2
          - 65/3\,\*\S(6)
          + 10\*\S(6)\*\eta
          + 10\*\S(7)
          - 56/3\,\*\Ss(1,1)\*D_1^2
          - 71/3\,\*\Ss(1,1)\*D_1^3
\\
&& \nonumber
          - 35\*\Ss(1,1)\*D_1^4
          - 20\*\Ss(1,1)\*D_1^5
          + 23/12\,\*\Ss(1,1)\*\eta
          - 587/8\,\*\Ss(1,1)\*\eta^2
          - 115/4\,\*\Ss(1,1)\*\eta^3
          - 109/3\,\*\Ss(1,1)\*\eta^4
\\
&& \nonumber
          - 11\*\Ss(1,1)\*\eta^5
          + 40\*\Ss(1,1)\*\zeta_3\*\eta
          - 16\*\Ss(1,1)\*\zeta_3\*\eta^2
          - 5552/81\,\*\Ss(1,2)
          + 184/3\,\*\Ss(1,2)\*D_1^2
          - 2\*\Ss(1,2)\*D_1^3
          + 6\*\Ss(1,2)\*D_1^4
\\
&& \nonumber
          - 383/6\,\*\Ss(1,2)\*\eta
          - 241/12\,\*\Ss(1,2)\*\eta^2
          - 137/6\,\*\Ss(1,2)\*\eta^3
          - 17/2\,\*\Ss(1,2)\*\eta^4
          + 890/27\,\*\Ss(1,3)
          - 22/3\,\*\Ss(1,3)\*D_1^2
\\
&& \nonumber
          + 53/3\,\*\Ss(1,3)\*\eta
          - 79/6\,\*\Ss(1,3)\*\eta^2
          - 11\*\Ss(1,3)\*\eta^3
          - 4\*\Ss(1,3)\*\zeta_3
          - 4745/72\,\*\Ss(1,4)
          + 14\*\Ss(1,4)\*D_1^2
          - 46/3\,\*\Ss(1,4)\*\eta
\\
&& \nonumber
          - \Ss(1,4)\*\eta^2
          + 70/3\,\*\Ss(1,5)
          - 4\*\Ss(1,5)\*\eta
          - 20\*\Ss(1,6)
          - 5552/81\,\*\Ss(2,1)
          + 184/3\,\*\Ss(2,1)\*D_1^2
          - 2\*\Ss(2,1)\*D_1^3
          + 6\*\Ss(2,1)\*D_1^4
\\
&& \nonumber
          - 383/6\,\*\Ss(2,1)\*\eta
          - 241/12\,\*\Ss(2,1)\*\eta^2
          - 137/6\,\*\Ss(2,1)\*\eta^3
          - 17/2\,\*\Ss(2,1)\*\eta^4
          + 9029/216\,\*\Ss(2,2)
          - 19\,\*\Ss(2,2)\*D_1^2
\\
&& \nonumber
          - 6\,\*\Ss(2,2)\*D_1^3
          - 38/3\,\*\Ss(2,2)\*\eta
          - 18\,\*\Ss(2,2)\*\eta^2
          - 9\*\Ss(2,2)\*\eta^3
          - 3635/72\,\*\Ss(2,3)
          + 12\*\Ss(2,3)\*D_1^2
          + 5/3\,\*\Ss(2,3)\*\eta
\\
&& \nonumber
          - 10\*\Ss(2,3)\*\eta^2
          + 31/2\,\*\Ss(2,4)
          + 3\*\Ss(2,4)\*\eta
          - 16\*\Ss(2,5)
          + 925/18\,\*\Ss(3,1)
          - 76/3\,\*\Ss(3,1)\*D_1^2
          - 8\*\Ss(3,1)\*D_1^3
\\
&& \nonumber
          - 1675/36\,\*\Ss(3,1)\*\eta
          - 21\*\Ss(3,1)\*\eta^2
          - 4\*\Ss(3,1)\*\eta^3
          - 4\*\Ss(3,1)\*\zeta_3
          - 5581/72\,\*\Ss(3,2)
          + 22\*\Ss(3,2)\*D_1^2
          - 53/3\,\*\Ss(3,2)\*\eta
\\
&& \nonumber
          - 16\*\Ss(3,2)\*\eta^2
          + 143/6\,\*\Ss(3,3)
          - 11\*\Ss(3,3)\*\eta
          - 14\*\Ss(3,4)
          - 6899/72\,\*\Ss(4,1)
          + 24\*\Ss(4,1)\*D_1^2
          - 74/3\,\*\Ss(4,1)\*\eta
\\
&& \nonumber
          - 11\*\Ss(4,1)\*\eta^2
          + 57/2\,\*\Ss(4,2)
          - 25\*\Ss(4,2)\*\eta
          - 26\*\Ss(4,3)
          + 63/2\,\*\Ss(5,1)
          - 23\*\Ss(5,1)\*\eta
          - 36\*\Ss(5,2)
          - 28\*\Ss(6,1)
\\
&& \nonumber
          - 12\*\Sss(1,1,1)\*D_1^4
          + 12\*\Sss(1,1,1)\*\eta^2
          + 24\*\Sss(1,1,1)\*\eta^3
          + 6\*\Sss(1,1,1)\*\eta^4
          + 18\*\Sss(1,1,2)\*\eta^2
          + 6\*\Sss(1,1,2)\*\eta^3
          - 20\*\Sss(1,1,3)\*\eta
\\
&& \nonumber
          + 8\*\Sss(1,1,3)\*\eta^2
          + 20/3\,\*\Sss(1,1,4)
          - 20\*\Sss(1,1,4)\*\eta
          + 8\*\Sss(1,1,5)
          + 18\*\Sss(1,2,1)\*\eta^2
          + 6\*\Sss(1,2,1)\*\eta^3
          + 134/3\,\*\Sss(1,2,2)
\\
&& \nonumber
          - 12\*\Sss(1,2,2)\*D_1^2
          + 12\*\Sss(1,2,2)\*\eta
          + 6\*\Sss(1,2,2)\*\eta^2
          - 22/3\,\*\Sss(1,2,3)
          - 6\*\Sss(1,2,4)
          + 1447/18\,\*\Sss(1,3,1)
          - 16\*\Sss(1,3,1)\*D_1^2
\\
&& \nonumber
          + 104/3\,\*\Sss(1,3,1)\*\eta
          - 6\*\Sss(1,3,1)\*\eta^2
          - 38/3\,\*\Sss(1,3,2)
          + 16\*\Sss(1,3,2)\*\eta
          + 22\*\Sss(1,3,3)
          - 56/3\,\*\Sss(1,4,1)
          + 12\*\Sss(1,4,1)\*\eta
\\
&& \nonumber
          + 50\*\Sss(1,4,2)
          + 46\*\Sss(1,5,1)
          + 18\*\Sss(2,1,1)\*\eta^2
          + 6\*\Sss(2,1,1)\*\eta^3
          + 134/3\,\*\Sss(2,1,2)
          - 12\*\Sss(2,1,2)\*D_1^2
          + 12\*\Sss(2,1,2)\*\eta
\\
&& \nonumber
          + 6\*\Sss(2,1,2)\*\eta^2
          - 22/3\,\*\Sss(2,1,3)
          - 6\*\Sss(2,1,4)
          + 134/3\,\*\Sss(2,2,1)
          - 12\*\Sss(2,2,1)\*D_1^2
          + 12\*\Sss(2,2,1)\*\eta
          + 6\*\Sss(2,2,1)\*\eta^2
\\
&& \nonumber
          - 13\*\Sss(2,2,2)
          + 6\*\Sss(2,2,2)\*\eta
          + 12\*\Sss(2,2,3)
          - 44/3\,\*\Sss(2,3,1)
          + 38\*\Sss(2,3,2)
          + 36\*\Sss(2,4,1)
          + 307/6\,\*\Sss(3,1,1)
          - 20\*\Sss(3,1,1)\*D_1^2
\\
&& \nonumber
          + 86/3\,\*\Sss(3,1,1)\*\eta
          + 16\*\Sss(3,1,1)\*\eta^2
          - 43/3\,\*\Sss(3,1,2)
          + 10\*\Sss(3,1,2)\*\eta
          + 14\*\Sss(3,1,3)
          - 43/3\,\*\Sss(3,2,1)
          + 10\*\Sss(3,2,1)\*\eta
\\
&& \nonumber
          + 24\*\Sss(3,2,2)
          + 22\*\Sss(3,3,1)
          - 37/3\,\*\Sss(4,1,1)
          + 26\*\Sss(4,1,1)\*\eta
          + 28\*\Sss(4,1,2)
          + 28\*\Sss(4,2,1)
          + 44\*\Sss(5,1,1)
          + 40\*\Ssss(1,1,1,4)
\\
&& \nonumber
          - 16/3\,\*\Ssss(1,1,3,1)
          + 16\*\Ssss(1,1,3,1)\*\eta
          - 32\*\Ssss(1,1,3,2)
          - 24\*\Ssss(1,1,4,1)
          - 12\*\Ssss(1,2,2,2)
          - 28/3\,\*\Ssss(1,3,1,1)
          - 16\*\Ssss(1,3,1,1)\*\eta
\\
&& \nonumber
          - 20\*\Ssss(1,3,1,2)
          - 20\*\Ssss(1,3,2,1)
          - 52\*\Ssss(1,4,1,1)
          - 12\*\Ssss(2,1,2,2)
          - 12\*\Ssss(2,2,1,2)
          - 12\*\Ssss(2,2,2,1)
          - 36\*\Ssss(2,3,1,1)
\\[-1mm]
&&
          - 12\*\Ssss(3,1,1,2)
          - 12\*\Ssss(3,1,2,1)
          - 12\*\Ssss(3,2,1,1)
          - 12\*\Ssss(4,1,1,1)
          - 32\*\Sssss(1,1,1,3,1)
          + 32\*\Sssss(1,1,3,1,1)
          \biggr)
\eea
and 
\bea
\label{eq:g3LncNF1}
{\lefteqn{
    \;\;\;\gamma_{\,\rm L,\,1}^{\:(3)}(N) \! = \!
}}
\nn
\\
&& \nonumber
         16 \* \biggl(
          - 353/48
          + 119917/864\,\*D_1^2
          + 15689/324\,\*D_1^3
          + 433/72\,\*D_1^4
          + 19/3\,\*D_1^5
          + 5/3\,\*D_1^6
\\[-1mm]
&& \nonumber
          - 112979/2592\,\*\eta
          + 13405/648\,\*\eta^2
          - 8045/1296\,\*\eta^3
          - 61/18\,\*\eta^4
          - 1/2\,\*\eta^5
          - 5/6\,\*\eta^6
          - 5\*\zeta_5
\\
&& \nonumber
          - 15\*\zeta_5\*\eta
          + 10\*\zeta_5\*\eta^2
          - 33/8\,\*\zeta_4
          - 11/4\,\*\zeta_4\*\eta
          + 235/16\,\*\zeta_3
          + 8/3\,\*\zeta_3\*D_1^2
          + 2\*\zeta_3\*D_1^3
          + 83/8\,\*\zeta_3\*\eta
\\
&& \nonumber
          + 3/2\,\*\zeta_3\*\eta^2
          + 2/3\,\*\zeta_3\*\eta^3
          - 39883/1296\,\*\S(1)
          + 19009/324\,\*\S(1)\*D_1^2
          + 77/9\,\*\S(1)\*D_1^3
          + 79/6\,\*\S(1)\*D_1^4
\\
&& \nonumber
          + 20/3\,\*\S(1)\*D_1^5
          - 19927/324\,\*\S(1)\*\eta
          + 1453/81\,\*\S(1)\*\eta^2
          - 7/24\,\*\S(1)\*\eta^3
          + 38/9\,\*\S(1)\*\eta^4
          + 3\*\S(1)\*\eta^5
\\
&& \nonumber
          + 10\*\S(1)\*\zeta_5
          + 11/2\,\*\S(1)\*\zeta_4
          - 317/12\,\*\S(1)\*\zeta_3
          + 4\*\S(1)\*\zeta_3\*D_1^2
          + 8/3\,\*\S(1)\*\zeta_3\*\eta
          - 8/3\,\*\S(1)\*\zeta_3\*\eta^2
          - 4\*\S(1)\*\zeta_3\*\eta^3
\\
&& \nonumber
          + 85175/2592\,\*\S(2)
          - 1873/54\,\*\S(2)\*D_1^2
          - 20/3\,\*\S(2)\*D_1^3
          - 4\*\S(2)\*D_1^4
          + 4943/648\,\*\S(2)\*\eta
          + 95/216\,\*\S(2)\*\eta^2
\\
&& \nonumber
          + 229/36\,\*\S(2)\*\eta^3
          + 25/6\,\*\S(2)\*\eta^4
          + 2/3\,\*\S(2)\*\zeta_3
          + 2\*\S(2)\*\zeta_3\*\eta
          - 4\*\S(2)\*\zeta_3\*\eta^2
          - 22247/648\,\*\S(3)
          + 241/18\,\*\S(3)\*D_1^2
\\
&& \nonumber
          + 2/3\,\*\S(3)\*D_1^3
          - 113/54\,\*\S(3)\*\eta
          + 37/18\,\*\S(3)\*\eta^2
          + 19/6\,\*\S(3)\*\eta^3
          - 8/3\,\*\S(3)\*\zeta_3
          + 725/24\,\*\S(4)
          - 16/3\,\*\S(4)\*D_1^2
\\
&& \nonumber
          - 73/36\,\*\S(4)\*\eta
          + 5/3\,\*\S(4)\*\eta^2
          - 46/3\,\*\S(5)
          + 8/3\,\*\S(5)\*\eta
          + 20/3\,\*\S(6)
          + 8/3\,\*\Ss(1,1)\*D_1^2
          + 8/3\,\*\Ss(1,1)\*D_1^3
          + 8\*\Ss(1,1)\*D_1^4
\\
&& \nonumber
          + 4/3\,\*\Ss(1,1)\*\eta
          + 9/4\,\*\Ss(1,1)\*\eta^2
          - 12\*\Ss(1,1)\*\eta^3
          - 14/3\,\*\Ss(1,1)\*\eta^4
          - 8\*\Ss(1,1)\*\zeta_3\*\eta
          + 8\*\Ss(1,1)\*\zeta_3\*\eta^2
          + 6673/324\,\*\Ss(1,2)
\\
&& \nonumber
          - 28/3\,\*\Ss(1,2)\*D_1^2
          + 28/3\,\*\Ss(1,2)\*\eta
          - 8/3\,\*\Ss(1,2)\*\eta^2
          - 8/3\,\*\Ss(1,2)\*\eta^3
          + 4\*\Ss(1,2)\*\zeta_3
          - 605/54\,\*\Ss(1,3)
          + 4/3\,\*\Ss(1,3)\*D_1^2
\\
&& \nonumber
          - 14/3\,\*\Ss(1,3)\*\eta
          - 1/3\,\*\Ss(1,3)\*\eta^2
          + 2\*\Ss(1,3)\*\eta^3
          + 181/18\,\*\Ss(1,4)
          + 10/3\,\*\Ss(1,4)\*\eta
          - 16/3\,\*\Ss(1,5)
          + 6673/324\,\*\Ss(2,1)
\\
&& \nonumber
          - 28/3\,\*\Ss(2,1)\*D_1^2
          + 28/3\,\*\Ss(2,1)\*\eta
          - 8/3\,\*\Ss(2,1)\*\eta^2
          - 8/3\,\*\Ss(2,1)\*\eta^3
          + 4\*\Ss(2,1)\*\zeta_3
          - 1021/54\,\*\Ss(2,2)
          + 4\*\Ss(2,2)\*D_1^2
\\
&& \nonumber
          + 2/3\,\*\Ss(2,2)\*\eta
          - 2\*\Ss(2,2)\*\eta^2
          + 181/18\,\*\Ss(2,3)
          - 8/3\,\*\Ss(2,3)\*\eta
          + 2\*\Ss(2,3)\*\eta^2
          - 2\*\Ss(2,4)
          - 479/18\,\*\Ss(3,1)
\\
&& \nonumber
          + 16/3\,\*\Ss(3,1)\*D_1^2
          + 59/9\,\*\Ss(3,1)\*\eta
          - 2\*\Ss(3,1)\*\eta^2
          - 2\*\Ss(3,1)\*\eta^3
          + 275/18\,\*\Ss(3,2)
          - 10/3\,\*\Ss(3,2)\*\eta
          - 22/3\,\*\Ss(3,3)
\\
&& \nonumber
          + 343/18\,\*\Ss(4,1)
          - 4/3\,\*\Ss(4,1)\*\eta
          - 2\*\Ss(4,1)\*\eta^2
          - 12\*\Ss(4,2)
          - 12\*\Ss(5,1)
          + 4\*\Sss(1,1,3)\*\eta
          - 4\*\Sss(1,1,3)\*\eta^2
          - 20/3\,\*\Sss(1,1,4)
\\
&& \nonumber
          - 20/3\,\*\Sss(1,2,2)
          + 4/3\*\Sss(1,2,3)
          - 94/9\,\*\Sss(1,3,1)
          - 20/3\,\*\Sss(1,3,1)\*\eta
          + 4\*\Sss(1,3,1)\*\eta^2
          + 20/3\,\*\Sss(1,3,2)
          + 20/3\,\*\Sss(1,4,1)
\\
&& \nonumber
          - 20/3\,\*\Sss(2,1,2)
          + 4/3\,\*\Sss(2,1,3)
          - 20/3\,\*\Sss(2,2,1)
          + 4\*\Sss(2,2,2)
          + 8/3\,\*\Sss(2,3,1)
          - 20/3\,\*\Sss(3,1,1)
          + 4/3\,\*\Sss(3,1,1)\*\eta
\\[-1mm]
&&
          + 16/3\,\*\Sss(3,1,2)
          + 16/3\,\*\Sss(3,2,1)
          + 28/3\,\*\Sss(4,1,1)
          + 16/3\,\*\Ssss(1,1,3,1)
          - 8/3\,\*\Ssss(1,3,1,1)
          \biggr)
\:\: .
\eea

\noindent
The large-$N$ limit of eq.~(\ref{eq:g3Lnc}) is of the form (\ref{eq:ntoinf})
with the large-$\nc$ cusp anomalous dimension 
\bea
\label{eq:A4Lnc}
  A_{L,4} &\!\!=\!& 
         \cf \* \nct \* \left(
            \frac{84278}{81}\,
          - \frac{88832}{81}\,\*\zeta_2
          + \frac{20992}{27}\,\*\zeta_3
          + 1804\*\zeta_4
          - \frac{352}{3}\,\*\zeta_2\*\zeta_3
          - 352\*\zeta_5
          - 32\*\zeta_3^2
          - 876\,\*\zeta_6 
\!
          \right)
\nonumber \\[1mm] && \mbox{\hspn}
       - \,\cf\*\ncs\,\*\nf \* \left(
            \frac{39883}{81}
          - \frac{26692}{81}\,\*\zeta_2
          + \frac{16252}{27}\,\*\zeta_3
          + \frac{440}{3}\,\*\zeta_4
          - \frac{256}{3}\,\*\zeta_2\*\zeta_3
          - 224\,\*\zeta_5
          \right)
\nonumber \\[1mm] && \mbox{\hspn}
        + \,\cf\*\nc\,\*\nfs \* \left(
            \frac{2119}{81}
          - \frac{608}{81}\,\*\zeta_2
          + \frac{1280}{27}\,\*\zeta_3
          - \frac{64}{3}\,\*\zeta_4
          \right)
        \,-\, \cf\*\nft \* \left(
            \frac{32}{81}
          - \frac{64}{27}\,\*\zeta_3
          \right)
\:\: . 
\eea
%
%
Our result for the (complete) $\nfs$ part was first presented at 
Loops$\,\&\,$Legs 2016, see ref.~\cite{Ruijl:2016pkm}, the rest in a Zurich 
seminar by one of us \cite{BRuijl6dec16}. Eq.~(\ref{eq:A4Lnc}) agrees with 
results of~refs.~\cite{Henn:2016men,Lee:2016ixa}, where this quantity was
obtained by computing the photon-quark form factor in the large-$\nc$ limit. 
The~lower-order coefficients can be found in eq.~(3.11) of 
ref.~\cite{Moch:2004pa}.

The one- to three-loop coefficients $B_{1,2,3}$ in eq.~(\ref{eq:xto1Lnc})
can be found, as coefficients of $\delta(1\!-\!x)$, in  eqs.~(4.5), (4.6)
and (4.9) of ref.~\cite{Moch:2004pa}. The four-loop coefficient in the 
large-$\nc$ limit reads
\bea
\label{eq:B4Lnc}
 B_{L,4} &\!\!=\!& 
         \cf \* \nct \* \bigg(
          - \frac{1379569}{5184}
          + \frac{24211}{27} \,\* \zeta_2
          - \frac{9803}{162} \,\* \zeta_3
          - \frac{9382}{9} \,\* \zeta_4
          + \frac{838}{9} \,\* \zeta_2 \* \zeta_3
          + 1002 \* \zeta_5
\nonumber \\ && \mbox{\hspp\hspp}
          + \frac{16}{3} \,\* \zeta_3^2
          + 135 \* \zeta_6
          - 80 \* \zeta_2 \* \zeta_5
          + 32 \* \zeta_3 \* \zeta_4
          - 560 \* \zeta_7
          \biggr)
\nonumber \\[1mm] && \mbox{\hspn}
       +\, \cf\*\ncs\*\,\nf \* \left(
            \frac{353}{3}
          - \frac{85175}{162} \,\* \zeta_2
          - \frac{137}{9} \,\* \zeta_3
          + \frac{16186}{27} \,\* \zeta_4
          - \frac{584}{9} \,\* \zeta_2 \* \zeta_3
          - \frac{248}{3} \,\* \zeta_5
          - \frac{16}{3} \,\* \zeta_3^2
          - 144 \* \zeta_6
\!
          \right)
\nonumber \\[1mm] && \mbox{\hspn}
       -\, \cf\*\nc\*\,\nfs \* \left(
            \frac{127}{18}
          - \frac{5036}{81} \,\* \zeta_2
          + \frac{932}{27} \,\* \zeta_3
          + \frac{1292}{27} \,\* \zeta_4
          - \frac{160}{9} \,\* \zeta_2 \* \zeta_3
          - \frac{32}{3} \,\* \zeta_5
          \right)
\nonumber \\[1mm] && \mbox{\hspn}
       -\, \cf\*\nft \* \left(
            \frac{131}{81}
          - \frac{32}{81} \,\* \zeta_2
          - \frac{304}{81} \,\* \zeta_3
          + \frac{32}{27} \,\* \zeta_4
          \right)
\:\: . 
\eea
%
%
The coefficients $B_n$ contain collinear contributions to the evolution 
kernels. With the help of the QCD corrections to the quark form factor in 
dimensional regularization, one can extract from them the universal eikonal 
anomalous dimension.
The latter governs the subleading infrared poles in gauge-theory amplitudes
and captures contributions from large-angle soft gluons
\cite{Ravindran:2004mb,Ravindran:2006cg,Dixon:2008gr}.

As mentioned above, the coefficients $C_n$ and $\widetilde{D}_n$ in 
eq.~(\ref{eq:xto1Lnc}) do not provide new information, but are functions of
lower-order quantities. They are given by 
\beq
\label{eq:CDofAB}
  C(\ars) \;=\; \left( \:\! A(\ars) \:\! \right)^{2^{}}
\;\; , \quad
  \widetilde{D}(\ars) \;=\; A(\ars) \cdot ( B(\ars)_{} - \beta(\ars)/\ars )
\:\: ,
\eeq
cf.~ref.~\cite{Dokshitzer:2005bf}, which leads to the four-loop relations
\beq
\label{eq:CD4ofAB}
  C_4 \;=\; A_2^{\,2^{}} \,+\, 2\,A_{1\,} A_3 
\;\; , \quad
  \widetilde{D}_4 \;=\; \sum_{k=1}^{3} A_k \cdot ( B_{4-k} - \beta_{3-k} ) 
\:\: . \qquad
\eeq
Using the results (\ref{eq:A4Lnc}) and (\ref{eq:B4Lnc}), it is actually now 
possible to predict $C_5$ and $\widetilde{D}_5$ for large $\nc$.

\pagebreak

\begin{figure}[htb]
\vspace{-1mm}
\centerline{\epsfig{file=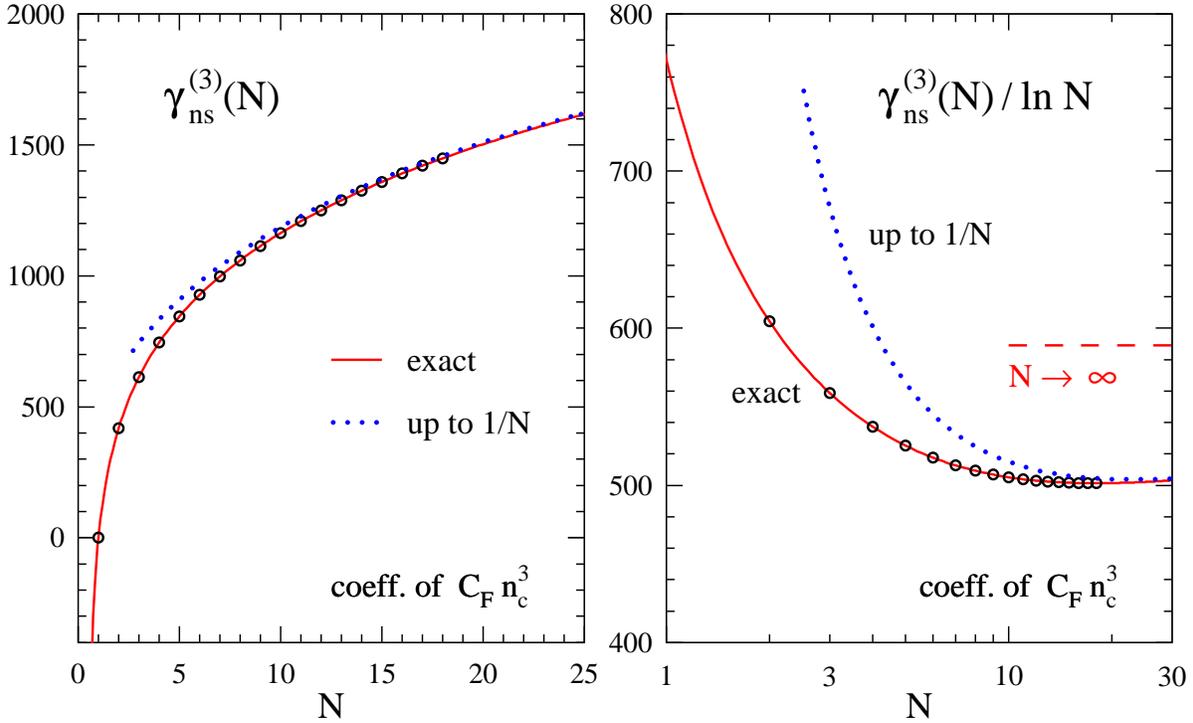,width=16.0cm,angle=0}}
\vspace{-1mm}
\caption{ \label{fig:g3Lnc0} \small
 The $\nfz$ part (\ref{eq:g3LncNF0}) of the anomalous dimensions 
 $\,\gamma_{\,\rm ns}^{\:(n)\pm}(N)\,$ in the large-$\nc$ limit, compared with 
 its large-$N$ expansion with all terms included in eq.~(\ref{eq:ntoinf}) 
 and, in the right panel, its asymptotic behaviour for $\,N \to \infty$. 
 The exact curve has been computed by via the $x$-space counterpart of 
 eq.~(\ref{eq:g3LncNF0}), see section~\ref{sec:Xres}.
 }
\vspace{1mm}
\end{figure}
 
\begin{figure}[p]
\vspace{-2mm}
\centerline{\epsfig{file=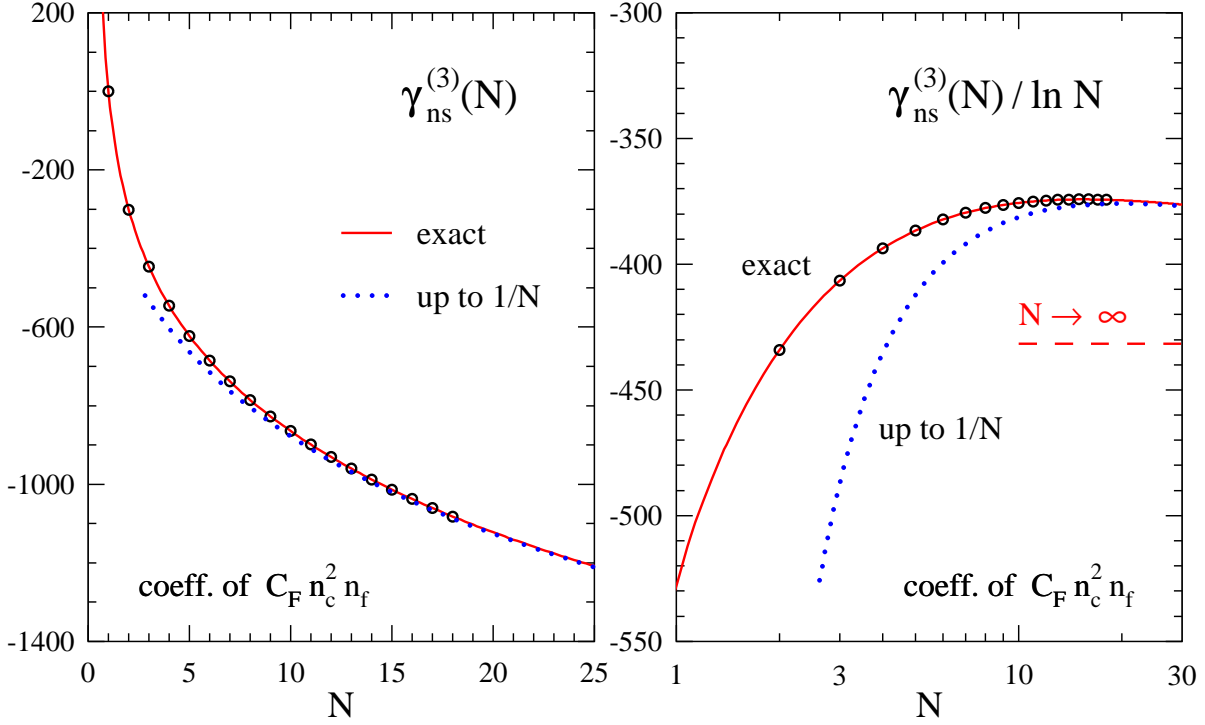,width=16.0cm,angle=0}}
\vspace{-1mm}
\caption{ \label{fig:g3Lnc1} \small
 As fig.~\ref{fig:g3Lnc0}, but for the $\nfo$ contribution (\ref{eq:g3LncNF1}). 
 The value of $\,\gamma_{\,\rm ns}^{\:(n)\pm}(N) / \ln N\,$ in the limit
 $\,N \to \infty\,$ is~given by the corresponding coefficient of $A_{L,4}$ 
 in the second line of eq.~(\ref{eq:A4Lnc}).
 }
\vspace{-2mm}
\end{figure}

The new functions (\ref{eq:g3LncNF0}) and (\ref{eq:g3LncNF1}) are shown in 
figs.~\ref{fig:g3Lnc0} and \ref{fig:g3Lnc1}, respectively, together with their 
large-$N$ approximation (\ref{eq:ntoinf}) with the coefficients given above.
In the right panels, the results are divided by $\ln\, N$, so for 
$N \to \infty$ the curves tend to constants given by the respective terms in 
the four-loop cusp anomalous dimension (\ref{eq:A4Lnc}).
 
The approach to this asymptotic behaviour is very slow: the $\nfz$ contribution
in fig.~\ref{fig:g3Lnc0} is 0.856 of its asymptotic result at $N=30$, yet it 
deviates by less than 10\% only from an $N$-value above $N = 500$. 
The corresponding numbers for the $\nfo$ part in fig.~\ref{fig:g3Lnc1} are
0.873 at $N=30$ and $N \simeq 185$ for a deviation by less than 10\%.
It might be interesting to note, on the other hand, that the corresponding
$n_{\!f}^{\,a}$ coefficient of $A_n$, here and in all lower-order cases
(in full QCD), falls in the interval spanned by the corresponding results 
for $\,\gamma_{\,\rm ns}^{\:(n)}(2) / \ln 2\,$ and 
$\,\gamma_{\,\rm ns}^{\:(n)}(4) / \ln 4$.

\bigskip

\begin{figure}[p]
\vspace{-2mm}
\centerline{\epsfig{file=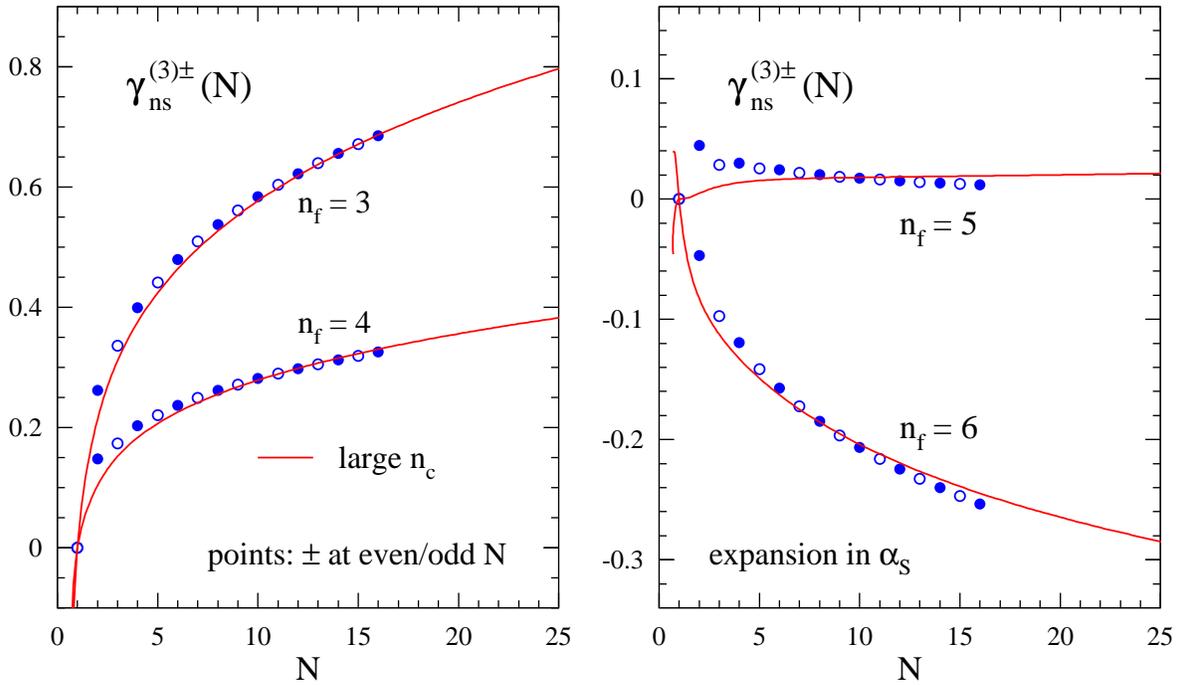,width=16.0cm,angle=0}}
\vspace{-1mm}
\caption{ \label{fig:g3Lnf3456} \small
 Our even-$N$ results for $\,\gamma_{\,\rm ns}^{\:(n)+}(N)$ and odd-$N$ 
 values for $\,\gamma_{\,\rm ns}^{\:(n)-}(N)$ at $\nf = 3, \ldots 6$, 
 compared with their common large-$\nc$ limit now known at all $N$.
 The results have been converted to an expansion in $\als$. 
 }
\vspace{-2mm}
\end{figure}

The results (\ref{gam+to16}) for $\,\gamma_{\,\rm ns}^{\:(3)+\,}$ (closed 
circles) and (\ref{gam-to15}) for $\,\gamma_{\,\rm ns}^{\:(3)-\,}$ (open
circles) are shown for the physically relevant values of $\nf$ in 
fig.~\ref{fig:g3Lnf3456}, together with the all-$N$ results in the large-$\nc$ 
limit. 
As at the previous orders in $\als$, there are 
cancellations between the $\nf$-independent and the \mbox{$\nf$-dependent} 
contributions, which are particularly pronounced here at $\nf = 5$.
For this number of light flavours, which is relevant for high-energy processes 
at the LHC, the large-$\nc$ result do not describe the (small) fourth-order QCD
contributions to the non-singlet evolution equations at the phenomenologically 
most relevant moments $N$ and momentum fractions $x$.
We therefore need to convert the calculated moments to practically usable 
constraints on the four-loop splitting functions $P_{\,\rm ns}^{\:(3)}(x)$.


\newpage

%
\setcounter{equation}{0}
\section{Results in $x\:\!$-$\:\!$space}
\label{sec:Xres}

The four-loop non-singlet splitting functions $P_{\rm ns}^{\,(3)}(x)$ 
are derived from the all-$N$ results for the corresponding anomalous 
dimensions by an inverse Mellin transformation that expresses these 
functions in terms of harmonic polylogarithms (HPLs). 
This transformation can be performed by an algebraic procedure~%
\cite{Remiddi:1999ew,Moch:1999eb} based on the fact that harmonic sums 
occur as coefficients of the Taylor expansion of HPLs.

For the convenience of the reader, we recall their basic definitions 
\cite{Remiddi:1999ew}. The lowest-weight ($w = 1$) functions $H_m(x)$ 
are given by
\beq
\label{eq:HPL1}
  H_0(x)       \:\: = \:\:\ln x \:\: , \quad\quad
  H_{\pm 1}(x) \:\: = \:\: \mp \, \ln (1 \mp x) \:\: .
\eeq
The higher-weight ($w \geq 2$) functions are recursively defined as
\beq
\label{eq:HPL2}
  H_{m_1,...,m_w}(x) \:\: = \:\:
  \left\{ 
    \begin{array}{cl}
      \displaystyle{ \frac{1}{w!}\,\ln^{\,w\!} x \:\: ,}
      & 
      \quad {\rm if} \:\:\: m^{}_1,...,m^{}_w \:=\: 0,\ldots ,0 
      \\[2ex]
      \displaystyle{ 
        \int_0^x \! dz\: f_{m_1}(z) \, H_{m_2,...,m_w}(z)
        \:\: , } 
      & 
      \quad {\rm otherwise}
    \end{array} \right.
\eeq
with
\beq
\label{eq:HPLf}
  f_0^{}(x)       \:\: = \:\: x^{\,-1} \:\: , \quad\quad
  f_{\pm 1}^{}(x) \:\: = \:\: (1 \mp x)^{-1} \;\; .
\eeq
For chains of indices `zero' we employ the abbreviated notation
\beq
\label{eq:HPP-abbr}
  H_{{\footnotesize \underbrace{0,\ldots ,0}_{\scriptstyle m} },\,
  \pm 1,\, {\footnotesize \underbrace{0,\ldots ,0}_{\scriptstyle n} },
  \, \pm 1,\, \ldots}(x) \:\: = \:\: H_{\pm (m+1),\,\pm (n+1),\, \ldots}(x)
\;\; .
\eeq
The argument $x$ will be suppressed in all results below, and we express the 
terms with $(1 \pm x)^{-1}$ in terms of the $x$-dependence of the leading-order 
splitting function $P_{\rm qq}^{\,(0)}$,
\bea
  \pqq(x) &\! =\! & 2\, (1 - x)^{-1} - 1 - x \, .
\eea

In this notation, the common large-$\nc$ limit of the functions 
$P_{\rm ns}^{\,(3)+}(x)$ and $P_{\rm ns}^{\,(3)-}(x)$ is given by 
\beq
\label{eq:P3Lnc}
P_{\rm ns,\,L}^{\,(3)}(x) \:\:=\:\:  \cf \left( 
  \nct\,     P_{\rm L,\,0}^{\,(3)}(x) 
  \:+\: \ncs \nf\, P_{\rm L,\,1}^{\,(3)}(x)
  \:+\: \nc \nfs\, P_{\rm L,\,2}^{\,(3)}(x)
  \:+\: \nft\,     P_{\rm L,\,3}^{\,(3)}(x) \right)
\eeq
with the new results
\bea
\label{eq:P3LncNF0}
{\lefteqn{
    P_{\rm L,\,0}^{\,(3)}(x) \; = \;
}}
\nn
\\
&& \nonumber
         \pqq(x)  \*  \biggl( \,
            \frac{42139}{81}
          - 160 \* \zeta_6
          - 204 \* \zeta_5
          + \frac{7666}{9} \,\* \zeta_4
          + \frac{10334}{27} \,\* \zeta_3
          + 16 \* \zts
          - \frac{44416}{81} \,\* \zeta_2
          - 72 \* \zeta_2 \* \zeta_3
\\
&& \nonumber
          + \frac{71591}{108} \,\* \Hh(0,0)
          + 504 \* \Hh(0,0) \* \zeta_4
          + \frac{212}{3} \,\* \Hh(0,0) \* \zeta_3
          - \frac{6899}{9} \,\* \Hh(0,0) \* \zeta_2
          + \frac{44416}{81} \,\* \Hh(1,0)
          + 680 \* \Hh(1,0) \,\* \zeta_4
\\
&& \nonumber
          + 128 \* \Hh(1,0) \,\* \zeta_3
          - \frac{5788}{9} \,\* \Hh(1,0) \* \zeta_2
          + 512 \* \Hh(1,1) \,\* \zeta_4
          + \frac{128}{3} \,\* \Hh(1,1) \* \zeta_3
          + 192 \* \Hh(1,2) \,\* \zeta_3
          + \frac{5788}{9} \,\* \Hh(1,3)
\\
&& \nonumber
          - 160 \* \Hh(1,3) \,\* \zeta_2
          + \frac{448}{3} \,\* \Hh(1,4)
          + 368 \* \Hh(1,5)
          + \frac{9029}{27} \,\* \Hh(2,0)
          + 32 \* \Hh(2,0) \* \zeta_3
          - \frac{352}{3} \,\* \Hh(2,0) \* \zeta_2
          + 192 \* \Hh(2,1) \* \zeta_3
\\
&& \nonumber
          + \frac{1072}{3} \,\* \Hh(2,2)
          - 96 \* \Hh(2,2) \* \zeta_2
          + \frac{352}{3} \,\* \Hh(2,3)
          + 288 \* \Hh(2,4)
          + \frac{5581}{9} \,\* \Hh(3,0)
          - 176 \* \Hh(3,0) \* \zeta_2
          + \frac{1228}{3} \,\* \Hh(3,1)
\\
&& \nonumber
          - 96 \* \Hh(3,1) \* \zeta_2
          + \frac{344}{3} \,\* \Hh(3,2)
          + 176 \* \Hh(3,3)
          + 228 \* \Hh(4,0)
          + \frac{296}{3} \,\* \Hh(4,1)
          + 224 \* \Hh(4,2)
          + 288 \* \Hh(5,0)
          + 352 \* \Hh(5,1)
\\
&& \nonumber
          + \frac{55291}{108} \,\* \Hhh(0,0,0)
          - 32 \* \Hhh(0,0,0) \,\* \zeta_3
          - 252 \* \Hhh(0,0,0) \,\* \zeta_2
          + \frac{7120}{27} \,\* \Hhh(1,0,0)
          - 16 \* \Hhh(1,0,0) \,\* \zeta_3
          - \frac{448}{3} \,\* \Hhh(1,0,0) \* \zeta_2
\\
&& \nonumber
          + 64 \* \Hhh(1,1,0) \* \zeta_3
          + \frac{128}{3} \,\* \Hhh(1,1,0) \* \zeta_2
          + 256 \* \Hhh(1,1,1) \* \zeta_3
          - \frac{128}{3} \,\* \Hhh(1,1,3)
          + 192 \* \Hhh(1,1,4)
          + \frac{1072}{3} \,\* \Hhh(1,2,0)
\\
&& \nonumber
          + \frac{304}{3} \,\* \Hhh(1,3,0)
          - \frac{224}{3} \,\* \Hhh(1,3,1)
          + 160 \* \Hhh(1,3,2)
          + 400 \* \Hhh(1,4,0)
          + 416 \* \Hhh(1,4,1)
          + \frac{3635}{9} \,\* \Hhh(2,0,0)
\\
&& \nonumber
          - 288 \* \Hhh(2,0,0) \* \zeta_2
          + \frac{1072}{3} \,\* \Hhh(2,1,0)
          + 104 \* \Hhh(2,2,0)
          + 96 \* \Hhh(2,2,2)
          + 304 \* \Hhh(2,3,0)
          + 288 \* \Hhh(2,3,1)
\\
&& \nonumber
          + \frac{572}{3} \,\* \Hhh(3,0,0)
          + \frac{344}{3} \,\* \Hhh(3,1,0)
          + 96 \* \Hhh(3,1,2)
          + 192 \* \Hhh(3,2,0)
          + 96 \* \Hhh(3,2,1)
          + 208 \* \Hhh(4,0,0)
          + 224 \* \Hhh(4,1,0)
\\
&& \nonumber
          + 96 \* \Hhh(4,1,1)
          + 454 \* \Hhhh(0,0,0,0)
          - 224 \* \Hhhh(0,0,0,0) \* \zeta_2
          + \frac{4745}{9} \* \,\Hhhh(1,0,0,0)
          - 368 \* \Hhhh(1,0,0,0) \* \zeta_2
          - 192 \* \Hhhh(1,1,0,0) \* \zeta_2
\\
&& \nonumber
          + 256 \* \Hhhh(1,1,1,0) \* \zeta_2
          - 256 \* \Hhhh(1,1,1,3)
          + 256 \* \Hhhh(1,1,3,0)
          + 256 \* \Hhhh(1,1,3,1)
          + \frac{176}{3} \,\* \Hhhh(1,2,0,0)
          + 96 \* \Hhhh(1,2,2,0)
\\
&& \nonumber
          + 176 \* \Hhhh(1,3,0,0)
          + 160 \* \Hhhh(1,3,1,0)
          + 124 \* \Hhhh(2,0,0,0)
          + \frac{176}{3} \,\* \Hhhh(2,1,0,0)
          + 96 \* \Hhhh(2,1,2,0)
          + 96 \* \Hhhh(2,2,0,0)
\\
&& \nonumber
          + 96 \* \Hhhh(2,2,1,0)
          + 112 \* \Hhhh(3,0,0,0)
          + 112 \* \Hhhh(3,1,0,0)
          + 96 \* \Hhhh(3,1,1,0)
          + \frac{520}{3} \,\* \Hhhhh(0,0,0,0,0)
          + \frac{560}{3} \,\* \Hhhhh(1,0,0,0,0)
\\
&& \nonumber
          - \frac{160}{3} \,\* \Hhhhh(1,1,0,0,0)
          - 48 \* \Hhhhh(1,2,0,0,0)
          + 128 \* \Hhhhh(2,0,0,0,0)
          - 48 \* \Hhhhh(2,1,0,0,0)
          + 80 \* \Hhhhhh(0,0,0,0,0,0)
\\
&& \nonumber
          + 160 \* \Hhhhhh(1,0,0,0,0,0)
          + 64 \* \Hhhhhh(1,1,0,0,0,0)
          - 320 \* \Hhhhhh(1,1,1,0,0,0)
          + \frac{24211}{54} \,\* \H(0)
          - 160 \* \H(0) \* \zeta_5
          + \frac{538}{3} \,\* \H(0) \* \zeta_4
\\
&& \nonumber
          + \frac{193}{3} \,\* \H(0) \* \zeta_3
          - \frac{3700}{9} \,\* \H(0) \* \zeta_2
          + 96 \* \H(0) \* \zeta_2 \* \zeta_3
          - 80 \* \H(1) \* \zeta_5
          - \frac{8}{3} \,\* \H(1) \* \zeta_4
          + \frac{644}{9} \,\* \H(1) \* \zeta_3
          + 96 \* \H(1) \* \zeta_2 \* \zeta_3
\\
&& \nonumber
          + \frac{44416}{81} \,\* \H(2)
          + 552 \* \H(2) \* \zeta_4
          + \frac{272}{3} \,\* \H(2) \* \zeta_3
          - \frac{1072}{3} \,\* \H(2) \* \zeta_2
          + \frac{3700}{9} \,\* \H(3)
          + 80 \* \H(3) \* \zeta_3
          - \frac{344}{3} \,\* \H(3) \* \zeta_2
\\
&& \nonumber
          + \frac{6899}{9} \,\* \H(4)
          - 224 \* \H(4) \* \zeta_2
          + 252 \* \H(5)
          + 224 \* \H(6)
          \biggr)
       + x  \*  \biggl(
          - \frac{184}{3} \,\* \zeta_5
          - \frac{80}{3} \,\* \zeta_2 \* \zeta_3
          + \frac{796}{3} \,\* \zeta_4
          - \frac{871}{3} \,\* \zeta_3
\\
&& \nonumber
          + \frac{68435}{54} \,\* \zeta_2
          - \frac{567245}{324} \,\* \Hh(0,0)
          - \frac{52}{3} \,\* \Hh(0,0) \* \zeta_3
          + \frac{875}{3} \,\* \Hh(0,0) \* \zeta_2
          - \frac{1015}{3} \,\* \Hh(2,0)
          - \frac{944}{3} \,\* \Hh(2,0) \* \zeta_2
\\
&& \nonumber
          + 254 \* \Hh(2,1)
          - \frac{116}{3} \,\* \Hh(2,2)
          + \frac{944}{3} \,\* \Hh(2,3)
          - \frac{1130}{3} \,\* \Hh(3,0)
          + \frac{40}{3} \,\* \Hh(3,1)
          - 96 \* \Hh(3,2)
          - \frac{1060}{3} \,\* \Hh(4,0)
          - 216 \* \Hh(4,1)
\\
&& \nonumber
          - \frac{33091}{36} \* \Hhh(0,0,0)
          + 428 \* \Hhh(0,0,0) \* \zeta_2
          - \frac{1223}{3} \,\* \Hhh(2,0,0)
          - \frac{116}{3} \,\* \Hhh(2,1,0)
          - 56 \* \Hhh(2,2,0)
          - 148 \* \Hhh(3,0,0)
\\
&& \nonumber
          - 96 \* \Hhh(3,1,0)
          - 550 \* \Hhhh(0,0,0,0)
          + \frac{152}{3} \,\* \Hhhh(2,0,0,0)
          - \frac{952}{3} \,\* \Hhhh(2,1,0,0)
          - 280 \* \Hhhhh(0,0,0,0,0)
          - \frac{237016}{81} \,\* \H(0)
 \\
&& \nonumber
         - 422 \* \H(0) \* \zeta_4
          + \frac{569}{3} \,\* \H(0) \* \zeta_3
          + \frac{937}{3} \,\* \H(0) \* \zeta_2
          - \frac{68435}{54} \,\* \H(2)
          + 208 \* \H(2) \* \zeta_3
          + \frac{116}{3} \,\* \H(2) \* \zeta_2
          - \frac{937}{3} \,\* \H(3)
\\
&& \nonumber
          + 96 \* \H(3) \* \zeta_2
          - \frac{875}{3} \,\* \H(4)
          - 428 \* \H(5)
          \biggr)
       + \xp1  \*  \biggl(
            360 \* \zeta_6
          + 64 \* \zts
          - 268 \* \Hh(0,0) \* \zeta_4
          - 80 \* \Hh(2,0) \* \zeta_3
\\
&& \nonumber
          - 160 \* \Hh(2,1) \* \zeta_3
          - 112 \* \Hh(2,4)
          + 112 \* \Hh(3,0) \* \zeta_2
          - 112 \* \Hh(3,3)
          - 88 \* \Hh(4,2)
          - 172 \* \Hh(5,0)
          - 176 \* \Hh(5,1)
\\[1mm]
&& \nonumber
          + 24 \* \Hhh(0,0,0) \* \zeta_3
          + 112 \* \Hhh(2,0,0) \* \zeta_2
          + 96 \* \Hhh(2,1,0) \* \zeta_2
          - 96 \* \Hhh(2,1,3)
          - 48 \* \Hhh(2,3,0)
          - 32 \* \Hhh(2,3,1)
          - 48 \* \Hhh(3,2,0)
\\[1mm]
&& \nonumber
          - 100 \* \Hhh(4,0,0)
          - 88 \* \Hhh(4,1,0)
          + 168 \* \Hhhh(0,0,0,0) \* \zeta_2
          + 64 \* \Hhhh(2,2,0,0)
          - 24 \* \Hhhh(3,0,0,0)
          + 64 \* \Hhhh(3,1,0,0)
\\[1mm]
&& \nonumber
          - 16 \* \Hhhhh(2,0,0,0,0)
          + 64 \* \Hhhhh(2,1,0,0,0)
          + 128 \* \Hhhhh(2,1,1,0,0)
          - 70 \* \Hhhhhh(0,0,0,0,0,0)
          + 40 \* \H(0) \* \zeta_5
          - 48 \* \H(0) \* \zeta_2 \* \zeta_3
\\[1mm]
&& \nonumber
          - 216 \* \H(2) \* \zeta_4
          - 112 \* \H(3) \* \zeta_3
          + 88 \* \H(4) \* \zeta_2
          - 168 \* \H(6)
          \biggr)
       + \xm1  \*  \biggl(
            \frac{1325809}{324}
          + \frac{43301}{27} \,\* \Hh(1,0)
\\
&& \nonumber
          - 256 \* \Hh(1,0) \* \zeta_3
          + \frac{560}{3} \,\* \Hh(1,0) \* \zeta_2
          + \frac{3224}{3} \,\* \Hh(1,1)
          - 448 \* \Hh(1,1) \* \zeta_3
          + \frac{448}{3} \,\* \Hh(1,2)
          - \frac{560}{3} \,\* \Hh(1,3)
          - 192 \* \Hh(1,4)
\\
&& \nonumber
          + \frac{3920}{9} \,\* \Hhh(1,0,0)
          + 192 \* \Hhh(1,0,0) \* \zeta_2
          + \frac{448}{3} \,\* \Hhh(1,1,0)
          + 704 \* \Hhh(1,1,0) \* \zeta_2
          - 704 \* \Hhh(1,1,3)
          + \frac{400}{3} \,\* \Hhh(1,2,0)
\\
&& \nonumber
          + 128 \* \Hhh(1,3,0)
          + 128 \* \Hhh(1,3,1)
          + \frac{256}{3} \,\* \Hhhh(1,0,0,0)
          + 352 \* \Hhhh(1,1,0,0)
          + 288 \* \Hhhh(1,2,0,0)
          + 32 \* \Hhhhh(1,0,0,0,0)
\\
&& \nonumber
          - 160 \* \Hhhhh(1,1,0,0,0)
          + 576 \* \Hhhhh(1,1,1,0,0)
          + \frac{41504}{27} \,\* \H(1)
          - 320 \* \H(1) \* \zeta_4
          - \frac{752}{3} \,\* \H(1) \* \zeta_3
          - \frac{448}{3} \,\* \H(1) \* \zeta_2
          \biggr)
\\
&& \nonumber
          - \frac{424}{3} \,\* \zeta_5
          + \frac{1364}{3} \,\* \zeta_4
          - 113 \* \zeta_3
          - \frac{11641}{6} \,\* \zeta_2
          + \frac{208}{3} \,\* \zeta_2 \* \zeta_3
          + \frac{451015}{324} \,\* \Hh(0,0)
          - \frac{100}{3} \,\* \Hh(0,0) \* \zeta_3
\\
&& \nonumber
          - 87 \* \Hh(0,0) \* \zeta_2
          + \frac{7715}{9} \,\* \Hh(2,0)
          + \frac{784}{3} \,\* \Hh(2,0) \* \zeta_2
          + \frac{1658}{3} \,\* \Hh(2,1)
          + 228 \* \Hh(2,2)
          - \frac{784}{3} \,\* \Hh(2,3)
          + \frac{758}{3} \,\* \Hh(3,0)
\\
&& \nonumber
          + \frac{1096}{3} \,\* \Hh(3,1)
          + 64 \* \Hh(3,2)
          + \frac{140}{3} \,\* \Hh(4,0)
          + 200 \* \Hh(4,1)
          + \frac{10465}{36} \,\* \Hhh(0,0,0)
          + 60 \* \Hhh(0,0,0) \* \zeta_2
          + \frac{809}{3} \,\* \Hhh(2,0,0)
\\
&& \nonumber
          + 228 \* \Hhh(2,1,0)
          + 40 \* \Hhh(2,2,0)
          + 28 \* \Hhh(3,0,0)
          + 64 \* \Hhh(3,1,0)
          - \frac{490}{3} \,\* \Hhhh(0,0,0,0)
          + \frac{8}{3} \,\* \Hhhh(2,0,0,0)
          + \frac{776}{3} \,\* \Hhhh(2,1,0,0)
\\
&& \nonumber
          - 120 \* \Hhhhh(0,0,0,0,0)
          + \frac{217325}{81} \,\* \H(0)
          + 258 \* \H(0) \* \zeta_4
          + \frac{41}{3} \* \H(0) \* \zeta_3
          - \frac{8521}{9} \,\* \H(0) \* \zeta_2
          + \frac{11641}{6} \,\* \H(2)
          - 176 \* \H(2) \* \zeta_3
\\
&& \nonumber
          - 228 \* \H(2) \* \zeta_2
          + \frac{8521}{9} \,\* \H(3)
          - 64 \* \H(3) \* \zeta_2
          + 87 \* \H(4)
          - 60 \* \H(5)
       - \delta \xm1  \*  \biggl(
            \frac{1379569}{5184}
          - \frac{24211}{27} \,\* \zeta_2
\\
&& \nonumber
          + \frac{9803}{162} \,\* \zeta_3
          + \frac{9382}{9} \,\* \zeta_4
          - \frac{838}{9} \,\* \zeta_2 \* \zeta_3
          - 1002 \* \zeta_5
          - \frac{16}{3} \,\* \zts
          - 135 \* \zeta_6
          + 80 \* \zeta_2 \* \zeta_5
          - 32 \* \zeta_3 \* \zeta_4
          + 560 \* \zeta_7
          \biggr)
\:\: ,
\\[2mm]
\eea

\vspace*{-5mm}
\noindent
and 
\bea
\label{eq:P3LncNF1}
{\lefteqn{
    P_{\rm L,\,1}^{\,(3)}(x) \;=\;
}}
\nn
\\
&& \nonumber
         \pqq(x)  \*  \biggl(
          - \frac{39883}{162}
          + 128 \* \zeta_5
          - \frac{680}{9} \,\* \zeta_4
          - \frac{8126}{27} \,\* \zeta_3
          + \frac{13346}{81} \,\* \zeta_2
          + 32 \* \zeta_2 \* \zeta_3
          - \frac{22247}{81} \,\* \Hh(0,0)
\\
&& \nonumber
          - \frac{128}{3} \,\* \Hh(0,0) \* \zeta_3
          + \frac{1372}{9} \,\* \Hh(0,0) \* \zeta_2
          - \frac{13346}{81} \,\* \Hh(1,0)
          - 64 \* \Hh(1,0) \* \zeta_3
          + \frac{752}{9} \,\* \Hh(1,0) \* \zeta_2
          - \frac{128}{3} \,\* \Hh(1,1) \* \zeta_3
\\
&& \nonumber
          - \frac{752}{9} \,\* \Hh(1,3)
          - \frac{160}{3} \,\* \Hh(1,4)
          - \frac{4084}{27} \,\* \Hh(2,0)
          + \frac{64}{3} \,\* \Hh(2,0) \* \zeta_2
          - \frac{160}{3} \,\* \Hh(2,2)
          - \frac{64}{3} \,\* \Hh(2,3)
          - \frac{1100}{9} \,\* \Hh(3,0)
\\
&& \nonumber
          - \frac{160}{3} \,\* \Hh(3,1)
          - \frac{128}{3} \,\* \Hh(3,2)
          - 96 \* \Hh(4,0)
          - \frac{224}{3} \,\* \Hh(4,1)
          - \frac{725}{3} \,\* \Hhh(0,0,0)
          + 96 \* \Hhh(0,0,0) \* \zeta_2
          - \frac{2420}{27} \,\* \Hhh(1,0,0)
\\
&& \nonumber
          + \frac{160}{3} \,\* \Hhh(1,0,0) \* \zeta_2
          - \frac{128}{3} \,\* \Hhh(1,1,0) \* \zeta_2
          + \frac{128}{3} \,\* \Hhh(1,1,3)
          - \frac{160}{3} \,\* \Hhh(1,2,0)
          - \frac{160}{3} \,\* \Hhh(1,3,0)
          - \frac{64}{3} \,\* \Hhh(1,3,1)
\\
&& \nonumber
          - \frac{724}{9} \,\* \Hhh(2,0,0)
          - \frac{160}{3} \,\* \Hhh(2,1,0)
          - 32 \* \Hhh(2,2,0)
          - \frac{176}{3} \,\* \Hhh(3,0,0)
          - \frac{128}{3} \,\* \Hhh(3,1,0)
          - \frac{368}{3} \,\* \Hhhh(0,0,0,0)
\\
&& \nonumber
          - \frac{724}{9} \,\* \Hhhh(1,0,0,0)
          - \frac{32}{3} \,\* \Hhhh(1,2,0,0)
          - 16 \* \Hhhh(2,0,0,0)
          - \frac{32}{3} \,\* \Hhhh(2,1,0,0)
          - \frac{160}{3} \,\* \Hhhhh(0,0,0,0,0)
          - \frac{128}{3} \,\* \Hhhhh(1,0,0,0,0)
\\
&& \nonumber
          + \frac{160}{3} \,\* \Hhhhh(1,1,0,0,0)
          - \frac{85175}{324} \* \H(0)
          - \frac{376}{3} \,\* \H(0) \* \zeta_4
          - 44 \* \H(0) \* \zeta_3
          + \frac{1916}{9} \,\* \H(0) \* \zeta_2
          - \frac{208}{3} \,\* \H(1) \* \zeta_4
          - \frac{208}{9} \,\* \H(1) \* \zeta_3
\:\:\: 
\\
&& \nonumber
          - \frac{13346}{81} \,\* \H(2)
          - \frac{224}{3} \,\* \H(2) \* \zeta_3
          + \frac{160}{3} \,\* \H(2) \* \zeta_2
          - \frac{1916}{9} \,\* \H(3)
          + \frac{128}{3} \,\* \H(3) \* \zeta_2
          - \frac{1372}{9} \,\* \H(4)
          - 96 \* \H(5)
          \biggr)
\\
&& \nonumber
       + x  \*  \biggl(
          - \frac{496}{3} \,\* \zeta_4
          + \frac{454}{3} \,\* \zeta_3
          - \frac{9752}{27} \,\* \zeta_2
          + \frac{44128}{81} \,\* \Hh(0,0)
          + \frac{28}{3} \,\* \Hh(0,0) \* \zeta_2
          + \frac{170}{3} \,\* \Hh(2,0)
          + \frac{320}{3} \,\* \Hh(2,0) \* \zeta_2
\\
&& \nonumber
          - 92 \* \Hh(2,1)
          + \frac{32}{3} \,\* \Hh(2,2)
          - \frac{320}{3} \,\* \Hh(2,3)
          + \frac{224}{3} \,\* \Hh(3,0)
          + \frac{32}{3} \,\* \Hh(3,1)
          + \frac{2339}{9} \,\* \Hhh(0,0,0)
          + \frac{428}{3} \,\* \Hhh(2,0,0)
\\
&& \nonumber
          + \frac{32}{3} \,\* \Hhh(2,1,0)
          + \frac{376}{3} \,\* \Hhhh(0,0,0,0)
          + \frac{304}{3} \,\* \Hhhh(2,1,0,0)
          + \frac{115273}{108} \,\* \H(0)
          - \frac{236}{3} \,\* \H(0) \* \zeta_3
          - \frac{74}{3} \,\* \H(0) \* \zeta_2
          + \frac{9752}{27} \,\* \H(2)
\\
&& \nonumber
          - \frac{32}{3} \,\* \H(2) \* \zeta_2
          + \frac{74}{3} \,\* \H(3)
          - \frac{28}{3} \,\* \H(4)
          \biggr)
       + \xp1  \*  \biggl(
          - 124 \* \zeta_6
          + \frac{88}{3} \,\* \zeta_5
          - 32 \* \zts
          - \frac{16}{3} \,\* \zeta_2 \* \zeta_3
          + \frac{64}{3} \,\* \Hh(0,0) \* \zeta_3
\\
&& \nonumber
          + 32 \* \Hh(2,0) \* \zeta_3
          + 64 \* \Hh(2,1) \* \zeta_3
          + 32 \* \Hh(2,4)
          - 32 \* \Hh(3,0) \* \zeta_2
          + 32 \* \Hh(3,3)
          + \frac{88}{3} \,\* \Hh(4,0)
          - 48 \* \Hhh(0,0,0) \* \zeta_2
\\
&& \nonumber
          - 32 \* \Hhh(2,0,0) \* \zeta_2
          - 64 \* \Hhh(2,1,0) \* \zeta_2
          + 64 \* \Hhh(2,1,3)
          + 8 \* \Hhh(3,0,0)
          - \frac{32}{3} \,\* \Hhhh(2,0,0,0)
          - 32 \* \Hhhh(2,2,0,0)
          - 32 \* \Hhhh(3,1,0,0)
\\
&& \nonumber
          + 40 \* \Hhhhh(0,0,0,0,0)
          - 64 \* \Hhhhh(2,1,1,0,0)
          + 16 \* \H(0) \* \zeta_4
          + 64 \* \H(2) \* \zeta_4
          + 32 \* \H(3) \* \zeta_3
          + 48 \* \H(5)
          \biggr)
       + \xm1  \*  \biggl(
          - \frac{136319}{81}
\\
&& \nonumber
          - \frac{12148}{27} \,\* \Hh(1,0)
          + 96 \* \Hh(1,0) \* \zeta_3
          - \frac{320}{3} \,\* \Hh(1,0) \* \zeta_2
          - \frac{872}{3} \,\* \Hh(1,1)
          + 192 \* \Hh(1,1) \* \zeta_3
          - \frac{64}{3} \,\* \Hh(1,2)
          + \frac{320}{3} \,\* \Hh(1,3)
\\
&& \nonumber
          + 96 \* \Hh(1,4)
          - \frac{1036}{9} \,\* \Hhh(1,0,0)
          - 96 \* \Hhh(1,0,0) \* \zeta_2
          - \frac{64}{3} \,\* \Hhh(1,1,0)
          - 192 \* \Hhh(1,1,0) \* \zeta_2
          + 192 \* \Hhh(1,1,3)
          - \frac{64}{3} \,\* \Hhh(1,2,0)
\\
&& \nonumber
          - \frac{16}{3} \,\* \Hhhh(1,0,0,0)
          - 128 \* \Hhhh(1,1,0,0)
          - 96 \* \Hhhh(1,2,0,0)
          - 192 \* \Hhhhh(1,1,1,0,0)
          - \frac{11416}{27} \,\* \H(1)
          + 192 \* \H(1) \* \zeta_4
\\
&& \nonumber
          + \frac{320}{3} \,\* \H(1) \* \zeta_3
          + \frac{64}{3} \,\* \H(1) \* \zeta_2
          + 96 \* \H(2) \* \zeta_3
          \biggr)
          - \frac{224}{3} \,\* \zeta_4
          - 38 \* \zeta_3
          + \frac{1616}{3} \,\* \zeta_2
          - \frac{38036}{81} \,\* \Hh(0,0)
          + 20 \* \Hh(0,0) \* \zeta_2
  \\
&& \nonumber
          - \frac{1954}{9} \,\* \Hh(2,0)
          - \frac{256}{3} \,\* \Hh(2,0) \* \zeta_2
          - \frac{404}{3} \,\* \Hh(2,1)
          - 32 \* \Hh(2,2)
          + \frac{256}{3} \,\* \Hh(2,3)
          - \frac{128}{3} \,\* \Hh(3,0)
          - \frac{160}{3} \,\* \Hh(3,1)
\\
&& \nonumber
          - \frac{473}{9} \,\* \Hhh(0,0,0)
          - \frac{116}{3} \,\* \Hhh(2,0,0)
          - 32 \* \Hhh(2,1,0)
          + \frac{152}{3} \,\* \Hhhh(0,0,0,0)
          - \frac{272}{3} \,\* \Hhhh(2,1,0,0)
          - \frac{340325}{324} \,\* \H(0)
\\
&& \nonumber
          - \frac{44}{3} \,\* \H(0) \* \zeta_3
          + \frac{2090}{9} \,\* \H(0) \* \zeta_2
          - \frac{1616}{3} \,\* \H(2)
          + 32 \* \H(2) \* \zeta_2
          - \frac{2090}{9} \* \H(3)
          - 20 \* \H(4)
       + \delta \xm1  \*  \biggl(
            \frac{353}{3}
\\
&&
          - \frac{85175}{162} \,\* \zeta_2
          - \frac{137}{9} \,\* \zeta_3
          + \frac{16186}{27} \,\* \zeta_4
          - \frac{584}{9} \,\* \zeta_2 \* \zeta_3
          - \frac{248}{3} \,\* \zeta_5
          - \frac{16}{3} \,\* \zts
          - 144 \* \zeta_6
          \biggr)
\:\: .
\eea
The $\nfs$ and $\nft$ contributions to eq.~(\ref{eq:P3Lnc}) are given by 
eqs.~(4.6) and (4.12) of ref.~\cite{Davies:2016jie}.
 
\vspace*{-2mm}
Disregarding terms that vanish for $x \!\ra\! 1$, the large-$x$ behaviour 
of $P_{\rm ns,\,L}^{\,(3)}(x)$ can be written as
\beq
\label{eq:xto1Lnc}
  P_{\,\rm ns,\,L}^{\,(n-1)}(x) \:\: = \:\;
        \frac{A_{L,n}}{(1-x)_+}
  \,+\, B_{L,n} \, \delta \xm1
  \,+\, C_{L,n} \, \ln(1-x)
  \,-\, A_{L,n} + \widetilde{D}_{L,n}
\eeq
in terms of the coefficients specified in eqs.~(\ref{eq:A4Lnc}), 
(\ref{eq:B4Lnc}) and (\ref{eq:CD4ofAB}) above. The numerical values of the
coefficients of the small-$x$ logarithms, $\ln^{\,k\!}x$ with $k=1,\ldots, 6$, 
can be read off from eq.~(\ref{eq:Pns3L12}) below. 
All six logarithms and the constant contribution for $x\to 0$ are required for 
a good approximation to the splitting functions at small $x$-values relevant
for collider physics.

In view of the length and complexity of the exact expressions
(\ref{eq:P3LncNF0}) and (\ref{eq:P3LncNF1}) it is useful to have at one's 
disposal also compact approximate representations involving, besides powers 
of $x$, only simple functions like the plus-distribution and the end-point 
logarithms
\bea
\label{eq:logs}
  {\cal D}_{\,0} \: = \: 1/(1\!-\!x)_+ \: ,
  \quad x_1 \: = \: 1\!-\!x \: ,
  \quad L_1 \: = \: \ln (1\!-\!x) \: ,
  \quad L_0 \: = \: \ln x \:\: .
\eea
Such approximations can be readily used in $N$-space evolution programmes, 
see, e.g., ref.~\cite{Vogt:2004ns}.
The results (\ref{eq:P3LncNF0}) and (\ref{eq:P3LncNF1}) can be parametrized
with a high accuracy (of 0.1\% or better) as
\bea
\label{eq:Pns3L12}
{\lefteqn{ \hspn\hspn
       \cf \nct\, P_{{\rm L},0}^{\,(3)}(x) 
 \,+\, \cf \ncs \nf\,P_{{\rm L},1}^{\,(3)}(x) 
 \:\:=\:\: 
}}
\nn
\\[2mm] && 
 21209.02 \* {\cal D}_{\,0} 
 + (25796.09 - 1.0) \* \delta(x_1) + 19069.80 \,\* L_1 - 29733.85
\nonumber \\[1mm] && \mbox{}
 \,+\, 25000 \,\* ( x_1 \*  ( 3.5242 
 + 8.3679 \,\*  x - 1.2395 \,\*  x^2 + 1.7423 \,\*  x^3 ) 
 + 11.916 \,\*  x \* L_0 
\nonumber \\[1mm] && \mbox{\hspp}
 - 0.2237 \,\*  x \* L_0^2 - 0.0129 \,\*  x \* L_0^3 
 + 11.937 \,\*  x_1 \* L_1 + 13.955  \,\*  L_0 \* L_1 ) 
\nonumber \\[1mm] && \mbox{}
 + 51671.33 \,\* L_0 
 + 17120.95 \,\* L_0^2 + 2863.226 \,\* L_0^3 + 297.8255 \,\* L_0^4 
 + 16 \,\* L_0^5 + 1/2\,\* L_0^6 
\nonumber \\[1mm] && \mbox{\hspn}
 + \nf \,\* \biggl(
 - 5179.372 \,\* {\cal D}_{\,0} 
 - ( 5818.637 + 0.35) \* \delta(x_1) 
 - 3079.761 \,\* L_1 + 8115.605
\nonumber \\ && \mbox{}
 + 2500 \,\*  ( x_1 \*  ( - 7.4077 
 + 4.5141 \,\*  x - 1.0069 \*  x^2
 + 0.7641 \,\*  x^3 ) + 8.4211 \,\*  x_1 \* L_1 
\nonumber \\ && \mbox{\hspp}
 + 7.5633 \,\*  L_0 \* L_1
 + 7.5236 \,\*  x \* L_0 + 0.2208 \,\*  x \* L_0^2 
 + 0.05712 \,\*  x \* L_0^3 )
\nonumber \\ && \mbox{}
 - 9239.374 \,\* L_0 - 2917.312 \* L_0^2 
 - 430.5308 \,\* L_0^3 - 36 \* L_0^4 - 4/3 \* L_0^5
 \biggr) 
\:\: .
\eea
Here the exact large-$x$ and small-$x$ coefficients have been rounded to 
seven significant figures.
The brackets multiplied by 25000 and 2500 have been obtained by fits to the 
exact expressions at $10^{\,-6} \leq x \leq 1 - 10^{\,-6}$.
The small shifts of the $\delta(1-x)$ fine-tune the accuracy of the resulting 
low moments and of the convolutions with the quark distributions.
The required evaluation of the HPLs has been performed using a weight-6
extension of ref.~\cite{Gehrmann:2001pz} and the program of 
ref.~\cite{ABRS-HPL}, which return identical results at the accuracy 
considered here.

\bigskip

\begin{figure}[p]
\vspace{-4mm}
\centerline{\epsfig{file=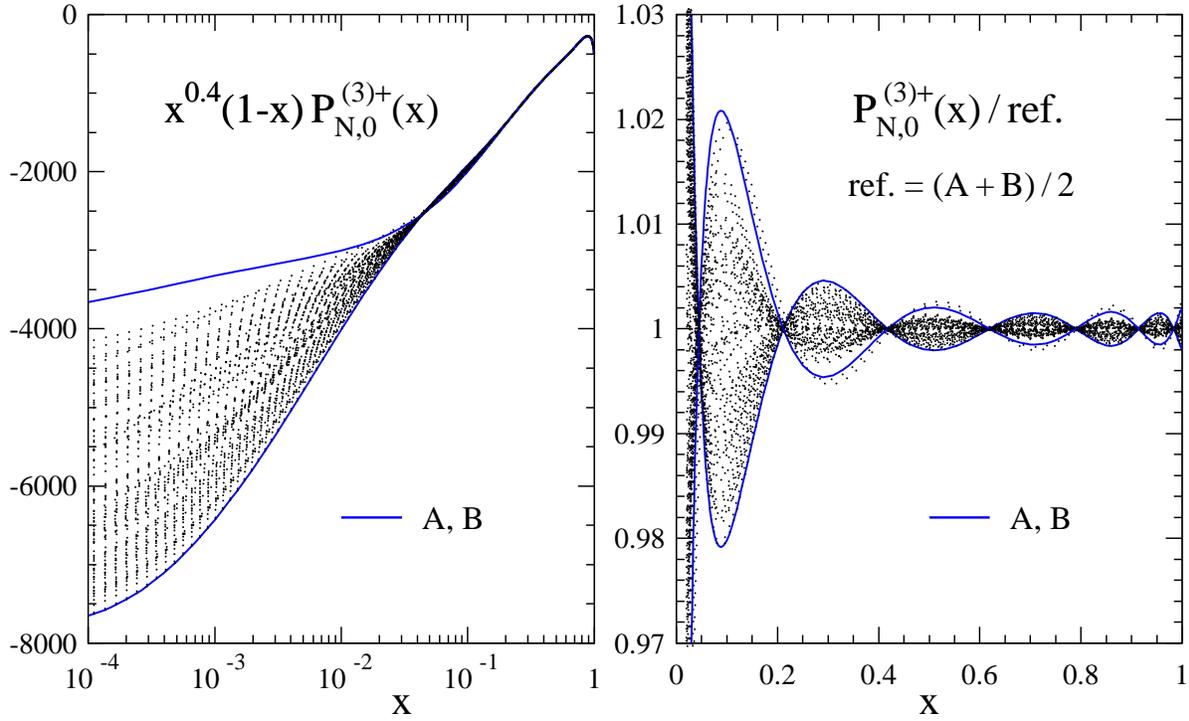,width=16.0cm,angle=0}}
\vspace{-2mm}
\caption{ \label{fig:P3Nnc0a} \small
 About 90 trial functions for the $\nf$-independent contribution to the non-%
 leading ($N$) large-$\nc$ part of splitting function $P_{\rm ns}^{\,(3)+}(x)$,
 multiplied by $x^{\,0.4}\xm1_{}$ for display purposes.
 The two functions chosen to represent the remaining uncertainty are denoted
 by $A$ and $B$ and shown by solid (blue) lines.
 }
\vspace{1mm}
\end{figure}

\begin{figure}[p]
\vspace{-2mm}
\centerline{\epsfig{file=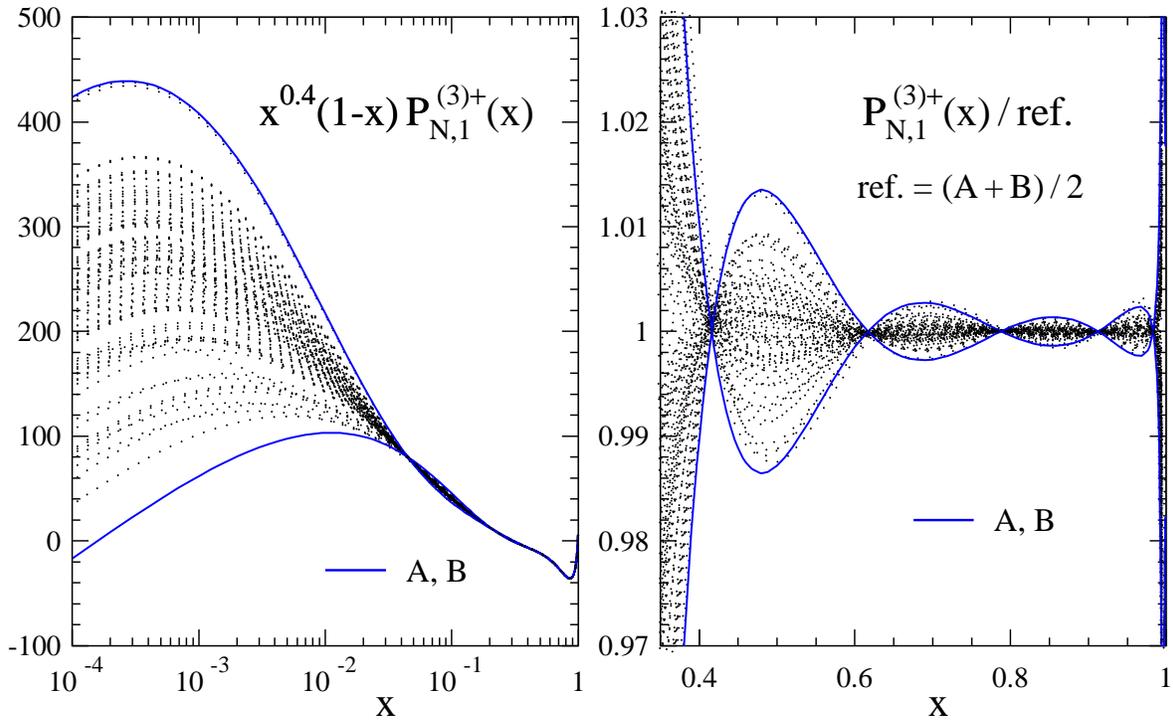,width=16.0cm,angle=0}}
\vspace{-2mm}
\caption{ \label{fig:P3Nnc1a} \small
 As fig.~\ref{fig:P3Nnc0a}, but for the $\nfo$ contribution.  The ratio in the 
 right panel is shown for a smaller $x$-range than in fig.~\ref{fig:P3Nnc0a} 
 due to a sign change of the function at $x \,\simeq\, 0.3$. 
 The large relative width of the uncertainty band close to $x=1$ is due to 
 another change of sign at $\xm1 \,\simeq\, 0.005$.
 }
\vspace{-1mm}
\end{figure}

For the corresponding non-leading contributions in the large-$\nc$ limit, 
denoted by the subscript $N$ in eqs.~(\ref{gam+to16}) and (\ref{gam-to15}) 
above, we are for now limited to approximations analogous to (but more accurate
than) those once constructed at three loops
\cite{vanNeerven:1999ca,vanNeerven:2000uj,vanNeerven:2000wp}.
For the $\nfz$ and $\nfo$ parts of $P_{\rm ns}^{\,(3)+}(x)$ we employ an 
ansatz consisting of 
\begin{itemize}
\item the two large-$x$ parameters $A_4$ and $B_4$, 
      cf.~eqs.~(\ref{eq:ntoinf}) and (\ref{eq:xto1Lnc}),
\item two of three suppressed large-$x$ logarithms 
      $\xm1\, \ln^{\,k\!}\xm1$, $k = 1,2,3$, 
\item one of ten two-parameter polynomials in $x$ that vanish for 
      $x \to 1$, 
\item two of the three unknown small-$x$ logarithms $\ln^{\,k\!}x$, 
      $k = 1,2,3$. 
\end{itemize}
The parameters of the 90 resulting trial functions are determined from the
eight available moments, and then two representatives as chosen that indicate
the remaining uncertainty. 
The result of this process is illustrated in figs.~\ref{fig:P3Nnc0a} and 
\ref{fig:P3Nnc1a}.


Supplementing the approximations $A$ and $B$ in figs.~\ref{fig:P3Nnc0a} and
\ref{fig:P3Nnc1a} by accurate parametrizations of the complete $\nfs$ results 
of ref.~\cite{Davies:2016jie} and the exact (but numerically truncated) $\nft$ 
expressions in a non-HPL notation, we obtain
\bea
\label{eq:Pqq3+A}
{\lefteqn{ \hspn\hspn
 P_{\rm ns,\,A}^{\,(3)+}(x) \; = \; 
         \cf \nct\, P_{{\rm L},0}^{\,(3)}(x) 
   \,+\, \cf \ncs \nf\,P_{{\rm L},1}^{\,(3)}(x)
   \,+\, P_{\,L\nf}^{\,(3)+}(x)
}}
\nn \\ && \mbox{}
%
 - 507.152 \,\* ( {\cal D}_{\,0} - 1) 
 - 2405.03 \,\* \delta(x_1) 
 - 1777.27 \,\*  L_1^2 \* x_1 - 204.183  \,\*  L_1^3 \* x_1 
\nonumber \\ && \mbox{}
 + 3948.16 \,\*  x_1 
 - 2464.61 \*  (2 \* x- x^2) \,\* x_1
 - 1839.44 \,\*  L_0^2 - 402.156 \,\*  L_0^3
\nonumber \\[1mm] && \mbox{}
 - 55.87553 \,\* L_0^4 - 2.831276 \,\* L_0^5 
 - 0.1488340 \,\* L_0^6 
 - 2601.749 - 2118.867 \,\* L_1
\nonumber \\[1mm] && \mbox{\hspn}
 + \nf \,\*  \biggl(
   7.33927 \,\* ( {\cal D}_{\,0} - 1)
 + 267.965 \,\* \delta(x_1) 
 - 143.813 \,\*  L_1 \* x_1 - 18.8803 \,\*  L_1^3 \* x_1 
\nonumber \\ && \mbox{}
 - (1116.34 - 1071.24 \,\* x) \,\*  x \,\* x_1
 - 59.3041 \,\*  L_0^2 - 8.4620 \,\* \,L_0^3
\nonumber \\ && \mbox{}
 + 4.658436 \,\* L_0^4 + 0.2798354 \,\* L_0^5
 + 312.1643 + 337.9310 \,\* L_1
 \biggr)
\eea
and 
\bea
\label{eq:Pqq3+B}
{\lefteqn{ \hspn\hspn
 P_{\rm ns,\,B}^{\,(3)+}(x) \; = \;
        \cf \nct\, P_{{\rm L},0}^{\,(3)}(x) 
  \,+\, \cf \ncs \nf\,P_{{\rm L},1}^{\,(3)}(x)
  \,+\, P_{\,L\nf}^{\,(3)+}(x)
}}
\nn \\ && \mbox{}
%
 - 505.209 \,\* ({\cal D}_{\,0} -1)
 - 2394.47 \,\* \delta(x_1) 
 - 173.936 \,\*  L_1^2 \* x_1 + 223.078 \,\* L_1^3 \* x_1 
\nonumber \\ && \mbox{}
 + (8698.39 - 10490.47 \* x) \,\*  x \,\* x_1
 + 1389.73 \,\*  L_0 + 189.576 \*  L_0^2
\nonumber \\[1mm] && \mbox{}
 - 55.87553 \,\* L_0^4 - 2.831276 \,\* L_0^5
 - 0.1488340 \,\* L_0^6 
 - 2601.749 - 2118.867 \,\* L_1
\nonumber \\[1mm] && \mbox{\hspn}
 + \nf \* \biggl(
   7.53662 \,\* ({\cal D}_{\,0} - 1)
 + 269.028 \,\* \delta(x_1) 
 - 745.573 \,\*  L_1 \* x_1 + 8.61438 \,\* L_1^3 \* x_1 
\nonumber \\ && \mbox{}
 - (690.151 + 656.386 \,\* x^2) \,\* x_1
 + 133.702 \,\*  L_0^2 + 34.0569 \,\*  L_0^3
\nonumber \\ && \mbox{}
 + 4.658437 \,\* L_0^4 + 0.2798354 \,\* L_0^5
 + 312.1643 + 337.9310 \,\* L_1
 \biggr)
\eea
with
\bea
\label{eq:PLnf+}
{\lefteqn{ \hspn\hspn
 P_{\,L\nf}^{\,(3)+}(x) \; = \;
}}
\nn \\ && \mbox{} \!\!\!
%
 \nfs \,\* \biggl(
   195.5772 \,\* {\cal D}_{\,0} + 26.68861 \,\* L_1 - 376.0092
 + ( 193.8554 + 0.0037) \* \delta(x_1)
\nonumber \\ && \mbox{}
 + 250 \,\*   ( x_1 \*  ( 3.0008 
 + 0.8619 \,\*  x - 0.12411 \*  x^2 
 + 0.31595 \,\*  x^3 ) - 0.37529 \,\*  x \* L_0 
\nonumber \\[1mm] && \mbox{}
 - 0.21684 \,\*  x \* L_0^2 - 0.02295 \,\*  x \* L_0^3 
 + ( 0.03394 \,\*  x_1 + 0.40431 \,\*  L_0 ) \,\* L_1 )
\nonumber \\ && \mbox{}
 + 393.0056 \,\* L_0 + 112.5705 \,\* L_0^2 
 + 16.52675 \,\* L_0^3 + 0.7901235 \,\* L_0^4
 \biggr)
\nonumber \\ && \mbox{\hspn}
 + \nft \,\* \biggl(
   3.272344 \,\* {\cal D}_{\,0} + 3.014982 \,\* \delta(x_1) 
 - 2.426296 - 0.8460488 \,\*  x 
\nonumber \\ && \mbox{}
 + ( 0.5267490 \,\*  x_1^{-1} 
 - 3.687243 + 3.160494 \,\*  x ) \,\*  L_0
 - ( 0.1316872 \,\*  ( 10\,\*x_1^{\,-1}+1) 
\nonumber \\ &&
 - 1.448560 \,\* x ) \,\* L_0^2
 - ( 0.2633745 \,\* x_1^{\,-1} 
   - 0.1316870 \,\*  (1+x) ) \,\*  L_0^3
          \biggr)
\:\: .
\eea
The case of $P_{\rm ns}^{\,(3)-}(x)$ can be treated in the same manner,
but taking into account that only its leading small-$x$ logarithm is
known up to now \cite{Blumlein:1995jp}. After careful consideration, the 
two approximations indicating the uncertainty band in this case are chosen as

\pagebreak
\vspace*{-1.2cm}
\bea
\label{eq:Pqq3-A}
{\lefteqn{ \hspn\hspn
 P_{\rm ns,\,A}^{\,(3)-}(x) \; = \;
         \cf \nct\, P_{{\rm L},0}^{\,(3)}(x) 
   \,+\, \cf \ncs \nf\,P_{{\rm L},1}^{\,(3)}(x)
   \,+\, P_{\,L\nf}^{\,(3)-}(x)
}}
\nn \\ && \mbox{}
%
 - 511.228 \,\* ({\cal D}_{\,0} -1 )
 - 2426.05 \,\* \delta(x_1)
 + 31897.82 \,\* L_1 \* x_1 + 4653.76 \,\* L_1^2 \* x_1
\nonumber \\[1mm] && \mbox{}
 + (5992.88 \,\* (1+2 \* x) + 31321.44 \,\* x^2) \* x_1 
 - 1618.07 \*  L_0 + 2.25480 \,\* L_0^3 
\nonumber \\[1mm] && \mbox{}
 + 0.4964335 \*  (L_0^6 + 6 \,\* L_0^5) - 2601.749 - 2118.867 \* L_1
\nonumber \\[1mm] && \mbox{\hspn}
 + \nf \,\*  \biggl(
   7.08645 \,\* ({\cal D}_{\,0}  - 1)
 + 266.669 \,\* \delta(x_1) 
 + 1856.63 \,\* L_1 \* x_1 + 440.17 \,\*  L_1^2 \* x_1 
\nonumber \\ && \mbox{}
 + (114.457 \,\*  (1+2 \* x) + 2570.73 \,\* x^2) \,\* x_1 
 - 127.012 \,\*  L_0^2 + 2.69618 \,\*  L_0^4 
\nonumber \\ && \mbox{}
 + 312.1643 + 337.9310 \,\* L_1
 \biggr)
\eea
and
\bea
\label{eq:Pqq3-B}
{\lefteqn{ \hspn\hspn
 P_{\rm ns,\,B}^{\,(3)-}(x) \; = \;
         \cf \nct\, P_{{\rm L},0}^{\,(3)}(x) 
   \,+\, \cf \ncs \nf\,P_{{\rm L},1}^{\,(3)}(x)
   \,+\, P_{\,L\nf}^{\,(3)-}(x)
}}
\nn \\ && \mbox{}
%
 - 502.481 \,\* ({\cal D}_{\,0} - 1)
 - 2380.255 \,\* \delta(x_1)
 - 3997.39 \,\*  L_1 \* x_1 + 511.567 \,\*  L_1^3 \* x_1
\nonumber \\ && \mbox{}
 + ( 4043.59 - 15386.6 \,\*  x) \*  \,x \,\* x_1 
 + 1532.96 \,\*  L_0^2 + 31.6023 \,\*  L_0^3 
\nonumber \\ && \mbox{}
 + 0.4964335 \,\*  (L_0^6 + 18 \,\* L_0^5)
 - 2601.749 - 2118.867 \,\* L_1
\nonumber \\ && \mbox{\hspn}
 + \nf \,\* \biggl(
   7.82077\* ({\cal D}_{\,0} - 1)
 + 270.468 \,\* \delta(x_1) 
 - 1360.04 \,\*  L_1 \* x_1 + 38.7337 \,\*  L_1^3 \* x_1 
\nonumber \\ && \mbox{}
 - (335.995 \,\*  (2+x) + 1605.91 \,\* x^2) \,\* x_1 
 - 9.76627 \,\* L_0^2 + 0.14218 \,\*  L_0^5 
\nonumber \\ && \mbox{}
 + 312.1643 + 337.9310 \* L_1
 \biggr)
\eea
with
\bea
\label{eq:PLnf-}
{\lefteqn{ \hspn\hspn
 P_{\,L\nf}^{\,(3)-}(x) \; = \;
}}
\nn \\ && \mbox{} \!\!\!
%
 \nfs \,\* \biggl(
   195.5772 \,\* {\cal D}_{\,0} + 26.68861 \,\* L_1 - 376.0092
 + ( 193.8554 + 0.0037) \* \delta(x_1)
\nonumber \\ && \mbox{}
 + 250 \,\*   ( x_1 \*  ( 3.2206 
 + 1.7507 \,\*  x + 0.13281 \,\* x^2
 + 0.45969 \,\*  x^3 ) + 1.5641 \,\*  x \* L_0 
\nonumber \\[1mm] && \mbox{}
 - 0.37902 \,\*  x \* L_0^2 - 0.03248 \,\*  x \* L_0^3 
 + ( 2.7511 \,\*  x_1 + 3.2709 \,\* L_0 ) \,\* L_1 )
\nonumber \\ && \mbox{}
 + 437.8810 \,\* L_0 + 128.2948 \,\* L_0^2 
 + 19.59945 \,\* L_0^3 + 0.9876543 \,\* L_0^4
 \biggr)
\nonumber \\ && \mbox{\hspn}
 + \nft \,\* \biggl(
   3.272344 \,\* {\cal D}_{\,0} + 3.014982 \,\* \delta(x_1) 
 - 2.426296 - 0.8460488 \,\*  x 
\nonumber \\ && \mbox{}
 + ( 0.5267490 \,\*  x_1^{-1} 
 - 3.687243 + 3.160494 \,\*  x ) \,\*  L_0
 - ( 0.1316872 \,\*  ( 10\,\*x_1^{\,-1}+1) 
\nonumber \\ &&
 - 1.448560 \,\* x ) \,\* L_0^2
 - ( 0.2633745 \,\* x_1^{\,-1} 
   - 0.1316870 \,\*  (1+x) ) \,\*  L_0^3
          \biggr)
\:\: .
\eea
The $\nft$ contribution to this last equation is the same as in 
eq.~(\ref{eq:PLnf+}).

\begin{figure}[p]
\vspace{-3mm}
\centerline{\epsfig{file=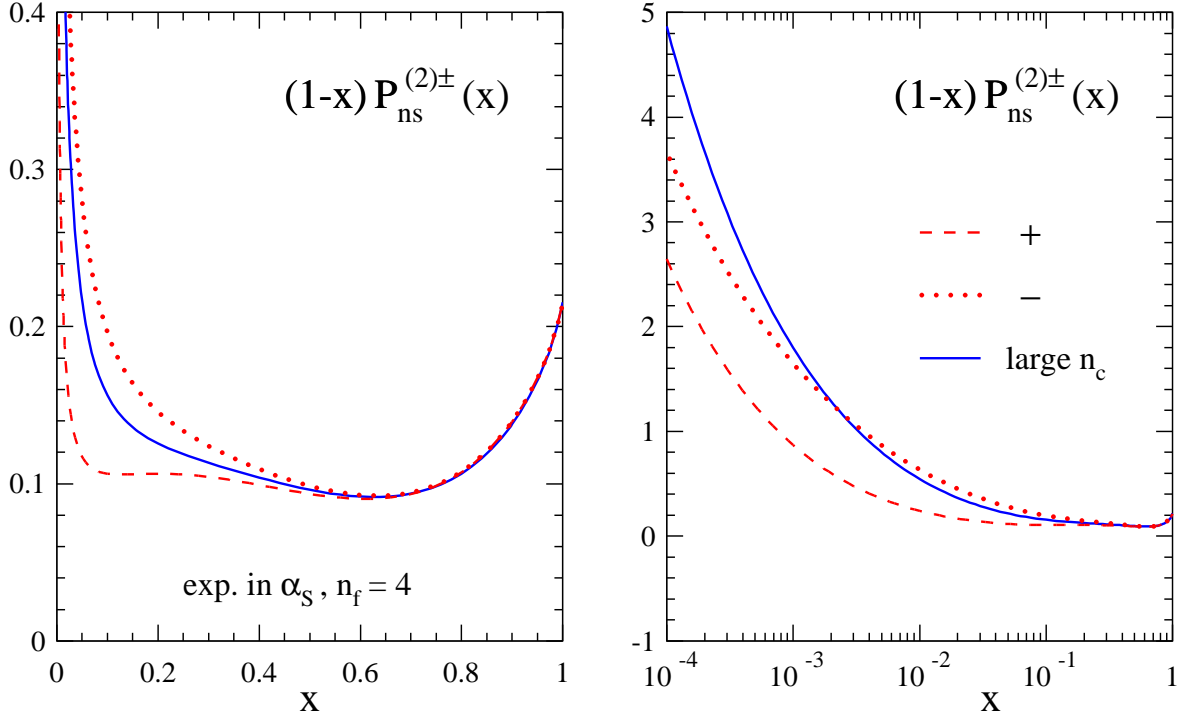,width=16.0cm,angle=0}}
\vspace{-1mm}
\caption{ \label{fig:P2nf4pm} \small
 The three-loop splitting functions $P_{\rm ns}^{\,(2)\pm}(x)$ and their
 large-$\nc$ limit for QCD with four flavours.
 }
\vspace{-1mm}
\end{figure}

\begin{figure}[p]
\vspace{-2mm}
\centerline{\epsfig{file=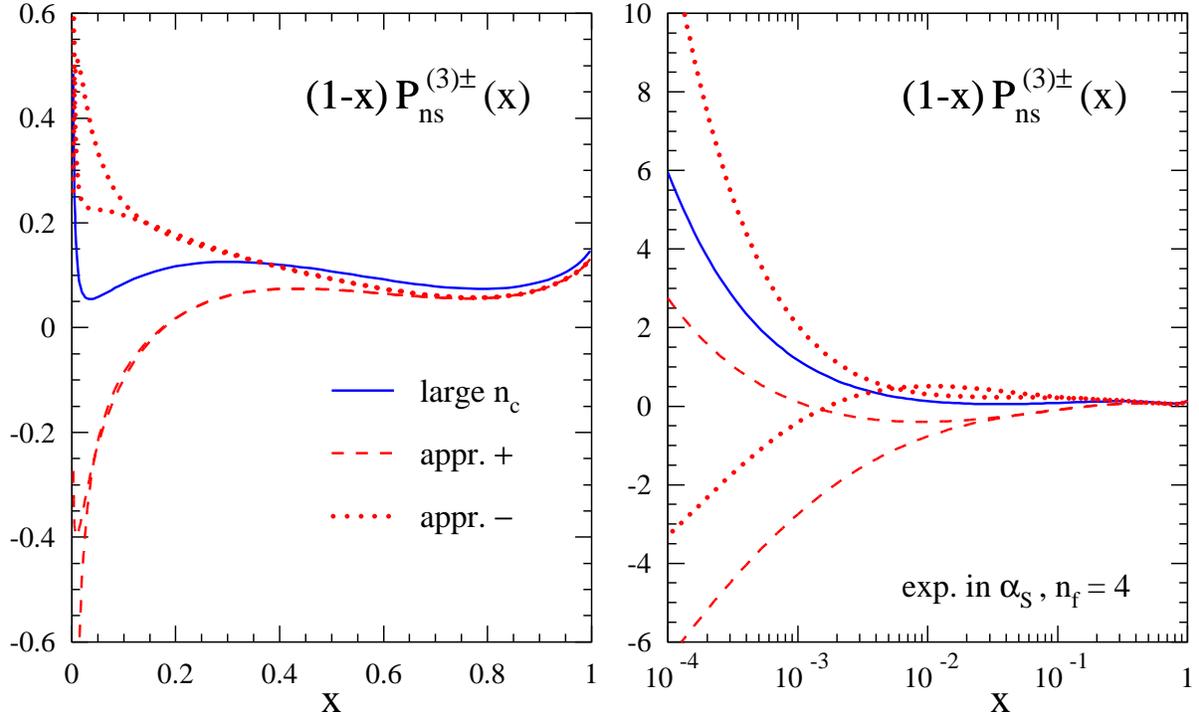,width=16.0cm,angle=0}}
\vspace{-1mm}
\caption{ \label{fig:P3nf4pm} \small
 The four-flavour uncertainty bands for the four-loop splitting functions
 $P_{\rm ns}^{\,(3)a}(x)$ generated by eqs.~(\ref{eq:Pqq3+A}) and 
 (\ref{eq:Pqq3+B}) for $a=+$ and by eqs.~(\ref{eq:Pqq3-A}) and 
 (\ref{eq:Pqq3-B}) for $a=-$, compared to their exact large-$\nc$ limit. 
 As in fig.~\ref{fig:P2nf4pm}, the curves are scaled to an expansion in 
 $\als$, and only the $x<1$ contributions are shown.
 }
\vspace{-1mm}
\end{figure}

Before we illustrate these results, it is useful to briefly recall the
behaviour of the corresponding third-order splitting functions. 
This is done in fig.~\ref{fig:P2nf4pm} for $\nf = 4$ flavours. 
The corresponding size and uncertainty bands of $P_{\rm ns}^{\,(3)+}(x)$
and $P_{\rm ns}^{\,(3)-}(x)$ are shown in fig.~\ref{fig:P3nf4pm}
together with their large-$\nc$ limit. The qualitative pattern and the
rough size of the corrections as coefficients of $\as(3)$ and $\as(4)$,
respectively, are comparable in the region of $x$ for which definite 
conclusions can be drawn. 


\pagebreak

The four-loop `sea' contribution $P_{\rm ns}^{\,(3)\rm s}(x)$ to the evolution 
of the total valence distribution is suppressed by two powers of $\xm1$ for 
$x \ra 1$, but its $\nfo$ part is completely unknown in the \mbox{small-$x$} 
limit.
In~this case, we use the nine odd moments (\ref{gamsto17}) with a suitably 
modified ansatz, in which the coefficient of $\ln^{\,6}x$ is varied `by hand' 
over a sufficiently wide range, and the coefficients of $\ln^{\,k}x$, 
$\,k=1,\ldots,5\,$, are all determined from the moments. 
In this manner we obtain
\beq
  P_{\rm ns, A/B}^{\,(3)\rm s}(x) \;=\; 
  \nf\, P_{1,\,\rm A/B}^{\,(3)\rm s}(x) + \nfs \, P_{2}^{\,(3)\rm s}(x)
\eeq
with
\bea
\label{eq:Pqq3s-A}
   P_{1,\,\rm A}^{\,(3)\rm s}(x) & \!=\! &
%
       60.40 \,\* x_1 \* L_1^2 + 4.685 \,\* x_1 \* L_1^3
       + x_1 \,\* x \*  ( 4989.2 - 1607.73 \,\*  x)
\nonumber \\ && \mbox{\hspn}
       + 3687.6 \,\*  L_0 + 3296.6 \,\*  L_0^2 + 1271.11 \,\*  L_0^3
       + 533.44  \,\*  L_0^4 + 97.27 \,\*   L_0^5 + 4 \,\*  L_0^6
\:\: , \quad \\[2mm]
\label{eq:Pqq3s-B}
   P_{1,\,\rm B}^{\,(3)\rm s}(x) & \!=\! &
%
       - 254.63 \,\*  x_1 \* L_1 - 0.28953 \,\* x_1 \* L_1^3
       + 1030.79 \,\*  x_1 \* x + 1266.77 \,\* x_1 \* (2-x^2) 
\nonumber \\ && \mbox{\hspn}
       + 2987.83  \,\*  L_0 + 273.05 \,\*  L_0^2 - 923.48 \,\*  L_0^3 
       - 236.76 \,\*  L_0^4 - 33.886 \,\*   L_0^5 - 4 \,\*  L_0^6 
\eea
and
\bea
\label{eq:Pqq3s2}
   P_{2}^{\,(3)\rm s}(x) & \!=\! &
%
     19.70002 \,\* x_1 \* L_1 - 3.435474 \,\* x_1 \* L_1^2  
   + 250 \,\* ( x_1 \*  ( -4.7656 + 1.6908 \,\*  x + 0.1703 \,\* x^2 ) 
\nonumber \\[1mm] && \mbox{\hspn}
   - 0.41652 \,\* x \* L_0 + 0.90777 \,\* x \* L_0^2 
   + 0.12478 \,\* x \* L_0^3 
   + 0.17155 \,\*  x_1 \* L_1 + 0.17191  \,\*  L_0 \* L_1 )
\nonumber \\[1mm] && \mbox{\hspn}
     -\, 647.3971 \,\* L_0 - 66.41219 \,\* L_0^2 
     -\, 5.353347 \,\* L_0^3 - 5.925926 \,\* L_0^4 - 0.3950617\, \* L_0^5
\:\: . \qquad
\eea
The last equation is a high-accuracy parametrization, constructed in the
same manner as eq.~(\ref{eq:Pns3L12}) above, of the exact result given in 
eq.~(4.11) of ref.~\cite{Davies:2016jie}.

\bigskip

The trial functions considered for all three cases lead to very similar 
predictions for the respective next moments, i.e., $N=18$ for 
$P_{\rm ns}^{\,(3)+}(x)$, and $N=17\:/\:19$ for $P_{\rm ns}^{\,(3)-/\rm s}(x)$.
The residual uncertainty at these $N$-values is a consequence of the width of 
the bands at large $x$, which in turn (recall figs.~\ref{fig:P3Nnc0a} and 
\ref{fig:P3Nnc1a}) is correlated with the uncertainties at smaller $x$. 
If the spread of the result $A$ and $B$ would underestimate the true remaining
uncertainties, then a comparison with additional analytic results at these
next values of $N$ should reveal a discrepancy.

In order to check this, we have extended the diagram computations of
the $\nfo$ parts of $P_{\rm ns}^{\,(3)-}(x)$ and $P_{\rm ns}^{\,(3)+}(x)$
to $N=17$ and $N=18$, respectively. The comparison of these results with
the Mellin-transformed $\nfo$ contributions to eqs.~(\ref{eq:Pqq3+A}), 
(\ref{eq:Pqq3+B}), (\ref{eq:Pqq3-A}) and (\ref{eq:Pqq3-B}) yields
\bea
  P_{\rm ns}^{\,(3)-}(N\!=\!17) &\! :\!&
  194.7126372_{\:\!B}  \; < \; 194.7126913_{\,\rm exact} \; < \; 194.7127561_A
\:\: ,
\nn \\
  P_{\rm ns}^{\,(3)+}(N\!=\!18) &\! :\!&
  195.8888792_{\:\!B}  \; < \; 195.8888857_{\,\rm exact} \; < \; 195.8888968_A 
\:\: .
\eea
Similar successful checks of our approximation procedure have been carried
out for the $\nfz$ parts of $P_{\rm ns}^{\,(3)\pm}(x)$ and the $\nfo$ part 
of $P_{\rm ns}^{\,(3)\rm s}(x)$ by deriving less accurate approximations 
using one fewer moment and comparing the results to the now unused highest 
calculated moments. 
As far as we can see from this and other checks, our approximation procedure, 
which is of course not mathematically rigorous, does not underestimate the 
remaining uncertainties.

\setcounter{equation}{0}
\section{Numerical implications}
\label{sec:numerics}

We are now ready to address two important applications of our new fourth-order 
results.
First, as already mentioned above, the large-$x\,/\,$large-$N$ limits of the 
splitting functions include coefficients that are relevant beyond the evolution
of parton distributions: the (light-like) four-loop cusp anomalous dimension 
$A_4$ and the $\delta\xm1$ coefficient $B_4$ for quark fields.
We are now able to provide approximate if rather accurate numerical results 
for these coefficients.
The obvious second application is a (further) improvement of the perturbative 
stability of the evolution of the  non-singlet quark distributions over a wide 
range in $x$.

The analytic large-$\nc$ expression for $A_4$ has been presented in 
eq.~(\ref{eq:A4Lnc}) above. Together with the approximate results
in eqs.~(\ref{eq:Pqq3+A}) and (\ref{eq:Pqq3+B}) and the known $\nfs$ 
and $\nft$ contributions, this yields
\beq
\label{eq:A4appr}
   A_4 \;=\;      20702(2)       \,-\, 5171.9(2) \,\nf 
             \,+\,195.5772\,\nfs \,+\, 3.272344 \,\nft
\eeq
in QCD with $\nf$ quark flavours. The numbers in brackets represent the 
uncertainty of the preceeding digit, for which we have increased the spread 
due to eqs.~(\ref{eq:Pqq3+A}) and (\ref{eq:Pqq3+B}) by a factor of~2. 
eq.~(\ref{eq:A4appr}) leads to
\beq
\label{eq:A4nf345}
  A_4 \;=\; 7035(2) \: ,\:\: 3353(2) \: ,\:\: 141(2)
  \quad \mbox{for} \quad \nf \; = \; 3\: ,\; 4\: ,\; 5 \:\: .
\eeq
For comparison: the corresponding [1/1] Pad\'e approximants used so far are
7849, 4313 and 1553 \cite{Moch:2005ba}. 
The agreement of the actual results with these approximants would be (much) 
better without the contributions of the quartic group invariant (see below).
A similar situation has been observed for the four-loop beta function in 
ref.~\cite{vanRitbergen:1997va}. The expansion of the cusp anomalous dimension,
now to the fourth order in $\als$, is given by the very benign series
\bea
\label{eq:Aqnum}
  A_q(\als,\,\nf\!=\!3) &\!=\!& 0.42441\:\als \:
  ( 1 \, + \, 0.72657\, \als \, + \, 0.73405\, \as(2) \,+\, 0.6647(2)\, \as(3)
  \, + \, \ldots )
\:\: , \nn \\
  A_q(\als,\,\nf\!=\!4) &\!=\!& 0.42441\:\als \:
  ( 1 \, + \, 0.63815\, \als \, + \, 0.50998\, \as(2) \ +\, 0.3168(2)\, \as(3)
  \, + \, \ldots )
\:\: ,\nn \\
  A_q(\als,\,\nf\!=\!5) &\!=\!& 0.42441\:\als \:
  ( 1 \, + \, 0.54973\, \als \, + \, 0.28403\, \as(2) \,+\, 0.0133(2)\, \as(3)
  \, + \, \ldots )
\:\: . \quad
\eea

The corresponding results for the four-loop coefficient $B_4$ in 
eqs.~(\ref{eq:ntoinf}) and (\ref{eq:xto1Lnc}) read
\beq
\label{eq:B4appr}
   B_4 \;=\;      23393(10)      \,-\, 5551(1) \,\nf
            \,+\, 193.8554\,\nfs \,+\, 3.014982\,\nft
\eeq
and
\bea
\label{eq:Bqnum}
  B_q(\als,\,\nf\!=\!3) &\!=\!& 0.31831\:\als \:
  ( 1 \, + \, 0.99712\, \als \, + \, 1.24116\, \as(2) \,+\, 1.0791(13)\, \as(3)
  \, + \, \ldots )
\:\: , \nn \\
  B_q(\als,\,\nf\!=\!4) &\!=\!& 0.31831\:\als \:
  ( 1 \, + \, 0.87192\, \als \, + \, 0.97833\, \as(2) \ +\, 0.5649(13)\, \as(3)
  \, + \, \ldots )
\:\: ,\nn \\
  B_q(\als,\,\nf\!=\!5) &\!=\!& 0.31831\:\als \:
  ( 1 \, + \, 0.74672\, \als \, + \, 0.71907\, \as(2) \,+\, 0.1085(13)\, \as(3)
  \, + \, \ldots )
\:\: . \quad
\eea
The dominant errors in eq.~(\ref{eq:A4appr}) and (\ref{eq:B4appr}) are those 
of the $\nf$-independent part; its relative uncertainty is $10^{\,-4}$ for
$A_4$ and about four times larger for $B_4$. Due to constraints by large-$N$ 
moments, the errors of $A_4$ and $B_4$ are fully correlated. The relative
uncertainties are larger for the physically relevant values of $\nf$, yet 
the accuracy in eqs.~(\ref{eq:Aqnum}) and \ref{eq:Bqnum}) should be amply
sufficient for phenomenological applications. 

\pagebreak

It may be interesting, for theoretical purposes, to consider the contributions
of the individual colour factors to $A_4$ and $B_4$. By repeating the 
approximation procedure of the previous sections separately for each colour 
factor, we arrive at the corresponding results collected in table~\ref{tab:AB}. 
Our results show that both quartic group invariants definitely contribute to 
the four-loop cusp anomalous dimension -- an issue that has attracted some 
interest, see, e.g., 
refs.~\cite{Gardi:2009qi,Becher:2009qa,Gardi:2009zv,Ahrens:2012qz,Boels:2017skl} 
-- which means that the so-called Casimir scaling between the quark and gluon 
cusp anomalous dimensions, $\,A_q \,=\, \cf/\ca\, A_g\,$, does not hold beyond 
three loops. A lower value, -113.66 after conversion to our notation, results
from assumptions made in ref.~\cite{Grozin:2015kna} for the coefficient of 
$\nf\,d_F^{\,abcd}d_F^{\,abcd}/N_R$.

\begin{table}[t!]
  \centering
  \renewcommand{\arraystretch}{1.2}
  \begin{tabular}{MMM}
\hline\\[-16pt]
   \mbox{colour factor} & A_4                  &  B_4     \\[1pt]
   \hline\\[-5mm] 
    \cff         &     0                       &  ~~~~197. \,\pm\, ~3. \\[0.5mm]
    \cft\, \ca   &     0                       &    ~-687. \,\pm\, 10. \\[0.5mm]
    \cfs \cas    &     0                       &  ~~~1219. \,\pm\, 12. \\[0.5mm]
    \cf\, \cat   & \phantom{-}  610.3 \,\pm\, 0.3 & ~~~295.6 \,\pm\,2.4\\[1mm]
    d_F^{\,abcd}d_A^{\,abcd}/N_R 
                 &    -507.5 \,\pm\, 6.0       &    ~-996. \,\pm\, 45. \\[0.5mm]
\hline\\[-5mm]
    \nf\, \cft   & -31.00 \pm 0.4     & \phantom{-0} 81.4 \pm 2.2      \\[0.5mm]
  \nf\,\cfs\,\ca & \phantom{-} 38.75 \,\pm\, 0.2  & -455.7 \,\pm\, 1.1 \\[0.5mm]
  \nf\,\cf\,\cas &  -440.65 \,\pm\, 0.2~~      &    -274.4 \,\pm\, 1.1 \\[0.5mm]
    \nf\,d_F^{\,abcd}d_F^{\,abcd}/N_R   
                 &  -123.90  \,\pm\, 0.2~~     &    -143.5 \,\pm\, 1.2 \\[1mm]
\hline\\[-5mm]
    \nfs\,\cfs   &   -21.31439         & -5.775288 \\
  \nfs\,\cf\,\ca & \phantom{-}58.36737 & \phantom{-}51.03056 \\
    \nft\,\cf    & \phantom{-}2.454258 & \phantom{-}2.261237 \\[1mm] 
  \hline
  \end{tabular}
  \caption{\small \label{tab:AB}
  Numerical results for the large-$x$ coefficients $A_4$ and $B_4$ for the 
  seven colour factors contributing to the $\nfz$ and $\nfo$ parts. 
  For completeness also the exactly known $\nfs$ and $\nft$ coefficients are 
  included.
  }
  \vspace*{-2mm}
\end{table}

We now turn to the effect of the four-loop splitting functions 
(\ref{eq:P3Lnc}) -- (\ref{eq:Pqq3s2}) on the evolution 
--~specifically the logarithmic derivatives 
  $\dot{q}_{\rm ns}^{\: i} \equiv d \ln q_{\rm ns}^{\: i}/ d\ln \mu_{\!f}^{\,2}$ 
where $\mu_{\!f}^{}$ is the factorization scale --
of the non-singlet combinations $q_{\,\rm ns}^{\,\pm, \rm v}(x,\mu_{\!f}^{\,2})$ 
of the quark and anti-quark distributions.
In all three cases we employ the same schematic, but characteristic model 
distribution
\beq
\label{eq:shape}
  xq_{\,\rm ns}^{\,\pm,\rm v}(x,\mu_{0}^{\,2}) \; = \;
  x^{\, 0.5} (1-x)^3 \:\: .
\eeq
This facilitates a direct comparison of effects of the various contributions
of the splitting functions. 
For the same reason the reference scale is specified by the order-independent 
value 
\beq
\label{eq:asref}
  \als (\mu_{0}^{\,2}) \; = \; 0.2
\eeq
for the strong coupling constant.
This value corresponds to $\mu_{0}^{\,2} \,\simeq\, 25\ldots 50$ GeV$^2$ for 
$\als (M_Z^{\, 2}) = 0.114 \ldots 0.120$ beyond the leading order. 
In this region of the physical scale $Q^2$ deep-inelastic scattering has been 
measured both at fixed-target experiments and, for much smaller $x$, at the 
{\it ep} collider HERA. 
Our default for the number of effectively massless flavours is $\nf =4$.

\newpage

The reliability of perturbative calculations can be assessed by the relative
size of the higher-order correction at a `nominal' value of the 
renormalization scale $\mu_r$, here $\mu_r = \mu_{\!f}$, and by investigating the 
stability of the results under variations of $\mu_r$. For $\mu_r \neq \mu_{\!f}$
the inverse Mellin transform, see eq.~(\ref{eq:Mtrf}), of the perturbative
expansion (\ref{eq:asexp}) in terms of $\ars \,=\, \als/(4\:\!\pi)$ has to be 
replaced by
\bea
  P_{\rm ns}^{\: i}(\mu_{\!f},\mu_r)
  &\! =\! & \quad
      \ars(\mu_r^2) \, P_{\rm ns}^{(\,0)}  \:\: + \:\:
      \ar(2)(\mu_r^2) \, \left( 
        P_{\rm ns}^{\,(1),i} 
      - \beta_0 \, L \, P_{\rm ns}^{\,(0)} 
      \right) 
\nn \\ & & \mbox{}\!\!\!
    + \: \ar(3)(\mu_r^2) \, \left( 
        P_{\rm ns}^{\,(2),i}
      - 2\beta_0 \, L \, P_{\rm ns}^{\,(1),i}
      - \left\{ \beta_1 \,L  - \beta_0^2 \,L^2 \right\} P_{\rm ns}^{\,(0)} 
      \right)
\nn \\  & & \mbox{}\!\!\!
    + \: \ar(4)(\mu_r^2) \, \left( 
         P_{\rm ns}^{\,(3),i} 
      - 3\beta_0 \, L \, P_{\rm ns}^{\,(2),i}
      - \left\{ 2\beta_1 \,L  - 3\beta_0^2 \,L^2 \right\} 
         P_{\rm ns}^{\,(1),i}
\right. \nn \\[-3mm]  & & \left. \mbox{\hspp}
       - \left\{ \beta_2 \,L - \frct{5}{2}\, \beta_1 \beta_0 \, L^2 
           + \beta_0^3 \, L^3 \right\} P_{\rm ns}^{\,(0)}
      \right)
      \quad \mbox{ with } \quad L = \ln \frac{\mu_{\!f}^{\,2}}{\mu_r^{\,2}}
\;\; .
\eea
For the \MSb\ expansion coefficients $\beta_k$ of the beta function of QCD to 
N$^3$LO see refs.~\cite{vanRitbergen:1997va,Czakon:2004bu} and references 
therein. 

\begin{figure}[p]
\vspace{-2mm}
\centerline{\epsfig{file=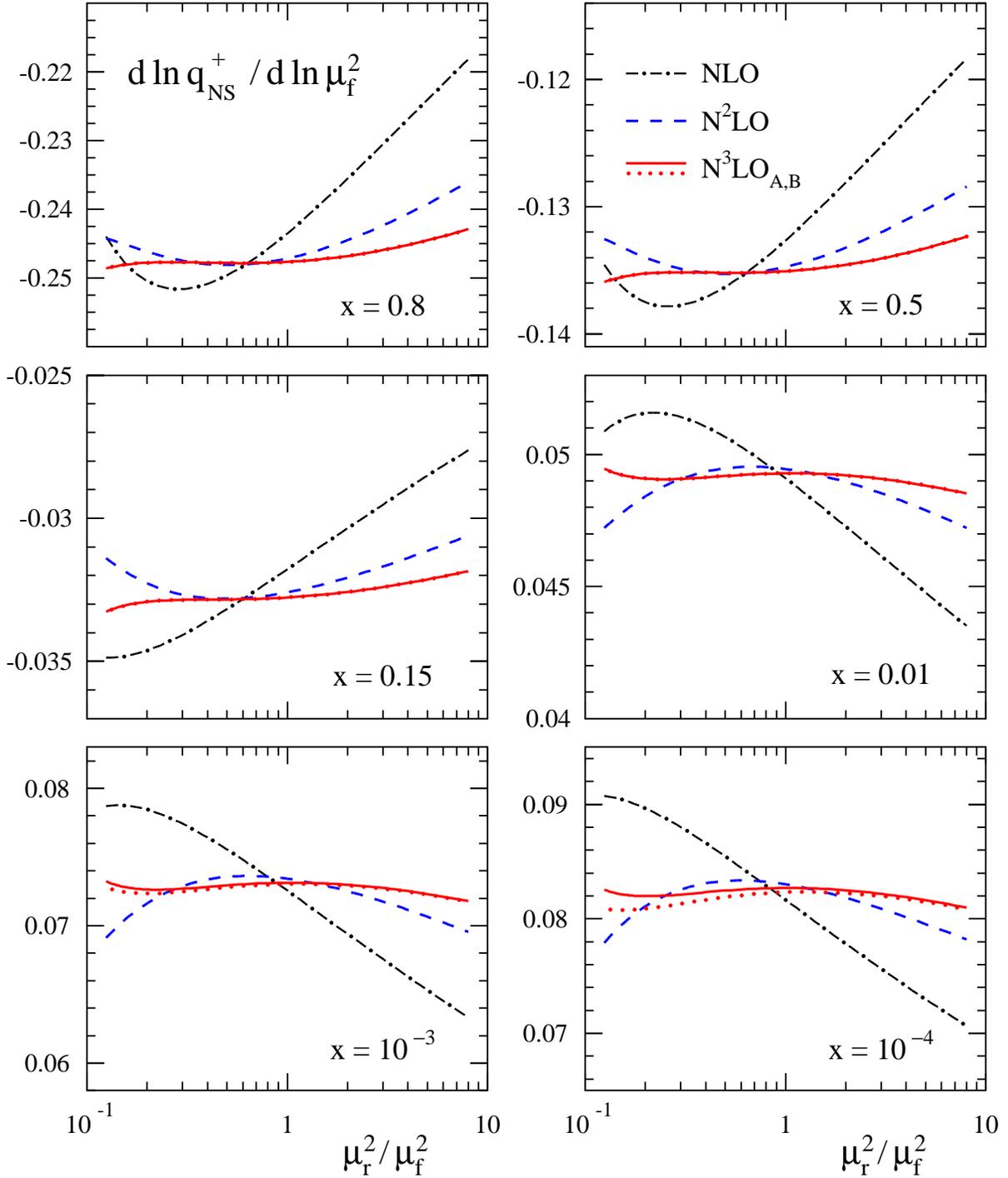,width=16cm,angle=0}}
\vspace{-2mm}
\caption{\small \label{pic:dqns+mur} 
 The dependence of the NLO, NNLO and N$^3$LO  predictions for 
 $\dot{q}_ {\rm ns}^{\, +} \:\equiv\: d \ln q_{\rm ns}^{\, +}/ d\ln\mu_{\!f}^2$ 
 on~the renormalization scale $\mu_r^{}$ at six typical values of $x$ for
 the initial conditions (\ref{eq:shape}) and (\ref{eq:asref}).
 The effect of the remaining uncertainty of the four-loop splitting function
 $P_{\rm ns}^{\,(3)+}$, indicated by the difference of the solid and dotted
 curves, is practically invisible except for the last two panel.
 }
\end{figure}

In fig.~\ref{pic:dqns+mur} the consequences of varying $\mu_r$ over the range
$\frac{1}{8}\,\mu_{\!f}^{\,2} \,\leq\, \mu_r^{\,2} \,\leq\, 8 \,\mu_{\!f}^{\,2}$ are 
displayed for $\dot{q}_{\rm ns}^{\, +}$ at six representative values of $x$ 
ranging from $x = 0.8$ to $x = 10^{-4}$. 
A clear improvement of the scale stability to N$^3$LO is found over this whole 
range. Due to the small size of the four-loop contributions and the 
`$x$-averaging' effect of the Mellin convolution given by
\beq
\label{eq:Mconv}
  [ \,P_{\rm ns}^{\:(n)} \otimes q_{\rm ns} ](x) \;=\;
  \int_x^1  \frac{dy}{y} \; P_{\rm ns}^{\:(n)} (y)\,
  q_{\rm ns}\bigg(\frac{x}{y}\bigg) 
\eeq
and its generalization for the plus-distribution contributions, the approximate
results of section 4 are applicable to lower values of $x$ than one might 
expect from fig.~\ref{fig:P3nf4pm}.

\begin{figure}[p]
\vspace{-4mm}
\centerline{\epsfig{file=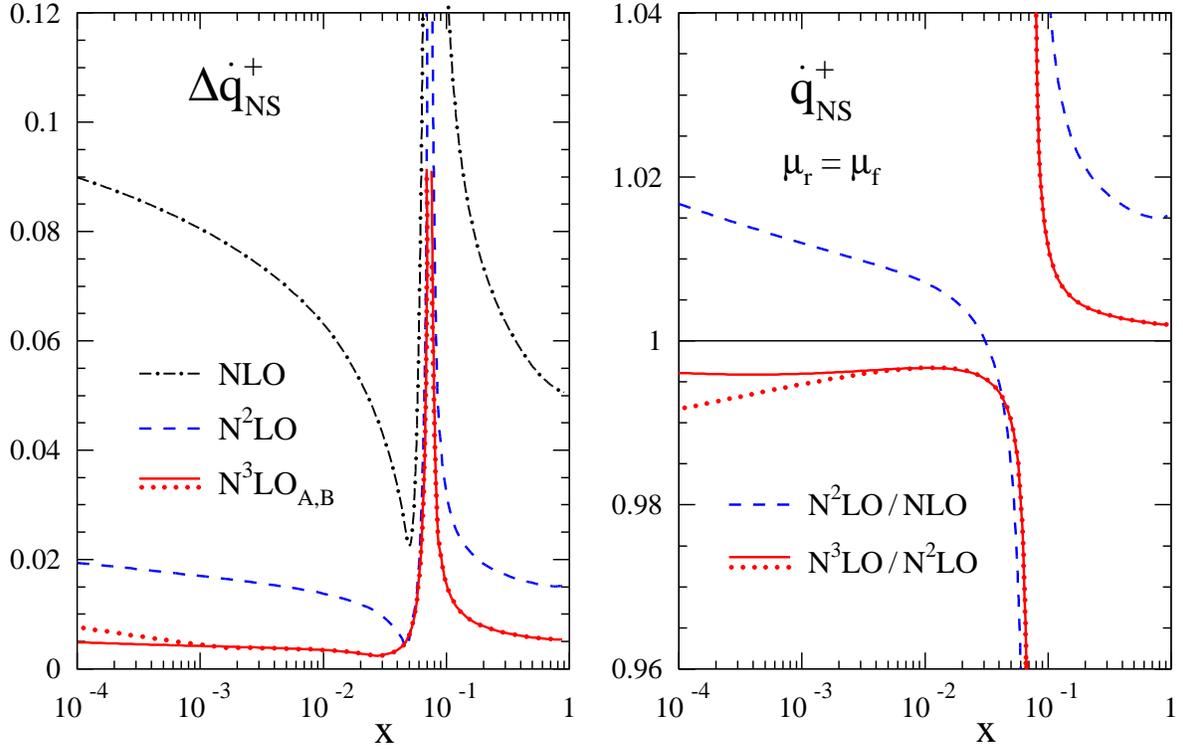,width=16cm,angle=0}}
\vspace{-2mm}
\caption{\small \label{pic:dqns+x}
 Left panel: the $\mu_r$-uncertainty of the NLO, N$^2$LO and N$^3$LO results 
 for the scale derivative of $q_{\rm ns}^{\,+}$ as obtained 
 from the quantity $\Delta \dot{q}_{\rm ns}^{\, i}$ in eq.~(\ref{eq:screl}).
 Right panel: the relative N$^2$LO and N$^3$LO corrections to the logarithmic 
 scale derivative of $q_{\,\rm ns}^{\,+}$ for the characteristic non-singlet 
 quark distribution (\ref{eq:shape}) at $\mu_r = \mu_f$.
 }
\end{figure}

\begin{figure}[p]
\vspace{-2mm}
\centerline{\epsfig{file=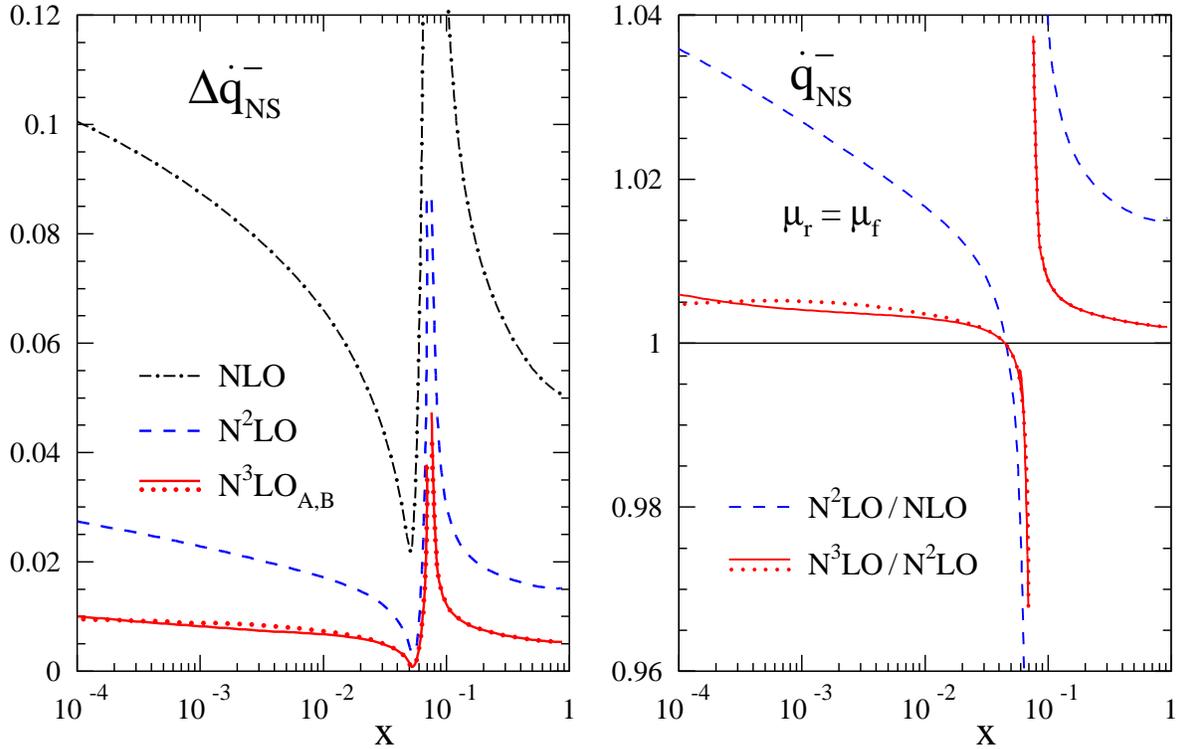,width=16cm,angle=0}}
\vspace{-2mm}
\caption{ \small \label{pic:dqns-x}
 As fig.~\ref{pic:dqns+x}, but for the scale derivative of 
 quark minus anti-quark flavour-differences $q_{\rm ns}^{\,-\,}.\quad\quad$
 }
\vspace{-1mm}
\end{figure}

\begin{figure}[t]
\vspace{-2mm}
\centerline{\epsfig{file=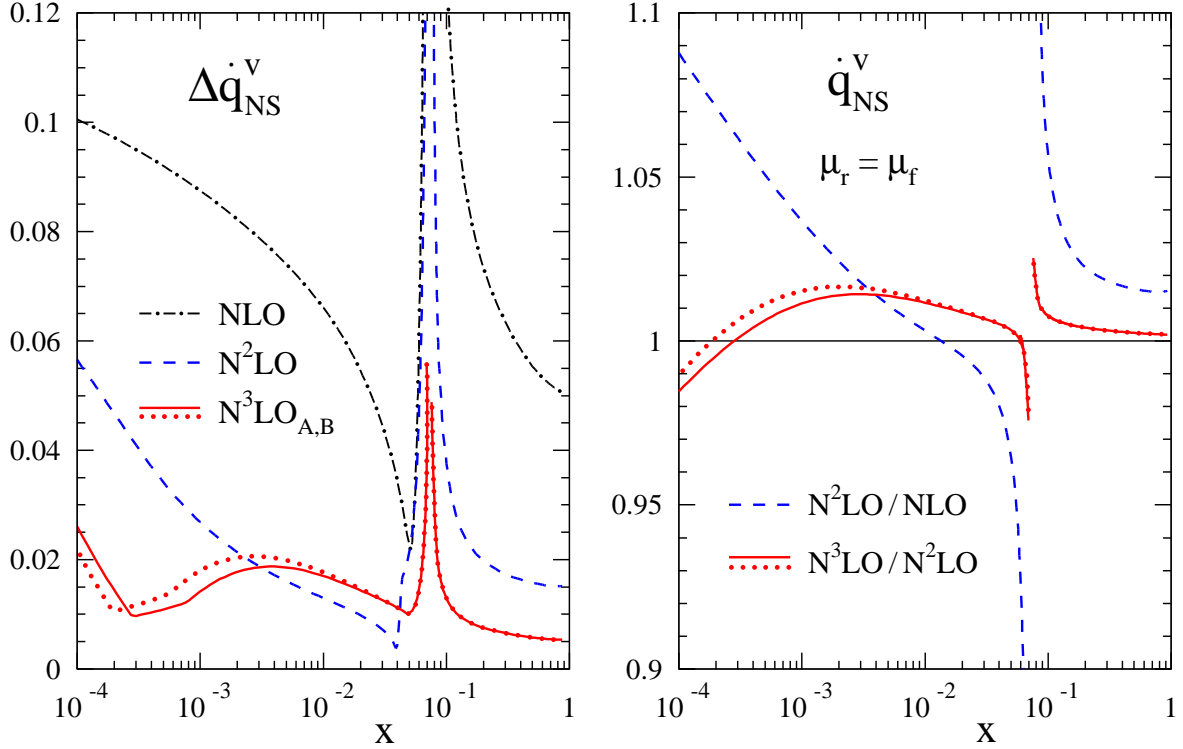,width=16cm,angle=0}}
\vspace{-2mm}
\caption{ \small \label{pic:dqnsvx}
 As figs.~\ref{pic:dqns+x} and \ref{pic:dqns-x}, but for the total 
(flavour-summed) valence quark distribution $q_{\rm ns}^{\,\rm v}$.
 Here and in figs.~\ref{pic:dqns+x} and \ref{pic:dqns-x} the spikes close to 
 $x = 0.07$ reflect the sign-changes of $\dot{q}_{\rm ns}^{\, i}$ and do not 
 constitute appreciable absolute corrections and uncertainties. All three 
 figures refer to $\,\nf=4\,$ and $\,\als(\mu_r^{\,2}\!=\!\mu_{\!f}^{\,2})
 \,=\, 0.2$.
 }
\vspace*{3mm}
\end{figure}

The relative scale uncertainties of the $\mu_r^{}$-average results, 
conventionally estimated by
\beq
\label{eq:screl}
 \Delta \,\dot{q}_{\rm ns}^{\, i} \; \equiv \;
 \frac{\max\, [ \,\dot{q}_{\rm ns}^{\, i}(x,\mu_r^{\,2} = \frac{1}{4}\,
 \mu_{\!f}^{\,2} \ldots 4\,\mu_{\!f}^{\,2})] - \min\, [ \,\dot{q}_{\rm ns}^{\, i} (x,
 \mu_r^2 = \frac{1}{4}\,\mu_{\!f}^{\,2} \ldots 4 \,\mu_{\!f}^{\,2})] }
 { 2\, |\, {\rm average}\, [ \,\dot{q}_{\rm ns}^{\, i}(x, \mu_r^{\,2} =
 \frac{1}{4}\,\mu_{\!f}^{\,2} \ldots 4 \,\mu_{\!f}^{\,2})]\, | }
\eeq
is shown in the left panels of figs.~\ref{pic:dqns+x}, \ref{pic:dqns-x} and
\ref{pic:dqnsvx} for all three cases $i = +,\,-$ and v. In the corresponding
right panels, the relative size of the N$^3$LO corrections at the scale
$\mu_r = \mu_{\!f}^{}$ are compared to the relative N$^2$LO effects. Both the
relative scale uncertainties and the relative corrections have a singularity
at about $\,x \simeq 0.07$ due to a sign change of the scaling violations 
$dq_{\rm ns}^{\:i}(x)/ d\ln \mu_{\!f}^{\,2}$.

Outside the region around $x=0.07$ where the $\mu_{\!f}^{}$-derivatives are
small, the remaining uncertainty of $\dot{q}_{\rm ns}^{\: +}$ is well below
1\% down to $x \simeq 10^{\,-3}$ and possibly, below. 
The size and $\mu_r$-variation of the NLO and NNLO contributions are somewhat 
larger for $\dot{q}_{\rm ns}^{\: -}$ at small $x$, yet neither the N$^3$LO 
correction nor its scale variation exceeds 1\% in the region shown in the plot. 

The case of $q_{\rm ns}^{\:\rm v}$, shown in fig.~\ref{pic:dqnsvx} is 
noticeably different beyond NLO due to the appearance of the 
$d^{abc}d_{abc} \,\nf$ contribution $P_{\rm ns}^{\:\rm s}$ which is negligible
and at large $x$ but large at small $x$ \cite{Moch:2004pa}: 
the difference of the NNLO curves of fig.~\ref{pic:dqns-x} and \ref{pic:dqnsvx}
-- note the different scales for the ordinate -- is~caused completely by this 
contribution due to our choice (\ref{eq:shape}) of the input quark distribution.
Also in this case our new N$^3$LO results leads to a considerable improvement 
and a remaining small-$x$ uncertainty of no more than about 2\% at 
$x \geq 10^{-4}$.

 
\setcounter{equation}{0}
\section{The time-like case}
\label{sec:timelike}

The differences between the initial-state (`space-like', $\sigma = -1$) and
the final-state (`time-like', $\sigma = 1$) splitting functions respectively
governing the evolution of the parton distributions and fragmentation functions
can be expressed in terms of lower-order quantities. At N$^3$LO they read
\bea
{\lefteqn{ \hspn\hspn
 \delta\, P^{\,(3)i}(x) \;\equiv\; P^{\,(3)i}_{{\rm ns},\,\sigma=1}(x)
 - P^{\,(3)i}_{{\rm ns},\,\sigma=-1}(x) \;=\;
}}
\nn \\[1mm] && \hspp
  2 \left\{ [\, \ln x \cdot \overline{P}_{\rm ns}^{\,(2)\,i} \,] 
  \otimes P_{\rm ns}^{\,(0)}  
    + [\, \ln x \cdot P_{\rm ns}^{\,(0)} \,]
      \otimes \overline{P}_{\rm ns}^{\,(2)\,i} 
    + [\, \ln x \cdot \overline{P}_{\rm ns}^{\,(1)\,i} \,]
      \otimes \overline{P}_{\rm ns}^{\,(1)\,i} 
\right\}
\nn\\ & & \mbox{}
  - 2\: P_{\rm ns}^{\,(0)} \otimes [\, \ln x \cdot P_{\rm ns}^{\,(0)} \,]^{\otimes 3} - 4\: [\, P_{\rm ns}^{\,(0)} \,]^{\otimes 2} \otimes [\, \ln x \cdot P_{\rm ns}^{\,(0)} \,] \otimes [\, \ln^2 x \cdot P_{\rm ns}^{\,(0)} \,]
\nn\\ & & \mbox{}
  - \frac{2}{3}\; [\, P_{\rm ns}^{\,(0)} \,]^{\otimes 3} \otimes [\, \ln^3 x \cdot P_{\rm ns}^{\,(0)} \,]
\:\: ,
\label{eq:dP3DMS}
\eea
where we have used the short-hand notations $A^{\otimes 2} \equiv A \otimes A$
etc for the Mellin convolutions, and $\overline{P}^{\,(n)\,i}$ stands for the 
average of the corresponding $\sigma=1$ and $\sigma=-1$ expansion coefficients,
normalized as in eq.~(\ref{eq:asexp}).
Eq.~(\ref{eq:dP3DMS}) has been derived in ref.~\cite{Mitov:2006ic} by
generalizing results in ref.~\cite{Dokshitzer:2005bf}; it is also a direct
consequence of eq.~(\ref{eq:selftune}) \cite{Basso:2006nk}.

The resulting rather lengthy explicit expressions can be found in appendix~%
\ref{sec:appC}. Here we present parametrizations in terms of powers of $x$ and 
the logarithms in eq.~(\ref{eq:logs}).
As above, their \mbox{small-$x$} and large-$x$ coefficients are exact up to 
their rounding to seven digits. 
The accuracy of the $n_{\!f}^{\,k}$ coefficients is better than 0.1\% except 
close to zeros. The three parametrizations are given by
\bea
\label{eq:DTSPqq3+}
{\lefteqn{
 \delta\, P^{\,(3)+}(x) \; = \;
}}
\nn \\ &&
%
 - 33901.87 \,\*  L_1 - 32392.47 
 + 25000 \,\*  (\, x_1 \*  ( 1.2960 + 1.7438 \,\*  x - 1.0943 \,\*  x^2
   - 0.44064 \,\*  x^3 ) 
\nonumber \\ && \mbox{}
   + x \* L_0 \,\*  (0.6440 + 0.8939 \,\*  L_0 + 0.21405 \,\*  L_0^2) 
   + L_1 \,\*  (2.0343 \,\*  x_1 + 0.35738 \,\*  L_0) ) - 10399.74 \,\* L_0 
\nonumber \\[1mm] && \mbox{}
 - 25718.24 \,\* L_0^2 - 5965.487 \,\* L_0^3 - 206.7846 \,\* L_0^4 
 + 4.213992 \,\* L_0^5 - 0.7023320 \,\* L_0^6
\nonumber \\ && \mbox{\hspn}
 + \nf \,\*  \biggl( 
          5483.660 \,\*  L_1 + 4975.255 
 + 250 \,\*  ( \, x_1 \*  ( -19.877 - 8.0977 \,\*  x + 12.335 \,\*  x^2
   + 8.1174 \,\*  x^3 ) 
\nonumber \\ && \mbox{}
   + x \* L_0 \,\*  (13.617 - 7.8856 \,\*  L_0 - 2.2491 \,\*  L_0^2) 
   + L_1  \,\*  (-20.171 \,\*  x_1 + 6.571 \,\*  L_0) ) 
 + 657.5425 \,\* L_0 
\nonumber \\ && \mbox{}
 + 3102.901 \,\* L_0^2 + 735.0891 \,\* L_0^3 + 45.82716 \,\* L_0^4 
 \biggr)
\nonumber \\ && \mbox{\hspn}
 + \nfs\, \*   \biggl(
 - 53.37723 \,\*  L_1 - 80.32433 
 + 50 \,\* ( \,x_1 \*  ( 1.6030 + 15.938 \,\*  x - 5.3145 \,\*  x^2
   + 1.8682 \,\*  x^3 ) 
\nonumber \\ && \mbox{}
   + x \* L_0 \,\*  (13.301 + 2.1060 \,\*  L_0 + 0.4375 \,\*  L_0^2) 
   + L_1  \,\*  (13.060 \,\*  x_1 + 11.023 \,\*  L_0) ) 
 - 9.550559 \,\* L_0 
\nonumber \\ && \mbox{}
 - 56.98805 \,\* L_0^2 - 21.59671 \,\* L_0^3 - 1.580247 \,\* L_0^4 
 \biggr)
\:\: , \\[2mm]
\label{eq:DTSPqq3-}
{\lefteqn{
 \delta\, P^{\,(3)-}(x) \; = \;
}}
\nn \\ &&
%
 - 33901.87 \,\*  L_1 - 32392.47 
 + 25000 \,\*  \,( x_1 \*  ( 1.2892 + 1.4892 \,\*  x - 1.4262 \,\*  x^2
   + 0.29374 \,\*  x^3 ) 
\nonumber \\ && \mbox{}
   + x \* L_0 \,\*  (1.1307 + 0.17484 \,\*  L_0 + 0.14894 \,\*  L_0^2) 
   + L_1  \,\*  (5.0547 \,\*  x_1 + 3.4824 \,\*  L_0) ) 
   - 7307.364 \,\* L_0 
\nonumber \\ && \mbox{}
   - 24617.82 \,\* L_0^2 - 7051.323 \,\* L_0^3
   - 665.0339 \,\* L_0^4 - 13.82716 \,\* L_0^5 + 1.035940 \,\* L_0^6
\nonumber \\ && \mbox{\hspn}
 + \nf\, \*  \biggl(
   5483.660 \,\*  L_1 + 4975.255 
 + 2500 \,\*  ( \,x_1 \*  ( - 1.9867 - 11.407 \,\*  x + 3.9156 \,\*  x^2
    - 1.6032 \*  x^3 ) 
\nonumber \\ && \mbox{}
   + x \* L_0 \,\*  ( - 11.069 - 1.1039 \,\*  L_0 - 0.2778 \,\*  L_0^2) 
   + L_1  \,\*  ( - 13.824 \,\* x_1 - 11.688 \,\* L_0) ) + 346.2303 \,\* L_0 
\nonumber \\ && \mbox{}
 + 2994.194 \,\* L_0^2 + 780.4122 \,\* L_0^3
 + 66.89712 \,\* L_0^4 + 1.580247 \,\* L_0^5
 \biggr)
\nonumber \\ && \mbox{\hspn}
 + \nfs\, \*  \biggl(
 - 53.37723 \,\*  L_1 - 80.32433 
 + 50 \,\*  (\, x_1 \*  ( 1.6030 + 15.938 \,\*  x - 5.3145 \,\*  x^2
   + 1.8682 \,\*  x^3 ) 
\nonumber \\ && \mbox{}
   + x \* L_0 \,\*  (13.301 + 2.1060 \,\*  L_0 + 0.4375 \,\*  L_0^2) 
   + L_1  \,\*  (13.060 \,\*  x_1 + 11.023 \,\*  L_0) )
 - 9.550559 \,\* L_0 
\nonumber \\ && \mbox{}
 - 56.98805 \,\* L_0^2 - 21.59671 \,\* L_0^3 - 1.580247 \,\* L_0^4 
 \biggr)
%
\:\: , \\[2mm]
\label{eq:DTSPqq3s}
{\lefteqn{
 \delta\, P^{\,(3)\rm s}(x) \; = \;
}}
\nn \\ && \mbox{\hspn} \phantom{-}
%
 \nf \*  \biggl(
  x_1^2 \* L_1 \*  ( 190.3189 - 27.48379 \,\*  L_1 )
  + 500 \,\*  ( \, x_1^2 \*  ( 0.0597 - 11.761 \,\* x + 3.0470 \,\* x^2
    - 0.8633 \,\* x^3 )
\nonumber \\ && \mbox{}
    + x \* L_0 \*  ( - 22.843 \,\*  x_1 - 10.899 \,\*  L_0
    - 3.1331 \,\*  L_0^2) + x_1 \* L_1 \*  (x_1 + L_0) \,\*  0.3835 )
  + 297.0894 \,\* x_1 \* L_0 
\nonumber \\ && \mbox{}
  - 949.8383 \,\* L_0^2 - 431.7969 \,\* L_0^3 
  + 52.21778 \,\* L_0^4 + 7.901235 \,\* L_0^5 - 1.580247 \,\* L_0^6
  \biggr)
\:\: .
\eea
 
\begin{figure}[p]
\vspace{-3mm}
\centerline{\epsfig{file=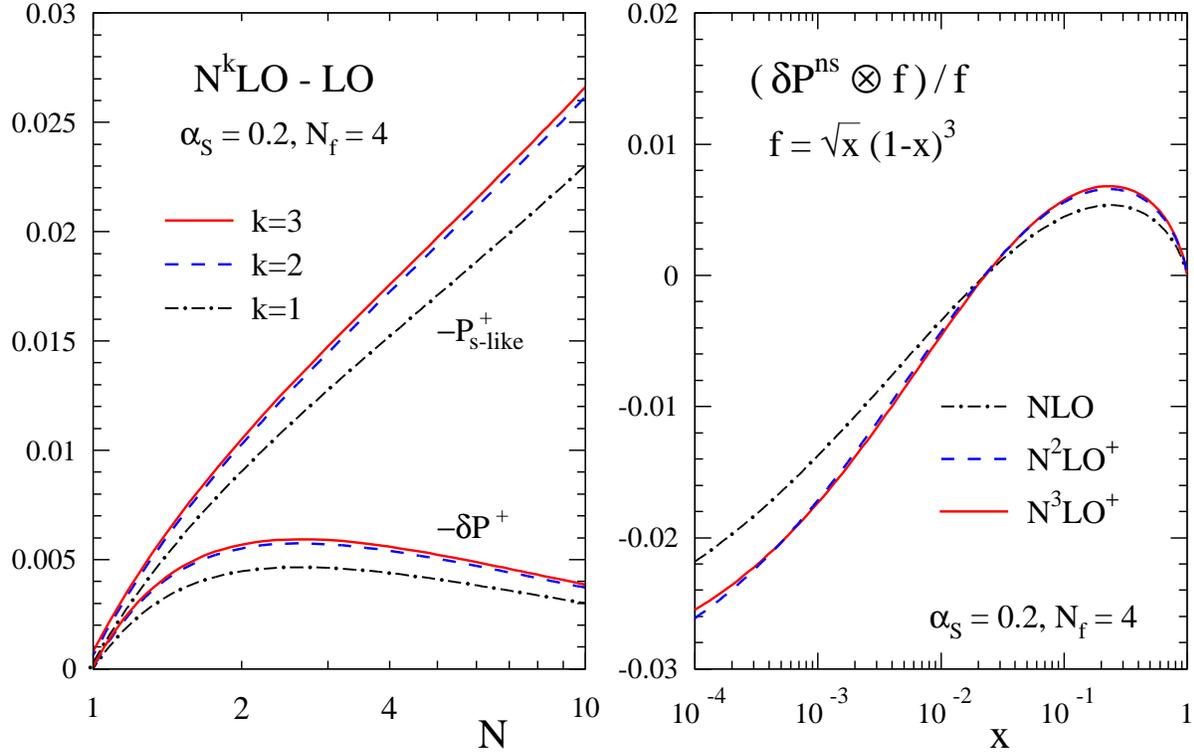,width=16cm,angle=0}}
\vspace{-2mm}
\caption{ \small \label{pic:dtsPns1}
 The perturbative expansion of the difference $\,\delta\, P^{\,+} \,=\, 
 P^{\,+}_{{\rm ns},\,\sigma=1} - P^{\,+}_{{\rm ns},\,\sigma=-1}$ of the 
 time-like ($\sigma=1$) and space-like ($\sigma=-1$) non-singlet$^+$ splitting 
 functions.
 Left: results in Mellin-space, compared to beyond-LO contributions in the
 space-like case. Right: convolution with the schematic input shape 
 (\ref{eq:shape}).
 }
\end{figure}

\begin{figure}[p]
\vspace{-1mm}
\centerline{\epsfig{file=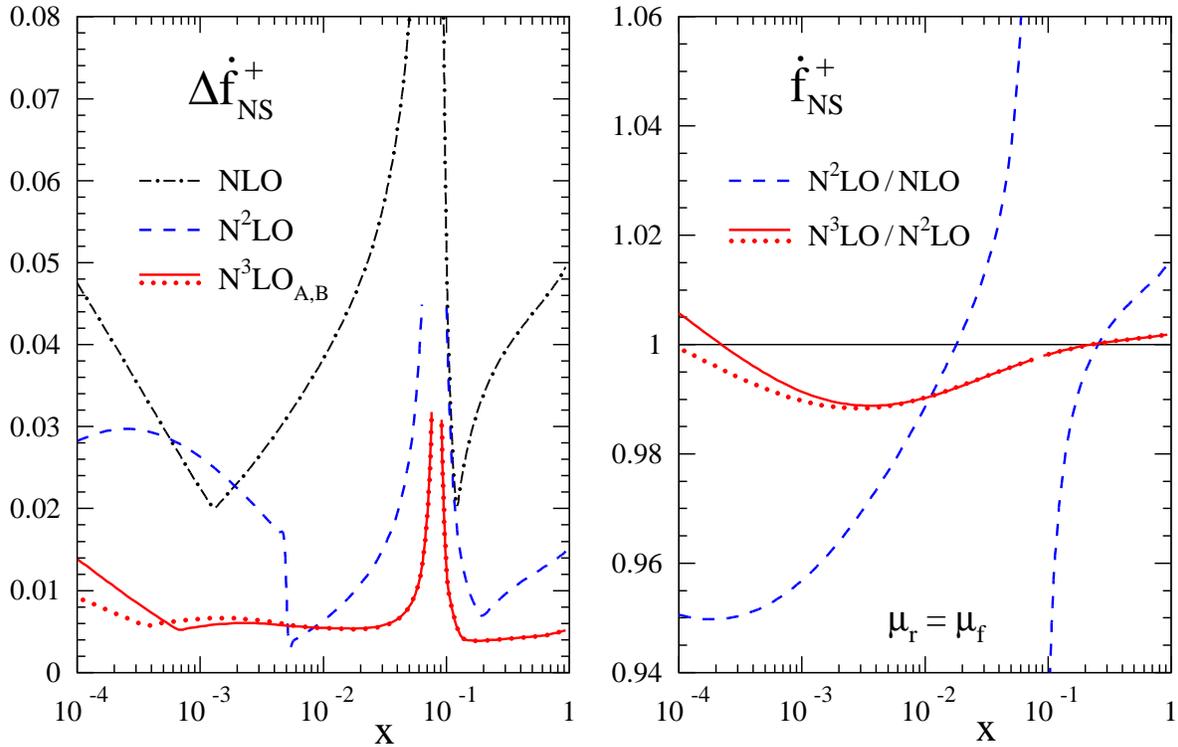,width=16cm,angle=0}}
\vspace{-3mm}
\caption{ \small \label{pic:dfns+T}
 As fig.~\ref{pic:dqns+x}, but for its time-like counterpart, the fragmentation 
 function $f_{\rm ns}^{\,+}$. 
 The same input (\ref{eq:shape}) and (\ref{eq:asref}) is used here, so all 
 differences to fig.~\ref{pic:dqns+x} are due to the different N$^{n>0}$LO 
 splitting functions.
 }
\vspace*{-1mm}
\end{figure}

The difference of the time-like and space-like splitting functions and the
resulting scale derivative are illustrated in figs.~\ref{pic:dtsPns1} and 
\ref{pic:dfns+T} for the most important case, NS$^+$.
The pattern is somewhat different in the time-like case, e.g., the N$^2$LO 
contribution is negative at small $x$. Yet also here the perturbative 
expansion is `perfectly' stable after including the N$^3$LO corrections, 
with a residual uncertainty of 1\% or less for $x > 10^{\,-4}$ at our 
(for the time-like case, low-scale) reference point.


%
\setcounter{equation}{0}
\section{Summary and outlook}
\label{sec:summ}
%

We have presented the four-loop corrections $P_{\rm ns}^{\,(3)\pm,\rm v}(x)$ 
to all three non-singlet splitting functions in perturbative QCD. 
Our results, which are partly approximate but 
sufficiently accurate for 
collider-physics applications, allow to set-up and solve the QCD evolution 
equations for flavour non-singlet (initial-state) parton distributions (PDFs) 
and (final-state) fragmentation functions (FFs) at N$^3$LO.
They thus provide a major step towards the consistent application of QCD 
factorization to theoretical predictions for N$^3$LO cross sections with
initial (final) state hadrons, as already obtained in 
refs.~\cite{Anastasiou:2015ema,Vermaseren:2005qc,Moch:2008fj,Davies:2016ruz,%
Dreyer:2016oyx}, which requires hard partonic cross section and PDFs (FFs) at 
the same order in renormalization-group improved perturbation theory.
 
The resulting logarithmic scale derivatives 
$\,d \ln q_{\,\rm ns}^{\,\pm,\,\rm v} / d\ln \mu_{\!f}^{\,2}\,$ exhibit a 
very good convergence of the perturbative expansion. 
Both the four-loop corrections and the N$^3$LO dependence on the 
renormalization scale $\mu_r^{}$ mostly amount to as little as 1\% or less 
(and maximally 2\%, for $P_{\rm ns}^{\,\rm v}(x)$ at small $x$) at momentum 
fractions $\,x > 10^{-4}\,$ for $\,\als(\mu_{\!f}^{\,2}) \,\simeq\, 0.2$.

Our results have been obtained via computations of fixed Mellin moments 
-- to $N=20$ for the diagrams contributing in the limit of a large number 
$\nc$ of colours, and $N=16$ otherwise -- 
for the QCD corrections to quark and anti-quark operator matrix elements 
(OMEs) up to four loops.
After~projection of all external spins and Lorentz indices, these OMEs reduce 
to four-loop massless propagator integrals which can be evaluated with the 
{\sc Forcer} program \cite{Ruijl:2017cxj} in the computer algebra system 
{\sc Form} \cite{Vermaseren:2000nd,Kuipers:2012rf,Tentyukov:2007mu,%
Ruijl:2017dtg}.
 
For the large-$\nc$ contributions to $P_{\rm ns}^{\,(3)\pm}$, these moments 
turn out to be sufficient for a reconstruction of the all-$N$ results in terms
of harmonic sums by solving systems of Diophantine equations. 
Additional knowledge about the limits for $x \to 0 $ and $x \to 1$, as well as 
the rephrasing (based on conformal symmetry) of the evolution equations in 
terms of a universal `reciprocity-respecting' evolution kernel 
\cite{Dokshitzer:2005bf,Dokshitzer:2006nm,Basso:2006nk}, have been instrumental
in this step. 
Beyond the large-$n_c$ contributions we have used the computed Mellin moments, 
again supplemented by endpoint constraints, to provide approximations for the 
four-loop splitting functions including $x$-dependent estimates of 
their residual uncertainties. The latter are very small, except in the region 
$x < 10^{-2}$. 
These small-$x$ uncertainties are subject to a further suppression in the 
actual evolution due to the convolution with the PDFs (or FFs). Due to this 
our results are found to be sufficiently precise for $x \gsim 10^{-4}$.

From the threshold limit $x \!\to\! 1$ we have been able to determine the 
complete cusp anomalous dimension $A_4$ for quarks at four loops.  Our results 
for the $\nfz$ and $\nfo$ parts beyond the \mbox{large-$\nc$} contribution
are numerical, and lead to a relative accuracy of $10^{-4}$ for these 
coefficients in QCD, which should be amply sufficient for phenomenological 
applications. The break-up of $A_4$ in terms of individual colour factors 
includes non-vanishing contributions with the quartic group invariants.
The exact results for $A_4$ of the present article and of 
ref.~\cite{Davies:2016jie} are in agreement with the form-factor calculations
in refs.~\cite{Henn:2016men,Lee:2016ixa} for the planar part and 
refs.~\cite{Grozin:2016ydd,Lee:2017mip} for the $\nfs$ contributions.
In particular, refs.~\cite{Lee:2016ixa} provided a quick confirmation of our
result for the hardest part, the $\nf$-independent contribution, and hence
of our (mathematically not completely rigorous) reconstruction of the all-$N$ 
expression of the large-$\nc$ non-singlet anomalous dimension.

The terms $B_4$ proportional to $\delta\xm1$ in the four-loop splitting
functions yield the universal eikonal anomalous dimension if properly combined
with information on infrared singularities from the QCD form factor. 
This is an important ingredient for extending the threshold resummation for
inclusive cross section to N$^4$LL accuracy, i.e. (next-to-)$^4$-leading 
logarithmic order.

\bigskip

In order to practically complete the QCD evolution equations at N$^3$LO, 
corresponding results are required for the singlet splitting functions at 
four loops, i.e., the pure-singlet quark-quark splitting function 
$P_{\,\rm ps}^{\,(3)}(x)$ and those involving gluons, 
$P_{\,\rm qg}^{\,(3)}(x)$, $P_{\,\rm gq}^{\,(3)}(x)$ and 
$P_{\,\rm gg}^{\,(3)}(x)$. All these quantities are currently unknown beyond 
the $N=2$ and $N=4$ moments presented in ref.~\cite{Ruijl:2016pkm}.
However, by following the approach of the present paper, it should be feasible 
to compute enough moments of the (theoretically much more complicated, see 
refs.~\cite{Hamberg:1991qt,Collins:1994ee}) corresponding flavour-singlet OMEs 
up to four loops with the {\sc Forcer} program to gather sufficient information
for first phenomenologically relevant approximations. 
We leave this topic to future research.

Incremental improvements in the flavour non-singlet sector can be obtained by 
calculating more moments, which is a hard problem within the present 
computational set-up for almost all colour factors, and by incorporating more 
external information, such as, e.g., a future exact result for the four-loop 
cusp anomalous dimension from calculations of the photon-quark form factor. 

A derivation of the exact expressions for the $\nf$-independent hardest 
parts will require, in addition, a much improved theoretical understanding.
In this context it may be interesting to note that the $\,\zeta_5\,$ part of
$\gamma_{\rm ns}^{\,(3)\pm}(N)$, which can be determined at all $N$ from the
presently available information (see appendix~D) 
includes a contribution
\beq
\label{eq:z5S1sq}
  -\,\frac{128}{3}\,\left\{ 3\,\cfs\,\cas - 2\,\cf\cat + 12\:\dfFAnc \right\}
   \:5\:\!\zeta_5^{}\, [S_1(N)]^2
\eeq
that vanishes in the large-$\nc$ limit.
The resulting $\ln^2 N$ large-$N$ behaviour needs to be compensated by 
non-$\zeta_5$ terms, and it is tempting to identify the $5\:\!\zeta_5$ in 
eq.~(\ref{eq:z5S1sq}) as the $\zeta_5$-`tail' of the function
\beq
     f(N) \;=\;
        5 \* \zeta_5
       - 2 \* \S(-5)
       + 4 \:\!\* \S(-2) \* \zeta_3
       - 4 \:\!\* \Ss(-2,-3)
       + 8 \* \Sss(-2,-2,1)
       + 4 \:\!\* \Ss(3,-2)
       - 4 \:\!\* \Ss(4,1)
       + 2 \* \S(5)
\:\: . 
\eeq 
This function first occurred multiplied with positive powers of $N$ in 
three-loop coefficient functions of DIS in ref.~\cite{Vermaseren:2005qc} and 
resurfaced, now multiplied with $[S_1(N)]^2$ as in eq.~(\ref{eq:z5S1sq}), as 
the `wrapping correction' in the anomalous dimensions in $\,{\cal N}=4$ 
maximally supersymmetric Yang-Mills theory \cite{Bajnok:2008qj}, where it is
crucial for obtaining the correct small-$x$ limit, see, e.g., 
ref.~\cite{Kotikov:2007cy}.
We thus hypothesize that eq.~(\ref{eq:z5S1sq}) represents the first glimpse 
of the wrapping corrections in an anomalous dimension in QCD.

\bigskip
\medskip


\noindent
{\sc Form} and {\sc Fortran} files with our results can be obtained from 
the preprint server http://arXiv.org by downloading the source of this article.
They are also available from the authors 
upon request.

\newpage

%
\subsection*{Acknowledgements}
S.M.~would like to thank J.~Gracey for useful discussions. We are grateful
to T. Gehrmann and J. Bl\"umlein for providing {\sc Fortran} codes for 
harmonic polylogarithms up to weight 6, and to P.~Marquard for converting of 
the coefficient of $\nf\,d_F^{\,abcd}d_F^{\,abcd}/N_R$ of the cusp anomalous 
dimension in ref.~\cite{Grozin:2015kna} to our notation.
This work has been supported by the {\it Deutsche Forschungsgemeinschaft} 
(DFG) grant MO~1801/1-2 and SFB 676 project A3, 
the {\it European Research Council}$\,$ (ERC) Advanced Grant 320651, 
{\it HEPGAME} 
and the UK {\it Science \& Technology Facilities Council}$\,$ (STFC) grant 
ST/L000431/1. 
We also are grateful for the opportunity to use most of the {\tt ulgqcd} 
computer cluster in Liverpool which was funded by the STFC grant ST/H008837/1.
The Feynman diagrams have been drawn with the packages {\sc Axodraw} 
\cite{Vermaseren:1994je} and {\sc Jaxo\-draw} \cite{Binosi:2003yf}.

\bigskip

\appendix
%
%
\renewcommand{\theequation}{\ref{sec:appA}.\arabic{equation}}
\setcounter{equation}{0}
\renewcommand{\thefigure}{\ref{sec:appA}.\arabic{figure}}
\setcounter{figure}{0}
\renewcommand{\thetable}{\ref{sec:appA}.\arabic{table}}
\setcounter{table}{0}
\section{Feynman rules}
\label{sec:appA}
Below we present the Feynman rules for vertices arising from insertions of 
the operator $O^{\,\rm ns}_{\{\mu^{\,}_1,...,\mu^{\,}_N\}}$
in eq.~(\ref{eq:loc-ops}). 
All momenta $p_i^{}$ are flowing into the operator vertex and we use
\begin{eqnarray}
q &=& \sum\limits_i p_i\, ,
\end{eqnarray}
where $q$ is the outgoing momentum flow through the operator. 
The free Lorentz indices of the operator 
$O^{\rm ns}_{\{\mu^{\,}_1,...,\mu^{\,}_N\}}$ are contracted with 
$$
\Delta_{\mu_1}\, \dots \,\Delta_{\mu_N}\, ,
$$
where the vector $\Delta$ fulfils $\Delta^2 = 0$.
We limit the derivation up to four additional gluons coupling to the operator, 
i.e., $n=6$ in fig.~\ref{fig:OMEng}. 
For Feynman rules with up to three additional gluons and zero momentum flow
through the operator, see also~ref.~\cite{Bierenbaum:2009mv} and references 
therein.

The expressions for unpolarized quark operators in 
eqs.~(\ref{eq:omeqq})--(\ref{eq:omeqqgggg})
are readily generalized to the polarized case by substituting 
$\Slash{\Delta}\,\to\, \Slash{\Delta}\gamma_5$.

\begin{tabular}[h]{ll}
\hspace*{-10mm}
\begin{minipage}{3cm}
\vspace*{5mm}
\begin{flushleft}
  \includegraphics[width=2.25\textwidth, angle=0]{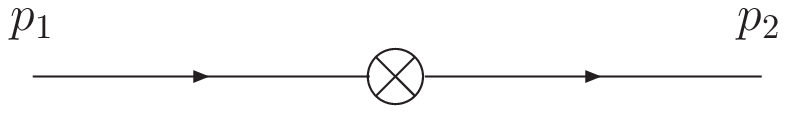}
\end{flushleft}
\end{minipage}
&
\begin{minipage}{13.0cm}
  \begin{eqnarray}
    \label{eq:omeqq}
    \Slash{\Delta}\, \left(\Delta \cdot p_2 \right)^{N-1}
  \end{eqnarray}
\end{minipage}
\end{tabular}

\begin{tabular}[h]{ll}
\hspace*{-10mm}
\begin{minipage}{3cm}
\vspace*{5mm}
\begin{flushleft}
  \includegraphics[width=2.25\textwidth, angle=0]{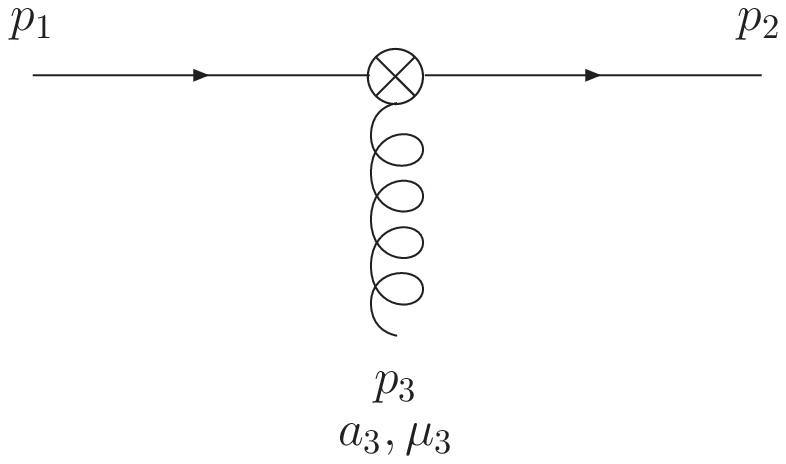}
\end{flushleft}
\end{minipage}
&
\begin{minipage}{13.0cm}
  \begin{eqnarray}
    \hspace*{20mm}
    \label{eq:omeqqg}
    - g t^{a_3} 
    \Slash{\Delta} \Delta^{\mu_3} 
    \sum\limits_{j_1=0}^{N-2} 
    (p_2\cdot \Delta)^{N-2-j_1} (q\cdot \Delta-p_1\cdot \Delta)^{j_1}
  \end{eqnarray}
\end{minipage}
\end{tabular}
\vspace{10mm}

\begin{tabular}[h]{ll}
\hspace*{-10mm}
\begin{minipage}{3cm}
\vspace*{-5mm}
\begin{flushleft}
  \includegraphics[width=2.25\textwidth, angle=0]{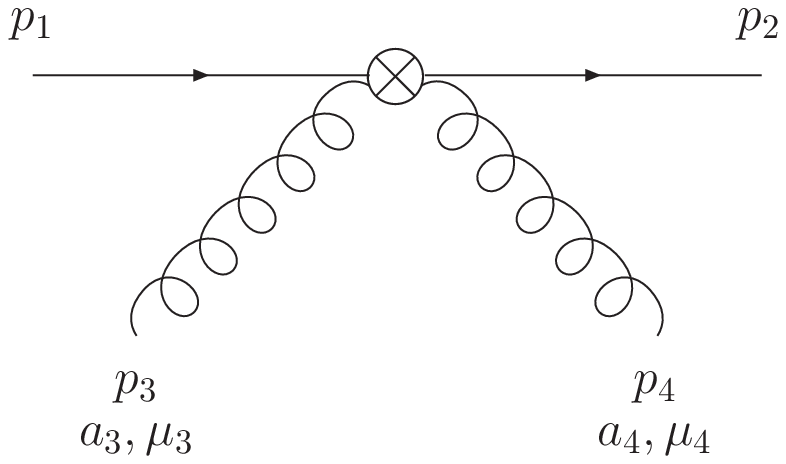}
\end{flushleft}
\end{minipage}
&
\begin{minipage}{13.0cm}
  \begin{eqnarray}
    \hspace*{20mm}
    \label{eq:omeqqgg}
    &&
    g^2 
    \Slash{\Delta} \Delta^{\mu_3} \Delta^{\mu_4} 
    \sum\limits_{j_1=0}^{N-3}
    \sum\limits_{j_2=0}^{j_1}
    (p_2\cdot \Delta)^{N-3-j_1} (q\cdot \Delta-p_1\cdot \Delta)^{j_2} 
    \nonumber
    \\[-1mm]
    &&
    \biggl(
    t^{a_3} t^{a_4} (q\cdot \Delta-p_1\cdot \Delta-p_3\cdot \Delta)^{j_1-j_2}
    \nonumber
    \\[-1mm]
    &&
    +t^{a_4} t^{a_3} (q\cdot \Delta-p_1\cdot \Delta-p_4\cdot \Delta)^{j_1-j_2}
    \biggr)
\end{eqnarray}
\end{minipage}
\end{tabular}
\vspace{3mm}

\begin{tabular}[h]{ll}
\hspace*{-10mm}
\begin{minipage}{3cm}
\vspace*{-5mm}
\begin{flushleft}
  \includegraphics[width=2.25\textwidth, angle=0]{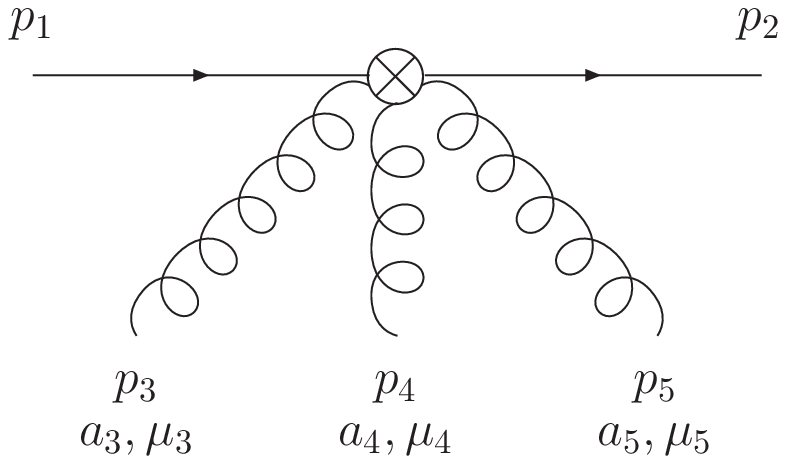}
\end{flushleft}
\end{minipage}
&
\begin{minipage}{13.0cm}
\end{minipage}
\end{tabular}
  \begin{eqnarray}
    \label{eq:omeqqggg}
    &&
    - g^3
    \Slash{\Delta} \Delta^{\mu_3} \Delta^{\mu_4} \Delta^{\mu_5} 
    \sum\limits_{j_1=0}^{N-4}
    \sum\limits_{j_2=0}^{j_1}
    \sum\limits_{j_3=0}^{j_2}
    (p_2\cdot \Delta)^{N-4-j_1} (q\cdot \Delta-p_1\cdot \Delta)^{j_3} 
    \nonumber
    \\
    &&
    \biggl( 
    t^{a_3} t^{a_4} t^{a_5} (q\cdot \Delta-p_1\cdot \Delta-p_3\cdot \Delta)^{j_2-j_3} (q\cdot \Delta-p_1\cdot \Delta-p_3\cdot \Delta-p_4\cdot \Delta)^{j_1-j_2}
    \nonumber
    \\[-1mm]
    &&
    + t^{a_3} t^{a_5} t^{a_4} (q\cdot \Delta-p_1\cdot \Delta-p_3\cdot \Delta)^{j_2-j_3} (q\cdot \Delta-p_1\cdot \Delta-p_3\cdot \Delta-p_5\cdot \Delta)^{j_1-j_2}
    \nonumber
    \\
    &&
    + t^{a_4} t^{a_3} t^{a_5} (q\cdot \Delta-p_1\cdot \Delta-p_4\cdot \Delta)^{j_2-j_3} (q\cdot \Delta-p_1\cdot \Delta-p_3\cdot \Delta-p_4\cdot \Delta)^{j_1-j_2}
    \nonumber
    \\
    &&
    + t^{a_4} t^{a_5} t^{a_3} (q\cdot \Delta-p_1\cdot \Delta-p_5\cdot \Delta)^{j_2-j_3} (q\cdot \Delta-p_1\cdot \Delta-p_3\cdot \Delta-p_5\cdot \Delta)^{j_1-j_2}
    \nonumber
    \\
    &&
    + t^{a_5} t^{a_3} t^{a_4} (q\cdot \Delta-p_1\cdot \Delta-p_4\cdot \Delta)^{j_2-j_3} (q\cdot \Delta-p_1\cdot \Delta-p_4\cdot \Delta-p_5\cdot \Delta)^{j_1-j_2}
    \nonumber
    \\
    &&
    + t^{a_5} t^{a_4} t^{a_3} (q\cdot \Delta-p_1\cdot \Delta-p_5\cdot \Delta)^{j_2-j_3} (q\cdot \Delta-p_1\cdot \Delta-p_4\cdot \Delta-p_5\cdot \Delta)^{j_1-j_2}
    \biggr)
  \end{eqnarray}
%

\begin{tabular}[h]{ll}
\hspace*{-10mm}
\begin{minipage}{3cm}
\vspace*{-5mm}
\begin{flushleft}
  \includegraphics[width=2.25\textwidth, angle=0]{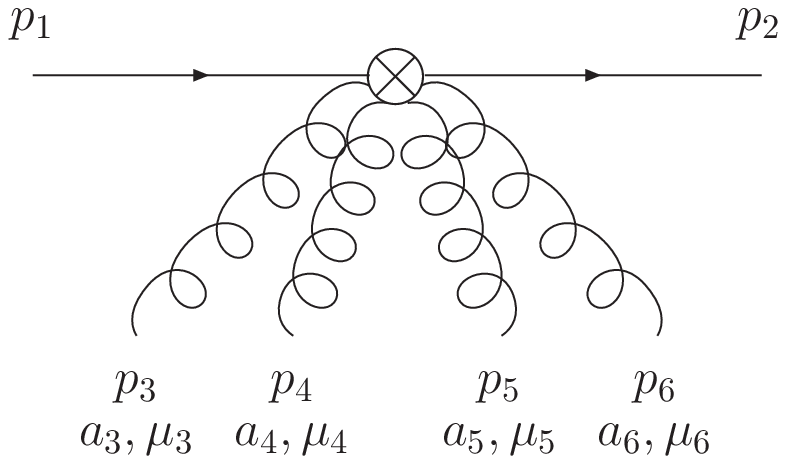}
\end{flushleft}
\end{minipage}
&
\begin{minipage}{13.0cm}
\end{minipage}
\end{tabular}

\begin{eqnarray}
    \label{eq:omeqqgggg}
    &&
    g^4 
    \Slash{\Delta} \Delta^{\mu_3} \Delta^{\mu_4} \Delta^{\mu_5} \Delta^{\mu_6} 
    \sum\limits_{j_1=0}^{N-5}
    \sum\limits_{j_2=0}^{j_1}
    \sum\limits_{j_3=0}^{j_2}
    \sum\limits_{j_4=0}^{j_3}
    (p_2\cdot \Delta)^{N-5-j_1} (q\cdot \Delta-p_1\cdot \Delta)^{j_4} 
    \nonumber
    \\
    &&
    \biggl( t^{a_3} t^{a_4} t^{a_5} t^{a_6} 
    (q\cdot \Delta-p_1\cdot \Delta-p_3\cdot \Delta)^{j_3-j_4} (q\cdot \Delta-p_1\cdot \Delta-p_3\cdot \Delta-p_4\cdot \Delta)^{j_2-j_3} 
    \nonumber
    \\[-1mm]
    &&\qquad (q\cdot \Delta-p_1\cdot \Delta-p_3\cdot \Delta-p_4\cdot \Delta-p_5\cdot \Delta)^{j_1-j_2}
    \nonumber
    \\
    &&
+ t^{a_3} t^{a_4} t^{a_6} t^{a_5} 
    (q\cdot \Delta-p_1\cdot \Delta-p_3\cdot \Delta)^{j_3-j_4} (q\cdot \Delta-p_1\cdot \Delta-p_3\cdot \Delta-p_4\cdot \Delta)^{j_2-j_3} 
    \nonumber
    \\
    &&\qquad (q\cdot \Delta-p_1\cdot \Delta-p_3\cdot \Delta-p_4\cdot \Delta-p_6\cdot \Delta)^{j_1-j_2}
    \nonumber
    \\
    &&
+ t^{a_3} t^{a_5} t^{a_4} t^{a_6} 
    (q\cdot \Delta-p_1\cdot \Delta-p_3\cdot \Delta)^{j_3-j_4} (q\cdot \Delta-p_1\cdot \Delta-p_3\cdot \Delta-p_5\cdot \Delta)^{j_2-j_3} 
    \nonumber
    \\
    &&\qquad (q\cdot \Delta-p_1\cdot \Delta-p_3\cdot \Delta-p_4\cdot \Delta-p_5\cdot \Delta)^{j_1-j_2}
    \nonumber
    \\
    &&
+ t^{a_3} t^{a_5} t^{a_6} t^{a_4} 
    (q\cdot \Delta-p_1\cdot \Delta-p_3\cdot \Delta)^{j_3-j_4} (q\cdot \Delta-p_1\cdot \Delta-p_3\cdot \Delta-p_5\cdot \Delta)^{j_2-j_3} 
    \nonumber
    \\
    &&\qquad (q\cdot \Delta-p_1\cdot \Delta-p_3\cdot \Delta-p_5\cdot \Delta-p_6\cdot \Delta)^{j_1-j_2}
    \nonumber
    \\
    &&
+ t^{a_3} t^{a_6} t^{a_4} t^{a_5} 
    (q\cdot \Delta-p_1\cdot \Delta-p_3\cdot \Delta)^{j_3-j_4} (q\cdot \Delta-p_1\cdot \Delta-p_3\cdot \Delta-p_6\cdot \Delta)^{j_2-j_3} 
    \nonumber
    \\
    &&\qquad (q\cdot \Delta-p_1\cdot \Delta-p_3\cdot \Delta-p_4\cdot \Delta-p_6\cdot \Delta)^{j_1-j_2}
    \nonumber
    \\
    &&
+ t^{a_3} t^{a_6} t^{a_5} t^{a_4} 
    (q\cdot \Delta-p_1\cdot \Delta-p_3\cdot \Delta)^{j_3-j_4} (q\cdot \Delta-p_1\cdot \Delta-p_3\cdot \Delta-p_6\cdot \Delta)^{j_2-j_3} 
    \nonumber
    \\
    &&\qquad (q\cdot \Delta-p_1\cdot \Delta-p_3\cdot \Delta-p_5\cdot \Delta-p_6\cdot \Delta)^{j_1-j_2}
    \nonumber
    \\
    &&
+ t^{a_4} t^{a_3} t^{a_5} t^{a_6} 
    (q\cdot \Delta-p_1\cdot \Delta-p_4\cdot \Delta)^{j_3-j_4} (q\cdot \Delta-p_1\cdot \Delta-p_3\cdot \Delta-p_4\cdot \Delta)^{j_2-j_3} 
    \nonumber
    \\
    &&\qquad (q\cdot \Delta-p_1\cdot \Delta-p_3\cdot \Delta-p_4\cdot \Delta-p_5\cdot \Delta)^{j_1-j_2}
    \nonumber
    \\
    &&
+ t^{a_4} t^{a_3} t^{a_6} t^{a_5} 
    (q\cdot \Delta-p_1\cdot \Delta-p_4\cdot \Delta)^{j_3-j_4} (q\cdot \Delta-p_1\cdot \Delta-p_3\cdot \Delta-p_4\cdot \Delta)^{j_2-j_3} 
    \nonumber
    \\
    &&\qquad (q\cdot \Delta-p_1\cdot \Delta-p_3\cdot \Delta-p_4\cdot \Delta-p_6\cdot \Delta)^{j_1-j_2}
    \nonumber
    \\
    &&
+ t^{a_4} t^{a_5} t^{a_3} t^{a_6} 
    (q\cdot \Delta-p_1\cdot \Delta-p_4\cdot \Delta)^{j_3-j_4} (q\cdot \Delta-p_1\cdot \Delta-p_4\cdot \Delta-p_5\cdot \Delta)^{j_2-j_3} 
    \nonumber
    \\
    &&\qquad (q\cdot \Delta-p_1\cdot \Delta-p_3\cdot \Delta-p_4\cdot \Delta-p_5\cdot \Delta)^{j_1-j_2}
    \nonumber
    \\
    &&
+ t^{a_4} t^{a_5} t^{a_6} t^{a_3} 
    (q\cdot \Delta-p_1\cdot \Delta-p_4\cdot \Delta)^{j_3-j_4} (q\cdot \Delta-p_1\cdot \Delta-p_4\cdot \Delta-p_5\cdot \Delta)^{j_2-j_3} 
    \nonumber
    \\
    &&\qquad (q\cdot \Delta-p_1\cdot \Delta-p_4\cdot \Delta-p_5\cdot \Delta-p_6\cdot \Delta)^{j_1-j_2}
    \nonumber
    \\
    &&
+ t^{a_4} t^{a_6} t^{a_3} t^{a_5} 
    (q\cdot \Delta-p_1\cdot \Delta-p_4\cdot \Delta)^{j_3-j_4} (q\cdot \Delta-p_1\cdot \Delta-p_4\cdot \Delta-p_6\cdot \Delta)^{j_2-j_3} 
    \nonumber
    \\
    &&\qquad (q\cdot \Delta-p_1\cdot \Delta-p_3\cdot \Delta-p_4\cdot \Delta-p_6\cdot \Delta)^{j_1-j_2}
    \nonumber
    \\
    &&
+ t^{a_4} t^{a_6} t^{a_5} t^{a_3} 
    (q\cdot \Delta-p_1\cdot \Delta-p_4\cdot \Delta)^{j_3-j_4} (q\cdot \Delta-p_1\cdot \Delta-p_4\cdot \Delta-p_6\cdot \Delta)^{j_2-j_3} 
    \nonumber
    \\
    &&\qquad (q\cdot \Delta-p_1\cdot \Delta-p_4\cdot \Delta-p_5\cdot \Delta-p_6\cdot \Delta)^{j_1-j_2}
    \nonumber
    \\
    &&
+ t^{a_5} t^{a_3} t^{a_4} t^{a_6} 
    (q\cdot \Delta-p_1\cdot \Delta-p_5\cdot \Delta)^{j_3-j_4} (q\cdot \Delta-p_1\cdot \Delta-p_3\cdot \Delta-p_5\cdot \Delta)^{j_2-j_3} 
    \nonumber
    \\
    &&\qquad (q\cdot \Delta-p_1\cdot \Delta-p_3\cdot \Delta-p_4\cdot \Delta-p_5\cdot \Delta)^{j_1-j_2}
    \nonumber
    \\
    &&
+ t^{a_5} t^{a_3} t^{a_6} t^{a_4} 
    (q\cdot \Delta-p_1\cdot \Delta-p_5\cdot \Delta)^{j_3-j_4} (q\cdot \Delta-p_1\cdot \Delta-p_3\cdot \Delta-p_5\cdot \Delta)^{j_2-j_3} 
    \nonumber
    \\
    &&\qquad (q\cdot \Delta-p_1\cdot \Delta-p_3\cdot \Delta-p_5\cdot \Delta-p_6\cdot \Delta)^{j_1-j_2}
    \nonumber
    \\
    &&
+ t^{a_5} t^{a_4} t^{a_3} t^{a_6} 
    (q\cdot \Delta-p_1\cdot \Delta-p_5\cdot \Delta)^{j_3-j_4} (q\cdot \Delta-p_1\cdot \Delta-p_4\cdot \Delta-p_5\cdot \Delta)^{j_2-j_3} 
    \nonumber
    \\
    &&\qquad (q\cdot \Delta-p_1\cdot \Delta-p_3\cdot \Delta-p_4\cdot \Delta-p_5\cdot \Delta)^{j_1-j_2}
    \nonumber
    \\
    &&
+ t^{a_5} t^{a_4} t^{a_6} t^{a_3} 
    (q\cdot \Delta-p_1\cdot \Delta-p_5\cdot \Delta)^{j_3-j_4} (q\cdot \Delta-p_1\cdot \Delta-p_4\cdot \Delta-p_5\cdot \Delta)^{j_2-j_3} 
    \nonumber
    \\
    &&\qquad (q\cdot \Delta-p_1\cdot \Delta-p_4\cdot \Delta-p_5\cdot \Delta-p_6\cdot \Delta)^{j_1-j_2}
    \nonumber
    \\
    &&
+ t^{a_5} t^{a_6} t^{a_3} t^{a_4} 
    (q\cdot \Delta-p_1\cdot \Delta-p_5\cdot \Delta)^{j_3-j_4} (q\cdot \Delta-p_1\cdot \Delta-p_5\cdot \Delta-p_6\cdot \Delta)^{j_2-j_3} 
    \nonumber
    \\
    &&\qquad (q\cdot \Delta-p_1\cdot \Delta-p_3\cdot \Delta-p_5\cdot \Delta-p_6\cdot \Delta)^{j_1-j_2}
    \nonumber
    \\
    &&
+ t^{a_5} t^{a_6} t^{a_4} t^{a_3} 
    (q\cdot \Delta-p_1\cdot \Delta-p_5\cdot \Delta)^{j_3-j_4} (q\cdot \Delta-p_1\cdot \Delta-p_5\cdot \Delta-p_6\cdot \Delta)^{j_2-j_3} 
    \nonumber
    \\
    &&\qquad (q\cdot \Delta-p_1\cdot \Delta-p_4\cdot \Delta-p_5\cdot \Delta-p_6\cdot \Delta)^{j_1-j_2}
    \nonumber
    \\
    &&
+ t^{a_6} t^{a_3} t^{a_4} t^{a_5} 
    (q\cdot \Delta-p_1\cdot \Delta-p_6\cdot \Delta)^{j_3-j_4} (q\cdot \Delta-p_1\cdot \Delta-p_3\cdot \Delta-p_6\cdot \Delta)^{j_2-j_3} 
    \nonumber
    \\
    &&\qquad (q\cdot \Delta-p_1\cdot \Delta-p_3\cdot \Delta-p_4\cdot \Delta-p_6\cdot \Delta)^{j_1-j_2}
    \nonumber
    \\
    &&
+ t^{a_6} t^{a_3} t^{a_5} t^{a_4} 
    (q\cdot \Delta-p_1\cdot \Delta-p_6\cdot \Delta)^{j_3-j_4} (q\cdot \Delta-p_1\cdot \Delta-p_3\cdot \Delta-p_6\cdot \Delta)^{j_2-j_3} 
    \nonumber
    \\
    &&\qquad (q\cdot \Delta-p_1\cdot \Delta-p_3\cdot \Delta-p_5\cdot \Delta-p_6\cdot \Delta)^{j_1-j_2}
    \nonumber
    \\
    &&
+ t^{a_6} t^{a_4} t^{a_3} t^{a_5} 
    (q\cdot \Delta-p_1\cdot \Delta-p_6\cdot \Delta)^{j_3-j_4} (q\cdot \Delta-p_1\cdot \Delta-p_4\cdot \Delta-p_6\cdot \Delta)^{j_2-j_3} 
    \nonumber
    \\
    &&\qquad (q\cdot \Delta-p_1\cdot \Delta-p_3\cdot \Delta-p_4\cdot \Delta-p_6\cdot \Delta)^{j_1-j_2}
    \nonumber
    \\
    &&
+ t^{a_6} t^{a_4} t^{a_5} t^{a_3} 
    (q\cdot \Delta-p_1\cdot \Delta-p_6\cdot \Delta)^{j_3-j_4} (q\cdot \Delta-p_1\cdot \Delta-p_4\cdot \Delta-p_6\cdot \Delta)^{j_2-j_3} 
    \nonumber
    \\
    &&\qquad (q\cdot \Delta-p_1\cdot \Delta-p_4\cdot \Delta-p_5\cdot \Delta-p_6\cdot \Delta)^{j_1-j_2}
    \nonumber
    \\
    &&
+ t^{a_6} t^{a_5} t^{a_3} t^{a_4} 
    (q\cdot \Delta-p_1\cdot \Delta-p_6\cdot \Delta)^{j_3-j_4} (q\cdot \Delta-p_1\cdot \Delta-p_5\cdot \Delta-p_6\cdot \Delta)^{j_2-j_3} 
    \nonumber
    \\
    &&\qquad (q\cdot \Delta-p_1\cdot \Delta-p_3\cdot \Delta-p_5\cdot \Delta-p_6\cdot \Delta)^{j_1-j_2}
    \nonumber
    \\
    &&
+ t^{a_6} t^{a_5} t^{a_4} t^{a_3} 
    (q\cdot \Delta-p_1\cdot \Delta-p_6\cdot \Delta)^{j_3-j_4} (q\cdot \Delta-p_1\cdot \Delta-p_5\cdot \Delta-p_6\cdot \Delta)^{j_2-j_3} 
    \nonumber
    \\
    &&\qquad (q\cdot \Delta-p_1\cdot \Delta-p_4\cdot \Delta-p_5\cdot \Delta-p_6\cdot \Delta)^{j_1-j_2} 
    \biggr)
\end{eqnarray}

\renewcommand{\theequation}{\ref{sec:appB}.\arabic{equation}}
\setcounter{equation}{0}
\renewcommand{\thefigure}{\ref{sec:appB}.\arabic{figure}}
\setcounter{figure}{0}
\renewcommand{\thetable}{\ref{sec:appB}.\arabic{table}}
\setcounter{table}{0}

\vspace*{-2mm}
\section{Mellin moments at four loops}
\label{sec:appB}
Here we present the anomalous dimensions at four loops for $1 \le N \le 16$. 
Obviously, $\gamma_{\rm ns}^{\,-}(N\!=\!1) = 0$ at all orders.
To fix our normalization, we write down the complete expression for $N=2$
including all lower orders. Recall that $\ars = \als/(4\pi)$.
\bea
{\lefteqn{
\gamma_{\rm ns}^{\,+}(N=2) \,=\,}}
\nonumber
\\
&& \nonumber
\ars \* \Biggl\{ 
        \frac{8}{3} \* \cf 
        \Biggr\}
+
\ars^2 \* \Biggl\{
        - \frac{112}{27} \* \cfs 
        + \frac{376}{27} \* \ca \* \cf 
        - \frac{64}{27} \* \nf \* \cf
        \Biggr\}
\\
&& \nonumber
+
\ars^3 \* \Biggl\{
          \cft \* \Biggl[  - \frac{560}{243} + \frac{128}{3}\*\zeta_3 \Biggr]
        + \ca\*\cfs \* \Biggl[  - \frac{8528}{243} - 64\*\zeta_3 \Biggr]
        + \cas\*\cf \* \Biggl[ \frac{20920}{243} + \frac{64}{3}\*\zeta_3 \Biggr]
\\
&& \nonumber
\hspace*{5mm}
        + \nf\*\cfs \* \Biggl[  - \frac{3412}{243} + \frac{64}{3}\*\zeta_3 \Biggr]
        + \nf\*\ca\*\cf \* \Biggl[  - \frac{3128}{243} - \frac{64}{3}\*\zeta_3 \Biggr]
        + \nfs\*\cf \* \Biggl[ - \frac{224}{243}  \Biggr]
        \Biggr\}
\\
&& \nonumber
+
\ars^4 \* \Biggl\{
          \cff \* \Biggl[ \frac{194392}{2187} + \frac{10880}{81}\*\zeta_3 - \frac{1280}{3}\*\zeta_5 \Biggr]
        + \ca\*\cft \* \Biggl[ \frac{238676}{2187} + \frac{31040}{81}\*\zeta_3 - \frac{704}{3}\*\zeta_4 + \frac{1280}{3}\*\zeta_5 \Biggr]
\\
&& \nonumber
\hspace*{5mm}
        + \cas\*\cfs \* \Biggl[  - \frac{1626064}{2187} - \frac{25744}{27}\*\zeta_3 + 352\*\zeta_4 + \frac{4480}{9}\*\zeta_5 \Biggr]
\\
&& \nonumber
\hspace*{5mm}
        + \cat\*\cf \* \Biggl[ \frac{1734130}{2187} + \frac{34936}{81}\*\zeta_3 - \frac{352}{3}\*\zeta_4 - \frac{12160}{27}\*\zeta_5 \Biggr]
\\
&& \nonumber
\hspace*{5mm}
        + \nf\*\cft \* \Biggl[ \frac{190912}{2187} - \frac{2528}{81}\*\zeta_3 + \frac{128}{3}\*\zeta_4 - \frac{640}{3}\*\zeta_5 \Biggr]
\\
&& \nonumber
\hspace*{5mm}
        + \nf\*\ca\*\cfs \* \Biggl[  - \frac{177748}{2187} + \frac{12928}{27}\*\zeta_3 - \frac{544}{3}\*\zeta_4 + \frac{320}{9}\*\zeta_5 \Biggr]
\\
&& \nonumber
\hspace*{5mm}
        + \nf\*\cas\*\cf \* \Biggl[  - \frac{53018}{243} - \frac{4040}{9}\*\zeta_3 + \frac{416}{3}\*\zeta_4 + \frac{4480}{27}\*\zeta_5 \Biggr]
        + \nfs\*\cfs \* \Biggl[ \frac{24944}{2187} - \frac{128}{3}\*\zeta_3 + \frac{64}{3}\*\zeta_4 \Biggr]
\\
&& \nonumber
\hspace*{5mm}
        + \nfs\*\ca\*\cf \* \Biggl[ \frac{6350}{729} + \frac{128}{3}\*\zeta_3 - \frac{64}{3}\*\zeta_4 \Biggr]
        + \nft\*\cf \* \Biggl[  - \frac{1024}{2187} + \frac{128}{81}\*\zeta_3 \Biggr]
\\
&& \nonumber
\hspace*{5mm}
        + \dfFAnc \* \Biggl[  - \frac{736}{9} + \frac{1984}{9}\*\zeta_3 + \frac{5120}{9}\*\zeta_5 \Biggr]
        + \nf\*\dfFFnc \* \Biggl[ \frac{832}{9} + \frac{1024}{9}\*\zeta_3 - \frac{2560}{9}\*\zeta_5 \Biggr]
        \Biggr\}
\:\: . \\
\eea

\noindent
The four-loop coefficient for the even moments $N = 4$ to $N=16$ are given by
\bea
{\lefteqn{
\gamma_{\rm ns}^{\,(3)+}(N=4) \,=\,}}
\nonumber
\\
&& \nonumber
\cff \* \Biggl[ \frac{3482407012657}{34992000000} + \frac{14504764}{50625}\*\zeta_3 - \frac{25136}{45}\*\zeta_5 \Biggr]
\\
&& \nonumber
   + \ca\*\cft \* \Biggl[ \frac{33802068299}{174960000} - \frac{10215349}{81000}\*\zeta_3 - \frac{15829}{75}\*\zeta_4
			  + \frac{33004}{45}\*\zeta_5 \Biggr]
\\
&& \nonumber
   + \cas\*\cfs \* \Biggl[ - \frac{1557367902137}{1749600000} - \frac{2497339}{16875}\*\zeta_3 
			   + \frac{15829}{50}\*\zeta_4 - \frac{1645}{9}\*\zeta_5 \Biggr]
\\
&& \nonumber
   + \cat\*\cf \* \Biggl[ \frac{49455970561}{43740000} + \frac{13461191}{81000}\*\zeta_3 - \frac{15829}{150}\*\zeta_4
			  - \frac{18646}{135}\*\zeta_5 \Biggr]
\\
&& \nonumber
   + \nf\*\cft \* \Biggl[ \frac{29581840417}{174960000} + \frac{820961}{10125}\*\zeta_3 + \frac{2878}{75}\*\zeta_4 
			  - \frac{1256}{3}\*\zeta_5 \Biggr]
\\
&& \nonumber
   + \nf\*\ca\*\cfs \* \Biggl[ - \frac{89325051233}{437400000} + \frac{4588639}{6750}\*\zeta_3 - \frac{21587}{75}\*\zeta_4
			       + \frac{628}{9}\*\zeta_5 \Biggr]
\\
&& \nonumber
   + \nf\*\cas\*\cf \* \Biggl[ - \frac{1796654459}{4860000} - \frac{5247961}{6750}\*\zeta_3 + \frac{18709}{75}\*\zeta_4
			       + \frac{43736}{135}\*\zeta_5 \Biggr]
\\
&& \nonumber
   + \nfs\*\cfs \* \Biggl[ \frac{5419760639}{218700000} - \frac{2146}{25}\*\zeta_3 + \frac{628}{15}\*\zeta_4 \Biggr]
   + \nfs\*\ca\*\cf \* \Biggl[ \frac{60167591}{3645000} + \frac{2146}{25}\*\zeta_3 - \frac{628}{15}\*\zeta_4 \Biggr]
\\
&& \nonumber
   + \nft\*\cf \* \Biggl[ - \frac{17813699}{21870000} + \frac{1256}{405}\*\zeta_3 \Biggr]
   + \dfFAnc \* \Biggl[ \frac{254713}{1350} + \frac{63568}{45}\*\zeta_3 - \frac{78868}{45}\*\zeta_5 \Biggr]
\\
&&
   + \nf\*\dfFFnc \* \Biggl[ \frac{16568}{135} + \frac{22552}{75}\*\zeta_3 - \frac{26912}{45}\*\zeta_5 \Biggr]
\:\: ,
\eea
\bea
{\lefteqn{
\gamma_{\rm ns}^{\,(3)+}(N=6) \,=\,}}
\nonumber
\\
&& \nonumber
\cff \* \Biggl[ \frac{287471623549488131}{1801088541000000} 
          + \frac{33026498018}{121550625}\*\zeta_3 - \frac{450712}{735}\*\zeta_5 \Biggr]
\\
&& \nonumber
   + \ca\*\cft \* \Biggl[ \frac{9809771626657}{476478450000} - \frac{9025033804}{121550625}\*\zeta_3
          - \frac{768482}{3675}\*\zeta_4 + \frac{621976}{735}\*\zeta_5 \Biggr]
\\
&& \nonumber
   + \cas\*\cfs \* \Biggl[ - \frac{588570119595401}{918922725000} + \frac{3466052671}{27011250}\*\zeta_3 
          + \frac{384241}{1225}\*\zeta_4 - \frac{1695382}{2205}\*\zeta_5 \Biggr]
\\
&& \nonumber
   + \cat\*\cf \* \Biggl[ \frac{64997866579309}{56010528000} + \frac{395253829}{27783000}\*\zeta_3 
          - \frac{384241}{3675}\*\zeta_4 + \frac{1378042}{6615}\*\zeta_5 \Biggr]
\\
&& \nonumber
       + \nf\*\cft \* \Biggl[ \frac{2697261071752787}{12864918150000} + \frac{2830802}{19845}\*\zeta_3 
          + \frac{139724}{3675}\*\zeta_4 - \frac{11344}{21}\*\zeta_5 \Biggr]
\\
&& \nonumber
       + \nf\*\ca\*\cfs \* \Biggl[  - \frac{160989027584717}{612615150000} + \frac{8830621}{11025}\*\zeta_3 
          - \frac{1301446}{3675}\*\zeta_4 + \frac{5672}{63}\*\zeta_5 \Biggr]
\\
&& \nonumber
       + \nf\*\cas\*\cf \* \Biggl[ - \frac{96176330975533}{210039480000} - \frac{1122625478}{1157625}\*\zeta_3
          + \frac{1161722}{3675}\*\zeta_4 + \frac{548152}{1323}\*\zeta_5 \Biggr]
\\
&& \nonumber
       + \nfs\*\cfs \* \Biggl[ \frac{5177594850433}{153153787500} - \frac{3697676}{33075}\*\zeta_3 
          + \frac{5672}{105}\*\zeta_4 \Biggr]
\\
&& \nonumber
       + \nfs\*\ca\*\cf \* \Biggl[ \frac{348632126303}{17503290000} + \frac{3697676}{33075}\*\zeta_3 
          - \frac{5672}{105}\*\zeta_4 \Biggr]
       + \nft\*\cf \* \Biggl[  - \frac{26381269339}{26254935000} + \frac{11344}{2835}\*\zeta_3 \Biggr]
\\
&& \nonumber
       + \dfFAnc \* \Biggl[ \frac{1816777297}{2646000} + \frac{48954263}{18375}\*\zeta_3 
          - \frac{8666912}{2205}\*\zeta_5 \Biggr]
\\
&&
       + \nf\*\dfFFnc \* \Biggl[ \frac{70029019}{330750} + \frac{4621504}{11025}\*\zeta_3 - \frac{379264}{441}\*\zeta_5 \Biggr]
\:\: ,
\eea
\bea
{\lefteqn{
\gamma_{\rm ns}^{\,(3)+}(N=8) \,=\,}}
\nonumber
\\
&& \nonumber
\cff \* \Biggl[ \frac{1526099947627950150458381}{8067032349014016000000} 
          + \frac{20349260276089}{78764805000}\*\zeta_3 - \frac{1804282}{2835}\*\zeta_5 \Biggr]
\\
&& \nonumber
       + \ca\*\cft \* \Biggl[  - \frac{1421752209961884815677}{19207219878604800000} 
          - \frac{3387289715833}{140026320000}\*\zeta_3 - \frac{27614477}{132300}\*\zeta_4 
          + \frac{17728727}{19845}\*\zeta_5 \Biggr]
\\
&& \nonumber
       + \cas\*\cfs \* \Biggl[  - \frac{136950111821419605151}{304876506009600000} 
          + \frac{81027197994671}{210039480000}\*\zeta_3 + \frac{27614477}{88200}\*\zeta_4 
          - \frac{4834637}{3780}\*\zeta_5 \Biggr]
\\
&& \nonumber
       + \cat\*\cf \* \Biggl[ \frac{1008933995789394973}{871075731456000} 
          - \frac{33568622483}{244944000}\*\zeta_3 - \frac{27614477}{264600}\*\zeta_4 + \frac{31576003}{59535}\*\zeta_5 \Biggr]
\\
&& \nonumber
       + \nf\*\cft \* \Biggl[ \frac{1136310990563809311301}{4801804969651200000} 
          + \frac{5612836699}{30005640}\*\zeta_3 + \frac{2510407}{66150}\*\zeta_4 - \frac{39532}{63}\*\zeta_5 \Biggr]
\\
&& \nonumber
       + \nf\*\ca\*\cfs \* \Biggl[  - \frac{11533221344476533811}{38109563251200000} 
          + \frac{88755158137}{100018800}\*\zeta_3 - \frac{17730227}{44100}\*\zeta_4 + \frac{19766}{189}\*\zeta_5 \Biggr]
\\
&& \nonumber
       + \nf\*\cas\*\cf \* \Biggl[  - \frac{189224450730372949}{362948221440000} 
          - \frac{1661823807613}{1500282000}\*\zeta_3 + \frac{48169867}{132300}\*\zeta_4 + \frac{813749}{1701}\*\zeta_5 \Biggr]
\\
&& \nonumber
       + \nfs\*\cfs \* \Biggl[ \frac{386628670506434189}{9527390812800000} 
          - \frac{77750543}{595350}\*\zeta_3 + \frac{19766}{315}\*\zeta_4 \Biggr]
\\
&& \nonumber
       + \nfs\*\ca\*\cf \* \Biggl[ \frac{2001327622434827}{90737055360000} + \frac{77750543}{595350}\*\zeta_3
          - \frac{19766}{315}\*\zeta_4 \Biggr]
\\
&& \nonumber
       + \nft\*\cf \* \Biggl[  - \frac{77182042631063}{68052791520000} + \frac{39532}{8505}\*\zeta_3 \Biggr]
\\
&& \nonumber
       + \dfFAnc \* \Biggl[ \frac{245446938529151}{216040608000} + \frac{964237792849}{250047000}\*\zeta_3
          - \frac{117093083}{19845}\*\zeta_5 \Biggr]
\\
&&
       + \nf\*\dfFFnc \* \Biggl[ \frac{1164731326841}{3857868000} + \frac{892046473}{1786050}\*\zeta_3 
          - \frac{605768}{567}\*\zeta_5 \Biggr]
\:\: ,
\eea
\bea
{\lefteqn{
\gamma_{\rm ns}^{\,(3)+}(N=10) \,=\,}}
\nonumber
\\
&& \nonumber
    \cff \* \Biggl[ \frac{2017383724760695233171991134991}{9825227427985488199296000000}
          + \frac{143667693462054187}{576597755002500}\*\zeta_3 
\\
&& \nonumber
\hspace*{5mm}
          - \frac{222343766}{343035}\*\zeta_5 \Biggr]
\\
&& \nonumber
    + \ca\*\cft \* \Biggl[  - \frac{1244204474536625986344420833}{9666693652091192640000000}
          + \frac{8470409428966531}{768797006670000}\*\zeta_3 
\\
&& \nonumber
\hspace*{5mm}
          - \frac{151796299}{727650}\*\zeta_4 
          + \frac{440406986}{480249}\*\zeta_5 \Biggr]
\\
&& \nonumber
       + \cas\*\cfs \* \Biggl[  - \frac{8344774766025843618923173}{27898105778040960000000} 
          + \frac{488048876510787149}{768797006670000}\*\zeta_3 
\\
&& \nonumber
\hspace*{5mm}
          + \frac{151796299}{485100}\*\zeta_4 
          - \frac{79369519}{45738}\*\zeta_5 \Biggr]
\\
&& \nonumber
       + \cat\*\cf \* \Biggl[ \frac{37809477198339250105213}{32937551095680000000} 
          - \frac{1032212221214159}{3630682440000}\*\zeta_3 - \frac{151796299}{1455300}\*\zeta_4 
\\
&& \nonumber
\hspace*{5mm}
          + \frac{2378996101}{2881494}\*\zeta_5 \Biggr]
\\
&& \nonumber
       + \nf\*\cft \* \Biggl[ \frac{273095498113782515546834719}{1063336301730031190400000} 
          + \frac{110938456177789}{499218835500}\*\zeta_3 
\\
&& \nonumber
\hspace*{5mm}
          + \frac{151796299}{4002075}\*\zeta_4 
          - \frac{482200}{693}\*\zeta_5 \Biggr]
\\
&& \nonumber
       + \nf\*\ca\*\cfs \* \Biggl[  - \frac{232521423145307204017937}{697452644451024000000} 
          + \frac{317697741600923}{332812557000}\*\zeta_3 
\\
&& \nonumber
\hspace*{5mm}
          - \frac{1172854799}{2668050}\*\zeta_4 
          + \frac{241100}{2079}\*\zeta_5 \Biggr]
\\
&& \nonumber
       + \nf\*\cas\*\cf \* \Biggl[  - \frac{3449950935218411700067}{6038551034208000000} 
          - \frac{168584409069107}{138671898750}\*\zeta_3 
\\
&& \nonumber
\hspace*{5mm}
          + \frac{3214971799}{8004150}\*\zeta_4 
          + \frac{181204484}{343035}\*\zeta_5 \Biggr]
\\
&& \nonumber
       + \nfs\*\cfs \* \Biggl[ \frac{17653381355656128339793}{383598954448063200000} 
          - \frac{5236700507}{36018675}\*\zeta_3 
          + \frac{48220}{693}\*\zeta_4 \Biggr]
\\
&& \nonumber
       + \nfs\*\ca\*\cf \* \Biggl[ \frac{7133251712396708693}{301927551710400000} 
          + \frac{5236700507}{36018675}\*\zeta_3 - \frac{48220}{693}\*\zeta_4 \Biggr]
\\
&& \nonumber
       + \nft\*\cf \* \Biggl[ - \frac{613551152411968391}{498180460322160000} + \frac{96440}{18711}\*\zeta_3 \Biggr]
\\
&& \nonumber
       + \dfFAnc \* \Biggl[ \frac{6387465101013091}{4149351360000} 
          + \frac{751245802206203}{151278435000}\*\zeta_3 - \frac{18465293144}{2401245}\*\zeta_5 \Biggr]
\\
&&
       + \nf\*\dfFFnc \* \Biggl[ \frac{51618947156153}{134469720000} + \frac{18626595374}{33350625}\*\zeta_3
          - \frac{141624128}{114345}\*\zeta_5 \Biggr]
\:\: ,
\eea
\bea
{\lefteqn{
\gamma_{\rm ns}^{\,(3)+}(N=12) \,=\,}}
\nonumber
\\
&& \nonumber
\cff \* \Biggl[ \frac{331516931678819616019516119238120540753}{1541296125734534205067061110080000000} 
			 + \frac{3996930795976177310551}{16468208480626402500}\*\zeta_3 
\\
&& \nonumber
\hspace*{5mm}
			 - \frac{37964613926}{57972915}\*\zeta_5 \Biggr]
\\
&& \nonumber
       + \ca\*\cft \* \Biggl[  - \frac{41810347217755725651032771751930697}{256626061560861506005171680000000} 
			       + \frac{98597799463449331819}{2744701413437733750}\*\zeta_3 
\\
&& \nonumber
\hspace*{5mm}
			       - \frac{25648239313}{122972850}\*\zeta_4 
          + \frac{34328650826}{36891855}\*\zeta_5 \Biggr]
\\
&& \nonumber
       + \cas\*\cfs \* \Biggl[  - \frac{238851091715148666377698033403459}{1367305023301293405288960000000} 
          + \frac{38330693033675619062839}{43915222615003740000}\*\zeta_3 
\\
&& \nonumber
\hspace*{5mm}
          + \frac{25648239313}{81981900}\*\zeta_4 - \frac{27699141079}{12882870}\*\zeta_5 \Biggr]
\\
&& \nonumber
       + \cat\*\cf \* \Biggl[ \frac{516560961693213572859380630401}{455313028072358776320000000} - \frac{3552780661279687493}{8356447859760000}\*\zeta_3
\\
&& \nonumber
\hspace*{5mm}
          - \frac{25648239313}{245945700}\*\zeta_4 + \frac{242831514391}{221351130}\*\zeta_5 \Biggr]
\\
&& \nonumber
\hspace*{5mm}
       + \nf\*\cft \* \Biggl[ \frac{14007724721120701350168397616773189}{51325212312172301201034336000000} 
          + \frac{275829519782497567}{1096783781593500}\*\zeta_3 
\\
&& \nonumber
\hspace*{5mm}
          + \frac{25648239313}{676350675}\*\zeta_4 
          - \frac{6774784}{9009}\*\zeta_5 \Biggr]
\\
&& \nonumber
       + \nf\*\ca\*\cfs \* \Biggl[  - \frac{12260374152873668292904144517329}{34182625582532335132224000000} 
				    + \frac{246056489027368057}{243729729243000}\*\zeta_3
\\
&& \nonumber
\hspace*{5mm}
          - \frac{212141105873}{450900450}\*\zeta_4 + \frac{3387392}{27027}\*\zeta_5 \Biggr]
\\
&& \nonumber
       + \nf\*\cas\*\cf \* \Biggl[  - \frac{581173376177155203686777699}{948568808484080784000000} 
				    - \frac{119245993023506038}{91398648466125}\*\zeta_3 
\\
&& \nonumber
\hspace*{5mm}
          + \frac{585126838993}{1352701350}\*\zeta_4 + \frac{599861228}{1054053}\*\zeta_5 \Biggr]
\\
&& \nonumber
\hspace*{5mm}
	 + \nfs\*\cfs \* \Biggl[ \frac{14406272473786523198816959409}{284855213187769459435200000}
          - \frac{191866603189}{1217431215}\*\zeta_3 + \frac{3387392}{45045}\*\zeta_4 \Biggr]
\\
&& \nonumber
       + \nfs\*\ca\*\cf \* \Biggl[ \frac{442446968888889719821403}{17785665159076514700000} 
          + \frac{191866603189}{1217431215}\*\zeta_3 - \frac{3387392}{45045}\*\zeta_4 \Biggr]
\\
&& \nonumber
       + \nft\*\cf \* \Biggl[  - \frac{18633947419926787652303}{14228532127261211760000} 
          + \frac{6774784}{1216215}\*\zeta_3 \Biggr]
\\
&& \nonumber
	 + \dfFAnc \* \Biggl[ \frac{55568956318895924814773}{29157574993747200000}
          + \frac{224835066938602217}{37496881422000}\*\zeta_3 - \frac{3785627672012}{405810405}\*\zeta_5 \Biggr]
\\
&&
      + \nf\*\dfFFnc \* \Biggl[ \frac{22664035135268075399}{49587712574400000} 
          + \frac{18402607883189}{30435780375}\*\zeta_3 - \frac{485441696}{351351}\*\zeta_5 \Biggr]
\:\: ,
\eea
\bea
{\lefteqn{
\gamma_{\rm ns}^{\,(3)+}(N=14) \,=\,}}
\nonumber
\\
&& \nonumber
\cff \* \Biggl[ \frac{97523736280058513278817419051773381749}{440370321638438344304874602880000000} 
          + \frac{1567683874499713548181}{6587283392250561000}\*\zeta_3 
\\
&& \nonumber
\hspace*{5mm}
          - \frac{267445675058}{405810405}\*\zeta_5 \Biggr]
\\
&& \nonumber
       + \ca\*\cft \* \Biggl[  - \frac{333627771877952197537388891905440907}{1796382430926030542036201760000000} 
          + \frac{118827371645736248177}{2195761130750187000}\*\zeta_3 
\\
&& \nonumber
\hspace*{5mm}
          - \frac{3663695353}{17567550}\*\zeta_4 + \frac{6927566278}{7378371}\*\zeta_5 \Biggr]
\\
&& \nonumber
       + \cas\*\cfs \* \Biggl[  - \frac{86656353251877447263713613209043}{1288422041187757247291520000000} 
          + \frac{6031813059800336882399}{5489402826875467500}\*\zeta_3 
\\
&& \nonumber
\hspace*{5mm}
          + \frac{3663695353}{11711700}\*\zeta_4 - \frac{45655020935}{18036018}\*\zeta_5 \Biggr]
\\
&& \nonumber
       + \cat\*\cf \* \Biggl[ \frac{3974468919372874040522232093269}{3549372014291342279040000000} 
			      - \frac{8803366113150246019}{15748690197240000}\*\zeta_3 
\\
&& \nonumber
\hspace*{5mm}
			      - \frac{3663695353}{35135100}\*\zeta_4 + \frac{59678850515}{44270226}\*\zeta_5 \Biggr]
\\
&& \nonumber
       + \nf\*\cft \* \Biggl[ \frac{1469298768736788933360199563898027}{5132521231217230120103433600000} 
          + \frac{303399887420914357}{1096783781593500}\*\zeta_3 
\\
&& \nonumber
\hspace*{5mm}
          + \frac{3663695353}{96621525}\*\zeta_4 
          - \frac{7204840}{9009}\*\zeta_5 \Biggr]
\\
&& \nonumber
       + \nf\*\ca\*\cfs \* \Biggl[  - \frac{22747053762883618602535741424929}{59819594769431586481392000000} 
				    + \frac{257384193913101469}{243729729243000}\*\zeta_3
\\
&& \nonumber
\hspace*{5mm}
          - \frac{31996728653}{64414350}\*\zeta_4 + \frac{3602420}{27027}\*\zeta_5 \Biggr]
\\
&& \nonumber
       + \nf\*\cas\*\cf \* \Biggl[  - \frac{15060401544121460041592815421}{23239935807859979208000000} 
				    - \frac{17663095878365457683}{12795810785257500}\*\zeta_3
\\
&& \nonumber
\hspace*{5mm}
          + \frac{88662795253}{193243050}\*\zeta_4 + \frac{131795548}{218295}\*\zeta_5 \Biggr]
\\
&& \nonumber
				    + \nfs\*\cfs \* \Biggl[ \frac{776154252048358965658599007}{14242760659388472971760000}
          - \frac{1022665028447}{6087156075}\*\zeta_3 + \frac{720484}{9009}\*\zeta_4 \Biggr]
\\
&& \nonumber
       + \nfs\*\ca\*\cf \* \Biggl[ \frac{10327818039687232051823761}{398398899563313929280000} 
          + \frac{1022665028447}{6087156075}\*\zeta_3 - \frac{720484}{9009}\*\zeta_4 \Biggr]
\\
&& \nonumber
       + \nft\*\cf \* \Biggl[  - \frac{3911903561688643214011}{2845706425452242352000} 
          + \frac{1440968}{243243}\*\zeta_3 \Biggr]
\\
&& \nonumber
       + \dfFAnc \* \Biggl[ \frac{1062557701742240655559903}{473810593648392000000} 
          + \frac{59328382591268105459}{8530540523505000}\*\zeta_3 
\\
&& \nonumber
\hspace*{5mm}
          - \frac{4399336453016}{405810405}\*\zeta_5 \Biggr]
\\
&& \nonumber
       + \nf\*\dfFFnc \* \Biggl[ \frac{144413233152647568760141}{276389512961562000000} 
          + \frac{4789034376385918}{7456766191875}\*\zeta_3 - \frac{1423427008}{945945}\*\zeta_5 \Biggr]
\:\: , \\
\eea
\bea
{\lefteqn{
\gamma_{\rm ns}^{\,(3)+}(N=16) \,=\,}}
\nonumber
\\
&& \nonumber
         \cff \* \Biggl[ 
       \frac{585091838928074604746949501506640014143607792143679}{2590529154801458855855422389944869371248640000000} 
\\
&& \nonumber
\hspace*{5mm}
       + 
       \frac{10317600287587322672417525639}{44014119696332728422480000}\*\zeta_3 
       - 
       \frac{22176170947759}{33508344870}\*\zeta_5 \Biggr]
\\
&& \nonumber
       + \ca\*\cft \* \Biggl[  
       -
       \frac{1377628380634900156963463567256914633249848729}{6830915722139930955540672272529162240000000} 
\\
&& \nonumber
\hspace*{5mm}
       + 
       \frac{15923437134502520764317446141}{234741971713774551586560000}\*\zeta_3 
       - 
       \frac{59290512768143}{284313229200}\*\zeta_4 + \frac{40278295293893}{42646984380}\*\zeta_5 \Biggr]
\\
&& \nonumber
       + \cas\*\cfs \* \Biggl[ 
       \frac{258914251298466366542816679113637834371}{9203567921633206758178924462080000000} 
\\
&& \nonumber
\hspace*{5mm}
       + 
       \frac{22014640304042236864687668923}{16767283693841039399040000}\*\zeta_3 
       + 
       \frac{59290512768143}{189542152800}\*\zeta_4 - \frac{1804393628665651}{625489104240}\*\zeta_5 \Biggr]
\\
&& \nonumber
       + \cat\*\cf \* \Biggl[ 
       \frac{14183627547279601657840082208324742517}{12843000517166864476020965376000000} 
\\
&& \nonumber
\hspace*{5mm}
       - \frac{2649527312649305104376683}{3862475881756882790400}\*\zeta_3 
       - \frac{59290512768143}{568626458400}\*\zeta_4 + \frac{101176031536771}{63970476570}\*\zeta_5 \Biggr]
\\
&& \nonumber
       + \nf\*\cft \* \Biggl[ 
       \frac{94403012063925170595517180859037562017787057}{317149658527925365792959784081711104000000} 
\\
&& \nonumber
\hspace*{5mm}
       + 
       \frac{25752552164686212840197}{86215979503501848000}\*\zeta_3 
       + \frac{59290512768143}{1563722760600}\*\zeta_4 - \frac{128839202}{153153}\*\zeta_5 \Biggr]
\\
&& \nonumber
       + \nf\*\ca\*\cfs \* \Biggl[  
       - \frac{2893023190248781306760837247560962327027}{7247809738286150322065903013888000000} 
\\
&& \nonumber
\hspace*{5mm}
       + 
       \frac{63010566480259446207079}{57477319669001232000}\*\zeta_3 
       - 
       \frac{541630986863623}{1042481840400}\*\zeta_4 + \frac{64419601}{459459}\*\zeta_5 \Biggr]
\\
&& \nonumber
       + \nf\*\cas\*\cf \* \Biggl[  
       - \frac{674736712195588046827798227902824541}{993803611447435941596860416000000} 
       - \frac{11083840592301151093813}{7663642622533497600}\*\zeta_3 
\\
&& \nonumber
\hspace*{5mm}
       + \frac{1506311935054583}{3127445521200}\*\zeta_4 + \frac{4569981743}{7209972}\*\zeta_5 \Biggr]
\\
&& \nonumber
       + \nfs\*\cfs \* \Biggl[ 
       \frac{1499767303942721479649857809345185131}{25885034779593394007378225049600000} 
       - \frac{2491999394100703}{14073504845400}\*\zeta_3 
\\
&& \nonumber
\hspace*{5mm}
       + \frac{64419601}{765765}\*\zeta_4 \Biggr]
\\
&& \nonumber
       + \nfs\*\ca\*\cf \* \Biggl[ 
       \frac{95217199819008483482793882523327}{3549298612312271219988787200000} 
       + \frac{2491999394100703}{14073504845400}\*\zeta_3 - \frac{64419601}{765765}\*\zeta_4 \Biggr]
\\
&& \nonumber
       + \nft\*\cf \* \Biggl[  
       - \frac{54396936195548207942967542719}{38028199417631477357022720000} + \frac{128839202}{20675655}\*\zeta_3 \Biggr]
\\
&& \nonumber
       + \dfFAnc \* \Biggl[ 
       \frac{13544672049501491403276515417}{5301437484764348928000000} 
       + 
       \frac{3791955378871486958507}{483002686294128000}\*\zeta_3 
\\
&& \nonumber
\hspace*{5mm}
       - 
       \frac{5744913623184647}{469116828180}\*\zeta_5 \Biggr]
\\
&& \nonumber
       + \nf\*\dfFFnc \* \Biggl[ 
       \frac{63048309341247319711503473}{108438494006543500800000} 
       + 
       \frac{1361755723179569}{2020810952160}\*\zeta_3 
\\
&& 
\hspace*{5mm}
       - \frac{12596460722}{7810803}\*\zeta_5 \Biggr]
\:\: .
\eea

The corresponding results for the odd moments $N = 3$ to $N = 15$ read
\bea
{\lefteqn{
\gamma_{\rm ns}^{\,(3)-}(N=3) \,=\,}}
\nonumber
\\
&& \nonumber
         \cff \* \Biggl[ \frac{341611945}{2239488} + \frac{24380}{81}\*\zeta_3 - \frac{2000}{3}\*\zeta_5 \Biggr]
\\
&& \nonumber
       + \cft\*\ca \* \Biggl[ \frac{40709323}{279936} - \frac{140057}{648}\*\zeta_3 - \frac{605}{3}\*\zeta_4 + \frac{2900}{3}\*\zeta_5 \Biggr]
\\
&& \nonumber
       + \cfs\*\cas \* \Biggl[  - \frac{503877829}{559872} - \frac{4843}{27}\*\zeta_3 + \frac{605}{2}\*\zeta_4 - \frac{1325}{9}\*\zeta_5 \Biggr]
\\
&& \nonumber
       + \cf\*\cat \* \Biggl[ \frac{72667541}{69984} + \frac{125219}{648}\*\zeta_3 - \frac{605}{6}\*\zeta_4 - \frac{5950}{27}\*\zeta_5 \Biggr]
\\
&& \nonumber
       + \cft\*\nf \* \Biggl[ \frac{38386673}{279936} + \frac{3493}{81}\*\zeta_3 + \frac{110}{3}\*\zeta_4 - \frac{1000}{3}\*\zeta_5 \Biggr]
\\
&& \nonumber
       + \cfs\*\ca\*\nf \* \Biggl[  - \frac{22941613}{139968} + \frac{31547}{54}\*\zeta_3 - \frac{715}{3}\*\zeta_4 + \frac{500}{9}\*\zeta_5 \Biggr]
\\
&& \nonumber
       + \cf\*\cas\*\nf \* \Biggl[  - \frac{2366971}{7776} - \frac{11483}{18}\*\zeta_3 + \frac{605}{3}\*\zeta_4 + \frac{7000}{27}\*\zeta_5 \Biggr]
\\
&& \nonumber
       + \cfs\*\nfs \* \Biggl[ \frac{1313443}{69984} - \frac{610}{9}\*\zeta_3 + \frac{100}{3}\*\zeta_4 \Biggr]
       + \cf\*\ca\*\nfs \* \Biggl[ \frac{79747}{5832} + \frac{610}{9}\*\zeta_3 - \frac{100}{3}\*\zeta_4 \Biggr]
\\
&& \nonumber
       + \cf\*\nft \* \Biggl[  - \frac{23587}{34992} + \frac{200}{81}\*\zeta_3 \Biggr]
       + \dfFAnc \* \Biggl[  - \frac{85}{3} + \frac{7600}{9}\*\zeta_3 - \frac{7300}{9}\*\zeta_5 \Biggr]
\\
&& 
       + \nf\*\dfFFnc \* \Biggl[ \frac{275}{3} + \frac{1960}{9}\*\zeta_3 - \frac{4000}{9}\*\zeta_5 \Biggr]
\:\: ,
\eea
\bea
{\lefteqn{
\gamma_{\rm ns}^{\,(3)-}(N=5) \,=\,}}
\nonumber
\\
&& \nonumber
 \cff \* \Biggl[ \frac{421526640437}{2187000000} + \frac{16940714}{50625}\*\zeta_3 - \frac{11032}{15}\*\zeta_5 \Biggr]
\\
&& \nonumber
       + \cft\*\ca \* \Biggl[ - \frac{27591809}{12150000} - \frac{2335808}{10125}\*\zeta_3 - \frac{15554}{75}\*\zeta_4 
          + \frac{16408}{15}\*\zeta_5 \Biggr]
\\
&& \nonumber
       + \cfs\*\cas \* \Biggl[ - \frac{151355831129}{218700000} + \frac{413533}{3750}\*\zeta_3 + \frac{7777}{25}\*\zeta_4
          - \frac{6398}{9}\*\zeta_5 \Biggr]
\\
&& \nonumber
       + \cf\*\cat \* \Biggl[ \frac{44149637147}{38880000} + \frac{4424851}{81000}\*\zeta_3 - \frac{7777}{75}\*\zeta_4 
          + \frac{13594}{135}\*\zeta_5 \Biggr]
\\
&& \nonumber
       + \nf\*\cft \* \Biggl[ \frac{20671110851}{109350000} + \frac{1181666}{10125}\*\zeta_3 + \frac{2828}{75}\*\zeta_4 
          - \frac{1456}{3}\*\zeta_5 \Biggr]
\\
&& \nonumber
       + \nf\*\cfs\*\ca \* \Biggl[ - \frac{8570931889}{36450000} + \frac{92921}{125}\*\zeta_3 - \frac{24262}{75}\*\zeta_4 
          + \frac{728}{9}\*\zeta_5 \Biggr]
\\
&& \nonumber
       + \nf\*\cf\*\cas \* \Biggl[ - \frac{36596231437}{87480000} - \frac{2976818}{3375}\*\zeta_3 + \frac{21434}{75}\*\zeta_4
          + \frac{50456}{135}\*\zeta_5 \Biggr]
\\
&& \nonumber
       + \nfs\*\cfs \* \Biggl[ \frac{269584781}{9112500} - \frac{67532}{675}\*\zeta_3 + \frac{728}{15}\*\zeta_4 \Biggr]
       + \nfs\*\cf\*\ca \* \Biggl[ \frac{44950469}{2430000} + \frac{67532}{675}\*\zeta_3 - \frac{728}{15}\*\zeta_4 \Biggr]
\\
&& \nonumber
       + \nft\*\cf \* \Biggl[ - \frac{10064827}{10935000} + \frac{1456}{405}\*\zeta_3 \Biggr]
       + \dfFAnc \* \Biggl[ \frac{38339}{80} + \frac{52857}{25}\*\zeta_3 - \frac{135968}{45}\*\zeta_5 \Biggr]
\\
&&
       + \nf\*\dfFFnc \* \Biggl[ \frac{74501}{450} + \frac{82432}{225}\*\zeta_3 - \frac{33152}{45}\*\zeta_5 \Biggr]
\:\: ,
\eea
\bea
{\lefteqn{
\gamma_{\rm ns}^{\,(3)-}(N=7) \,=\,}}
\nonumber
\\
&& \nonumber
\cff \* \Biggl[ \frac{89260315086226821967}{409847703552000000} 
          + \frac{107378934083}{324135000}\*\zeta_3 - \frac{555362}{735}\*\zeta_5 \Biggr]
\\
&& \nonumber
       + \ca\*\cft \* \Biggl[  - \frac{2731357455838412101}{26347352371200000} 
          - \frac{976102618619}{5186160000}\*\zeta_3 - \frac{1020151}{4900}\*\zeta_4 + \frac{832501}{735}\*\zeta_5 \Biggr]
\\
&& \nonumber
       + \cas\*\cfs \* \Biggl[  - \frac{1842858791204823727}{3763907481600000} 
          + \frac{964491706751}{2593080000}\*\zeta_3 + \frac{3060453}{9800}\*\zeta_4 - \frac{3607811}{2940}\*\zeta_5 \Biggr]
\\
&& \nonumber
       + \cat\*\cf \* \Biggl[ \frac{6856528500444857}{5974456320000} - \frac{42180835897}{444528000}\*\zeta_3
          - \frac{1020151}{9800}\*\zeta_4 + \frac{942569}{2205}\*\zeta_5 \Biggr]
\\
&& \nonumber
       + \nf\*\cft \* \Biggl[ \frac{1462431578723251501}{6586838092800000} + \frac{44264401}{264600}\*\zeta_3
          + \frac{92741}{2450}\*\zeta_4 - \frac{4108}{7}\*\zeta_5 \Biggr]
\\
&& \nonumber
       + \nf\*\cfs\*\ca \* \Biggl[ - \frac{132713980134736771}{470488435200000} 
          + \frac{447543373}{529200}\*\zeta_3 - \frac{1859803}{4900}\*\zeta_4 + \frac{2054}{21}\*\zeta_5 \Biggr]
\\
&& \nonumber
       + \nf\*\cas\*\cf \* \Biggl[ - \frac{6617534246169487}{13442526720000} 
          - \frac{57952538719}{55566000}\*\zeta_3 + \frac{1674321}{4900}\*\zeta_4 + \frac{197849}{441}\*\zeta_5 \Biggr]
\\
&& \nonumber
       + \nfs\*\cfs \* \Biggl[ \frac{1463487948290143}{39207369600000} - \frac{8059127}{66150}\*\zeta_3 
          + \frac{2054}{35}\*\zeta_4 \Biggr]
\\
&& \nonumber
       + \nfs\*\cf\*\ca \* \Biggl[ \frac{23660663137019}{1120210560000} + \frac{8059127}{66150}\*\zeta_3 
          - \frac{2054}{35}\*\zeta_4 \Biggr]
\\
&& \nonumber
       + \nft\*\cf \* \Biggl[  - \frac{902896393223}{840157920000} + \frac{4108}{945}\*\zeta_3 \Biggr]
\\
&& \nonumber
       + \dfFAnc \* \Biggl[ \frac{40269598361}{42336000} + \frac{885141073}{264600}\*\zeta_3 
          - \frac{3727489}{735}\*\zeta_5 \Biggr]
\\
&&
       + \nf\*\dfFFnc \* \Biggl[ \frac{584326699}{2268000} + \frac{30610507}{66150}\*\zeta_3 
          - \frac{142568}{147}\*\zeta_5 \Biggr]
\:\: ,
\eea
\bea
{\lefteqn{
\gamma_{\rm ns}^{\,(3)-}(N=9) \,=\,}}
\nonumber
\\
&& \nonumber
\cff \* \Biggl[ \frac{117290741389735897476581}{504189521813376000000} 
          + \frac{12756265517567}{39382402500}\*\zeta_3 - \frac{2166406}{2835}\*\zeta_5 \Biggr]
\\
&& \nonumber
       + \ca\*\cft \* \Biggl[  - \frac{1414388556845197343069}{8574651731520000000} 
          - \frac{2630697602423}{17503290000}\*\zeta_3 - \frac{13785409}{66150}\*\zeta_4 + \frac{4562074}{3969}\*\zeta_5 \Biggr]
\\
&& \nonumber
       + \cas\*\cfs \* \Biggl[  - \frac{624559062347065092853}{1905478162560000000} 
          + \frac{32816667064709}{52509870000}\*\zeta_3 + \frac{13785409}{44100}\*\zeta_4 - \frac{91553}{54}\*\zeta_5 \Biggr]
\\
&& \nonumber
       + \cat\*\cf \* \Biggl[ \frac{44405787094076715779}{38887309440000000} 
          - \frac{7312706287799}{30005640000}\*\zeta_3 - \frac{13785409}{132300}\*\zeta_4 
          + \frac{17432173}{23814}\*\zeta_5 \Biggr]
\\
&& \nonumber
       + \nf\*\cft \* \Biggl[ \frac{29494412200734623467}{120045124241280000} 
          + \frac{77319514799}{375070500}\*\zeta_3 + \frac{1253219}{33075}\*\zeta_4 - \frac{41800}{63}\*\zeta_5 \Biggr]
\\
&& \nonumber
       + \nf\*\ca\*\cfs \* \Biggl[  - \frac{15104254419980130497}{47636954064000000} 
          + \frac{230513485753}{250047000}\*\zeta_3 - \frac{9299719}{22050}\*\zeta_4 + \frac{20900}{189}\*\zeta_5 \Biggr]
\\
&& \nonumber
       + \nf\*\cas\*\cf \* \Biggl[  - \frac{2486710179097323617}{4536852768000000} 
          - \frac{1091690689753}{937676250}\*\zeta_3 + \frac{25392719}{66150}\*\zeta_4 + \frac{4292332}{8505}\*\zeta_5 \Biggr]
\\
&& \nonumber
       + \nfs\*\cfs \* \Biggl[ \frac{103302930942446363}{2381847703200000} 
          - \frac{41192947}{297675}\*\zeta_3 + \frac{4180}{63}\*\zeta_4 \Biggr]
\\
&& \nonumber
       + \nfs\*\ca\*\cf \* \Biggl[ \frac{5199820982878583}{226842638400000} + \frac{41192947}{297675}\*\zeta_3
          - \frac{4180}{63}\*\zeta_4 \Biggr]
\\
&& \nonumber
       + \nft\*\cf \* \Biggl[  - \frac{40350728956471}{34026395760000} + \frac{8360}{1701}\*\zeta_3 \Biggr]
\\
&& \nonumber
       + \dfFAnc \* \Biggl[ \frac{2967981583758917}{2160406080000} 
          + \frac{5621430297923}{1250235000}\*\zeta_3 - \frac{137680664}{19845}\*\zeta_5 \Biggr]
\\
&&
       + \nf\*\dfFFnc \* \Biggl[ \frac{92904062646461}{270050760000} 
          + \frac{11854602254}{22325625}\*\zeta_3 - \frac{3281344}{2835}\*\zeta_5 \Biggr]
\:\: ,
\eea
\bea
{\lefteqn{
\gamma_{\rm ns}^{\,(3)-}(N=11) \,=\,}}
\nonumber
\\
&& \nonumber
\cff \* \Biggl[ \frac{5942479928802007050870338007469}{24563068569963720498240000000}
           + \frac{182985306975827551}{576597755002500}\*\zeta_3 - \frac{263657654}{343035}\*\zeta_5 \Biggr]
\\
&& \nonumber
       + \ca\*\cft \* \Biggl[  - \frac{140643955245651718928769629}{690478118006513760000000}
          - \frac{11683631163531071}{96099625833750}\*\zeta_3 - \frac{151689577}{727650}\*\zeta_4 
\\
&& \nonumber
\hspace*{5mm}
          + \frac{2781069694}{2401245}\*\zeta_5 \Biggr]
\\
&& \nonumber
       + \cas\*\cfs \* \Biggl[  - \frac{65080732936319895430465693}{334777269336491520000000}
          + \frac{1334072255443149799}{1537594013340000}\*\zeta_3 + \frac{151689577}{485100}\*\zeta_4 
\\
&& \nonumber
\hspace*{5mm}
          - \frac{23116873}{10890}\*\zeta_5 \Biggr]
\\
&& \nonumber
       + \cat\*\cf \* \Biggl[ \frac{1934103459065790006173}{1711041615360000000}
           - \frac{937411447527013}{2420454960000}\*\zeta_3 - \frac{151689577}{1455300}\*\zeta_4 
\\
&& \nonumber
\hspace*{5mm}
           + \frac{14593135349}{14407470}\*\zeta_5 \Biggr]
\\
&& \nonumber
       + \nf\*\cft \* \Biggl[ \frac{2807323127469698463268574221}{10633363017300311904000000}
          + \frac{118758665137891}{499218835500}\*\zeta_3 + \frac{151689577}{4002075}\*\zeta_4 
\\
&& \nonumber
\hspace*{5mm}
          - \frac{502528}{693}\*\zeta_5 \Biggr]
\\
&& \nonumber
       + \nf\*\ca\*\cfs \* \Biggl[  - \frac{2888600005383327025035503}{8369431733412288000000} 
          + \frac{109012601223061}{110937519000}\*\zeta_3 - \frac{1215792617}{2668050}\*\zeta_4 
\\
&& \nonumber
\hspace*{5mm}
          + \frac{251264}{2079}\*\zeta_5 \Biggr]
\\
&& \nonumber
       + \nf\*\cas\*\cf \* \Biggl[  - \frac{1791290424092116907479}{3019275517104000000} 
          - \frac{52497666373939}{41601569625}\*\zeta_3 + \frac{3343998697}{8004150}\*\zeta_4
\\
&& \nonumber
\hspace*{5mm}
          + \frac{37705252}{68607}\*\zeta_5 \Biggr]
\\
&& \nonumber
       + \nfs\*\cfs \* \Biggl[ \frac{37084682543660792132933}{767197908896126400000} 
          - \frac{1093228621}{7203735}\*\zeta_3 + \frac{251264}{3465}\*\zeta_4 \Biggr]
\\
&& \nonumber
       + \nfs\*\ca\*\cf \* \Biggl[ \frac{31274848451808887}{1286623089675000} 
          + \frac{1093228621}{7203735}\*\zeta_3 - \frac{251264}{3465}\*\zeta_4 \Biggr]
\\
&& \nonumber
       + \nft\*\cf \* \Biggl[  - \frac{633953354507891423}{498180460322160000} 
          + \frac{502528}{93555}\*\zeta_3 \Biggr]
\\
&& \nonumber
       + \dfFAnc \* \Biggl[ \frac{27544072006307986519}{15684548140800000} 
           + \frac{112240841713507}{20170458000}\*\zeta_3 - \frac{20755381868}{2401245}\*\zeta_5 \Biggr]
\\
&&
       + \nf\*\dfFFnc \* \Biggl[ \frac{550922957081846869}{1307045678400000} 
          + \frac{104968283381}{180093375}\*\zeta_3 - \frac{30026464}{22869}\*\zeta_5 \Biggr]
\:\: ,
\eea
\bea
{\lefteqn{
\gamma_{\rm ns}^{\,(3)-}(N=13) \,=\,}}
\nonumber
\\
&& \nonumber
       + \cff \* \Biggl[ \frac{109254370623053143691943402173119101653}{440370321638438344304874602880000000} 
		  + \frac{10279814538281025664097}{32936416961252805000}\*\zeta_3 
\\
&& \nonumber
\hspace*{5mm}
		  - \frac{312960229682}{405810405}\*\zeta_5 \Biggr]
\\
&& \nonumber
       + \ca\*\cft \* \Biggl[  - \frac{137433149672175525416627043272450693}{598794143642010180678733920000000} 
		      - \frac{122111455215270203911}{1219867294861215000}\*\zeta_3 
\\
&& \nonumber
\hspace*{5mm}
		      - \frac{3662719609}{17567550}\*\zeta_4 
		      + \frac{42916930526}{36891855}\*\zeta_5 \Biggr]
\\
&& \nonumber
       + \cas\*\cfs \* \Biggl[  - \frac{448216384754306083916951583828253}{5583162178480281404929920000000} 
			+ \frac{6030472603025623852733}{5489402826875467500}\*\zeta_3 
\\
&& \nonumber
\hspace*{5mm}
			+ \frac{3662719609}{11711700}\*\zeta_4 
			- \frac{680613539689}{270540270}\*\zeta_5 \Biggr]
\\
&& \nonumber
       + \cat\*\cf \* \Biggl[ \frac{43589208948057316412721986370967}{39043092157204765069440000000} 
		     - \frac{29267635901342586703}{55836265244760000}\*\zeta_3 
\\
&& \nonumber
\hspace*{5mm}
		     - \frac{3662719609}{35135100}\*\zeta_4 
		     + \frac{281672586271}{221351130}\*\zeta_5 \Biggr]
\\
&& \nonumber
       + \nf\*\cft \* \Biggl[ \frac{7156992636760583450168754611387207}{25662606156086150600517168000000} 
		     + \frac{290443295483160277}{1096783781593500}\*\zeta_3 
\\
&& \nonumber
\hspace*{5mm}
		     + \frac{3662719609}{96621525}\*\zeta_4
		     - \frac{6997864}{9009}\*\zeta_5 \Biggr]
\\
&& \nonumber
       + \nf\*\ca\*\cfs \* \Biggl[  - \frac{22054474323190222766265253462577}{59819594769431586481392000000} 
			 + \frac{755458878766886551}{731189187729000}\*\zeta_3 
\\
&& \nonumber
\hspace*{5mm}
			 - \frac{31181819789}{64414350}\*\zeta_4 
			 + \frac{3498932}{27027}\*\zeta_5 \Biggr]
\\
&& \nonumber
       + \nf\*\cas\*\cf \* \Biggl[  - \frac{44016652745476286989105451999}{69719807423579937624000000} 
			 - \frac{10316777282946572197}{7677486471154500}\*\zeta_3
\\
&& \nonumber
\hspace*{5mm}
			 + \frac{86220020149}{193243050}\*\zeta_4 
			 + \frac{4331692556}{7378371}\*\zeta_5 \Biggr]
\\
&& \nonumber
       + \nfs\*\cfs \* \Biggl[ \frac{3742938630150542702886432311}{71213803296942364858800000}
		       - \frac{198433044499}{1217431215}\*\zeta_3 + \frac{3498932}{45045}\*\zeta_4 \Biggr]
\\
&& \nonumber
       + \nfs\*\ca\*\cf \* \Biggl[ \frac{50680931177006728293019381}{1991994497816569646400000}
			+ \frac{198433044499}{1217431215}\*\zeta_3 - \frac{3498932}{45045}\*\zeta_4 \Biggr]
\\
&& \nonumber
       + \nft\*\cf \* \Biggl[  - \frac{19115924965400169467303}{14228532127261211760000}
		      + \frac{6997864}{1216215}\*\zeta_3 \Biggr]
\\
&& \nonumber
       + \dfFAnc \* \Biggl[ \frac{107463452886020580200393}{51025756239057600000} 
		       + \frac{2581697503167399049}{393717254931000}\*\zeta_3 
\\
&& \nonumber
\hspace*{5mm}
		      - \frac{4144734743192}{405810405}\*\zeta_5 \Biggr]
\\
&& \nonumber
       + \nf\*\dfFFnc \* \Biggl[ \frac{6259112433726496309379}{12756439059764400000} 
			  + \frac{2793224880481058}{4474059715125}\*\zeta_3 - \frac{3554796992}{2459457}\*\zeta_5 \Biggr]
\:\: , \\
\eea
\bea
{\lefteqn{
\gamma_{\rm ns}^{\,(3)-}(N=15) \,=\,}}
\nonumber
\\
&& \nonumber
  \cff \* \Biggl[ \frac{1593418359971838203424246724210060722526639}{6313148931008652103954682306887680000000} 
		  + \frac{259680235193827374263}{843172274208071808}\*\zeta_3 
\\
&& \nonumber
\hspace*{5mm}
                  - \frac{89608395343}{115945830}\*\zeta_5 \Biggr]
\\
&& \nonumber
       + \ca\*\cft \* \Biggl[  - \frac{910674811575654880535423035274865836261}{3678991218536510550090141204480000000} 
		      - \frac{47105526692823873500839}{562114849472047872000}\*\zeta_3 
\\
&& \nonumber
\hspace*{5mm}
		      - \frac{205125530543}{983782800}\*\zeta_4 
		      + \frac{34430154145}{29513484}\*\zeta_5 \Biggr]
\\
&& \nonumber
       + \cas\*\cfs \* \Biggl[ \frac{24794744696072980162780614700810631}{1225105300877958891138908160000000} 
		       + \frac{264518190199865408922107}{200755303382874240000}\*\zeta_3 
\\
&& \nonumber
\hspace*{5mm}
		       + \frac{205125530543}{655855200}\*\zeta_4 
		       - \frac{415521270781}{144288144}\*\zeta_5 \Biggr]
\\
&& \nonumber
       + \cat\*\cf \* \Biggl[ \frac{11008793137870484173113743839424629}{9995031592244419857776640000000} 
		     - \frac{4285776141034052935649}{6551455122051840000}\*\zeta_3 
\\
&& \nonumber
\hspace*{5mm}
		     - \frac{205125530543}{1967565600}\*\zeta_4 
		     + \frac{67011783863}{44270226}\*\zeta_5 \Biggr]
\\
&& \nonumber
       + \nf\*\cft \* \Biggl[ \frac{765704555141235451117809578252282141}{2627850870383221821492958003200000} 
		     + \frac{5058205287659325397}{17548540505496000}\*\zeta_3 
\\
&& \nonumber
\hspace*{5mm}
		     + \frac{205125530543}{5410805400}\*\zeta_4 - \frac{7397890}{9009}\*\zeta_5 \Biggr]
\\
&& \nonumber
       + \nf\*\ca\*\cfs \* \Biggl[  - \frac{5957585274731644785588178652001289}{15313816260974486139236352000000} 
			 + \frac{4197787411208679709}{3899675667888000}\*\zeta_3 
\\
&& \nonumber
\hspace*{5mm}
			 - \frac{1834288866343}{3607203600}\*\zeta_4 + \frac{3698945}{27027}\*\zeta_5
			 \Biggr]
\\
&& \nonumber
       + \nf\*\cas\*\cf \* \Biggl[  - \frac{718538849608998968737940767607}{1081713375784028123136000000} 
			 - \frac{165455940798017850551}{116990270036640000}\*\zeta_3 
\\
&& \nonumber
\hspace*{5mm}
			 + \frac{5092615537943}{10821610800}\*\zeta_4 + \frac{77250091}{124740}\*\zeta_5 \Biggr]
\\
&& \nonumber
       + \nfs\*\cfs \* \Biggl[ \frac{205033222941376179044590449727}{3646146728803449080770560000} 
		       - \frac{8409151022911}{48697248600}\*\zeta_3 
		       + \frac{739789}{9009}\*\zeta_4 \Biggr]
\\
&& \nonumber
       + \nfs\*\ca\*\cf \* \Biggl[ \frac{673351343605311635893107289}{25497529572052091473920000}
			+ \frac{8409151022911}{48697248600}\*\zeta_3 - \frac{739789}{9009}\*\zeta_4 \Biggr]
\\
&& \nonumber
       + \nft\*\cf \* \Biggl[  - \frac{1022483569869843897775}{728500844915774042112} 
		      + \frac{1479578}{243243}\*\zeta_3 \Biggr]
\\
&& \nonumber
       + \dfFAnc \* \Biggl[ \frac{328171179053663132153783483}{135079092880123392000000} 
		       + \frac{145906081836810245387}{19498378339440000}\*\zeta_3 
\\
&& \nonumber
\hspace*{5mm}
		       - \frac{18940495337639}{1623241620}\*\zeta_5 \Biggr]
\\
&& \nonumber
       + \nf\*\dfFFnc \* \Biggl[ \frac{34218686926433684318010149}{61911250903389888000000}
			  + \frac{22453298164820123}{34088074020000}\*\zeta_3 
\\
&&
\hspace*{5mm}
                          - \frac{210864634}{135135}\*\zeta_5 \Biggr]
\:\: ,
\eea
and 
\bea
{\lefteqn{
\gamma_{\rm ns}^{\,(3){\rm s}}(N=3) \,=\,}}
\nonumber
\\
&& \nonumber
      \nf\* \cf\* \dabcnc \* \Biggl[ - \frac{72365}{216} + \frac{3200}{9}\*\zeta_3 \Biggr]
\\
&& 
    + \nf\* \ca\* \dabcnc \* \Biggl[ \frac{60629}{648} - \frac{1010}{9}\*\zeta_3 + \frac{200}{9}\*\zeta_5 \Biggr]
    + \nfs\* \dabcnc \* \Biggl[ - \frac{1879}{324} \Biggr]
\:\: ,
\eea
\bea
{\lefteqn{
\gamma_{\rm ns}^{\,(3){\rm s}}(N=5) \,=\,}}
\nonumber
\\
&& \nonumber
      \nf\* \cf\* \dabcnc \* \Biggl[ - \frac{1110626839}{5467500} + \frac{133952}{675}\*\zeta_3 \Biggr]
\\
&& 
    + \nf\* \ca\* \dabcnc \* \Biggl[ \frac{11071396}{273375} - \frac{48608}{675}\*\zeta_3 + \frac{1792}{45}\*\zeta_5 \Biggr]
    + \nfs\* \dabcnc \* \Biggl[ \frac{150143}{273375} \Biggr]
\:\: ,
\eea
\bea
{\lefteqn{
\gamma_{\rm ns}^{\,(3){\rm s}}(N=7) \,=\,}}
\nonumber
\\
&& \nonumber
      \nf\* \cf\* \dabcnc \* \Biggl[ - \frac{779352339134399}{5601052800000} + \frac{116051}{882}\*\zeta_3 \Biggr]
\\
&& \nonumber
    + \nf\* \ca\* \dabcnc \* \Biggl[ \frac{4572181575853}{240045120000} - \frac{3093067}{52920}\*\zeta_3 
	       + \frac{2250}{49}\*\zeta_5 \Biggr]
\\
&& 
	     + \nfs\* \dabcnc \* \Biggl[ \frac{133768649869}{120022560000} \Biggr]
\:\: ,
\eea
\bea
{\lefteqn{
\gamma_{\rm ns}^{\,(3){\rm s}}(N=9) \,=\,}}
\nonumber
\\
&& \nonumber
      \nf\* \cf\* \dabcnc \* \Biggl[  - \frac{1021276699865745697}{9924365430000000} 
          + \frac{85388204}{893025}\*\zeta_3 \Biggr]
\\
&& \nonumber
    + \nf\* \ca\* \dabcnc \* \Biggl[ \frac{2755119043881017}{315059220000000} 
          - \frac{1155263626}{22325625}\*\zeta_3 + \frac{19712}{405}\*\zeta_5 \Biggr]
\\
&& 
    + \nfs\* \dabcnc \* \Biggl[ \frac{767593737326}{738420046875} \Biggr]
\:\: ,
\eea
\bea
{\lefteqn{
\gamma_{\rm ns}^{\,(3){\rm s}}(N=11) \,=\,}}
\nonumber
\\
&& \nonumber
      \nf\* \cf\* \dabcnc \* \Biggl[  - \frac{111603773330166434882669}{1394905288902048000000} + \frac{39731371264}{540280125}\*\zeta_3 \Biggr]
\\
&& \nonumber
    + \nf\* \ca\* \dabcnc \* \Biggl[ \frac{4824168874878027761}{1646877554784000000} 
          - \frac{103259298727}{2161120500}\*\zeta_3 + \frac{18200}{363}\*\zeta_5 \Biggr]
\\
&& 
    + \nfs\* \dabcnc \* \Biggl[ \frac{40350238520199037}{45289132756560000} \Biggr]
\:\: ,
\eea
\bea
{\lefteqn{
\gamma_{\rm ns}^{\,(3){\rm s}}(N=13) \,=\,}}
\nonumber
\\
&& \nonumber
         \nf\* \cf\* \dabcnc \* \Biggl[  - \frac{18532792410668780136669601961}{287594205622267242699000000} 
         + \frac{87687308676848}{1491353238375}\*\zeta_3 \Biggr]
\\
&& \nonumber
       + \nf\* \ca\* \dabcnc \* \Biggl[  - \frac{3391753143539038458111467}{4692679345817880417000000} 
       - \frac{202143595520348}{4474059715125}\*\zeta_3 
          + \frac{422400}{8281}\*\zeta_5 \Biggr]
\\
&& 
    + \nfs\* \dabcnc \* \Biggl[ \frac{8444176392147052173457}{11173046061471143850000} \Biggr]
\:\: ,
\eea
\bea
{\lefteqn{
\gamma_{\rm ns}^{\,(3){\rm s}}(N=15) \,=\,}}
\nonumber
\\
&& \nonumber
         \nf\* \cf\* \dabcnc \* \Biggl[
         - \frac{12364402852607089809412352833493}{232027519105674032412672000000} + \frac{89928183421837}{1859349492000}\*\zeta_3 
         \Biggr]
\\
&& \nonumber
       + \nf\* \ca\* \dabcnc \* \Biggl[ 
       - \frac{6470906684625313369880498347}{2028212579595052730880000000} - \frac{17737091969947579}{409056888240000}\*\zeta_3
          + \frac{1547}{30}\*\zeta_5
       \Biggr]
\\
&& 
       + \nfs\* \dabcnc \* \Biggl[ \frac{1437544924718401819943003}{2228805032522035968000000} \Biggr]
\:\: .
\eea

\renewcommand{\theequation}{\ref{sec:appC}.\arabic{equation}}
\setcounter{equation}{0}
\renewcommand{\thefigure}{\ref{sec:appC}.\arabic{figure}}
\setcounter{figure}{0}
\renewcommand{\thetable}{\ref{sec:appC}.\arabic{table}}
\setcounter{table}{0}
\section{Time-like splitting function}
\label{sec:appC}
Here we present the difference between the space- and time-like non-singlet 
splitting functions at four loops, defined by $\delta\, P^{\,(3)\pm}(x) 
\,\equiv\, P^{\,(3)\pm}_{\sigma=1}(x) - P^{\,(3)\pm}_{\sigma=-1}(x)$.
The expression for $\delta\, P^{\,(3)+}(x)$ reads
\bea
\lefteqn{ 
\delta\, P^{\,(3)+}(x) \, = \,
}
\nonumber \\ && \mbox{\hspn} \phantom{+}\nonumber
         16\*\cff \* \biggl\{
         (1+x) \* \biggl(
          - 792\*\H(-2)\*\zeta_3
          - 364\*\H(-2)\*\zeta_2
          - 64\*\H(3)\*\zeta_3
          + 12\*\Hh(-4,0)
          + 768\*\Hh(-2,-1)\*\zeta_2
          - 28\*\Hh(-2,0)
\\
&& \nonumber
          + 304\*\Hh(-2,2)
          + 768\*\Hh(-1,-2)\*\zeta_2
          - 792\*\Hh(-1,0)\*\zeta_3
          - 364\*\Hh(-1,0)\*\zeta_2
          + 608\*\Hh(-1,3)
          + 1584\*\Hh(-1,4)
          - 32\*\Hh(2,0)\*\zeta_3
\\
&& \nonumber
          - 324\*\Hh(2,2)
          + 120\*\Hh(2,4)
          - 648\*\Hh(3,1)
          + 112\*\Hh(3,3)
          + 40\*\Hh(4,2)
          - 32\*\Hh(5,1)
          + 192\*\Hhh(-4,-1,0)
          + 64\*\Hhh(-3,-2,0)
\\
&& \nonumber
          - 192\*\Hhh(-3,-1,0)
          - 192\*\Hhh(-2,-2,0)
          - 120\*\Hhh(-2,-1,0)
          - 672\*\Hhh(-2,-1,2)
          + 384\*\Hhh(-2,2,1)
          - 192\*\Hhh(-1,-3,0)
\\
&& \nonumber
          - 120\*\Hhh(-1,-2,0)
          - 672\*\Hhh(-1,-2,2)
          + 768\*\Hhh(-1,-1,0)\*\zeta_2
          - 1344\*\Hhh(-1,-1,3)
          - 56\*\Hhh(-1,0,0)
          - 1280\*\Hhh(-1,0,0)\*\zeta_2
\\
&& \nonumber
          + 304\*\Hhh(-1,2,0)
          + 384\*\Hhh(-1,2,2)
          + 912\*\Hhh(-1,3,0)
          + 768\*\Hhh(-1,3,1)
          - 64\*\Hhh(2,0,0)\*\zeta_2
          - 324\*\Hhh(2,1,0)
          + 56\*\Hhh(2,3,0)
\\
&& \nonumber
          + 48\*\Hhh(3,2,0)
          + 76\*\Hhh(4,0,0)
          + 40\*\Hhh(4,1,0)
          + 192\*\Hhhh(-2,-1,-1,0)
          - 624\*\Hhhh(-2,-1,0,0)
          + 192\*\Hhhh(-1,-2,-1,0)
\\
&& \nonumber
          - 624\*\Hhhh(-1,-2,0,0)
          + 192\*\Hhhh(-1,-1,-2,0)
          - 240\*\Hhhh(-1,-1,0,0)
          - 672\*\Hhhh(-1,-1,2,0)
          + 636\*\Hhhh(-1,0,0,0)
\\
&& \nonumber
          + 384\*\Hhhh(-1,2,0,0)
          + 384\*\Hhhh(-1,2,1,0)
          + 24\*\Hhhh(2,2,0,0)
          + 116\*\Hhhh(3,0,0,0)
          + 48\*\Hhhh(3,1,0,0)
          + 384\*\Hhhhh(-1,-1,-1,0,0)
\\
&& \nonumber
          - 1296\*\Hhhhh(-1,-1,0,0,0)
          + 704\*\Hhhhh(-1,0,0,0,0)
          + 112\*\Hhhhh(2,0,0,0,0)
          + 24\*\Hhhhh(2,1,0,0,0)
          \biggr)
       + (1-x) \* \biggl(
            288\*\H(-3)\*\zeta_3
          + \frct{9}{2}\*\H(0)
\\
&& \nonumber
          - 393\*\H(2)
          - 56\*\H(2)\*\zeta_3
          - 76\*\H(2)\*\zeta_2
          + 96\*\H(4)\*\zeta_2
          - 384\*\Hh(-3,-1)\*\zeta_2
          - 192\*\Hh(-2,-2)\*\zeta_2
          + 144\*\Hh(-2,0)\*\zeta_3
\\
&& \nonumber
          - 288\*\Hh(-2,4)
          - 393\*\Hh(1,0)
          - 56\*\Hh(1,0)\*\zeta_3
          - 76\*\Hh(1,0)\*\zeta_2
          - 560\*\Hh(1,2)
          - 96\*\Hh(1,2)\*\zeta_2
          - 96\*\Hh(1,4)
          - 560\*\Hh(2,1)
\\
&& \nonumber
          - 96\*\Hh(2,1)\*\zeta_2
          - 64\*\Hh(2,3)
          + 384\*\Hhh(-3,-1,2)
          + 192\*\Hhh(-2,-2,2)
          - 192\*\Hhh(-2,-1,0)\*\zeta_2
          + 384\*\Hhh(-2,-1,3)
\\
&& \nonumber
          + 256\*\Hhh(-2,0,0)\*\zeta_2
          - 96\*\Hhh(-2,3,0)
          - 560\*\Hhh(1,0,0)
          - 64\*\Hhh(1,0,0)\*\zeta_2
          - 560\*\Hhh(1,1,0)
          - 96\*\Hhh(1,1,0)\*\zeta_2
          - 64\*\Hhh(1,3,0)
\\
&& \nonumber
          - 48\*\Hhh(2,2,0)
          + 96\*\Hhhh(-2,-2,0,0)
          + 192\*\Hhhh(-2,-1,2,0)
          + 168\*\Hhhh(1,0,0,0)
          + 288\*\Hhhhh(-2,-1,0,0,0)
          - 64\*\Hhhhh(-2,0,0,0,0)
\\
&& \nonumber
          - 32\*\Hhhhh(1,0,0,0,0)
          + 96\*\Hhhhh(1,1,0,0,0)
          \biggr)
       - (20 + 264\*x) \* \H(0)\*\zeta_5
       - (28 - 268\*x) \* \Hh(-3,0)
       - (32 + 96\*x) \* \Hh(3,0)\*\zeta_2
\\
&& \nonumber
       - (32 + 224\*x) \* \Hh(-5,0)
       + \biggl(\frct{69}{2} + \frct{1641}{2}\*x\biggr) \* \Hh(0,0)
       - \biggl(\frct{77}{2} - \frct{725}{2}\*x\biggr) \* \Hhh(0,0,0)
       - (46 + 306\*x) \* \Hh(0,0)\*\zeta_4
\\
&& \nonumber
       + (48 - 1240\*x) \* \Hhhh(0,0,0,0)
       + (61 + 839\*x) \* \H(0)\*\zeta_3
       + (70 + 310\*x) \* \Hhhhhh(0,0,0,0,0,0)
       + (72 + 792\*x) \* \Hh(0,0)\*\zeta_3
\\
&& \nonumber
       + (80 + 48\*x) \* \Hh(5,0)
       - (88 - 72\*x) \* \Hh(2,0)\*\zeta_2
       - (98 + 82\*x) \* \Hhhh(2,0,0,0)
       - (110 + 130\*x) \* \Hhhhh(0,0,0,0,0)
\\
&& \nonumber
       - (112 - 80\*x) \* (
           \H(3)\*\zeta_2
          + \Hhh(-3,2,0)
          )
       - (120 + 168\*x) \* \Hhh(0,0,0)\*\zeta_3
       + (126 + 906\*x) \* \Hhh(0,0,0)\*\zeta_2
\\
&& \nonumber
       - (144 + 112\*x) \* \Hhhh(0,0,0,0)\*\zeta_2
       + (168 - 56\*x) \* \H(0)\*\zeta_2\*\zeta_3
       - (176 + 208\*x) \* (
            \Hh(3,2)
          + \Hhh(3,1,0)
          )
\\
&& \nonumber
       + (184 + 480\*x) \* \Hhh(-2,0,0)
       + (188 - 376\*x) \* \H(0)\*\zeta_2
       - \biggl(\frct{403}{2} + \frct{851}{2}\*x\biggr) \* \H(0)\*\zeta_4
       + (240 + 80\*x) \* \H(6)
\\
&& \nonumber
       - (244 + 276\*x) \* \Hhh(3,0,0)
       + (248 + 1284\*x) \* \Hh(0,0)\*\zeta_2
       - (260 + 388\*x) \* \Hhh(2,0,0)
       - (264 + 904\*x) \* \H(5)
\\
&& \nonumber
       - (280 + 88\*x) \* \Hhh(-4,0,0)
       + (280 + 600\*x) \* \Hhh(-3,0,0)
       - (296 - 56\*x) \* \Hhhh(-3,0,0,0)
       + (320 - 64\*x) \* \Hhhh(-3,-1,0,0)
\\
&& \nonumber
       - (336 - 240\*x) \* \Hh(-4,2)
       - (342 + 550\*x) \* \Hh(4,0)
       + (352 + 992\*x) \* \Hh(-3,2)
       + (368 + 688\*x) \* \Hhh(-2,2,0)
\\
&& \nonumber
       - (378 + 1506\*x) \* \H(4)
       - (400 - 720\*x) \* \H(3)
       + (400 - 304\*x) \* \Hh(-3,0)\*\zeta_2
       - (416 - 352\*x) \* \Hh(-3,3)
\\
&& \nonumber
       + (432 - 144\*x) \* \H(-4)\*\zeta_2
       - (432 + 720\*x) \* \Hh(4,1)
       - (442 + 834\*x) \* \Hh(3,0)
       - (448 + 1088\*x) \* \H(-3)\*\zeta_2
\\
&& \nonumber
       - (480 - 640\*x) \* \Hh(2,0)
       + (584 + 1064\*x) \* \Hhhh(-2,0,0,0)
       - (864 + 1184\*x) \* \Hh(-2,0)\*\zeta_2
       + (880 + 1520\*x) \* \Hh(-2,3)
\\
&& \nonumber
       + \pqq( - x) \* \biggl(
          - 456\*\H(-4)\*\zeta_2
          - 1248\*\H(-3)\*\zeta_3
          + 192\*\H(-3)\*\zeta_2
          + 910\*\H(-2)\*\zeta_4
          + 96\*\H(-2)\*\zeta_3
          + 36\*\H(-2)\*\zeta_2
\\
&& \nonumber
          - 244\*\H(0)\*\zeta_5
          + \frct{45}{2}\*\H(0)\*\zeta_4
          - 27\*\H(0)\*\zeta_3
          - \frct{39}{2}\*\H(0)\*\zeta_2
          - 308\*\H(0)\*\zeta_2\*\zeta_3
          + 24\*\H(3)
          - 48\*\H(3)\*\zeta_3
          + 72\*\H(3)\*\zeta_2
          + 54\*\H(4)
\\
&& \nonumber
          - 120\*\H(4)\*\zeta_2
          + 48\*\H(5)
          - 160\*\H(6)
          - 192\*\Hh(-5,0)
          + 36\*\Hh(-4,0)
          + 528\*\Hh(-4,2)
          + 1152\*\Hh(-3,-1)\*\zeta_2
\\
&& \nonumber
          - 1240\*\Hh(-3,0)\*\zeta_2
          - 192\*\Hh(-3,2)
          + 1312\*\Hh(-3,3)
          + 1152\*\Hh(-2,-2)\*\zeta_2
          + 1248\*\Hh(-2,-1)\*\zeta_3
          - 192\*\Hh(-2,-1)\*\zeta_2
\\
&& \nonumber
          - 15\*\Hh(-2,0)
          - 960\*\Hh(-2,0)\*\zeta_3
          + 216\*\Hh(-2,0)\*\zeta_2
          - 36\*\Hh(-2,2)
          - 16\*\Hh(-2,2)\*\zeta_2
          - 240\*\Hh(-2,3)
          + 1712\*\Hh(-2,4)
\\
&& \nonumber
          + 1152\*\Hh(-1,-3)\*\zeta_2
          + 1248\*\Hh(-1,-2)\*\zeta_3
          - 192\*\Hh(-1,-2)\*\zeta_2
          + 910\*\Hh(-1,0)\*\zeta_4
          + 96\*\Hh(-1,0)\*\zeta_3
          + 36\*\Hh(-1,0)\*\zeta_2
\\
&& \nonumber
          - 72\*\Hh(-1,3)
          - 32\*\Hh(-1,3)\*\zeta_2
          - 144\*\Hh(-1,4)
          + 1088\*\Hh(-1,5)
          - 18\*\Hh(0,0)
          - 612\*\Hh(0,0)\*\zeta_4
          - 36\*\Hh(0,0)\*\zeta_2
          + 12\*\Hh(2,0)
\\
&& \nonumber
          - 24\*\Hh(2,0)\*\zeta_3
          + 36\*\Hh(2,0)\*\zeta_2
          - 96\*\Hh(2,2)\*\zeta_2
          + 18\*\Hh(3,0)
          - 136\*\Hh(3,0)\*\zeta_2
          - 192\*\Hh(3,1)\*\zeta_2
          - 48\*\Hh(3,2)
          - 60\*\Hh(4,0)
\\
&& \nonumber
          - 144\*\Hh(4,1)
          - 96\*\Hh(4,2)
          - 224\*\Hh(5,0)
          - 384\*\Hh(5,1)
          + 144\*\Hhh(-4,-1,0)
          + 96\*\Hhh(-4,0,0)
          + 48\*\Hhh(-3,-2,0)
\\
&& \nonumber
          - 1152\*\Hhh(-3,-1,2)
          - 72\*\Hhh(-3,0,0)
          + 560\*\Hhh(-3,2,0)
          + 768\*\Hhh(-3,2,1)
          + 48\*\Hhh(-2,-3,0)
          - 1152\*\Hhh(-2,-2,2)
\\
&& \nonumber
          - 1152\*\Hhh(-2,-1,-1)\*\zeta_2
          + 1664\*\Hhh(-2,-1,0)\*\zeta_2
          + 192\*\Hhh(-2,-1,2)
          - 2112\*\Hhh(-2,-1,3)
          - 18\*\Hhh(-2,0,0)
\\
&& \nonumber
          - 1408\*\Hhh(-2,0,0)\*\zeta_2
          - 48\*\Hhh(-2,2,0)
          + 96\*\Hhh(-2,2,1)
          + 384\*\Hhh(-2,2,2)
          + 1056\*\Hhh(-2,3,0)
          + 1152\*\Hhh(-2,3,1)
\\
&& \nonumber
          + 144\*\Hhh(-1,-4,0)
          - 1152\*\Hhh(-1,-3,2)
          - 1152\*\Hhh(-1,-2,-1)\*\zeta_2
          + 1664\*\Hhh(-1,-2,0)\*\zeta_2
          + 192\*\Hhh(-1,-2,2)
\\
&& \nonumber
          - 2112\*\Hhh(-1,-2,3)
          - 1152\*\Hhh(-1,-1,-2)\*\zeta_2
          + 1248\*\Hhh(-1,-1,0)\*\zeta_3
          - 192\*\Hhh(-1,-1,0)\*\zeta_2
          + 384\*\Hhh(-1,-1,3)
\\
&& \nonumber
          - 2880\*\Hhh(-1,-1,4)
          - 30\*\Hhh(-1,0,0)
          - 672\*\Hhh(-1,0,0)\*\zeta_3
          + 240\*\Hhh(-1,0,0)\*\zeta_2
          - 36\*\Hhh(-1,2,0)
          - 16\*\Hhh(-1,2,0)\*\zeta_2
\\
&& \nonumber
          + 96\*\Hhh(-1,2,2)
          + 48\*\Hhh(-1,3,0)
          + 192\*\Hhh(-1,3,1)
          + 384\*\Hhh(-1,3,2)
          + 848\*\Hhh(-1,4,0)
          + 1152\*\Hhh(-1,4,1)
          + \frct{81}{2}\*\Hhh(0,0,0)
\\
&& \nonumber
          - 72\*\Hhh(0,0,0)\*\zeta_2
          - 96\*\Hhh(2,0,0)\*\zeta_2
          - 96\*\Hhh(2,1,0)\*\zeta_2
          - 48\*\Hhh(3,0,0)
          - 48\*\Hhh(3,1,0)
          - 48\*\Hhh(4,0,0)
          - 96\*\Hhh(4,1,0)
\\
&& \nonumber
          - 480\*\Hhhh(-3,-1,0,0)
          + 448\*\Hhhh(-3,0,0,0)
          - 576\*\Hhhh(-2,-2,0,0)
          + 1152\*\Hhhh(-2,-1,-1,2)
          + 96\*\Hhhh(-2,-1,0,0)
\\
&& \nonumber
          - 960\*\Hhhh(-2,-1,2,0)
          - 768\*\Hhhh(-2,-1,2,1)
          - 144\*\Hhhh(-2,0,0,0)
          + 352\*\Hhhh(-2,2,0,0)
          + 384\*\Hhhh(-2,2,1,0)
          - 480\*\Hhhh(-1,-3,0,0)
\\
&& \nonumber
          + 1152\*\Hhhh(-1,-2,-1,2)
          + 96\*\Hhhh(-1,-2,0,0)
          - 960\*\Hhhh(-1,-2,2,0)
          - 768\*\Hhhh(-1,-2,2,1)
          + 1152\*\Hhhh(-1,-1,-2,2)
\\
&& \nonumber
          - 1152\*\Hhhh(-1,-1,-1,0)\*\zeta_2
          + 2304\*\Hhhh(-1,-1,-1,3)
          + 2176\*\Hhhh(-1,-1,0,0)\*\zeta_2
          + 192\*\Hhhh(-1,-1,2,0)
          - 768\*\Hhhh(-1,-1,2,2)
\\
&& \nonumber
          - 1728\*\Hhhh(-1,-1,3,0)
          - 1536\*\Hhhh(-1,-1,3,1)
          - 54\*\Hhhh(-1,0,0,0)
          - 960\*\Hhhh(-1,0,0,0)\*\zeta_2
          + 96\*\Hhhh(-1,2,0,0)
          + 96\*\Hhhh(-1,2,1,0)
\\
&& \nonumber
          + 320\*\Hhhh(-1,3,0,0)
          + 384\*\Hhhh(-1,3,1,0)
          + 64\*\Hhhh(0,0,0,0)\*\zeta_2
          + 48\*\Hhhh(3,0,0,0)
          + 576\*\Hhhhh(-2,-1,-1,0,0)
\\
&& \nonumber
          - 1120\*\Hhhhh(-2,-1,0,0,0)
          + 496\*\Hhhhh(-2,0,0,0,0)
          + 576\*\Hhhhh(-1,-2,-1,0,0)
          - 1120\*\Hhhhh(-1,-2,0,0,0)
          + 576\*\Hhhhh(-1,-1,-2,0,0)
\\
&& \nonumber
          + 1152\*\Hhhhh(-1,-1,-1,2,0)
          + 288\*\Hhhhh(-1,-1,0,0,0)
          - 768\*\Hhhhh(-1,-1,2,0,0)
          - 768\*\Hhhhh(-1,-1,2,1,0)
          - 96\*\Hhhhh(-1,2,0,0,0)
\\
&& \nonumber
          - 60\*\Hhhhh(0,0,0,0,0)
          + 1728\*\Hhhhhh(-1,-1,-1,0,0,0)
          - 1024\*\Hhhhhh(-1,-1,0,0,0,0)
          - 80\*\Hhhhhh(-1,0,0,0,0,0)
          + 192\*\Hhhhhh(0,0,0,0,0,0)
          \biggr)
\\
&& \nonumber
       + \pqq(x) \* \biggl(
          - 264\*\H(-4)\*\zeta_2
          - 336\*\H(-3)\*\zeta_3
          + 72\*\H(-3)\*\zeta_2
          + \frct{29}{8}\*\H(0)
          + 144\*\H(0)\*\zeta_5
          - \frct{45}{2}\*\H(0)\*\zeta_4
          + 25\*\H(0)\*\zeta_3
\\
&& \nonumber
          + \frct{9}{2}\*\H(0)\*\zeta_2
          - 124\*\H(0)\*\zeta_2\*\zeta_3
          - 66\*\H(2)\*\zeta_4
          + 24\*\H(2)\*\zeta_3
          - 64\*\H(3)\*\zeta_3
          + 15\*\H(4)
          + 24\*\H(4)\*\zeta_2
          + 96\*\H(5)
          - 320\*\H(6)
\\
&& \nonumber
          + 192\*\Hh(-5,0)
          - 36\*\Hh(-4,0)
          + 48\*\Hh(-4,2)
          + 384\*\Hh(-3,-1)\*\zeta_2
          - 280\*\Hh(-3,0)\*\zeta_2
          + 128\*\Hh(-3,3)
          + 192\*\Hh(-2,-2)\*\zeta_2
\\
&& \nonumber
          - 168\*\Hh(-2,0)\*\zeta_3
          + 36\*\Hh(-2,0)\*\zeta_2
          + 144\*\Hh(-2,4)
          + 76\*\Hh(0,0)\*\zeta_4
          - 28\*\Hh(0,0)\*\zeta_2
          + 32\*\Hh(1,-3)\*\zeta_2
          - 66\*\Hh(1,0)\*\zeta_4
\\
&& \nonumber
          + 24\*\Hh(1,0)\*\zeta_3
          - 64\*\Hh(1,2)\*\zeta_3
          + 144\*\Hh(1,4)
          - 256\*\Hh(1,5)
          + 16\*\Hh(2,-2)\*\zeta_2
          - 32\*\Hh(2,0)\*\zeta_3
          - 64\*\Hh(2,1)\*\zeta_3
          + 96\*\Hh(2,3)
\\
&& \nonumber
          - 256\*\Hh(2,4)
          - 13\*\Hh(3,0)
          + 72\*\Hh(3,0)\*\zeta_2
          + 24\*\Hh(3,2)
          - 224\*\Hh(3,3)
          + 60\*\Hh(4,0)
          - 72\*\Hh(4,1)
          - 224\*\Hh(4,2)
          - 288\*\Hh(5,0)
\\
&& \nonumber
          - 320\*\Hh(5,1)
          - 432\*\Hhh(-4,-1,0)
          + 336\*\Hhh(-4,0,0)
          - 336\*\Hhh(-3,-2,0)
          + 144\*\Hhh(-3,-1,0)
          - 192\*\Hhh(-3,-1,2)
\\
&& \nonumber
          - 96\*\Hhh(-3,0,0)
          + 16\*\Hhh(-3,2,0)
          - 192\*\Hhh(-2,-3,0)
          + 72\*\Hhh(-2,-2,0)
          - 96\*\Hhh(-2,-2,2)
          + 192\*\Hhh(-2,-1,0)\*\zeta_2
\\
&& \nonumber
          - 192\*\Hhh(-2,-1,3)
          - 192\*\Hhh(-2,0,0)\*\zeta_2
          + 48\*\Hhh(-2,3,0)
          + \frct{45}{4}\*\Hhh(0,0,0)
          + 160\*\Hhh(0,0,0)\*\zeta_3
          + 36\*\Hhh(0,0,0)\*\zeta_2
\\
&& \nonumber
          + 144\*\Hhh(1,-4,0)
          + 96\*\Hhh(1,-3,0)
          - 128\*\Hhh(1,-3,2)
          + 16\*\Hhh(1,-2,0)\*\zeta_2
          - 128\*\Hhh(1,-2,3)
          - 64\*\Hhh(1,1,0)\*\zeta_3
          - 192\*\Hhh(1,1,4)
\\
&& \nonumber
          + 32\*\Hhh(1,2,0)\*\zeta_2
          - 128\*\Hhh(1,2,3)
          + 96\*\Hhh(1,3,0)
          - 160\*\Hhh(1,3,2)
          - 304\*\Hhh(1,4,0)
          - 288\*\Hhh(1,4,1)
          + 48\*\Hhh(2,-3,0)
\\
&& \nonumber
          + 48\*\Hhh(2,-2,0)
          - 64\*\Hhh(2,-2,2)
          - 49\*\Hhh(2,0,0)
          + 112\*\Hhh(2,0,0)\*\zeta_2
          + 32\*\Hhh(2,1,0)\*\zeta_2
          - 128\*\Hhh(2,1,3)
          + 72\*\Hhh(2,2,0)
\\
&& \nonumber
          - 96\*\Hhh(2,2,2)
          - 240\*\Hhh(2,3,0)
          - 160\*\Hhh(2,3,1)
          + 120\*\Hhh(3,0,0)
          + 24\*\Hhh(3,1,0)
          - 96\*\Hhh(3,1,2)
          - 192\*\Hhh(3,2,0)
          - 96\*\Hhh(3,2,1)
\\
&& \nonumber
          - 240\*\Hhh(4,0,0)
          - 224\*\Hhh(4,1,0)
          - 96\*\Hhh(4,1,1)
          + 384\*\Hhhh(-3,-1,-1,0)
          - 576\*\Hhhh(-3,-1,0,0)
          + 336\*\Hhhh(-3,0,0,0)
\\
&& \nonumber
          + 192\*\Hhhh(-2,-2,-1,0)
          - 336\*\Hhhh(-2,-2,0,0)
          + 192\*\Hhhh(-2,-1,-2,0)
          + 144\*\Hhhh(-2,-1,0,0)
          - 96\*\Hhhh(-2,-1,2,0)
\\
&& \nonumber
          - 108\*\Hhhh(-2,0,0,0)
          - 70\*\Hhhh(0,0,0,0)
          + 256\*\Hhhh(0,0,0,0)\*\zeta_2
          - 192\*\Hhhh(1,-3,-1,0)
          + 128\*\Hhhh(1,-3,0,0)
          - 96\*\Hhhh(1,-2,-2,0)
\\
&& \nonumber
          + 96\*\Hhhh(1,-2,0,0)
          - 64\*\Hhhh(1,-2,2,0)
          - 93\*\Hhhh(1,0,0,0)
          + 144\*\Hhhh(1,0,0,0)\*\zeta_2
          + 64\*\Hhhh(1,1,0,0)\*\zeta_2
          - 128\*\Hhhh(1,1,3,0)
\\
&& \nonumber
          + 96\*\Hhhh(1,2,0,0)
          - 96\*\Hhhh(1,2,2,0)
          - 160\*\Hhhh(1,3,0,0)
          - 160\*\Hhhh(1,3,1,0)
          - 96\*\Hhhh(2,-2,-1,0)
          + 16\*\Hhhh(2,-2,0,0)
          + 180\*\Hhhh(2,0,0,0)
\\
&& \nonumber
          + 96\*\Hhhh(2,1,0,0)
          - 96\*\Hhhh(2,1,2,0)
          - 96\*\Hhhh(2,2,0,0)
          - 96\*\Hhhh(2,2,1,0)
          - 176\*\Hhhh(3,0,0,0)
          - 96\*\Hhhh(3,1,0,0)
          - 96\*\Hhhh(3,1,1,0)
\\
&& \nonumber
          + 384\*\Hhhhh(-2,-1,-1,0,0)
          - 432\*\Hhhhh(-2,-1,0,0,0)
          + 192\*\Hhhhh(-2,0,0,0,0)
          + 180\*\Hhhhh(0,0,0,0,0)
          - 192\*\Hhhhh(1,-2,-1,0,0)
\\
&& \nonumber
          + 48\*\Hhhhh(1,-2,0,0,0)
          + 144\*\Hhhhh(1,0,0,0,0)
          + 144\*\Hhhhh(1,1,0,0,0)
          - 16\*\Hhhhh(1,2,0,0,0)
          - 144\*\Hhhhh(2,0,0,0,0)
          - 16\*\Hhhhh(2,1,0,0,0)
\\
&& \nonumber
          - 272\*\Hhhhhh(0,0,0,0,0,0)
          - 240\*\Hhhhhh(1,0,0,0,0,0)
          - 64\*\Hhhhhh(1,1,0,0,0,0)
          + 192\*\Hhhhhh(1,1,1,0,0,0)
          \biggr)
          \biggr\}
\\
&& \mbox{\hspn} \nonumber
       + 16\,\*\cft\*\ca \* \biggl\{
         (1+x) \* \biggl(
            868\*\H(-2)\*\zeta_3
          + \frct{874}{3}\*\H(-2)\*\zeta_2
          - 32\*\H(3)\*\zeta_3
          - 864\*\Hh(-2,-1)\*\zeta_2
          + 238\*\Hh(-2,0)
          - 288\*\Hh(-2,2)
\\
&& \nonumber
          - 864\*\Hh(-1,-2)\*\zeta_2
          + 868\*\Hh(-1,0)\*\zeta_3
          + \frct{874}{3}\*\Hh(-1,0)\*\zeta_2
          - 576\*\Hh(-1,3)
          - 1704\*\Hh(-1,4)
          - 16\*\Hh(2,0)\*\zeta_3
          + \frct{1016}{3}\*\Hh(2,2)
\\
&& \nonumber
          - 16\*\Hh(2,2)\*\zeta_2
          - 144\*\Hh(2,4)
          + \frct{2032}{3}\*\Hh(3,1)
          - 32\*\Hh(3,1)\*\zeta_2
          - 96\*\Hh(3,3)
          + 48\*\Hh(4,2)
          + 192\*\Hh(5,1)
          + \frct{20}{3}\*\Hhh(-2,-1,0)
\\
&& \nonumber
          + 752\*\Hhh(-2,-1,2)
          - 384\*\Hhh(-2,2,1)
          + 224\*\Hhh(-1,-3,0)
          + \frct{20}{3}\*\Hhh(-1,-2,0)
          + 752\*\Hhh(-1,-2,2)
          - 864\*\Hhh(-1,-1,0)\*\zeta_2
\\
&& \nonumber
          + 1504\*\Hhh(-1,-1,3)
          + 476\*\Hhh(-1,0,0)
          + 1376\*\Hhh(-1,0,0)\*\zeta_2
          - 288\*\Hhh(-1,2,0)
          - 384\*\Hhh(-1,2,2)
          - 952\*\Hhh(-1,3,0)
\\
&& \nonumber
          - 768\*\Hhh(-1,3,1)
          - 64\*\Hhh(2,-3,0)
          + 48\*\Hhh(2,0,0)\*\zeta_2
          + \frct{1016}{3}\*\Hhh(2,1,0)
          - 16\*\Hhh(2,1,0)\*\zeta_2
          - 48\*\Hhh(2,3,0)
          - 64\*\Hhh(3,-2,0)
\\
&& \nonumber
          + 72\*\Hhh(4,0,0)
          + 48\*\Hhh(4,1,0)
          - 224\*\Hhhh(-2,-1,-1,0)
          - 224\*\Hhhh(-1,-2,-1,0)
          + 712\*\Hhhh(-1,-2,0,0)
          - 224\*\Hhhh(-1,-1,-2,0)
\\
&& \nonumber
          + \frct{40}{3}\*\Hhhh(-1,-1,0,0)
          + 752\*\Hhhh(-1,-1,2,0)
          - 266\*\Hhhh(-1,0,0,0)
          - 384\*\Hhhh(-1,2,0,0)
          - 384\*\Hhhh(-1,2,1,0)
          - 64\*\Hhhh(2,-2,0,0)
\\
&& \nonumber
          + 16\*\Hhhh(2,2,0,0)
          + 24\*\Hhhh(3,0,0,0)
          + 32\*\Hhhh(3,1,0,0)
          - 448\*\Hhhhh(-1,-1,-1,0,0)
          + 1464\*\Hhhhh(-1,-1,0,0,0)
          - 864\*\Hhhhh(-1,0,0,0,0)
\\
&& \nonumber
          + 48\*\Hhhhh(2,1,0,0,0)
          \biggr)
       + (1-x) \* \biggl(
          - 392\*\H(-3)\*\zeta_3
          + \frct{373}{12}\*\H(0)
          + \frct{1373}{3}\*\H(2)
          + 196\*\H(2)\*\zeta_3
          + 78\*\H(2)\*\zeta_2
\\
&& \nonumber
          + 512\*\Hh(-3,-1)\*\zeta_2
          + 256\*\Hh(-2,-2)\*\zeta_2
          - 196\*\Hh(-2,0)\*\zeta_3
          + 360\*\Hh(-2,4)
          + \frct{1373}{3}\*\Hh(1,0)
          + 196\*\Hh(1,0)\*\zeta_3
\\
&& \nonumber
          + 78\*\Hh(1,0)\*\zeta_2
          + \frct{1672}{3}\*\Hh(1,2)
          + 112\*\Hh(1,2)\*\zeta_2
          + 96\*\Hh(1,4)
          + \frct{1672}{3}\*\Hh(2,1)
          + 112\*\Hh(2,1)\*\zeta_2
          + 32\*\Hh(2,3)
\\
&& \nonumber
          - 480\*\Hhh(-3,-1,2)
          - 32\*\Hhh(-2,-3,0)
          - 240\*\Hhh(-2,-2,2)
          + 256\*\Hhh(-2,-1,0)\*\zeta_2
          - 480\*\Hhh(-2,-1,3)
          - 320\*\Hhh(-2,0,0)\*\zeta_2
\\
&& \nonumber
          + 120\*\Hhh(-2,3,0)
          + 128\*\Hhh(1,-3,0)
          + 572\*\Hhh(1,0,0)
          + 112\*\Hhh(1,0,0)\*\zeta_2
          + \frct{1672}{3}\*\Hhh(1,1,0)
          + 112\*\Hhh(1,1,0)\*\zeta_2
          + 32\*\Hhh(1,3,0)
\\
&& \nonumber
          + 64\*\Hhh(2,-2,0)
          + 64\*\Hhhh(-3,-1,-1,0)
          + 32\*\Hhhh(-2,-2,-1,0)
          - 168\*\Hhhh(-2,-2,0,0)
          + 32\*\Hhhh(-2,-1,-2,0)
\\
&& \nonumber
          - 240\*\Hhhh(-2,-1,2,0)
          + 96\*\Hhhh(0,0,0,0)\*\zeta_2
          + 128\*\Hhhh(1,-2,0,0)
          - 204\*\Hhhh(1,0,0,0)
          - 96\*\Hhhh(1,2,0,0)
          - 96\*\Hhhh(2,1,0,0)
\\
&& \nonumber
          + 64\*\Hhhhh(-2,-1,-1,0,0)
          - 408\*\Hhhhh(-2,-1,0,0,0)
          + 160\*\Hhhhh(-2,0,0,0,0)
          - 256\*\Hhhhh(1,0,0,0,0)
          - 288\*\Hhhhh(1,1,0,0,0)
          \biggr)
\\
&& \nonumber
       + \biggl(6 + \frct{1550}{3}\*x\biggr) \* \Hhhh(0,0,0,0)
       + \biggl(\frct{22}{3} + \frct{290}{3}\*x\biggr) \* \Hh(-3,0)
       + (8 + 72\*x) \* \Hh(3,0)\*\zeta_2
       + \biggl(\frct{31}{3} - 927\*x\biggr) \* \Hh(0,0)
\\
&& \nonumber
       + (32 + 96\*x) \* \Hh(-5,0)
       + (50 + 88\*x) \* \H(0)\*\zeta_5
       + (54 - 90\*x) \* \Hhh(0,0,0)\*\zeta_3
       - (62 - 354\*x) \* \Hhhh(2,0,0,0)
\\
&& \nonumber
       - \biggl(\frct{196}{3} + \frct{3004}{3}\*x\biggr) \* \Hh(0,0)\*\zeta_3
       + (72 + 128\*x) \* \Hh(5,0)
       - \biggl(\frct{463}{6} + \frct{4201}{6}\*x\biggr) \* \H(0)\*\zeta_3
       - \biggl(\frct{244}{3} - 8\*x\biggr) \* \Hhh(-2,0,0)
\\
&& \nonumber
       + \biggl(\frct{335}{4} - \frct{3877}{4}\*x\biggr) \* \Hhh(0,0,0)
       - (100 + 32\*x) \* \Hh(-4,0)
       + (112 - 128\*x) \* \Hhh(-3,2,0)
       - (112 + 16\*x) \* \Hhh(-3,-2,0)
\\
&& \nonumber
       - (120 - 160\*x) \* \H(6)
       - (120 - 72\*x) \* \H(4)\*\zeta_2
       - \biggl(\frct{249}{2} - \frct{939}{2}\*x\biggr) \* \H(0)\*\zeta_2
       + (126 - 98\*x) \* \Hh(2,0)\*\zeta_2
\\
&& \nonumber
       + (126 + 234\*x) \* \Hh(0,0)\*\zeta_4
       - (132 - 156\*x) \* \H(0)\*\zeta_2\*\zeta_3
       + \biggl(\frct{400}{3} + \frct{2960}{3}\*x\biggr) \* \H(5)
       + (140 - 84\*x) \* \H(3)\*\zeta_2
\\
&& \nonumber
       + (144 + 240\*x) \* (
            \Hh(3,2)
          + \Hhh(3,1,0)
          )
       - (144 + 912\*x) \* \Hhh(0,0,0)\*\zeta_2
       + \biggl(\frct{436}{3} + \frct{1300}{3}\*x\biggr) \* \Hhh(3,0,0)
\\
&& \nonumber
       + \biggl(\frct{983}{6} + \frct{728}{3}\*x\biggr) \* \H(0)\*\zeta_4
       - (240 + 144\*x) \* \Hhh(-4,-1,0)
       + \biggl(\frct{748}{3} + \frct{1820}{3}\*x\biggr) \* \Hh(4,0)
\\
&& \nonumber
       + (252 + 196\*x) \* \Hhh(-2,-2,0)
       + (278 + 414\*x) \* \Hhh(2,0,0)
       + (280 + 168\*x) \* \Hhh(-3,-1,0)
\\
&& \nonumber
       - \biggl(\frct{860}{3} + 1078\*x\biggr) \* \Hh(0,0)\*\zeta_2
       + (304 - 72\*x) \* \Hhh(-4,0,0)
       + (336 - 384\*x) \* \Hh(-4,2)
       - \biggl(\frct{1072}{3} + \frct{3440}{3}\*x\biggr) \* \Hh(-3,2)
\\
&& \nonumber
       + (368 - 200\*x) \* \Hhhh(-3,0,0,0)
       - \biggl(\frct{1112}{3} + \frct{2296}{3}\*x\biggr) \* \Hhh(-2,2,0)
       - (432 - 176\*x) \* \Hhhh(-3,-1,0,0)
\\
&& \nonumber
       + (432 + 720\*x) \* \Hh(4,1)
       - (444 + 620\*x) \* \Hhh(-3,0,0)
       + (452 + 1472\*x) \* \H(4)
       - (456 - 312\*x) \* \H(-4)\*\zeta_2
\\
&& \nonumber
       + (464 - 496\*x) \* \Hh(-3,3)
       + (465 - 679\*x) \* \H(3)
       - (472 - 424\*x) \* \Hh(-3,0)\*\zeta_2
       + \biggl(\frct{1490}{3} + \frct{2510}{3}\*x\biggr) \* \Hh(3,0)
\\
&& \nonumber
       + \biggl(\frct{1492}{3} + \frct{3692}{3}\*x\biggr) \* \H(-3)\*\zeta_2
       + \biggl(\frct{1037}{2} - \frct{1251}{2}\*x\biggr) \* \Hh(2,0)
       + (768 + 656\*x) \* \Hhhh(-2,-1,0,0)
\\
&& \nonumber
       - (782 + 1114\*x) \* \Hhhh(-2,0,0,0)
       - \biggl(\frct{2776}{3} + \frct{5144}{3}\*x\biggr) \* \Hh(-2,3)
       + \biggl(\frct{2810}{3} + \frct{3910}{3}\*x\biggr) \* \Hh(-2,0)\*\zeta_2
\\
&& \nonumber
       - 48\*x \* \Hhhhhh(0,0,0,0,0,0)
       + 140\*x \* \Hhhhh(0,0,0,0,0)
       + \pqq( - x) \* \biggl(
            660\*\H(-4)\*\zeta_2
          + 1456\*\H(-3)\*\zeta_3
          - 288\*\H(-3)\*\zeta_2
\\
&& \nonumber
          - 835\*\H(-2)\*\zeta_4
          - \frct{136}{3}\*\H(-2)\*\zeta_3
          - \frct{1258}{3}\*\H(-2)\*\zeta_2
          + 36\*\H(0)\*\zeta_5
          + \frct{65}{12}\*\H(0)\*\zeta_4
          + \frct{2155}{6}\*\H(0)\*\zeta_3
          - \frct{79}{4}\*\H(0)\*\zeta_2
\\
&& \nonumber
          + 390\*\H(0)\*\zeta_2\*\zeta_3
          - 36\*\H(3)
          + 72\*\H(3)\*\zeta_3
          - 60\*\H(3)\*\zeta_2
          - 361\*\H(4)
          + 156\*\H(4)\*\zeta_2
          - 168\*\H(5)
          + 320\*\H(6)
          + 96\*\Hh(-5,0)
\\
&& \nonumber
          + 194\*\Hh(-4,0)
          - 696\*\Hh(-4,2)
          - 1344\*\Hh(-3,-1)\*\zeta_2
          + \frct{940}{3}\*\Hh(-3,0)
          + 1468\*\Hh(-3,0)\*\zeta_2
          + \frct{688}{3}\*\Hh(-3,2)
\\
&& \nonumber
          - 1584\*\Hh(-3,3)
          - 1344\*\Hh(-2,-2)\*\zeta_2
          - 1456\*\Hh(-2,-1)\*\zeta_3
          + \frct{512}{3}\*\Hh(-2,-1)\*\zeta_2
          - \frct{151}{2}\*\Hh(-2,0)
          + 1184\*\Hh(-2,0)\*\zeta_3
\\
&& \nonumber
          - \frct{956}{3}\*\Hh(-2,0)\*\zeta_2
          + \frct{722}{3}\*\Hh(-2,2)
          + 8\*\Hh(-2,2)\*\zeta_2
          + \frct{872}{3}\*\Hh(-2,3)
          - 2024\*\Hh(-2,4)
          - 1344\*\Hh(-1,-3)\*\zeta_2
\\
&& \nonumber
          - 1456\*\Hh(-1,-2)\*\zeta_3
          + \frct{512}{3}\*\Hh(-1,-2)\*\zeta_2
          - 835\*\Hh(-1,0)\*\zeta_4
          - \frct{136}{3}\*\Hh(-1,0)\*\zeta_3
          - \frct{1258}{3}\*\Hh(-1,0)\*\zeta_2
          + \frct{1444}{3}\*\Hh(-1,3)
\\
&& \nonumber
          + 16\*\Hh(-1,3)\*\zeta_2
          + 184\*\Hh(-1,4)
          - 1376\*\Hh(-1,5)
          + 27\*\Hh(0,0)
          + 454\*\Hh(0,0)\*\zeta_4
          + \frct{316}{3}\*\Hh(0,0)\*\zeta_3
          + \frct{1138}{3}\*\Hh(0,0)\*\zeta_2
\\
&& \nonumber
          - 18\*\Hh(2,0)
          + 36\*\Hh(2,0)\*\zeta_3
          - 30\*\Hh(2,0)\*\zeta_2
          + 112\*\Hh(2,2)\*\zeta_2
          - \frct{361}{3}\*\Hh(3,0)
          + 164\*\Hh(3,0)\*\zeta_2
          + 224\*\Hh(3,1)\*\zeta_2
          + \frct{248}{3}\*\Hh(3,2)
\\
&& \nonumber
          + 38\*\Hh(4,0)
          + 248\*\Hh(4,1)
          + 112\*\Hh(4,2)
          + 288\*\Hh(5,0)
          + 448\*\Hh(5,1)
          - 72\*\Hhh(-4,-1,0)
          - 264\*\Hhh(-4,0,0)
          - 24\*\Hhh(-3,-2,0)
\\
&& \nonumber
          - \frct{352}{3}\*\Hhh(-3,-1,0)
          + 1344\*\Hhh(-3,-1,2)
          + 420\*\Hhh(-3,0,0)
          - 680\*\Hhh(-3,2,0)
          - 896\*\Hhh(-3,2,1)
          - 24\*\Hhh(-2,-3,0)
\\
&& \nonumber
          - \frct{352}{3}\*\Hhh(-2,-2,0)
          + 1344\*\Hhh(-2,-2,2)
          + 1344\*\Hhh(-2,-1,-1)\*\zeta_2
          - \frct{1072}{3}\*\Hhh(-2,-1,0)
          - 1920\*\Hhh(-2,-1,0)\*\zeta_2
\\
&& \nonumber
          - \frct{512}{3}\*\Hhh(-2,-1,2)
          + 2464\*\Hhh(-2,-1,3)
          + \frct{1705}{3}\*\Hhh(-2,0,0)
          + 1648\*\Hhh(-2,0,0)\*\zeta_2
          + \frct{184}{3}\*\Hhh(-2,2,0)
          - \frct{496}{3}\*\Hhh(-2,2,1)
\\
&& \nonumber
          - 448\*\Hhh(-2,2,2)
          - 1232\*\Hhh(-2,3,0)
          - 1344\*\Hhh(-2,3,1)
          - 72\*\Hhh(-1,-4,0)
          - \frct{352}{3}\*\Hhh(-1,-3,0)
          + 1344\*\Hhh(-1,-3,2)
\\
&& \nonumber
          + 1344\*\Hhh(-1,-2,-1)\*\zeta_2
          - \frct{1072}{3}\*\Hhh(-1,-2,0)
          - 1920\*\Hhh(-1,-2,0)\*\zeta_2
          - \frct{512}{3}\*\Hhh(-1,-2,2)
          + 2464\*\Hhh(-1,-2,3)
\\
&& \nonumber
          + 1344\*\Hhh(-1,-1,-2)\*\zeta_2
          - 1456\*\Hhh(-1,-1,0)\*\zeta_3
          + \frct{512}{3}\*\Hhh(-1,-1,0)\*\zeta_2
          - \frct{1024}{3}\*\Hhh(-1,-1,3)
          + 3360\*\Hhh(-1,-1,4)
\\
&& \nonumber
          - 151\*\Hhh(-1,0,0)
          + 912\*\Hhh(-1,0,0)\*\zeta_3
          - \frct{1048}{3}\*\Hhh(-1,0,0)\*\zeta_2
          + \frct{722}{3}\*\Hhh(-1,2,0)
          + 8\*\Hhh(-1,2,0)\*\zeta_2
          - \frct{496}{3}\*\Hhh(-1,2,2)
\\
&& \nonumber
          - \frct{136}{3}\*\Hhh(-1,3,0)
          - \frct{992}{3}\*\Hhh(-1,3,1)
          - 448\*\Hhh(-1,3,2)
          - 1016\*\Hhh(-1,4,0)
          - 1344\*\Hhh(-1,4,1)
          + \frct{345}{4}\*\Hhh(0,0,0)
\\
&& \nonumber
          - 192\*\Hhh(0,0,0)\*\zeta_3
          + 304\*\Hhh(0,0,0)\*\zeta_2
          + 112\*\Hhh(2,0,0)\*\zeta_2
          + 112\*\Hhh(2,1,0)\*\zeta_2
          + \frct{160}{3}\*\Hhh(3,0,0)
          + \frct{248}{3}\*\Hhh(3,1,0)
          + 64\*\Hhh(4,0,0)
\\
&& \nonumber
          + 112\*\Hhh(4,1,0)
          + 624\*\Hhhh(-3,-1,0,0)
          - 664\*\Hhhh(-3,0,0,0)
          + 672\*\Hhhh(-2,-2,0,0)
          - 1344\*\Hhhh(-2,-1,-1,2)
\\
&& \nonumber
          - 320\*\Hhhh(-2,-1,0,0)
          + 1120\*\Hhhh(-2,-1,2,0)
          + 896\*\Hhhh(-2,-1,2,1)
          + 572\*\Hhhh(-2,0,0,0)
          - 416\*\Hhhh(-2,2,0,0)
\\
&& \nonumber
          - 448\*\Hhhh(-2,2,1,0)
          + 624\*\Hhhh(-1,-3,0,0)
          - 1344\*\Hhhh(-1,-2,-1,2)
          - 320\*\Hhhh(-1,-2,0,0)
          + 1120\*\Hhhh(-1,-2,2,0)
\\
&& \nonumber
          + 896\*\Hhhh(-1,-2,2,1)
          - 1344\*\Hhhh(-1,-1,-2,2)
          + 1344\*\Hhhh(-1,-1,-1,0)\*\zeta_2
          - 2688\*\Hhhh(-1,-1,-1,3)
          - \frct{2144}{3}\*\Hhhh(-1,-1,0,0)
\\
&& \nonumber
          - 2496\*\Hhhh(-1,-1,0,0)\*\zeta_2
          - \frct{512}{3}\*\Hhhh(-1,-1,2,0)
          + 896\*\Hhhh(-1,-1,2,2)
          + 2016\*\Hhhh(-1,-1,3,0)
          + 1792\*\Hhhh(-1,-1,3,1)
\\
&& \nonumber
          + 765\*\Hhhh(-1,0,0,0)
          + 1200\*\Hhhh(-1,0,0,0)\*\zeta_2
          - \frct{320}{3}\*\Hhhh(-1,2,0,0)
          - \frct{496}{3}\*\Hhhh(-1,2,1,0)
          - 384\*\Hhhh(-1,3,0,0)
          - 448\*\Hhhh(-1,3,1,0)
\\
&& \nonumber
          - \frct{700}{3}\*\Hhhh(0,0,0,0)
          - 224\*\Hhhh(0,0,0,0)\*\zeta_2
          - 48\*\Hhhh(3,0,0,0)
          - 672\*\Hhhhh(-2,-1,-1,0,0)
          + 1360\*\Hhhhh(-2,-1,0,0,0)
\\
&& \nonumber
          - 744\*\Hhhhh(-2,0,0,0,0)
          - 672\*\Hhhhh(-1,-2,-1,0,0)
          + 1360\*\Hhhhh(-1,-2,0,0,0)
          - 672\*\Hhhhh(-1,-1,-2,0,0)
\\
&& \nonumber
          - 1344\*\Hhhhh(-1,-1,-1,2,0)
          - 608\*\Hhhhh(-1,-1,0,0,0)
          + 896\*\Hhhhh(-1,-1,2,0,0)
          + 896\*\Hhhhh(-1,-1,2,1,0)
          + 544\*\Hhhhh(-1,0,0,0,0)
\\
&& \nonumber
          + 96\*\Hhhhh(-1,2,0,0,0)
          - 310\*\Hhhhh(0,0,0,0,0)
          - 2016\*\Hhhhhh(-1,-1,-1,0,0,0)
          + 1408\*\Hhhhhh(-1,-1,0,0,0,0)
          - 200\*\Hhhhhh(-1,0,0,0,0,0)
\\
&& \nonumber
          - 48\*\Hhhhhh(0,0,0,0,0,0)
          \biggr)
       + \pqq(x) \* \biggl(
            84\*\H(-4)\*\zeta_2
          + 280\*\H(-3)\*\zeta_3
          + 36\*\H(-3)\*\zeta_2
          + \frct{587}{48}\*\H(0)
          - 26\*\H(0)\*\zeta_5
          + \frct{149}{4}\*\H(0)\*\zeta_4
\\
&& \nonumber
          - \frct{200}{3}\*\H(0)\*\zeta_3
          - \frct{229}{4}\*\H(0)\*\zeta_2
          - 6\*\H(0)\*\zeta_2\*\zeta_3
          + 313\*\H(2)\*\zeta_4
          - 44\*\H(2)\*\zeta_3
          + 320\*\H(3)\*\zeta_3
          + 226\*\H(4)
          + 12\*\H(4)\*\zeta_2
\\
&& \nonumber
          + \frct{640}{3}\*\H(5)
          + 40\*\H(6)
          - 96\*\Hh(-5,0)
          - 90\*\Hh(-4,0)
          + 120\*\Hh(-4,2)
          - 384\*\Hh(-3,-1)\*\zeta_2
          + 36\*\Hh(-3,0)
          + 156\*\Hh(-3,0)\*\zeta_2
\\
&& \nonumber
          - 96\*\Hh(-3,2)
          + 64\*\Hh(-3,3)
          - 192\*\Hh(-2,-2)\*\zeta_2
          + 140\*\Hh(-2,0)\*\zeta_3
          + 18\*\Hh(-2,0)\*\zeta_2
          - 96\*\Hh(-2,3)
          - 24\*\Hh(-2,4)
\\
&& \nonumber
          + \frct{11}{4}\*\Hh(0,0)
          + 34\*\Hh(0,0)\*\zeta_4
          - \frct{356}{3}\*\Hh(0,0)\*\zeta_3
          - \frct{464}{3}\*\Hh(0,0)\*\zeta_2
          - 272\*\Hh(1,-3)\*\zeta_2
          + 313\*\Hh(1,0)\*\zeta_4
          - 44\*\Hh(1,0)\*\zeta_3
\\
&& \nonumber
          + 320\*\Hh(1,2)\*\zeta_3
          - 232\*\Hh(1,4)
          + 64\*\Hh(1,5)
          - 136\*\Hh(2,-2)\*\zeta_2
          + 352\*\Hh(2,0)\*\zeta_3
          + \frct{52}{3}\*\Hh(2,0)\*\zeta_2
          + 320\*\Hh(2,1)\*\zeta_3
\\
&& \nonumber
          - \frct{232}{3}\*\Hh(2,3)
          + 112\*\Hh(2,4)
          + \frct{226}{3}\*\Hh(3,0)
          + 4\*\Hh(3,0)\*\zeta_2
          + 48\*\Hh(3,2)
          + 64\*\Hh(3,3)
          + \frct{508}{3}\*\Hh(4,0)
          + 144\*\Hh(4,1)
\\
&& \nonumber
          + 32\*\Hh(4,2)
          + 72\*\Hh(5,0)
          + 128\*\Hh(5,1)
          + 408\*\Hhh(-4,-1,0)
          - 192\*\Hhh(-4,0,0)
          + 360\*\Hhh(-3,-2,0)
          - 120\*\Hhh(-3,-1,0)
\\
&& \nonumber
          + 160\*\Hhh(-3,-1,2)
          - 48\*\Hhh(-3,0,0)
          + 104\*\Hhh(-3,2,0)
          + 128\*\Hhh(-3,2,1)
          + 224\*\Hhh(-2,-3,0)
          - 60\*\Hhh(-2,-2,0)
\\
&& \nonumber
          + 80\*\Hhh(-2,-2,2)
          - 192\*\Hhh(-2,-1,0)\*\zeta_2
          + 160\*\Hhh(-2,-1,3)
          + 36\*\Hhh(-2,0,0)
          + 128\*\Hhh(-2,0,0)\*\zeta_2
          - 48\*\Hhh(-2,2,0)
\\
&& \nonumber
          + 64\*\Hhh(-2,2,2)
          + 56\*\Hhh(-2,3,0)
          + 128\*\Hhh(-2,3,1)
          - \frct{597}{4}\*\Hhh(0,0,0)
          + 120\*\Hhh(0,0,0)\*\zeta_3
          - 250\*\Hhh(0,0,0)\*\zeta_2
          + 120\*\Hhh(1,-4,0)
\\
&& \nonumber
          - 240\*\Hhh(1,-3,0)
          + 320\*\Hhh(1,-3,2)
          - 136\*\Hhh(1,-2,0)\*\zeta_2
          + 320\*\Hhh(1,-2,3)
          + 384\*\Hhh(1,0,0)\*\zeta_3
          + \frct{104}{3}\*\Hhh(1,0,0)\*\zeta_2
\\
&& \nonumber
          + 320\*\Hhh(1,1,0)\*\zeta_3
          + 192\*\Hhh(1,1,4)
          + 64\*\Hhh(1,2,3)
          - \frct{232}{3}\*\Hhh(1,3,0)
          + 128\*\Hhh(1,3,2)
          + 208\*\Hhh(1,4,0)
          + 384\*\Hhh(1,4,1)
\\
&& \nonumber
          + 168\*\Hhh(2,-3,0)
          - 120\*\Hhh(2,-2,0)
          + 160\*\Hhh(2,-2,2)
          + \frct{400}{3}\*\Hhh(2,0,0)
          + 8\*\Hhh(2,0,0)\*\zeta_2
          + 64\*\Hhh(2,1,3)
          + 96\*\Hhh(2,3,0)
\\
&& \nonumber
          + 128\*\Hhh(2,3,1)
          + 128\*\Hhh(3,-2,0)
          + 48\*\Hhh(3,0,0)
          + 48\*\Hhh(3,1,0)
          - 16\*\Hhh(4,0,0)
          + 32\*\Hhh(4,1,0)
          - 448\*\Hhhh(-3,-1,-1,0)
\\
&& \nonumber
          + 576\*\Hhhh(-3,-1,0,0)
          - 224\*\Hhhh(-3,0,0,0)
          - 224\*\Hhhh(-2,-2,-1,0)
          + 376\*\Hhhh(-2,-2,0,0)
          - 224\*\Hhhh(-2,-1,-2,0)
\\
&& \nonumber
          - 120\*\Hhhh(-2,-1,0,0)
          + 80\*\Hhhh(-2,-1,2,0)
          + 18\*\Hhhh(-2,0,0,0)
          + 64\*\Hhhh(-2,2,0,0)
          + 64\*\Hhhh(-2,2,1,0)
          + \frct{748}{3}\*\Hhhh(0,0,0,0)
\\
&& \nonumber
          - 64\*\Hhhh(0,0,0,0)\*\zeta_2
          + 96\*\Hhhh(1,-3,-1,0)
          + 192\*\Hhhh(1,-3,0,0)
          + 48\*\Hhhh(1,-2,-2,0)
          - 240\*\Hhhh(1,-2,0,0)
          + 160\*\Hhhh(1,-2,2,0)
\\
&& \nonumber
          + 400\*\Hhhh(1,0,0,0)
          + 24\*\Hhhh(1,0,0,0)\*\zeta_2
          + 256\*\Hhhh(1,1,-3,0)
          + 64\*\Hhhh(1,1,3,0)
          + 128\*\Hhhh(1,2,-2,0)
          - \frct{176}{3}\*\Hhhh(1,2,0,0)
\\
&& \nonumber
          - 64\*\Hhhh(1,3,0,0)
          + 128\*\Hhhh(1,3,1,0)
          + 48\*\Hhhh(2,-2,-1,0)
          + 184\*\Hhhh(2,-2,0,0)
          - 8\*\Hhhh(2,0,0,0)
          + 128\*\Hhhh(2,1,-2,0)
\\
&& \nonumber
          - \frct{176}{3}\*\Hhhh(2,1,0,0)
          - 192\*\Hhhh(2,2,0,0)
          - 176\*\Hhhh(3,0,0,0)
          - 192\*\Hhhh(3,1,0,0)
          - 448\*\Hhhhh(-2,-1,-1,0,0)
          + 456\*\Hhhhh(-2,-1,0,0,0)
\\
&& \nonumber
          - 160\*\Hhhhh(-2,0,0,0,0)
          + 310\*\Hhhhh(0,0,0,0,0)
          + 96\*\Hhhhh(1,-2,-1,0,0)
          + 168\*\Hhhhh(1,-2,0,0,0)
          + 144\*\Hhhhh(1,0,0,0,0)
\\
&& \nonumber
          + 256\*\Hhhhh(1,1,-2,0,0)
          - 176\*\Hhhhh(1,1,0,0,0)
          - 192\*\Hhhhh(1,1,2,0,0)
          - 416\*\Hhhhh(1,2,0,0,0)
          - 192\*\Hhhhh(1,2,1,0,0)
          - 296\*\Hhhhh(2,0,0,0,0)
\\
&& \nonumber
          - 416\*\Hhhhh(2,1,0,0,0)
          - 192\*\Hhhhh(2,1,1,0,0)
          + 48\*\Hhhhhh(0,0,0,0,0,0)
          - 200\*\Hhhhhh(1,0,0,0,0,0)
          - 512\*\Hhhhhh(1,1,0,0,0,0)
\\
&& \nonumber
          - 576\*\Hhhhhh(1,1,1,0,0,0)
          \biggr)
          \biggr\}
\\
&& \mbox{\hspn} \nonumber
       + \frct{16}{3}\,\*\cfs\*\cas \* \biggl\{
         (1+x) \* \biggl(
          - 708\*\H(-2)\*\zeta_3
          - 164\*\H(-2)\*\zeta_2
          + 120\*\H(3)\*\zeta_3
          + 720\*\Hh(-2,-1)\*\zeta_2
          - 336\*\Hh(-2,0)
\\
&& \nonumber
          + 204\*\Hh(-2,2)
          + 720\*\Hh(-1,-2)\*\zeta_2
          - 708\*\Hh(-1,0)\*\zeta_3
          - 164\*\Hh(-1,0)\*\zeta_2
          + 408\*\Hh(-1,3)
          + 1368\*\Hh(-1,4)
          + 60\*\Hh(2,0)\*\zeta_3
\\
&& \nonumber
          - 336\*\Hh(2,2)
          + 24\*\Hh(2,2)\*\zeta_2
          + 162\*\Hh(2,4)
          - 672\*\Hh(3,1)
          + 48\*\Hh(3,1)\*\zeta_2
          - 144\*\Hh(3,2)
          + 108\*\Hh(3,3)
          - 432\*\Hh(4,1)
\\
&& \nonumber
          - 36\*\Hh(4,2)
          - 144\*\Hh(5,1)
          + 80\*\Hhh(-2,-1,0)
          - 624\*\Hhh(-2,-1,2)
          + 288\*\Hhh(-2,2,1)
          - 192\*\Hhh(-1,-3,0)
          + 80\*\Hhh(-1,-2,0)
\\
&& \nonumber
          - 624\*\Hhh(-1,-2,2)
          + 720\*\Hhh(-1,-1,0)\*\zeta_2
          - 1248\*\Hhh(-1,-1,3)
          - 672\*\Hhh(-1,0,0)
          - 1104\*\Hhh(-1,0,0)\*\zeta_2
          + 204\*\Hhh(-1,2,0)
\\
&& \nonumber
          + 288\*\Hhh(-1,2,2)
          + 744\*\Hhh(-1,3,0)
          + 576\*\Hhh(-1,3,1)
          + 96\*\Hhh(2,-3,0)
          - 36\*\Hhh(2,0,0)\*\zeta_2
          - 336\*\Hhh(2,1,0)
          + 24\*\Hhh(2,1,0)\*\zeta_2
\\
&& \nonumber
          + 54\*\Hhh(2,3,0)
          + 96\*\Hhh(3,-2,0)
          - 144\*\Hhh(3,1,0)
          - 99\*\Hhh(4,0,0)
          - 36\*\Hhh(4,1,0)
          + 192\*\Hhhh(-2,-1,-1,0)
          + 192\*\Hhhh(-1,-2,-1,0)
\\
&& \nonumber
          - 600\*\Hhhh(-1,-2,0,0)
          + 192\*\Hhhh(-1,-1,-2,0)
          + 160\*\Hhhh(-1,-1,0,0)
          - 624\*\Hhhh(-1,-1,2,0)
          - 78\*\Hhhh(-1,0,0,0)
\\
&& \nonumber
          + 288\*\Hhhh(-1,2,0,0)
          + 288\*\Hhhh(-1,2,1,0)
          + 96\*\Hhhh(2,-2,0,0)
          - 42\*\Hhhh(2,2,0,0)
          - 99\*\Hhhh(3,0,0,0)
          - 84\*\Hhhh(3,1,0,0)
\\
&& \nonumber
          + 384\*\Hhhhh(-1,-1,-1,0,0)
          - 1224\*\Hhhhh(-1,-1,0,0,0)
          + 768\*\Hhhhh(-1,0,0,0,0)
          - 72\*\Hhhhh(2,0,0,0,0)
          - 126\*\Hhhhh(2,1,0,0,0)
          \biggr)
\\
&& \nonumber
       + (1-x) \* \biggl(
            360\*\H(-4)\*\zeta_2
          + 372\*\H(-3)\*\zeta_3
          - \frct{16997}{27}\*\H(0)
          - \frct{9800}{9}\*\H(2)
          - 276\*\H(2)\*\zeta_3
          + 40\*\H(2)\*\zeta_2
          - 24\*\Hh(-5,0)
\\
&& \nonumber
          - 480\*\Hh(-3,-1)\*\zeta_2
          + 408\*\Hh(-3,0)\*\zeta_2
          - 240\*\Hh(-2,-2)\*\zeta_2
          + 186\*\Hh(-2,0)\*\zeta_3
          - 324\*\Hh(-2,4)
          - \frct{9800}{9}\*\Hh(1,0)
\\
&& \nonumber
          - 276\*\Hh(1,0)\*\zeta_3
          + 40\*\Hh(1,0)\*\zeta_2
          - 528\*\Hh(1,2)
          - 96\*\Hh(1,2)\*\zeta_2
          - 144\*\Hh(1,4)
          - 528\*\Hh(2,1)
          - 96\*\Hh(2,1)\*\zeta_2
          - 48\*\Hh(2,3)
\\
&& \nonumber
          + 432\*\Hhh(-3,-1,2)
          + 48\*\Hhh(-2,-3,0)
          + 216\*\Hhh(-2,-2,2)
          - 240\*\Hhh(-2,-1,0)\*\zeta_2
          + 432\*\Hhh(-2,-1,3)
          + 288\*\Hhh(-2,0,0)\*\zeta_2
\\
&& \nonumber
          - 108\*\Hhh(-2,3,0)
          - 192\*\Hhh(1,-3,0)
          - \frct{2200}{3}\*\Hhh(1,0,0)
          - 96\*\Hhh(1,0,0)\*\zeta_2
          - 528\*\Hhh(1,1,0)
          - 96\*\Hhh(1,1,0)\*\zeta_2
          - 48\*\Hhh(1,3,0)
\\
&& \nonumber
          - 96\*\Hhh(2,-2,0)
          - 96\*\Hhhh(-3,-1,-1,0)
          - 48\*\Hhhh(-2,-2,-1,0)
          + 180\*\Hhhh(-2,-2,0,0)
          - 48\*\Hhhh(-2,-1,-2,0)
\\
&& \nonumber
          + 216\*\Hhhh(-2,-1,2,0)
          - 192\*\Hhhh(1,-2,0,0)
          + 228\*\Hhhh(1,0,0,0)
          + 144\*\Hhhh(1,2,0,0)
          + 144\*\Hhhh(2,1,0,0)
          - 96\*\Hhhhh(-2,-1,-1,0,0)
\\
&& \nonumber
          + 396\*\Hhhhh(-2,-1,0,0,0)
          - 192\*\Hhhhh(-2,0,0,0,0)
          + 384\*\Hhhhh(1,0,0,0,0)
          + 432\*\Hhhhh(1,1,0,0,0)
          \biggr)
       + (10 - 346\*x) \* \Hh(-3,0)
\\
&& \nonumber
       - (16 + 372\*x) \* \Hhh(-2,0,0)
       - (\frct{87}{2} + \frct{21}{2}\*x) \* \Hh(0,0)\*\zeta_4
       - (45 + 105\*x) \* \Hh(5,0)
       + \biggl(\frct{123}{2} + \frct{1001}{2}\*x\biggr) \* \H(0)\*\zeta_3
\\
&& \nonumber
       + (66 + 930\*x) \* \Hh(0,0)\*\zeta_3
       - (75 - 51\*x) \* \H(0)\*\zeta_5
       - (84 - 132\*x) \* \Hhh(-3,2,0)
       + (90 - 174\*x) \* \H(0)\*\zeta_2\*\zeta_3
\\
&& \nonumber
       + \biggl(\frct{207}{2} + \frct{1053}{2}\*x\biggr) \* \Hhh(0,0,0)\*\zeta_2
       - (108 - 156\*x) \* \Hhhh(0,0,0,0)\*\zeta_2
       + (108 - 36\*x) \* \H(4)\*\zeta_2
       - (117 - 75\*x) \* \Hh(2,0)\*\zeta_2
\\
&& \nonumber
       - (117 + 405\*x) \* \Hhh(3,0,0)
       + (120 - 24\*x) \* \Hhh(-3,-2,0)
       + (121 + 197\*x) \* \Hhhh(0,0,0,0)
       - \biggl(\frct{755}{6} - \frct{11119}{6}\*x\biggr) \* \Hhh(0,0,0)
\\
&& \nonumber
       + (135 - 165\*x) \* \H(6)
       + \biggl(\frct{273}{2} - \frct{975}{2}\*x\biggr) \* \Hhhh(2,0,0,0)
       - (138 - 54\*x) \* \H(3)\*\zeta_2
       - (138 + 650\*x) \* \H(5)
\\
&& \nonumber
       + (141 + 39\*x) \* \Hh(-4,0)
       - \biggl(\frct{1421}{8} - \frct{753}{8}\*x\biggr) \* \H(0)\*\zeta_4
       + (216 + 72\*x) \* \Hhh(-4,-1,0)
       - (234 + 150\*x) \* \Hhh(-2,-2,0)
\\
&& \nonumber
       - (246 - 174\*x) \* \Hhh(-4,0,0)
       - \biggl(\frct{501}{2} + \frct{757}{2}\*x\biggr) \* \Hh(4,0)
       - (252 - 396\*x) \* \Hh(-4,2)
       - (260 + 412\*x) \* \Hhh(2,0,0)
\\
&& \nonumber
       + (272 + 976\*x) \* \Hh(-3,2)
       - (276 + 108\*x) \* \Hhh(-3,-1,0)
       + (280 + 632\*x) \* \Hhh(-2,2,0)
       + (288 + 446\*x) \* \Hh(0,0)\*\zeta_2
\\
&& \nonumber
       + \biggl(\frct{974}{3} - \frct{2234}{3}\*x\biggr) \* \H(0)\*\zeta_2
       - (330 - 258\*x) \* \Hhhh(-3,0,0,0)
       - (384 - 480\*x) \* \Hh(-3,3)
\\
&& \nonumber
       + (408 - 216\*x) \* \Hhhh(-3,-1,0,0)
       - (410 + 1030\*x) \* \H(-3)\*\zeta_2
       - (432 + 1188\*x) \* \H(4)
       + (456 + 480\*x) \* \Hhh(-3,0,0)
\\
&& \nonumber
       - (480 + 732\*x) \* \Hh(3,0)
       - \biggl(\frct{1948}{3} - \frct{2452}{3}\*x\biggr) \* \H(3)
       - (684 + 516\*x) \* \Hhhh(-2,-1,0,0)
       - \biggl(\frct{2074}{3} - \frct{2326}{3}\*x\biggr) \* \Hh(2,0)
\\
&& \nonumber
       + (728 + 1432\*x) \* \Hh(-2,3)
       + (735 + 873\*x) \* \Hhhh(-2,0,0,0)
       - (757 + 1067\*x) \* \Hh(-2,0)\*\zeta_2
\\
&& \nonumber
       - \biggl(\frct{20899}{27} - \frct{51353}{27}\*x\biggr) \* \Hh(0,0)
       - 108\*x \* \Hhhhhh(0,0,0,0,0,0)
       - 75\*x \* \Hhhhh(0,0,0,0,0)
       - 48\*x \* \Hh(3,0)\*\zeta_2
       + 252\*x \* \Hhh(0,0,0)\*\zeta_3
\\
&& \nonumber
       + \pqq( - x) \* \biggl(
          - 648\*\H(-4)\*\zeta_2
          - 1248\*\H(-3)\*\zeta_3
          + 288\*\H(-3)\*\zeta_2
          + 570\*\H(-2)\*\zeta_4
          - 4\*\H(-2)\*\zeta_3
          + 602\*\H(-2)\*\zeta_2
\\
&& \nonumber
          + 129\*\H(0)\*\zeta_5
          - 25\*\H(0)\*\zeta_4
          - \frct{1037}{2}\*\H(0)\*\zeta_3
          + \frct{177}{4}\*\H(0)\*\zeta_2
          - 354\*\H(0)\*\zeta_2\*\zeta_3
          + 36\*\H(3)
          - 72\*\H(3)\*\zeta_3
          + 36\*\H(3)\*\zeta_2
\\
&& \nonumber
          + 501\*\H(4)
          - 144\*\H(4)\*\zeta_2
          + 216\*\H(5)
          - 360\*\H(6)
          - 318\*\Hh(-4,0)
          + 648\*\Hh(-4,2)
          + 1152\*\Hh(-3,-1)\*\zeta_2
          - 470\*\Hh(-3,0)
\\
&& \nonumber
          - 1272\*\Hh(-3,0)\*\zeta_2
          - 200\*\Hh(-3,2)
          + 1392\*\Hh(-3,3)
          + 1152\*\Hh(-2,-2)\*\zeta_2
          + 1248\*\Hh(-2,-1)\*\zeta_3
          - 112\*\Hh(-2,-1)\*\zeta_2
\\
&& \nonumber
          + \frct{249}{2}\*\Hh(-2,0)
          - 1056\*\Hh(-2,0)\*\zeta_3
          + 316\*\Hh(-2,0)\*\zeta_2
          - 334\*\Hh(-2,2)
          - 256\*\Hh(-2,3)
          + 1752\*\Hh(-2,4)
\\
&& \nonumber
          + 1152\*\Hh(-1,-3)\*\zeta_2
          + 1248\*\Hh(-1,-2)\*\zeta_3
          - 112\*\Hh(-1,-2)\*\zeta_2
          + 570\*\Hh(-1,0)\*\zeta_4
          - 4\*\Hh(-1,0)\*\zeta_3
          + 602\*\Hh(-1,0)\*\zeta_2
\\
&& \nonumber
          - 668\*\Hh(-1,3)
          - 168\*\Hh(-1,4)
          + 1248\*\Hh(-1,5)
          - 27\*\Hh(0,0)
          - 222\*\Hh(0,0)\*\zeta_4
          - 158\*\Hh(0,0)\*\zeta_3
          - 542\*\Hh(0,0)\*\zeta_2
\\
&& \nonumber
          + 18\*\Hh(2,0)
          - 36\*\Hh(2,0)\*\zeta_3
          + 18\*\Hh(2,0)\*\zeta_2
          - 96\*\Hh(2,2)\*\zeta_2
          + 167\*\Hh(3,0)
          - 144\*\Hh(3,0)\*\zeta_2
          - 192\*\Hh(3,1)\*\zeta_2
          - 88\*\Hh(3,2)
\\
&& \nonumber
          - 12\*\Hh(4,0)
          - 264\*\Hh(4,1)
          - 96\*\Hh(4,2)
          - 264\*\Hh(5,0)
          - 384\*\Hh(5,1)
          + 324\*\Hhh(-4,0,0)
          + 176\*\Hhh(-3,-1,0)
\\
&& \nonumber
          - 1152\*\Hhh(-3,-1,2)
          - 576\*\Hhh(-3,0,0)
          + 600\*\Hhh(-3,2,0)
          + 768\*\Hhh(-3,2,1)
          + 176\*\Hhh(-2,-2,0)
          - 1152\*\Hhh(-2,-2,2)
\\
&& \nonumber
          - 1152\*\Hhh(-2,-1,-1)\*\zeta_2
          + 536\*\Hhh(-2,-1,0)
          + 1632\*\Hhh(-2,-1,0)\*\zeta_2
          + 112\*\Hhh(-2,-1,2)
          - 2112\*\Hhh(-2,-1,3)
\\
&& \nonumber
          - 839\*\Hhh(-2,0,0)
          - 1416\*\Hhh(-2,0,0)\*\zeta_2
          - 56\*\Hhh(-2,2,0)
          + 176\*\Hhh(-2,2,1)
          + 384\*\Hhh(-2,2,2)
          + 1056\*\Hhh(-2,3,0)
\\
&& \nonumber
          + 1152\*\Hhh(-2,3,1)
          + 176\*\Hhh(-1,-3,0)
          - 1152\*\Hhh(-1,-3,2)
          - 1152\*\Hhh(-1,-2,-1)\*\zeta_2
          + 536\*\Hhh(-1,-2,0)
\\
&& \nonumber
          + 1632\*\Hhh(-1,-2,0)\*\zeta_2
          + 112\*\Hhh(-1,-2,2)
          - 2112\*\Hhh(-1,-2,3)
          - 1152\*\Hhh(-1,-1,-2)\*\zeta_2
\\
&& \nonumber
          + 1248\*\Hhh(-1,-1,0)\*\zeta_3
          - 112\*\Hhh(-1,-1,0)\*\zeta_2
          + 224\*\Hhh(-1,-1,3)
          - 2880\*\Hhh(-1,-1,4)
          + 249\*\Hhh(-1,0,0)
\\
&& \nonumber
          - 864\*\Hhh(-1,0,0)\*\zeta_3
          + 344\*\Hhh(-1,0,0)\*\zeta_2
          - 334\*\Hhh(-1,2,0)
          + 176\*\Hhh(-1,2,2)
          + 32\*\Hhh(-1,3,0)
          + 352\*\Hhh(-1,3,1)
\\
&& \nonumber
          + 384\*\Hhh(-1,3,2)
          + 888\*\Hhh(-1,4,0)
          + 1152\*\Hhh(-1,4,1)
          - \frct{639}{4}\*\Hhh(0,0,0)
          + 288\*\Hhh(0,0,0)\*\zeta_3
          - 402\*\Hhh(0,0,0)\*\zeta_2
\\
&& \nonumber
          - 96\*\Hhh(2,0,0)\*\zeta_2
          - 96\*\Hhh(2,1,0)\*\zeta_2
          - 44\*\Hhh(3,0,0)
          - 88\*\Hhh(3,1,0)
          - 60\*\Hhh(4,0,0)
          - 96\*\Hhh(4,1,0)
          - 576\*\Hhhh(-3,-1,0,0)
\\
&& \nonumber
          + 660\*\Hhhh(-3,0,0,0)
          - 576\*\Hhhh(-2,-2,0,0)
          + 1152\*\Hhhh(-2,-1,-1,2)
          + 408\*\Hhhh(-2,-1,0,0)
          - 960\*\Hhhh(-2,-1,2,0)
\\
&& \nonumber
          - 768\*\Hhhh(-2,-1,2,1)
          - 750\*\Hhhh(-2,0,0,0)
          + 360\*\Hhhh(-2,2,0,0)
          + 384\*\Hhhh(-2,2,1,0)
          - 576\*\Hhhh(-1,-3,0,0)
\\
&& \nonumber
          + 1152\*\Hhhh(-1,-2,-1,2)
          + 408\*\Hhhh(-1,-2,0,0)
          - 960\*\Hhhh(-1,-2,2,0)
          - 768\*\Hhhh(-1,-2,2,1)
          + 1152\*\Hhhh(-1,-1,-2,2)
\\
&& \nonumber
          - 1152\*\Hhhh(-1,-1,-1,0)\*\zeta_2
          + 2304\*\Hhhh(-1,-1,-1,3)
          + 1072\*\Hhhh(-1,-1,0,0)
          + 2112\*\Hhhh(-1,-1,0,0)\*\zeta_2
          + 112\*\Hhhh(-1,-1,2,0)
\\
&& \nonumber
          - 768\*\Hhhh(-1,-1,2,2)
          - 1728\*\Hhhh(-1,-1,3,0)
          - 1536\*\Hhhh(-1,-1,3,1)
          - 1107\*\Hhhh(-1,0,0,0)
          - 1080\*\Hhhh(-1,0,0,0)\*\zeta_2
\\
&& \nonumber
          + 88\*\Hhhh(-1,2,0,0)
          + 176\*\Hhhh(-1,2,1,0)
          + 336\*\Hhhh(-1,3,0,0)
          + 384\*\Hhhh(-1,3,1,0)
          + 350\*\Hhhh(0,0,0,0)
          + 288\*\Hhhh(0,0,0,0)\*\zeta_2
\\
&& \nonumber
          + 36\*\Hhhh(3,0,0,0)
          + 576\*\Hhhhh(-2,-1,-1,0,0)
          - 1200\*\Hhhhh(-2,-1,0,0,0)
          + 744\*\Hhhhh(-2,0,0,0,0)
          + 576\*\Hhhhh(-1,-2,-1,0,0)
\\
&& \nonumber
          - 1200\*\Hhhhh(-1,-2,0,0,0)
          + 576\*\Hhhhh(-1,-1,-2,0,0)
          + 1152\*\Hhhhh(-1,-1,-1,2,0)
          + 696\*\Hhhhh(-1,-1,0,0,0)
\\
&& \nonumber
          - 768\*\Hhhhh(-1,-1,2,0,0)
          - 768\*\Hhhhh(-1,-1,2,1,0)
          - 816\*\Hhhhh(-1,0,0,0,0)
          - 72\*\Hhhhh(-1,2,0,0,0)
          + 510\*\Hhhhh(0,0,0,0,0)
\\
&& \nonumber
          + 1728\*\Hhhhhh(-1,-1,-1,0,0,0)
          - 1344\*\Hhhhhh(-1,-1,0,0,0,0)
          + 360\*\Hhhhhh(-1,0,0,0,0,0)
          - 72\*\Hhhhhh(0,0,0,0,0,0)
          \biggr)
\\
&& \nonumber
       + \pqq(x) \* \biggl(
            72\*\H(-4)\*\zeta_2
          - 168\*\H(-3)\*\zeta_3
          - 108\*\H(-3)\*\zeta_2
          + \frct{10537}{144}\*\H(0)
          - 9\*\H(0)\*\zeta_5
          - \frct{163}{2}\*\H(0)\*\zeta_4
          + \frct{242}{3}\*\H(0)\*\zeta_3
\\
&& \nonumber
          + \frct{4879}{36}\*\H(0)\*\zeta_2
          + 66\*\H(0)\*\zeta_2\*\zeta_3
          - \frct{11104}{27}\*\H(2)
          - 642\*\H(2)\*\zeta_4
          - 4\*\H(2)\*\zeta_3
          + 268\*\H(2)\*\zeta_2
          - \frct{3560}{9}\*\H(3)
          - 480\*\H(3)\*\zeta_3
\\
&& \nonumber
          + 88\*\H(3)\*\zeta_2
          - \frct{947}{2}\*\H(4)
          - 264\*\H(5)
          + 162\*\Hh(-4,0)
          - 216\*\Hh(-4,2)
          + 288\*\Hh(-3,-1)\*\zeta_2
          - 54\*\Hh(-3,0)
          - 24\*\Hh(-3,0)\*\zeta_2
\\
&& \nonumber
          + 144\*\Hh(-3,2)
          - 192\*\Hh(-3,3)
          + 144\*\Hh(-2,-2)\*\zeta_2
          - 84\*\Hh(-2,0)\*\zeta_3
          - 54\*\Hh(-2,0)\*\zeta_2
          + 144\*\Hh(-2,3)
          - 72\*\Hh(-2,4)
\\
&& \nonumber
          - \frct{23959}{108}\*\Hh(0,0)
          - 306\*\Hh(0,0)\*\zeta_4
          + 134\*\Hh(0,0)\*\zeta_3
          + \frct{1571}{3}\*\Hh(0,0)\*\zeta_2
          + 384\*\Hh(1,-3)\*\zeta_2
          - \frct{11104}{27}\*\Hh(1,0)
\\
&& \nonumber
          - 642\*\Hh(1,0)\*\zeta_4
          - 4\*\Hh(1,0)\*\zeta_3
          + 268\*\Hh(1,0)\*\zeta_2
          - 480\*\Hh(1,2)\*\zeta_3
          + 348\*\Hh(1,4)
          - 192\*\Hh(1,5)
          + 192\*\Hh(2,-2)\*\zeta_2
\\
&& \nonumber
          - \frct{3560}{9}\*\Hh(2,0)
          - 528\*\Hh(2,0)\*\zeta_3
          + 44\*\Hh(2,0)\*\zeta_2
          - 480\*\Hh(2,1)\*\zeta_3
          + 116\*\Hh(2,3)
          - 192\*\Hh(2,4)
          - \frct{477}{2}\*\Hh(3,0)
          - 44\*\Hh(3,2)
\\
&& \nonumber
          - 96\*\Hh(3,3)
          - 198\*\Hh(4,0)
          - 132\*\Hh(4,1)
          - 48\*\Hh(4,2)
          - 96\*\Hh(5,0)
          - 192\*\Hh(5,1)
          - 288\*\Hhh(-4,-1,0)
          + 36\*\Hhh(-4,0,0)
\\
&& \nonumber
          - 288\*\Hhh(-3,-2,0)
          + 72\*\Hhh(-3,-1,0)
          - 96\*\Hhh(-3,-1,2)
          + 144\*\Hhh(-3,0,0)
          - 168\*\Hhh(-3,2,0)
          - 192\*\Hhh(-3,2,1)
\\
&& \nonumber
          - 192\*\Hhh(-2,-3,0)
          + 36\*\Hhh(-2,-2,0)
          - 48\*\Hhh(-2,-2,2)
          + 144\*\Hhh(-2,-1,0)\*\zeta_2
          - 96\*\Hhh(-2,-1,3)
          - 54\*\Hhh(-2,0,0)
\\
&& \nonumber
          - 48\*\Hhh(-2,0,0)\*\zeta_2
          + 72\*\Hhh(-2,2,0)
          - 96\*\Hhh(-2,2,2)
          - 120\*\Hhh(-2,3,0)
          - 192\*\Hhh(-2,3,1)
          - \frct{1733}{6}\*\Hhh(0,0,0)
          - 288\*\Hhh(0,0,0)\*\zeta_3
\\
&& \nonumber
          + 438\*\Hhh(0,0,0)\*\zeta_2
          - 288\*\Hhh(1,-4,0)
          + 288\*\Hhh(1,-3,0)
          - 384\*\Hhh(1,-3,2)
          + 192\*\Hhh(1,-2,0)\*\zeta_2
          - 384\*\Hhh(1,-2,3)
\\
&& \nonumber
          - \frct{3560}{9}\*\Hhh(1,0,0)
          - 576\*\Hhh(1,0,0)\*\zeta_3
          - 480\*\Hhh(1,1,0)\*\zeta_3
          - 288\*\Hhh(1,1,4)
          - 96\*\Hhh(1,2,3)
          + 116\*\Hhh(1,3,0)
          - 192\*\Hhh(1,3,2)
\\
&& \nonumber
          - 336\*\Hhh(1,4,0)
          - 576\*\Hhh(1,4,1)
          - 288\*\Hhh(2,-3,0)
          + 144\*\Hhh(2,-2,0)
          - 192\*\Hhh(2,-2,2)
          - \frct{609}{2}\*\Hhh(2,0,0)
          - 96\*\Hhh(2,1,3)
\\
&& \nonumber
          - 144\*\Hhh(2,3,0)
          - 192\*\Hhh(2,3,1)
          - 192\*\Hhh(3,-2,0)
          - 72\*\Hhh(3,0,0)
          - 44\*\Hhh(3,1,0)
          + 60\*\Hhh(4,0,0)
          - 48\*\Hhh(4,1,0)
\\
&& \nonumber
          + 384\*\Hhhh(-3,-1,-1,0)
          - 432\*\Hhhh(-3,-1,0,0)
          + 84\*\Hhhh(-3,0,0,0)
          + 192\*\Hhhh(-2,-2,-1,0)
          - 312\*\Hhhh(-2,-2,0,0)
\\
&& \nonumber
          + 192\*\Hhhh(-2,-1,-2,0)
          + 72\*\Hhhh(-2,-1,0,0)
          - 48\*\Hhhh(-2,-1,2,0)
          + 54\*\Hhhh(-2,0,0,0)
          - 96\*\Hhhh(-2,2,0,0)
          - 96\*\Hhhh(-2,2,1,0)
\\
&& \nonumber
          - 592\*\Hhhh(0,0,0,0)
          - 384\*\Hhhh(1,-3,0,0)
          + 288\*\Hhhh(1,-2,0,0)
          - 192\*\Hhhh(1,-2,2,0)
          - \frct{1343}{2}\*\Hhhh(1,0,0,0)
          - 384\*\Hhhh(1,1,-3,0)
\\
&& \nonumber
          - 96\*\Hhhh(1,1,3,0)
          - 192\*\Hhhh(1,2,-2,0)
          + 60\*\Hhhh(1,2,0,0)
          + 96\*\Hhhh(1,3,0,0)
          - 192\*\Hhhh(1,3,1,0)
          - 288\*\Hhhh(2,-2,0,0)
          - 26\*\Hhhh(2,0,0,0)
\\
&& \nonumber
          - 192\*\Hhhh(2,1,-2,0)
          + 60\*\Hhhh(2,1,0,0)
          + 288\*\Hhhh(2,2,0,0)
          + 300\*\Hhhh(3,0,0,0)
          + 288\*\Hhhh(3,1,0,0)
          + 384\*\Hhhhh(-2,-1,-1,0,0)
\\
&& \nonumber
          - 360\*\Hhhhh(-2,-1,0,0,0)
          + 96\*\Hhhhh(-2,0,0,0,0)
          - 510\*\Hhhhh(0,0,0,0,0)
          - 288\*\Hhhhh(1,-2,0,0,0)
          - 200\*\Hhhhh(1,0,0,0,0)
\\
&& \nonumber
          - 384\*\Hhhhh(1,1,-2,0,0)
          + 180\*\Hhhhh(1,1,0,0,0)
          + 288\*\Hhhhh(1,1,2,0,0)
          + 624\*\Hhhhh(1,2,0,0,0)
          + 288\*\Hhhhh(1,2,1,0,0)
          + 456\*\Hhhhh(2,0,0,0,0)
\\
&& \nonumber
          + 624\*\Hhhhh(2,1,0,0,0)
          + 288\*\Hhhhh(2,1,1,0,0)
          + 72\*\Hhhhhh(0,0,0,0,0,0)
          + 360\*\Hhhhhh(1,0,0,0,0,0)
          + 768\*\Hhhhhh(1,1,0,0,0,0)
\\
&& \nonumber
          + 864\*\Hhhhhh(1,1,1,0,0,0)
          \biggr)
          \biggr\}
\\
&& \mbox{\hspn} \nonumber
       + \frct{16}{3}\,\*\cft\*\nf \* \biggl\{
         (1+x) \* \biggl(
            32\*\H(-2)\*\zeta_2
          - 60\*\H(4)
          - 120\*\Hh(-2,0)
          + 32\*\Hh(-1,0)\*\zeta_2
          + 40\*\Hh(0,0)\*\zeta_3
          - 32\*\Hh(2,2)
          - 56\*\Hh(3,0)
\\
&& \nonumber
          - 64\*\Hh(3,1)
          + 64\*\Hhh(-2,-1,0)
          + 64\*\Hhh(-1,-2,0)
          - 240\*\Hhh(-1,0,0)
          - 36\*\Hhh(2,0,0)
          - 32\*\Hhh(2,1,0)
          - 16\*\Hhh(3,0,0)
\\
&& \nonumber
          + 128\*\Hhhh(-1,-1,0,0)
          - 192\*\Hhhh(-1,0,0,0)
          - 24\*\Hhhh(2,0,0,0)
          \biggr)
       + (1-x) \* \biggl(
            32\*\H(-3)\*\zeta_2
          + \frct{35}{2}\*\H(0)
          - 20\*\H(2)
          + 32\*\H(5)
\\
&& \nonumber
          + 48\*\Hh(-4,0)
          - 32\*\Hh(-3,2)
          + 16\*\Hh(-2,0)\*\zeta_2
          - 32\*\Hh(-2,3)
          - 20\*\Hh(1,0)
          - 64\*\Hh(1,2)
          - 64\*\Hh(2,1)
          + 8\*\Hh(4,0)
          + 48\*\Hhh(-3,0,0)
\\
&& \nonumber
          - 16\*\Hhh(-2,2,0)
          - 72\*\Hhh(1,0,0)
          - 64\*\Hhh(1,1,0)
          + 24\*\Hhhh(-2,0,0,0)
          \biggr)
       - (4 + 28\*x) \* \H(0)\*\zeta_3
       - (12 - 340\*x) \* \Hhhh(0,0,0,0)
\\
&& \nonumber
       - (16 + 176\*x) \* \Hh(-3,0)
       - \biggl(\frct{39}{2} - \frct{825}{2}\*x\biggr) \* \Hhh(0,0,0)
       + (20 + 28\*x) \* \H(0)\*\zeta_4
       - \biggl(\frct{51}{2} - \frct{29}{2}\*x\biggr) \* \Hh(0,0)
 \\
&& \nonumber
      - (27 + 123\*x) \* \H(0)\*\zeta_2
       + (32 - 48\*x) \* \Hh(0,0)\*\zeta_2
       - (42 - 102\*x) \* \H(3)
       - (57 - 87\*x) \* \Hh(2,0)
\\
&& \nonumber
       - (80 + 240\*x) \* \Hhh(-2,0,0)
       + 120\*x \* \Hhhhh(0,0,0,0,0)
       + \pqq( - x) \* \biggl(
          - 80\*\H(-2)\*\zeta_3
          + 184\*\H(-2)\*\zeta_2
          - 5\*\H(0)\*\zeta_4
          - 158\*\H(0)\*\zeta_3
\\
&& \nonumber 
         + 21\*\H(0)\*\zeta_2
          + 156\*\H(4)
          - 96\*\Hh(-4,0)
          - 136\*\Hh(-3,0)
          + 32\*\Hh(-3,2)
          + 64\*\Hh(-2,-1)\*\zeta_2
          + 42\*\Hh(-2,0)
          - 16\*\Hh(-2,0)\*\zeta_2
\\
&& \nonumber
          - 104\*\Hh(-2,2)
          + 64\*\Hh(-2,3)
          + 64\*\Hh(-1,-2)\*\zeta_2
          - 80\*\Hh(-1,0)\*\zeta_3
          + 184\*\Hh(-1,0)\*\zeta_2
          - 208\*\Hh(-1,3)
          + 96\*\Hh(-1,4)
\\
&& \nonumber
          + 8\*\Hh(0,0)\*\zeta_3
          - 172\*\Hh(0,0)\*\zeta_2
          + 52\*\Hh(3,0)
          - 32\*\Hh(3,2)
          - 24\*\Hh(4,0)
          - 96\*\Hh(4,1)
          + 64\*\Hhh(-3,-1,0)
          - 144\*\Hhh(-3,0,0)
\\
&& \nonumber
          + 64\*\Hhh(-2,-2,0)
          + 160\*\Hhh(-2,-1,0)
          - 64\*\Hhh(-2,-1,2)
          - 244\*\Hhh(-2,0,0)
          + 32\*\Hhh(-2,2,0)
          + 64\*\Hhh(-2,2,1)
          + 64\*\Hhh(-1,-3,0)
\\
&& \nonumber
          + 160\*\Hhh(-1,-2,0)
          - 64\*\Hhh(-1,-2,2)
          + 64\*\Hhh(-1,-1,0)\*\zeta_2
          - 128\*\Hhh(-1,-1,3)
          + 84\*\Hhh(-1,0,0)
          - 32\*\Hhh(-1,0,0)\*\zeta_2
\\
&& \nonumber
          - 104\*\Hhh(-1,2,0)
          + 64\*\Hhh(-1,2,2)
          + 64\*\Hhh(-1,3,0)
          + 128\*\Hhh(-1,3,1)
          - 63\*\Hhh(0,0,0)
          - 48\*\Hhh(0,0,0)\*\zeta_2
          - 16\*\Hhh(3,0,0)
\\
&& \nonumber
          - 32\*\Hhh(3,1,0)
          + 96\*\Hhhh(-2,-1,0,0)
          - 168\*\Hhhh(-2,0,0,0)
          + 96\*\Hhhh(-1,-2,0,0)
          + 320\*\Hhhh(-1,-1,0,0)
          - 64\*\Hhhh(-1,-1,2,0)
\\
&& \nonumber
          - 324\*\Hhhh(-1,0,0,0)
          + 32\*\Hhhh(-1,2,0,0)
          + 64\*\Hhhh(-1,2,1,0)
          + 112\*\Hhhh(0,0,0,0)
          + 96\*\Hhhhh(-1,-1,0,0,0)
          - 192\*\Hhhhh(-1,0,0,0,0)
\\
&& \nonumber
          + 120\*\Hhhhh(0,0,0,0,0)
          \biggr)
       + \pqq(x) \* \biggl(
          - \frct{313}{8}\*\H(0)
          - 15\*\H(0)\*\zeta_4
          + 56\*\H(0)\*\zeta_3
          + 15\*\H(0)\*\zeta_2
          + 55\*\H(2)
          - 96\*\H(2)\*\zeta_3
          + 12\*\H(3)
\\
&& \nonumber
          - 120\*\H(4)
          - 64\*\H(5)
          + \frct{89}{2}\*\Hh(0,0)
          - 88\*\Hh(0,0)\*\zeta_3
          + 92\*\Hh(0,0)\*\zeta_2
          + 55\*\Hh(1,0)
          - 96\*\Hh(1,0)\*\zeta_3
          + 48\*\Hh(1,4)
          + 12\*\Hh(2,0)
\\
&& \nonumber
          - 16\*\Hh(2,0)\*\zeta_2
          + 16\*\Hh(2,3)
          - 40\*\Hh(3,0)
          - 40\*\Hh(4,0)
          + 81\*\Hhh(0,0,0)
          + 48\*\Hhh(0,0,0)\*\zeta_2
          + 12\*\Hhh(1,0,0)
          - 32\*\Hhh(1,0,0)\*\zeta_2
\\
&& \nonumber
          + 16\*\Hhh(1,3,0)
          - 52\*\Hhh(2,0,0)
          - 112\*\Hhhh(0,0,0,0)
          - 156\*\Hhhh(1,0,0,0)
          + 32\*\Hhhh(1,2,0,0)
          + 24\*\Hhhh(2,0,0,0)
          + 32\*\Hhhh(2,1,0,0)
\\
&& \nonumber
          - 120\*\Hhhhh(0,0,0,0,0)
          + 96\*\Hhhhh(1,1,0,0,0)
          \biggl)
          \biggr\}
\\
&& \mbox{\hspn} \nonumber
       + \frct{16}{3}\,\*\cfs\*\ca\*\nf \* \biggl\{
         (1+x) \* \biggl(
          - 16\*\H(-2)\*\zeta_2
          + 36\*\H(4)
          + 60\*\Hh(-2,0)
          - 16\*\Hh(-1,0)\*\zeta_2
          - 36\*\Hh(0,0)\*\zeta_3
          + 24\*\Hh(2,2)
\\
&& \nonumber
          + 36\*\Hh(3,0)
          + 48\*\Hh(3,1)
          - 32\*\Hhh(-2,-1,0)
          - 32\*\Hhh(-1,-2,0)
          + 120\*\Hhh(-1,0,0)
          + 18\*\Hhh(0,0,0)\*\zeta_2
          + 24\*\Hhh(2,1,0)
          + 12\*\Hhh(3,0,0)
\\
&& \nonumber
          - 64\*\Hhhh(-1,-1,0,0)
          + 96\*\Hhhh(-1,0,0,0)
          + 18\*\Hhhh(2,0,0,0)
          \biggr)
       + (1-x) \* \biggl(
          - 16\*\H(-3)\*\zeta_2
          + \frct{9161}{54}\*\H(0)
          + \frct{1846}{9}\*\H(2)
\\
&& \nonumber
          - 16\*\H(2)\*\zeta_2
          - 24\*\Hh(-4,0)
          + 16\*\Hh(-3,2)
          - 8\*\Hh(-2,0)\*\zeta_2
          + 16\*\Hh(-2,3)
          + \frct{1846}{9}\*\Hh(1,0)
          - 16\*\Hh(1,0)\*\zeta_2
          + 48\*\Hh(1,2)
\\
&& \nonumber
          + 48\*\Hh(2,1)
          - 24\*\Hhh(-3,0,0)
          + 8\*\Hhh(-2,2,0)
          + \frct{344}{3}\*\Hhh(1,0,0)
          + 48\*\Hhh(1,1,0)
          - 12\*\Hhhh(-2,0,0,0)
          - 12\*\Hhhh(1,0,0,0)
          \biggr)
\\
&& \nonumber
       - (6 - 2\*x) \* \Hh(4,0)
       + (8 + 88\*x) \* \Hh(-3,0)
       - (14 + 18\*x) \* \H(0)\*\zeta_4
       + (18 - 10\*x) \* \H(0)\*\zeta_3
       + (20 + 28\*x) \* \Hhh(2,0,0)
\\
&& \nonumber
       - (24 - 64\*x) \* \Hh(0,0)\*\zeta_2
       - (24 - 8\*x) \* \H(5)
       + \biggl(\frct{77}{3} - \frct{1276}{3}\*x\biggr) \* \Hhh(0,0,0)
       + (40 + 120\*x) \* \Hhh(-2,0,0)
\\
&& \nonumber
       - (44 + 204\*x) \* \Hhhh(0,0,0,0)
       - \biggl(\frct{145}{3} - \frct{379}{3}\*x\biggr) \* \H(0)\*\zeta_2
       + \biggl(\frct{290}{3} - \frct{398}{3}\*x\biggr) \* \H(3)
       + \biggl(\frct{317}{3} - \frct{371}{3}\*x\biggr) \* \Hh(2,0)
\\
&& \nonumber
       + \biggl(\frct{6623}{27} - \frct{8821}{27}\*x\biggr) \* \Hh(0,0)
       - 60\*x \* \Hhhhh(0,0,0,0,0)
       + \pqq( - x) \* \biggl(
            40\*\H(-2)\*\zeta_3
          - 92\*\H(-2)\*\zeta_2
          + \frct{5}{2}\*\H(0)\*\zeta_4
          + 79\*\H(0)\*\zeta_3
\\
&& \nonumber
          - \frct{21}{2}\*\H(0)\*\zeta_2
          - 78\*\H(4)
          + 48\*\Hh(-4,0)
          + 68\*\Hh(-3,0)
          - 16\*\Hh(-3,2)
          - 32\*\Hh(-2,-1)\*\zeta_2
          - 21\*\Hh(-2,0)
          + 8\*\Hh(-2,0)\*\zeta_2
\\
&& \nonumber
          + 52\*\Hh(-2,2)
          - 32\*\Hh(-2,3)
          - 32\*\Hh(-1,-2)\*\zeta_2
          + 40\*\Hh(-1,0)\*\zeta_3
          - 92\*\Hh(-1,0)\*\zeta_2
          + 104\*\Hh(-1,3)
          - 48\*\Hh(-1,4)
\\
&& \nonumber
          - 4\*\Hh(0,0)\*\zeta_3
          + 86\*\Hh(0,0)\*\zeta_2
          - 26\*\Hh(3,0)
          + 16\*\Hh(3,2)
          + 12\*\Hh(4,0)
          + 48\*\Hh(4,1)
          - 32\*\Hhh(-3,-1,0)
          + 72\*\Hhh(-3,0,0)
\\
&& \nonumber
          - 32\*\Hhh(-2,-2,0)
          - 80\*\Hhh(-2,-1,0)
          + 32\*\Hhh(-2,-1,2)
          + 122\*\Hhh(-2,0,0)
          - 16\*\Hhh(-2,2,0)
          - 32\*\Hhh(-2,2,1)
          - 32\*\Hhh(-1,-3,0)
\\
&& \nonumber
          - 80\*\Hhh(-1,-2,0)
          + 32\*\Hhh(-1,-2,2)
          - 32\*\Hhh(-1,-1,0)\*\zeta_2
          + 64\*\Hhh(-1,-1,3)
          - 42\*\Hhh(-1,0,0)
          + 16\*\Hhh(-1,0,0)\*\zeta_2
\\
&& \nonumber
          + 52\*\Hhh(-1,2,0)
          - 32\*\Hhh(-1,2,2)
          - 32\*\Hhh(-1,3,0)
          - 64\*\Hhh(-1,3,1)
          + \frct{63}{2}\*\Hhh(0,0,0)
          + 24\*\Hhh(0,0,0)\*\zeta_2
          + 8\*\Hhh(3,0,0)
          + 16\*\Hhh(3,1,0)
\\
&& \nonumber
          - 48\*\Hhhh(-2,-1,0,0)
          + 84\*\Hhhh(-2,0,0,0)
          - 48\*\Hhhh(-1,-2,0,0)
          - 160\*\Hhhh(-1,-1,0,0)
          + 32\*\Hhhh(-1,-1,2,0)
          + 162\*\Hhhh(-1,0,0,0)
\\
&& \nonumber
          - 16\*\Hhhh(-1,2,0,0)
          - 32\*\Hhhh(-1,2,1,0)
          - 56\*\Hhhh(0,0,0,0)
          - 48\*\Hhhhh(-1,-1,0,0,0)
          + 96\*\Hhhhh(-1,0,0,0,0)
          - 60\*\Hhhhh(0,0,0,0,0)
          \biggr)
\\
&& \nonumber
       + \pqq(x) \* \biggl(
          - \frct{239}{36}\*\H(0)
          + \frct{91}{2}\*\H(0)\*\zeta_4
          - \frct{121}{3}\*\H(0)\*\zeta_3
          - \frct{967}{18}\*\H(0)\*\zeta_2
          + \frct{2594}{27}\*\H(2)
          + 88\*\H(2)\*\zeta_3
          - 40\*\H(2)\*\zeta_2
\\
&& \nonumber
          + \frct{1156}{9}\*\H(3)
          - 16\*\H(3)\*\zeta_2
          + 95\*\H(4)
          + 48\*\H(5)
          + \frct{965}{27}\*\Hh(0,0)
          + 76\*\Hh(0,0)\*\zeta_3
          - \frct{292}{3}\*\Hh(0,0)\*\zeta_2
          + \frct{2594}{27}\*\Hh(1,0)
\\
&& \nonumber
          + 88\*\Hh(1,0)\*\zeta_3
          - 40\*\Hh(1,0)\*\zeta_2
          - 24\*\Hh(1,4)
          + \frct{1156}{9}\*\Hh(2,0)
          - 8\*\Hh(2,0)\*\zeta_2
          - 8\*\Hh(2,3)
          + 61\*\Hh(3,0)
          + 8\*\Hh(3,2)
          + 36\*\Hh(4,0)
\\
&& \nonumber
          + 24\*\Hh(4,1)
          + \frct{769}{6}\*\Hhh(0,0,0)
          - 60\*\Hhh(0,0,0)\*\zeta_2
          + \frct{1156}{9}\*\Hhh(1,0,0)
          - 8\*\Hhh(1,3,0)
          + 73\*\Hhh(2,0,0)
          + 8\*\Hhh(3,1,0)
          + 144\*\Hhhh(0,0,0,0)
\\
&& \nonumber
          + 131\*\Hhhh(1,0,0,0)
          - 24\*\Hhhh(1,2,0,0)
          - 28\*\Hhhh(2,0,0,0)
          - 24\*\Hhhh(2,1,0,0)
          + 60\*\Hhhhh(0,0,0,0,0)
          - 16\*\Hhhhh(1,0,0,0,0)
          - 72\*\Hhhhh(1,1,0,0,0)
          \biggr)
          \biggr\}
\\
&& \mbox{\hspn} \nonumber
       + \frct{16}{81}\,\*\cfs\*\nfs \* \biggl\{
         108\* (1+x) \* \Hhhh(0,0,0,0)
       + (1-x) \* \biggl(
          - 260\*\H(0)
          + 72\*\H(0)\*\zeta_2
          - 276\*\H(2)
          - 144\*\H(3)
          - 276\*\Hh(1,0)
\\
&& \nonumber
          - 144\*\Hh(2,0)
          - 144\*\Hhh(1,0,0)
          \biggr)
       - (90 - 450\*x) \* \Hhh(0,0,0)
       - (466 - 398\*x) \* \Hh(0,0)
       + \pqq(x) \* \biggl(
          - \frct{159}{4}\*\H(0)
          - 36\*\H(0)\*\zeta_3
\\
&& \nonumber
          + 120\*\H(0)\*\zeta_2
          - 76\*\H(2)
          - 240\*\H(3)
          - 108\*\H(4)
          + 23\*\Hh(0,0)
          + 72\*\Hh(0,0)\*\zeta_2
          - 76\*\Hh(1,0)
          - 240\*\Hh(2,0)
          - 108\*\Hh(3,0)
\\
&&
          - 279\*\Hhh(0,0,0)
          - 240\*\Hhh(1,0,0)
          - 108\*\Hhh(2,0,0)
          - 216\*\Hhhh(0,0,0,0)
          - 108\*\Hhhh(1,0,0,0)
          \biggr)
          \biggr\}
\:\: .
\eea
The most compact representation of $\delta\, P^{\,(3)-}(x)$ is via its
difference to $\delta\, P^{\,(3)+}(x)$,
\bea
\lefteqn{
\delta\, P^{\,(3)+}(x) - \delta\, P^{\,(3)-}(x) \, = \,
} 
\nonumber \\ && \mbox{\hspn} \phantom{+}\nonumber
         16\*\cft\*(\ca-2\*\cf) \* \biggl\{
         (1+x) \* \biggl(
          - 208\*\H(-2)\*\zeta_3
          + \frct{40}{3}\*\H(-2)\*\zeta_2
          + 192\*\Hh(-2,-1)\*\zeta_2
          - 376\*\Hh(-2,0)
          + 96\*\Hh(-2,2)
\\
&& \nonumber
          + 192\*\Hh(-1,-2)\*\zeta_2
          - 208\*\Hh(-1,0)\*\zeta_3
          + \frct{40}{3}\*\Hh(-1,0)\*\zeta_2
          + 192\*\Hh(-1,3)
          + 480\*\Hh(-1,4)
          - \frct{400}{3}\*\Hh(2,2)
          - \frct{800}{3}\*\Hh(3,1)
\\
&& \nonumber
          + \frct{656}{3}\*\Hhh(-2,-1,0)
          - 192\*\Hhh(-2,-1,2)
          + 128\*\Hhh(-2,2,1)
          + \frct{656}{3}\*\Hhh(-1,-2,0)
          - 192\*\Hhh(-1,-2,2)
          + 192\*\Hhh(-1,-1,0)\*\zeta_2
\\
&& \nonumber
          - 384\*\Hhh(-1,-1,3)
          - 752\*\Hhh(-1,0,0)
          - 320\*\Hhh(-1,0,0)\*\zeta_2
          + 96\*\Hhh(-1,2,0)
          + 128\*\Hhh(-1,2,2)
          + 288\*\Hhh(-1,3,0)
\\
&& \nonumber
          + 256\*\Hhh(-1,3,1)
          - \frct{224}{3}\*\Hhh(2,0,0)
          - \frct{400}{3}\*\Hhh(2,1,0)
          - 96\*\Hhhh(-1,-2,0,0)
          + \frct{1312}{3}\*\Hhhh(-1,-1,0,0)
          - 192\*\Hhhh(-1,-1,2,0)
\\
&& \nonumber
          - 536\*\Hhhh(-1,0,0,0)
          + 128\*\Hhhh(-1,2,0,0)
          + 128\*\Hhhh(-1,2,1,0)
          - 288\*\Hhhhh(-1,-1,0,0,0)
          + 384\*\Hhhhh(-1,0,0,0,0)
          \biggr)
\\
&& \nonumber
       + (1-x) \* \biggl(
            192\*\H(-4)\*\zeta_2
          + 208\*\H(-3)\*\zeta_3
          + 347\*\H(0)
          - 206\*\H(0)\*\zeta_5
          + 64\*\H(0)\*\zeta_2\*\zeta_3
          - \frct{452}{3}\*\H(2)
          - 24\*\H(2)\*\zeta_3
\\
&& \nonumber
          - 60\*\H(2)\*\zeta_2
          + 120\*\H(6)
          - 128\*\Hh(-5,0)
          - 144\*\Hh(-4,2)
          - 256\*\Hh(-3,-1)\*\zeta_2
          + 192\*\Hh(-3,0)\*\zeta_2
          - 192\*\Hh(-3,3)
\\
&& \nonumber
          - 128\*\Hh(-2,-2)\*\zeta_2
          + 104\*\Hh(-2,0)\*\zeta_3
          - 144\*\Hh(-2,4)
          - 152\*\Hh(0,0)\*\zeta_4
          - \frct{452}{3}\*\Hh(1,0)
          - 24\*\Hh(1,0)\*\zeta_3
          - 60\*\Hh(1,0)\*\zeta_2
\\
&& \nonumber
          - \frct{800}{3}\*\Hh(1,2)
          - 32\*\Hh(1,2)\*\zeta_2
          - \frct{800}{3}\*\Hh(2,1)
          - 32\*\Hh(2,1)\*\zeta_2
          + 24\*\Hh(5,0)
          + 96\*\Hhh(-4,-1,0)
          - 184\*\Hhh(-4,0,0)
\\
&& \nonumber
          + 96\*\Hhh(-3,-2,0)
          - 208\*\Hhh(-3,-1,0)
          + 192\*\Hhh(-3,-1,2)
          - 48\*\Hhh(-3,2,0)
          + 64\*\Hhh(-2,-3,0)
          - 104\*\Hhh(-2,-2,0)
\\
&& \nonumber
          + 96\*\Hhh(-2,-2,2)
          - 128\*\Hhh(-2,-1,0)\*\zeta_2
          + 192\*\Hhh(-2,-1,3)
          + 128\*\Hhh(-2,0,0)\*\zeta_2
          - 48\*\Hhh(-2,3,0)
          - 192\*\Hhh(0,0,0)\*\zeta_3
\\
&& \nonumber
          - \frct{448}{3}\*\Hhh(1,0,0)
          - 32\*\Hhh(1,0,0)\*\zeta_2
          - \frct{800}{3}\*\Hhh(1,1,0)
          - 32\*\Hhh(1,1,0)\*\zeta_2
          - 128\*\Hhhh(-3,-1,-1,0)
          + 224\*\Hhhh(-3,-1,0,0)
\\
&& \nonumber
          - 216\*\Hhhh(-3,0,0,0)
          - 64\*\Hhhh(-2,-2,-1,0)
          + 144\*\Hhhh(-2,-2,0,0)
          - 64\*\Hhhh(-2,-1,-2,0)
          + 96\*\Hhhh(-2,-1,2,0)
\\
&& \nonumber
          - 160\*\Hhhh(0,0,0,0)\*\zeta_2
          - 128\*\Hhhhh(-2,-1,-1,0,0)
          + 240\*\Hhhhh(-2,-1,0,0,0)
          - 192\*\Hhhhh(-2,0,0,0,0)
          + 192\*\Hhhhhh(0,0,0,0,0,0)
          \biggr)
\\
&& \nonumber
     + \biggl(12 + \frct{3292}{3}\*x\biggr) \* \Hhhh(0,0,0,0)
       - (32 + 96\*x) \* (
            \Hh(3,2)
          + \Hhh(3,0,0)
          + \Hhh(3,1,0)
          )
       - (40 - 24\*x) \* \Hh(2,0)\*\zeta_2
\\
&& \nonumber
       + (42 - 186\*x) \* \Hh(0,0)\*\zeta_2
       - (48 - 16\*x) \* \H(3)\*\zeta_2
     + \biggl(\frct{148}{3} - \frct{2164}{3}\*x\biggr) \* \Hh(-3,0)
     - \biggl(60 - \frct{716}{3}\*x\biggr) \* \H(3)
\\
&& \nonumber
       - (96 + 288\*x) \* \Hh(4,1)
     - \biggl(\frct{314}{3} - 194\*x\biggr) \* \Hh(2,0)
     - \biggl(\frct{344}{3} + \frct{664}{3}\*x\biggr) \* \Hh(4,0)
     - \biggl(\frct{364}{3} + 140\*x\biggr) \* \Hh(3,0)
\\
&& \nonumber
     - \biggl(\frct{388}{3} + 900\*x\biggr) \* \Hhh(-2,0,0)
       - (140 + 196\*x) \* \H(4)
     + \biggl(\frct{1061}{6} - \frct{161}{6}\*x\biggr) \* \H(0)\*\zeta_4
     + \biggl(\frct{544}{3} + \frct{416}{3}\*x\biggr) \* \Hhh(-2,2,0)
\\
&& \nonumber
     - \biggl(216 + \frct{748}{3}\*x\biggr) \* \H(0)\*\zeta_2
     + \biggl(\frct{704}{3} + \frct{448}{3}\*x\biggr) \* \Hh(-3,2)
     - \biggl(\frct{800}{3} + \frct{928}{3}\*x\biggr) \* \H(5)
     + \biggl(279 - \frct{661}{3}\*x\biggr) \* \H(0)\*\zeta_3
\\
&& \nonumber
       - (304 - 112\*x) \* \Hhhh(-2,-1,0,0)
     - \biggl(\frct{988}{3} + \frct{548}{3}\*x\biggr) \* \Hh(-2,0)\*\zeta_2
     - \biggl(\frct{1016}{3} + \frct{136}{3}\*x\biggr) \* \H(-3)\*\zeta_2
\\
&& \nonumber
     + \biggl(\frct{1184}{3} + \frct{928}{3}\*x\biggr) \* \Hh(-2,3)
       + (404 - 260\*x) \* \Hh(-4,0)
       + (408 + 72\*x) \* \Hh(0,0)\*\zeta_3
       + (420 + 60\*x) \* \Hhh(0,0,0)\*\zeta_2
\\
&& \nonumber
     + \biggl(\frct{1541}{3} + \frct{1565}{3}\*x\biggr) \* \Hh(0,0)
       + (524 + 826\*x) \* \Hhh(0,0,0)
       - (600 - 40\*x) \* \Hhhhh(0,0,0,0,0)
       + (604 - 124\*x) \* \Hhhh(-2,0,0,0)
\\
&& \nonumber
       + (608 - 320\*x) \* \Hhh(-3,0,0)
       + \pqq( - x) \* \biggl(
          - 408\*\H(-4)\*\zeta_2
          - 416\*\H(-3)\*\zeta_3
          + 192\*\H(-3)\*\zeta_2
          - 150\*\H(-2)\*\zeta_4
\\
&& \nonumber
          - \frct{304}{3}\*\H(-2)\*\zeta_3
          + \frct{2300}{3}\*\H(-2)\*\zeta_2
          + 416\*\H(0)\*\zeta_5
          - \frct{335}{6}\*\H(0)\*\zeta_4
          - \frct{1993}{3}\*\H(0)\*\zeta_3
          + \frct{157}{2}\*\H(0)\*\zeta_2
          - 164\*\H(0)\*\zeta_2\*\zeta_3
\\
&& \nonumber
          + 24\*\H(3)
          - 48\*\H(3)\*\zeta_3
          - 24\*\H(3)\*\zeta_2
          + 614\*\H(4)
          - 72\*\H(4)\*\zeta_2
          + 240\*\H(5)
          - 320\*\H(6)
          + 192\*\Hh(-5,0)
          - 460\*\Hh(-4,0)
\\
&& \nonumber
          + 336\*\Hh(-4,2)
          + 384\*\Hh(-3,-1)\*\zeta_2
          - \frct{1880}{3}\*\Hh(-3,0)
          - 456\*\Hh(-3,0)\*\zeta_2
          - \frct{224}{3}\*\Hh(-3,2)
          + 544\*\Hh(-3,3)
\\
&& \nonumber
          + 384\*\Hh(-2,-2)\*\zeta_2
          + 416\*\Hh(-2,-1)\*\zeta_3
          + \frct{128}{3}\*\Hh(-2,-1)\*\zeta_2
          + 181\*\Hh(-2,0)
          - 448\*\Hh(-2,0)\*\zeta_3
          + \frct{616}{3}\*\Hh(-2,0)\*\zeta_2
\\
&& \nonumber
          - \frct{1228}{3}\*\Hh(-2,2)
          + 16\*\Hh(-2,2)\*\zeta_2
          - \frct{304}{3}\*\Hh(-2,3)
          + 624\*\Hh(-2,4)
          + 384\*\Hh(-1,-3)\*\zeta_2
          + 416\*\Hh(-1,-2)\*\zeta_3
\\
&& \nonumber
          + \frct{128}{3}\*\Hh(-1,-2)\*\zeta_2
          - 150\*\Hh(-1,0)\*\zeta_4
          - \frct{304}{3}\*\Hh(-1,0)\*\zeta_3
          + \frct{2300}{3}\*\Hh(-1,0)\*\zeta_2
          - \frct{2456}{3}\*\Hh(-1,3)
          + 32\*\Hh(-1,3)\*\zeta_2
\\
&& \nonumber
          - 80\*\Hh(-1,4)
          + 576\*\Hh(-1,5)
          - 18\*\Hh(0,0)
          + 316\*\Hh(0,0)\*\zeta_4
          - \frct{632}{3}\*\Hh(0,0)\*\zeta_3
          - \frct{2060}{3}\*\Hh(0,0)\*\zeta_2
          + 12\*\Hh(2,0)
          - 24\*\Hh(2,0)\*\zeta_3
\\
&& \nonumber
          - 12\*\Hh(2,0)\*\zeta_2
          - 32\*\Hh(2,2)\*\zeta_2
          + \frct{614}{3}\*\Hh(3,0)
          - 56\*\Hh(3,0)\*\zeta_2
          - 64\*\Hh(3,1)\*\zeta_2
          - \frct{208}{3}\*\Hh(3,2)
          + 44\*\Hh(4,0)
          - 208\*\Hh(4,1)
\\
&& \nonumber
          - 32\*\Hh(4,2)
          - 128\*\Hh(5,0)
          - 128\*\Hh(5,1)
          - 144\*\Hhh(-4,-1,0)
          + 336\*\Hhh(-4,0,0)
          - 48\*\Hhh(-3,-2,0)
          + \frct{704}{3}\*\Hhh(-3,-1,0)
\\
&& \nonumber
          - 384\*\Hhh(-3,-1,2)
          - 696\*\Hhh(-3,0,0)
          + 240\*\Hhh(-3,2,0)
          + 256\*\Hhh(-3,2,1)
          - 48\*\Hhh(-2,-3,0)
          + \frct{704}{3}\*\Hhh(-2,-2,0)
\\
&& \nonumber
          - 384\*\Hhh(-2,-2,2)
          - 384\*\Hhh(-2,-1,-1)\*\zeta_2
          + \frct{2144}{3}\*\Hhh(-2,-1,0)
          + 512\*\Hhh(-2,-1,0)\*\zeta_2
          - \frct{128}{3}\*\Hhh(-2,-1,2)
\\
&& \nonumber
          - 704\*\Hhh(-2,-1,3)
          - \frct{3302}{3}\*\Hhh(-2,0,0)
          - 480\*\Hhh(-2,0,0)\*\zeta_2
          - \frct{80}{3}\*\Hhh(-2,2,0)
          + \frct{416}{3}\*\Hhh(-2,2,1)
          + 128\*\Hhh(-2,2,2)
\\
&& \nonumber
          + 352\*\Hhh(-2,3,0)
          + 384\*\Hhh(-2,3,1)
          - 144\*\Hhh(-1,-4,0)
          + \frct{704}{3}\*\Hhh(-1,-3,0)
          - 384\*\Hhh(-1,-3,2)
          - 384\*\Hhh(-1,-2,-1)\*\zeta_2
\\
&& \nonumber
          + \frct{2144}{3}\*\Hhh(-1,-2,0)
          + 512\*\Hhh(-1,-2,0)\*\zeta_2
          - \frct{128}{3}\*\Hhh(-1,-2,2)
          - 704\*\Hhh(-1,-2,3)
          - 384\*\Hhh(-1,-1,-2)\*\zeta_2
\\
&& \nonumber
          + 416\*\Hhh(-1,-1,0)\*\zeta_3
          + \frct{128}{3}\*\Hhh(-1,-1,0)\*\zeta_2
          - \frct{256}{3}\*\Hhh(-1,-1,3)
          - 960\*\Hhh(-1,-1,4)
          + 362\*\Hhh(-1,0,0)
          - 480\*\Hhh(-1,0,0)\*\zeta_3
\\
&& \nonumber
          + \frct{656}{3}\*\Hhh(-1,0,0)\*\zeta_2
          - \frct{1228}{3}\*\Hhh(-1,2,0)
          + 16\*\Hhh(-1,2,0)\*\zeta_2
          + \frct{416}{3}\*\Hhh(-1,2,2)
          - \frct{16}{3}\*\Hhh(-1,3,0)
          + \frct{832}{3}\*\Hhh(-1,3,1)
\\
&& \nonumber
          + 128\*\Hhh(-1,3,2)
          + 336\*\Hhh(-1,4,0)
          + 384\*\Hhh(-1,4,1)
          - \frct{507}{2}\*\Hhh(0,0,0)
          + 384\*\Hhh(0,0,0)\*\zeta_3
          - 464\*\Hhh(0,0,0)\*\zeta_2
\\
&& \nonumber
          - 32\*\Hhh(2,0,0)\*\zeta_2
          - 32\*\Hhh(2,1,0)\*\zeta_2
          - \frct{32}{3}\*\Hhh(3,0,0)
          - \frct{208}{3}\*\Hhh(3,1,0)
          - 32\*\Hhh(4,0,0)
          - 32\*\Hhh(4,1,0)
          - 288\*\Hhhh(-3,-1,0,0)
\\
&& \nonumber
          + 432\*\Hhhh(-3,0,0,0)
          - 192\*\Hhhh(-2,-2,0,0)
          + 384\*\Hhhh(-2,-1,-1,2)
          + 448\*\Hhhh(-2,-1,0,0)
          - 320\*\Hhhh(-2,-1,2,0)
\\
&& \nonumber
          - 256\*\Hhhh(-2,-1,2,1)
          - 856\*\Hhhh(-2,0,0,0)
          + 128\*\Hhhh(-2,2,0,0)
          + 128\*\Hhhh(-2,2,1,0)
          - 288\*\Hhhh(-1,-3,0,0)
\\
&& \nonumber
          + 384\*\Hhhh(-1,-2,-1,2)
          + 448\*\Hhhh(-1,-2,0,0)
          - 320\*\Hhhh(-1,-2,2,0)
          - 256\*\Hhhh(-1,-2,2,1)
          + 384\*\Hhhh(-1,-1,-2,2)
\\
&& \nonumber
          - 384\*\Hhhh(-1,-1,-1,0)\*\zeta_2
          + 768\*\Hhhh(-1,-1,-1,3)
          + \frct{4288}{3}\*\Hhhh(-1,-1,0,0)
          + 640\*\Hhhh(-1,-1,0,0)\*\zeta_2
          - \frct{128}{3}\*\Hhhh(-1,-1,2,0)
\\
&& \nonumber
          - 256\*\Hhhh(-1,-1,2,2)
          - 576\*\Hhhh(-1,-1,3,0)
          - 512\*\Hhhh(-1,-1,3,1)
          - 1422\*\Hhhh(-1,0,0,0)
          - 480\*\Hhhh(-1,0,0,0)\*\zeta_2
\\
&& \nonumber
          + \frct{64}{3}\*\Hhhh(-1,2,0,0)
          + \frct{416}{3}\*\Hhhh(-1,2,1,0)
          + 128\*\Hhhh(-1,3,0,0)
          + 128\*\Hhhh(-1,3,1,0)
          + \frct{1400}{3}\*\Hhhh(0,0,0,0)
          + 320\*\Hhhh(0,0,0,0)\*\zeta_2
\\
&& \nonumber
          + 192\*\Hhhhh(-2,-1,-1,0,0)
          - 480\*\Hhhhh(-2,-1,0,0,0)
          + 496\*\Hhhhh(-2,0,0,0,0)
          + 192\*\Hhhhh(-1,-2,-1,0,0)
          - 480\*\Hhhhh(-1,-2,0,0,0)
\\
&& \nonumber
          + 192\*\Hhhhh(-1,-1,-2,0,0)
          + 384\*\Hhhhh(-1,-1,-1,2,0)
          + 640\*\Hhhhh(-1,-1,0,0,0)
          - 256\*\Hhhhh(-1,-1,2,0,0)
          - 256\*\Hhhhh(-1,-1,2,1,0)
\\
&& \nonumber
          - 1088\*\Hhhhh(-1,0,0,0,0)
          + 740\*\Hhhhh(0,0,0,0,0)
          + 576\*\Hhhhhh(-1,-1,-1,0,0,0)
          - 768\*\Hhhhhh(-1,-1,0,0,0,0)
          + 560\*\Hhhhhh(-1,0,0,0,0,0)
\\
&& \nonumber
          - 288\*\Hhhhhh(0,0,0,0,0,0)
          \biggr)
       \biggr\}
\\
&& \mbox{\hspn} \nonumber
       + \frct{16}{3}\,\*\cfs\*(\ca-2\*\cf)^2 \* \biggl\{
       (1+x) \* \biggl(
          - 1248\*\H(-2)\*\zeta_3
          - 316\*\H(-2)\*\zeta_2
          + 1152\*\Hh(-2,-1)\*\zeta_2
          - 834\*\Hh(-2,0)
\\
&& \nonumber
          + 552\*\Hh(-2,2)
          + 1152\*\Hh(-1,-2)\*\zeta_2
          - 1248\*\Hh(-1,0)\*\zeta_3
          - 316\*\Hh(-1,0)\*\zeta_2
          + 1104\*\Hh(-1,3)
          + 2880\*\Hh(-1,4)
\\
&& \nonumber
          - 560\*\Hh(2,2)
          - 1120\*\Hh(3,1)
          + 472\*\Hhh(-2,-1,0)
          - 1152\*\Hhh(-2,-1,2)
          + 768\*\Hhh(-2,2,1)
          + 472\*\Hhh(-1,-2,0)
\\
&& \nonumber
          - 1152\*\Hhh(-1,-2,2)
          + 1152\*\Hhh(-1,-1,0)\*\zeta_2
          - 2304\*\Hhh(-1,-1,3)
          - 1668\*\Hhh(-1,0,0)
          - 2112\*\Hhh(-1,0,0)\*\zeta_2
\\
&& \nonumber
          + 552\*\Hhh(-1,2,0)
          + 768\*\Hhh(-1,2,2)
          + 1728\*\Hhh(-1,3,0)
          + 1536\*\Hhh(-1,3,1)
          - 560\*\Hhh(2,1,0)
          - 576\*\Hhhh(-1,-2,0,0)
\\
&& \nonumber
          + 944\*\Hhhh(-1,-1,0,0)
          - 1152\*\Hhhh(-1,-1,2,0)
          - 408\*\Hhhh(-1,0,0,0)
          + 768\*\Hhhh(-1,2,0,0)
          + 768\*\Hhhh(-1,2,1,0)
          + 72\*\Hhhh(2,0,0,0)
\\
&& \nonumber
          - 1728\*\Hhhhh(-1,-1,0,0,0)
          + 1344\*\Hhhhh(-1,0,0,0,0)
          \biggr)
       + (1-x) \* \biggl(
            720\*\H(-4)\*\zeta_2
          + 744\*\H(-3)\*\zeta_3
          + \frct{2445}{2}\*\H(0)
\\
&& \nonumber
          - 126\*\H(0)\*\zeta_5
          + 264\*\H(0)\*\zeta_2\*\zeta_3
          - 766\*\H(2)
          - 72\*\H(2)\*\zeta_3
          - 156\*\H(2)\*\zeta_2
          - 192\*\H(3)\*\zeta_2
          + 144\*\H(4)\*\zeta_2
          + 300\*\H(6)
\\
&& \nonumber
          - 48\*\Hh(-5,0)
          + 390\*\Hh(-4,0)
          - 648\*\Hh(-4,2)
          - 960\*\Hh(-3,-1)\*\zeta_2
          + 816\*\Hh(-3,0)\*\zeta_2
          - 864\*\Hh(-3,3)
          - 480\*\Hh(-2,-2)\*\zeta_2
\\
&& \nonumber
          + 372\*\Hh(-2,0)\*\zeta_3
          - 648\*\Hh(-2,4)
          - 33\*\Hh(0,0)\*\zeta_4
          - 766\*\Hh(1,0)
          - 72\*\Hh(1,0)\*\zeta_3
          - 156\*\Hh(1,0)\*\zeta_2
          - 1120\*\Hh(1,2)
\\
&& \nonumber
          - 192\*\Hh(1,2)\*\zeta_2
          - 192\*\Hh(2,0)\*\zeta_2
          - 1120\*\Hh(2,1)
          - 192\*\Hh(2,1)\*\zeta_2
          + 48\*\Hh(3,0)\*\zeta_2
          + 60\*\Hh(5,0)
          + 144\*\Hhh(-4,-1,0)
\\
&& \nonumber
          - 420\*\Hhh(-4,0,0)
          + 144\*\Hhh(-3,-2,0)
          - 168\*\Hhh(-3,-1,0)
          + 864\*\Hhh(-3,-1,2)
          - 216\*\Hhh(-3,2,0)
          + 96\*\Hhh(-2,-3,0)
\\
&& \nonumber
          - 84\*\Hhh(-2,-2,0)
          + 432\*\Hhh(-2,-2,2)
          - 480\*\Hhh(-2,-1,0)\*\zeta_2
          + 864\*\Hhh(-2,-1,3)
          + 576\*\Hhh(-2,0,0)\*\zeta_2
          - 216\*\Hhh(-2,3,0)
\\
&& \nonumber
          - 252\*\Hhh(0,0,0)\*\zeta_3
          - 944\*\Hhh(1,0,0)
          - 192\*\Hhh(1,0,0)\*\zeta_2
          - 1120\*\Hhh(1,1,0)
          - 192\*\Hhh(1,1,0)\*\zeta_2
          - 192\*\Hhhh(-3,-1,-1,0)
\\
&& \nonumber
          + 624\*\Hhhh(-3,-1,0,0)
          - 588\*\Hhhh(-3,0,0,0)
          - 96\*\Hhhh(-2,-2,-1,0)
          + 360\*\Hhhh(-2,-2,0,0)
          - 96\*\Hhhh(-2,-1,-2,0)
\\
&& \nonumber
          + 432\*\Hhhh(-2,-1,2,0)
          - 264\*\Hhhh(0,0,0,0)\*\zeta_2
          + 144\*\Hhhh(1,0,0,0)
          - 192\*\Hhhhh(-2,-1,-1,0,0)
          + 792\*\Hhhhh(-2,-1,0,0,0)
\\
&& \nonumber
          - 384\*\Hhhhh(-2,0,0,0,0)
          + 108\*\Hhhhhh(0,0,0,0,0,0)
          \biggr)
       - (144 + 528\*x) \* \Hhh(3,0,0)
       + (162 + 1158\*x) \* \Hh(0,0)\*\zeta_2
\\
&& \nonumber
       - (174 + 956\*x) \* \H(0)\*\zeta_2
       - (192 + 576\*x) \* (
            \Hh(3,2)
          + \Hhh(3,1,0)
          )
       - (196 + 1100\*x) \* \Hh(-3,0)
\\
&& \nonumber
     - \biggl(\frct{799}{4} + \frct{3761}{4}\*x\biggr) \* \H(0)\*\zeta_4
       + (324 + 428\*x) \* \Hhhh(0,0,0,0)
       - (332 + 1236\*x) \* \Hhh(-2,0,0)
       - (424 + 520\*x) \* \Hhh(2,0,0)
\\
&& \nonumber
       - (424 + 1352\*x) \* \Hh(4,0)
       + (447 + 469\*x) \* \H(0)\*\zeta_3
       - (544 + 1952\*x) \* \H(5)
       - (552 + 2256\*x) \* \H(4)
\\
&& \nonumber
       + (576 + 1152\*x) \* \Hh(0,0)\*\zeta_3
       - (576 + 1728\*x) \* \Hh(4,1)
       - (645 + 75\*x) \* \Hhhhh(0,0,0,0,0)
       - (648 - 1240\*x) \* \H(3)
\\
&& \nonumber
       - (656 + 1224\*x) \* \Hh(3,0)
       + (657 + 1503\*x) \* \Hhh(0,0,0)\*\zeta_2
       - (744 + 408\*x) \* \Hhhh(-2,-1,0,0)
       - (796 - 1092\*x) \* \Hh(2,0)
\\
&& \nonumber
       + (800 + 1120\*x) \* \Hhh(-2,2,0)
       + (832 + 1472\*x) \* \Hh(-3,2)
       - (916 + 1388\*x) \* \H(-3)\*\zeta_2
       + (936 + 216\*x) \* \Hhh(-3,0,0)
\\
&& \nonumber
       + (1209 + 1869\*x) \* \Hhh(0,0,0)
       + (1350 + 1050\*x) \* \Hhhh(-2,0,0,0)
       + (1486 + 2578\*x) \* \Hh(0,0)
\\
&& \nonumber
       - (1514 + 1750\*x) \* \Hh(-2,0)\*\zeta_2
       + (1792 + 2432\*x) \* \Hh(-2,3)
       + \pqq( - x) \* \biggl(
          - 1296\*\H(-4)\*\zeta_2
          - 2496\*\H(-3)\*\zeta_3
\\
&& \nonumber
          + 576\*\H(-3)\*\zeta_2
          + 1140\*\H(-2)\*\zeta_4
          - 8\*\H(-2)\*\zeta_3
          + 1204\*\H(-2)\*\zeta_2
          + 258\*\H(0)\*\zeta_5
          - 50\*\H(0)\*\zeta_4
          - 1037\*\H(0)\*\zeta_3
\\
&& \nonumber
          + \frct{177}{2}\*\H(0)\*\zeta_2
          - 708\*\H(0)\*\zeta_2\*\zeta_3
          + 72\*\H(3)
          - 144\*\H(3)\*\zeta_3
          + 72\*\H(3)\*\zeta_2
          + 1002\*\H(4)
          - 288\*\H(4)\*\zeta_2
          + 432\*\H(5)
\\
&& \nonumber
          - 720\*\H(6)
          - 636\*\Hh(-4,0)
          + 1296\*\Hh(-4,2)
          + 2304\*\Hh(-3,-1)\*\zeta_2
          - 940\*\Hh(-3,0)
          - 2544\*\Hh(-3,0)\*\zeta_2
          - 400\*\Hh(-3,2)
\\
&& \nonumber
          + 2784\*\Hh(-3,3)
          + 2304\*\Hh(-2,-2)\*\zeta_2
          + 2496\*\Hh(-2,-1)\*\zeta_3
          - 224\*\Hh(-2,-1)\*\zeta_2
          + 249\*\Hh(-2,0)
          - 2112\*\Hh(-2,0)\*\zeta_3
\\
&& \nonumber
          + 632\*\Hh(-2,0)\*\zeta_2
          - 668\*\Hh(-2,2)
          - 512\*\Hh(-2,3)
          + 3504\*\Hh(-2,4)
          + 2304\*\Hh(-1,-3)\*\zeta_2
          + 2496\*\Hh(-1,-2)\*\zeta_3
\\
&& \nonumber
          - 224\*\Hh(-1,-2)\*\zeta_2
          + 1140\*\Hh(-1,0)\*\zeta_4
          - 8\*\Hh(-1,0)\*\zeta_3
          + 1204\*\Hh(-1,0)\*\zeta_2
          - 1336\*\Hh(-1,3)
          - 336\*\Hh(-1,4)
\\
&& \nonumber
          + 2496\*\Hh(-1,5)
          - 54\*\Hh(0,0)
          - 444\*\Hh(0,0)\*\zeta_4
          - 316\*\Hh(0,0)\*\zeta_3
          - 1084\*\Hh(0,0)\*\zeta_2
          + 36\*\Hh(2,0)
          - 72\*\Hh(2,0)\*\zeta_3
\\
&& \nonumber
          + 36\*\Hh(2,0)\*\zeta_2
          - 192\*\Hh(2,2)\*\zeta_2
          + 334\*\Hh(3,0)
          - 288\*\Hh(3,0)\*\zeta_2
          - 384\*\Hh(3,1)\*\zeta_2
          - 176\*\Hh(3,2)
          - 24\*\Hh(4,0)
          - 528\*\Hh(4,1)
\\
&& \nonumber
          - 192\*\Hh(4,2)
          - 528\*\Hh(5,0)
          - 768\*\Hh(5,1)
          + 648\*\Hhh(-4,0,0)
          + 352\*\Hhh(-3,-1,0)
          - 2304\*\Hhh(-3,-1,2)
          - 1152\*\Hhh(-3,0,0)
\\
&& \nonumber
          + 1200\*\Hhh(-3,2,0)
          + 1536\*\Hhh(-3,2,1)
          + 352\*\Hhh(-2,-2,0)
          - 2304\*\Hhh(-2,-2,2)
          - 2304\*\Hhh(-2,-1,-1)\*\zeta_2
\\
&& \nonumber
          + 1072\*\Hhh(-2,-1,0)
          + 3264\*\Hhh(-2,-1,0)\*\zeta_2
          + 224\*\Hhh(-2,-1,2)
          - 4224\*\Hhh(-2,-1,3)
          - 1678\*\Hhh(-2,0,0)
\\
&& \nonumber
          - 2832\*\Hhh(-2,0,0)\*\zeta_2
          - 112\*\Hhh(-2,2,0)
          + 352\*\Hhh(-2,2,1)
          + 768\*\Hhh(-2,2,2)
          + 2112\*\Hhh(-2,3,0)
          + 2304\*\Hhh(-2,3,1)
\\
&& \nonumber
          + 352\*\Hhh(-1,-3,0)
          - 2304\*\Hhh(-1,-3,2)
          - 2304\*\Hhh(-1,-2,-1)\*\zeta_2
          + 1072\*\Hhh(-1,-2,0)
          + 3264\*\Hhh(-1,-2,0)\*\zeta_2
\\
&& \nonumber
          + 224\*\Hhh(-1,-2,2)
          - 4224\*\Hhh(-1,-2,3)
          - 2304\*\Hhh(-1,-1,-2)\*\zeta_2
          + 2496\*\Hhh(-1,-1,0)\*\zeta_3
          - 224\*\Hhh(-1,-1,0)\*\zeta_2
\\
&& \nonumber
          + 448\*\Hhh(-1,-1,3)
          - 5760\*\Hhh(-1,-1,4)
          + 498\*\Hhh(-1,0,0)
          - 1728\*\Hhh(-1,0,0)\*\zeta_3
          + 688\*\Hhh(-1,0,0)\*\zeta_2
          - 668\*\Hhh(-1,2,0)
\\
&& \nonumber
          + 352\*\Hhh(-1,2,2)
          + 64\*\Hhh(-1,3,0)
          + 704\*\Hhh(-1,3,1)
          + 768\*\Hhh(-1,3,2)
          + 1776\*\Hhh(-1,4,0)
          + 2304\*\Hhh(-1,4,1)\\
&& \nonumber
          - \frct{639}{2}\*\Hhh(0,0,0)
          + 576\*\Hhh(0,0,0)\*\zeta_3
          - 804\*\Hhh(0,0,0)\*\zeta_2
          - 192\*\Hhh(2,0,0)\*\zeta_2
          - 192\*\Hhh(2,1,0)\*\zeta_2
          - 88\*\Hhh(3,0,0)
          - 176\*\Hhh(3,1,0)
\\
&& \nonumber
          - 120\*\Hhh(4,0,0)
          - 192\*\Hhh(4,1,0)
          - 1152\*\Hhhh(-3,-1,0,0)
          + 1320\*\Hhhh(-3,0,0,0)
          - 1152\*\Hhhh(-2,-2,0,0)
          + 2304\*\Hhhh(-2,-1,-1,2)
\\
&& \nonumber
          + 816\*\Hhhh(-2,-1,0,0)
          - 1920\*\Hhhh(-2,-1,2,0)
          - 1536\*\Hhhh(-2,-1,2,1)
          - 1500\*\Hhhh(-2,0,0,0)
          + 720\*\Hhhh(-2,2,0,0)
\\
&& \nonumber
          + 768\*\Hhhh(-2,2,1,0)
          - 1152\*\Hhhh(-1,-3,0,0)
          + 2304\*\Hhhh(-1,-2,-1,2)
          + 816\*\Hhhh(-1,-2,0,0)
          - 1920\*\Hhhh(-1,-2,2,0)
\\
&& \nonumber
          - 1536\*\Hhhh(-1,-2,2,1)
          + 2304\*\Hhhh(-1,-1,-2,2)
          - 2304\*\Hhhh(-1,-1,-1,0)\*\zeta_2
          + 4608\*\Hhhh(-1,-1,-1,3)
\\
&& \nonumber
          + 2144\*\Hhhh(-1,-1,0,0)
          + 4224\*\Hhhh(-1,-1,0,0)\*\zeta_2
          + 224\*\Hhhh(-1,-1,2,0)
          - 1536\*\Hhhh(-1,-1,2,2)
          - 3456\*\Hhhh(-1,-1,3,0)
\\
&& \nonumber
          - 3072\*\Hhhh(-1,-1,3,1)
          - 2214\*\Hhhh(-1,0,0,0)
          - 2160\*\Hhhh(-1,0,0,0)\*\zeta_2
          + 176\*\Hhhh(-1,2,0,0)
          + 352\*\Hhhh(-1,2,1,0)
\\
&& \nonumber
          + 672\*\Hhhh(-1,3,0,0)
          + 768\*\Hhhh(-1,3,1,0)
          + 700\*\Hhhh(0,0,0,0)
          + 576\*\Hhhh(0,0,0,0)\*\zeta_2
          + 72\*\Hhhh(3,0,0,0)
          + 1152\*\Hhhhh(-2,-1,-1,0,0)
\\
&& \nonumber
          - 2400\*\Hhhhh(-2,-1,0,0,0)
          + 1488\*\Hhhhh(-2,0,0,0,0)
          + 1152\*\Hhhhh(-1,-2,-1,0,0)
          - 2400\*\Hhhhh(-1,-2,0,0,0)
\\
&& \nonumber
          + 1152\*\Hhhhh(-1,-1,-2,0,0)
          + 2304\*\Hhhhh(-1,-1,-1,2,0)
          + 1392\*\Hhhhh(-1,-1,0,0,0)
          - 1536\*\Hhhhh(-1,-1,2,0,0)
\\
&& \nonumber
          - 1536\*\Hhhhh(-1,-1,2,1,0)
          - 1632\*\Hhhhh(-1,0,0,0,0)
          - 144\*\Hhhhh(-1,2,0,0,0)
          + 1020\*\Hhhhh(0,0,0,0,0)
          + 3456\*\Hhhhhh(-1,-1,-1,0,0,0)
\\
&& \nonumber
          - 2688\*\Hhhhhh(-1,-1,0,0,0,0)
          + 720\*\Hhhhhh(-1,0,0,0,0,0)
          - 144\*\Hhhhhh(0,0,0,0,0,0)
          \biggr)
       \biggr\}
\\
&& \mbox{\hspn} \nonumber
       + \frct{16}{3}\,\*\cfs\*(\ca-2\*\cf) \*\nf \* \biggl\{
         (1+x) \* \biggl(
          - 32\*\H(-2)\*\zeta_2
          + 120\*\Hh(-2,0)
          - 32\*\Hh(-1,0)\*\zeta_2
          + 32\*\Hh(2,2)
          + 64\*\Hh(3,1)
\\
&& \nonumber
          - 64\*\Hhh(-2,-1,0)
          - 64\*\Hhh(-1,-2,0)
          + 240\*\Hhh(-1,0,0)
          + 16\*\Hhh(2,0,0)
          + 32\*\Hhh(2,1,0)
          - 128\*\Hhhh(-1,-1,0,0)
\\
&& \nonumber
          + 192\*\Hhhh(-1,0,0,0)
          \biggr)
       + (1-x) \* \biggl(
          - 32\*\H(-3)\*\zeta_2
          - 201\*\H(0)
          + 4\*\H(0)\*\zeta_4
          - 68\*\H(2)
          - 24\*\H(4)
          - 32\*\H(5)
          - 48\*\Hh(-4,0)
\\
&& \nonumber
          + 32\*\Hh(-3,2)
          - 16\*\Hh(-2,0)\*\zeta_2
          + 32\*\Hh(-2,3)
          - 68\*\Hh(1,0)
          + 64\*\Hh(1,2)
          + 64\*\Hh(2,1)
          - 8\*\Hh(4,0)
          - 48\*\Hhh(-3,0,0)
\\
&& \nonumber
          + 16\*\Hhh(-2,2,0)
          + 32\*\Hhh(1,0,0)
          + 64\*\Hhh(1,1,0)
          - 24\*\Hhhh(-2,0,0,0)
          + 60\*\Hhhhh(0,0,0,0,0)
          \biggr)
       + (8 + 24\*x) \* \Hh(3,0)
\\
&& \nonumber
       + (16 + 176\*x) \* \Hh(-3,0)
       + (24 + 72\*x) \* \Hh(0,0)\*\zeta_2
       - (32 + 96\*x) \* \Hh(2,0)
       - (36 - 68\*x) \* \H(0)\*\zeta_3
\\
&& \nonumber
       - (48 + 272\*x) \* \Hhhh(0,0,0,0)
       + (80 + 240\*x) \* \Hhh(-2,0,0)
       - (96 + 160\*x) \* \H(3)
       + (108 + 140\*x) \* \H(0)\*\zeta_2
\\
&& \nonumber
       - (228 + 348\*x) \* \Hhh(0,0,0)
       - (244 + 28\*x) \* \Hh(0,0)
       + \pqq( - x) \* \biggl(
            80\*\H(-2)\*\zeta_3
          - 184\*\H(-2)\*\zeta_2
          + 5\*\H(0)\*\zeta_4
\\
&& \nonumber
          + 158\*\H(0)\*\zeta_3
          - 21\*\H(0)\*\zeta_2
          - 156\*\H(4)
          + 96\*\Hh(-4,0)
          + 136\*\Hh(-3,0)
          - 32\*\Hh(-3,2)
          - 64\*\Hh(-2,-1)\*\zeta_2
          - 42\*\Hh(-2,0)
\\
&& \nonumber
          + 16\*\Hh(-2,0)\*\zeta_2
          + 104\*\Hh(-2,2)
          - 64\*\Hh(-2,3)
          - 64\*\Hh(-1,-2)\*\zeta_2
          + 80\*\Hh(-1,0)\*\zeta_3
          - 184\*\Hh(-1,0)\*\zeta_2
          + 208\*\Hh(-1,3)
\\
&& \nonumber
          - 96\*\Hh(-1,4)
          - 8\*\Hh(0,0)\*\zeta_3
          + 172\*\Hh(0,0)\*\zeta_2
          - 52\*\Hh(3,0)
          + 32\*\Hh(3,2)
          + 24\*\Hh(4,0)
          + 96\*\Hh(4,1)
          - 64\*\Hhh(-3,-1,0)
\\
&& \nonumber
          + 144\*\Hhh(-3,0,0)
          - 64\*\Hhh(-2,-2,0)
          - 160\*\Hhh(-2,-1,0)
          + 64\*\Hhh(-2,-1,2)
          + 244\*\Hhh(-2,0,0)
          - 32\*\Hhh(-2,2,0)
          - 64\*\Hhh(-2,2,1)
\\
&& \nonumber
          - 64\*\Hhh(-1,-3,0)
          - 160\*\Hhh(-1,-2,0)
          + 64\*\Hhh(-1,-2,2)
          - 64\*\Hhh(-1,-1,0)\*\zeta_2
          + 128\*\Hhh(-1,-1,3)
          - 84\*\Hhh(-1,0,0)
\\
&& \nonumber
          + 32\*\Hhh(-1,0,0)\*\zeta_2
          + 104\*\Hhh(-1,2,0)
          - 64\*\Hhh(-1,2,2)
          - 64\*\Hhh(-1,3,0)
          - 128\*\Hhh(-1,3,1)
          + 63\*\Hhh(0,0,0)
          + 48\*\Hhh(0,0,0)\*\zeta_2
\\
&& \nonumber
          + 16\*\Hhh(3,0,0)
          + 32\*\Hhh(3,1,0)
          - 96\*\Hhhh(-2,-1,0,0)
          + 168\*\Hhhh(-2,0,0,0)
          - 96\*\Hhhh(-1,-2,0,0)
          - 320\*\Hhhh(-1,-1,0,0)
\\
&& \nonumber
          + 64\*\Hhhh(-1,-1,2,0)
          + 324\*\Hhhh(-1,0,0,0)
          - 32\*\Hhhh(-1,2,0,0)
          - 64\*\Hhhh(-1,2,1,0)
          - 112\*\Hhhh(0,0,0,0)
          - 96\*\Hhhhh(-1,-1,0,0,0)
\\
&& 
          + 192\*\Hhhhh(-1,0,0,0,0)
          - 120\*\Hhhhh(0,0,0,0,0)
          \biggr)
       \biggr\}
\:\: .
\eea
A difference between the time-like and space-like case appears for the 
quantities $P_{\rm ns}^{\,(3)\rm s}$ for the first time at the four-loop 
level with
\bea
\lefteqn{
\delta\, P^{\,(3)\rm s}(x) \, = \,
}
\nonumber \\ && \mbox{\hspn} \nonumber
       - \frct{16}{3}\* \nf\* \cf\* \dabcnc \* \biggl\{
       (1+x) \* \biggl(
          - 336 \* \H(-2) \* \zeta_3
          + 648 \* \H(-2) \* \zeta_2
          + 384 \* \H(3) \* \zeta_3
          - 576 \* \H(4) \* \zeta_2
          - 328 \* \Hh(-2,0)
\\
&& \nonumber
          - 256 \* \Hh(-2,2)
          - 336 \* \Hh(-1,0) \* \zeta_3
          + 648 \* \Hh(-1,0) \* \zeta_2
          - 512 \* \Hh(-1,3)
          + 1440 \* \Hh(-1,4)
          + 192 \* \Hh(2,0) \* \zeta_3
          + 1312 \* \Hh(2,2)
\\
&& \nonumber
          - 384 \* \Hh(2,2) \* \zeta_2
          - 288 \* \Hh(2,4)
          - 384 \* \Hh(3,0) \* \zeta_2
          + 2624 \* \Hh(3,1)
          - 768 \* \Hh(3,1) \* \zeta_2
          - 192 \* \Hh(3,3)
          + 576 \* \Hh(4,2)
          + 2304 \* \Hh(5,1)
\\
&& \nonumber
          + 784 \* \Hhh(-2,-1,0)
          - 192 \* \Hhh(-2,-1,2)
          + 768 \* \Hhh(-2,2,1)
          + 384 \* \Hhh(-1,-3,0)
          + 784 \* \Hhh(-1,-2,0)
          - 192 \* \Hhh(-1,-2,2)
\\
&& \nonumber
          - 384 \* \Hhh(-1,-1,3)
          - 656 \* \Hhh(-1,0,0)
          - 768 \* \Hhh(-1,0,0) \* \zeta_2
          - 256 \* \Hhh(-1,2,0)
          + 768 \* \Hhh(-1,2,2)
          + 1248 \* \Hhh(-1,3,0)
\\
&& \nonumber
          + 1536 \* \Hhh(-1,3,1)
          - 192 \* \Hhh(2,0,0) \* \zeta_2
          + 1312 \* \Hhh(2,1,0)
          - 384 \* \Hhh(2,1,0) \* \zeta_2
          - 96 \* \Hhh(2,3,0)
          + 432 \* \Hhh(4,0,0)
          + 576 \* \Hhh(4,1,0)
\\
&& \nonumber
          - 384 \* \Hhhh(-2,-1,-1,0)
          - 384 \* \Hhhh(-1,-2,-1,0)
          + 480 \* \Hhhh(-1,-2,0,0)
          - 384 \* \Hhhh(-1,-1,-2,0)
          + 1568 \* \Hhhh(-1,-1,0,0)
\\
&& \nonumber
          - 192 \* \Hhhh(-1,-1,2,0)
          - 1560 \* \Hhhh(-1,0,0,0)
          + 768 \* \Hhhh(-1,2,0,0)
          + 768 \* \Hhhh(-1,2,1,0)
          - 96 \* \Hhhh(2,2,0,0)
          - 336 \* \Hhhh(3,0,0,0)
\\
&& \nonumber
          - 192 \* \Hhhh(3,1,0,0)
          - 768 \* \Hhhhh(-1,-1,-1,0,0)
          + 288 \* \Hhhhh(-1,-1,0,0,0)
          + 384 \* \Hhhhh(-1,0,0,0,0)
          - 384 \* \Hhhhh(2,0,0,0,0)
\\
&& \nonumber
          - 288 \* \Hhhhh(2,1,0,0,0)
          \biggr)
       + (1-x)  \*  \biggl(
            1152 \* \H(-4) \* \zeta_2
          + 1344 \* \H(-3) \* \zeta_3
          + 3232 \* \H(0)
          + 4684 \* \H(2)
          + 288 \* \H(2) \* \zeta_3
\\
&& \nonumber
          - 392 \* \H(2) \* \zeta_2
          - 384 \* \Hh(-5,0)
          - 576 \* \Hh(-4,2)
          - 1536 \* \Hh(-3,-1) \* \zeta_2
          + 1152 \* \Hh(-3,0) \* \zeta_2
          - 768 \* \Hh(-3,3)
\\
&& \nonumber
          - 768 \* \Hh(-2,-2) \* \zeta_2
          + 672 \* \Hh(-2,0) \* \zeta_3
          - 576 \* \Hh(-2,4)
          + 4684 \* \Hh(1,0)
          + 288 \* \Hh(1,0) \* \zeta_3
          - 392 \* \Hh(1,0) \* \zeta_2
          + 2912 \* \Hh(1,2)
\\
&& \nonumber
          - 432 \* \Hh(1,4)
          + 2912 \* \Hh(2,1)
          - 144 \* \Hh(2,3)
          + 1152 \* \Hhh(-4,-1,0)
          - 1056 \* \Hhh(-4,0,0)
          + 1152 \* \Hhh(-3,-2,0)
\\
&& \nonumber
          + 768 \* \Hhh(-3,-1,2)
          - 192 \* \Hhh(-3,2,0)
          + 768 \* \Hhh(-2,-3,0)
          + 384 \* \Hhh(-2,-2,2)
          - 768 \* \Hhh(-2,-1,0) \* \zeta_2
          + 768 \* \Hhh(-2,-1,3)
\\
&& \nonumber
          + 768 \* \Hhh(-2,0,0) \* \zeta_2
          - 192 \* \Hhh(-2,3,0)
          + 2912 \* \Hhh(1,0,0)
          + 2912 \* \Hhh(1,1,0)
          - 144 \* \Hhh(1,3,0)
          - 1536 \* \Hhhh(-3,-1,-1,0)
\\
&& \nonumber
          + 1920 \* \Hhhh(-3,-1,0,0)
          - 1248 \* \Hhhh(-3,0,0,0)
          - 768 \* \Hhhh(-2,-2,-1,0)
          + 1344 \* \Hhhh(-2,-2,0,0)
          - 768 \* \Hhhh(-2,-1,-2,0)
\\
&& \nonumber
          + 384 \* \Hhhh(-2,-1,2,0)
          - 396 \* \Hhhh(1,0,0,0)
          - 144 \* \Hhhh(1,2,0,0)
          - 144 \* \Hhhh(2,1,0,0)
          - 1536 \* \Hhhhh(-2,-1,-1,0,0)
\\
&& \nonumber
          + 1728 \* \Hhhhh(-2,-1,0,0,0)
          - 768 \* \Hhhhh(-2,0,0,0,0)
          - 576 \* \Hhhhh(1,0,0,0,0)
          - 432 \* \Hhhhh(1,1,0,0,0)
          \biggr)
\\
&& \nonumber
       - (96 - 1056 \* x + 1024 \* x^2)  \* \Hhhh(-2,-1,0,0)
       + (96 - 864 \* x - 512 \* x^2)  \*  (
            \Hh(3,2)
          + \Hhh(3,1,0)
          )
\\
&& \nonumber
       + (96 - 96 \* x + 512 \* x^2)  \*  \Hhh(1,0,0) \* \zeta_2
       + (96 + 192 \* x + 512 \* x^2)  \*  \Hh(2,0) \* \zeta_2
       + (96 + 480 \* x + 512 \* x^2)  \*  \H(3) \* \zeta_2
\\
&& \nonumber
       + (96 + 672 \* x - 512 \* x^2)  \*  \Hhh(-2,-2,0)
       - (144 - 240 \* x)  \*  \H(0) \* \zeta_5
       - (192 - 960 \* x + 1024 \* x^2)  \*  \Hhh(-3,-1,0)
\\
&& \nonumber
       - (192 - 192 \* x - 512 \* x^2)  \*  (
            \Hh(1,2) \* \zeta_2
          + \Hh(2,1) \* \zeta_2
          + \Hhh(1,1,0) \* \zeta_2
          )
       + (288 - 2592 \* x - 1536 \* x^2)  \*  \Hh(4,1)
\\
&& \nonumber
       + (288 - 576 \* x + 768 \* x^2)  \*  \Hh(-4,0)
       - (288 - 432 \* x)  \*  \Hhhh(2,0,0,0)
       + (372 - 1680 \* x - 736 \* x^2)  \*  \H(0) \* \zeta_4
\\
&& \nonumber
       - (400 + 1168 \* x)  \*  \Hh(-3,0)
       - (480 + 1280 \* x^2)  \*  \Hhhhh(0,0,0,0,0)
       + (576 - 960 \* x + 1536 \* x^2)  \*  \Hhh(-3,0,0)
\\
&& \nonumber
       + (576 - 192 \* x + 1024 \* x^2)  \*  \Hh(-3,2)
       - (576 + 1152 \* x + 2048 \* x^2)  \*  \H(5)
       - (976 + 1504 \* x)  \*  \H(0) \* \zeta_3
\\
&& \nonumber
       + (672 - 768 \* x + 1536 \* x^2)  \*  \Hhhh(-2,0,0,0)
       - (672 - 672 \* x + 1536 \* x^2)  \*  \H(-3) \* \zeta_2
       + (1184 + 1856 \* x)  \*  \Hhhh(0,0,0,0)
\\
&& \nonumber
       + (672 + 288 \* x + 512 \* x^2)  \*  \Hhh(-2,2,0)
       - (720 + 48 \* x + 768 \* x^2)  \*  \Hh(-2,0) \* \zeta_2
       - (2280 + 1256 \* x)  \*  \Hh(0,0) \* \zeta_2
\\
&& \nonumber
       + (720 + 864 \* x + 2304 \* x^2)  \*  \Hhh(0,0,0) \* \zeta_2
       + (768 + 1792 \* x^2)  \*  \Hh(0,0) \* \zeta_3
       - (920 + 1688 \* x)  \*  \Hhh(-2,0,0)
\\
&& \nonumber
       + (1056 + 288 \* x)  \*  \H(0) \* \zeta_2 \* \zeta_3
       + (1056 + 288 \* x + 1024 \* x^2)  \*  \Hh(-2,3)
       + (1180 + 1444 \* x)  \*  \Hhh(2,0,0)
\\
&& \nonumber
       + (1488 + 1296 \* x)  \*  \Hh(5,0)
       + (1680 + 720 \* x)  \*  \H(6)
       - (1728 + 576 \* x)  \*  \Hhhh(0,0,0,0) \* \zeta_2
       + (2252 + 1940 \* x)  \*  \Hh(3,0)
\\
&& \nonumber
       + (2280 + 1524 \* x)  \*  \Hhh(0,0,0)
       + (2376 + 2040 \* x)  \*  \Hh(0,0) \* \zeta_4
       - (2724 - 516 \* x)  \*  \H(0) \* \zeta_2
       + (2820 + 1884 \* x)  \*  \H(4)
\\
&& \nonumber
       + (3232 - 6136 \* x)  \*  \Hh(0,0)
       + (3852 - 1972 \* x)  \*  \Hh(2,0)
       + (4792 - 1032 \* x)  \*  \H(3)
       - (1584 \* x + 1280 \* x^2)  \*  \Hh(4,0)
\\
&& \nonumber
       - (672 \* x + 512 \* x^2)  \* \Hhh(3,0,0)
     + \biggl(\frct{1}{x}+x^2\biggr) \ \*  \biggl(
          - 896 \* \H(-2) \* \zeta_3
          + 768 \* \Hh(-2,-1) \* \zeta_2
          - 640 \* \Hh(-2,0) \* \zeta_2
          + 512 \* \Hh(-2,3)
\\
&& \nonumber
          + 768 \* \Hh(-1,-2) \* \zeta_2
          - 896 \* \Hh(-1,0) \* \zeta_3
          + 1536 \* \Hh(-1,4)
          - 256 \* \Hh(1,2) \* \zeta_2
          - 128 \* \Hh(2,0) \* \zeta_2
          - 256 \* \Hh(2,1) \* \zeta_2
\\
&& \nonumber
          - 256 \* \Hhh(-2,-2,0)
          - 512 \* \Hhh(-2,-1,2)
          + 256 \* \Hhh(-2,2,0)
          + 512 \* \Hhh(-2,2,1)
          - 512 \* \Hhh(-1,-3,0)
          - 512 \* \Hhh(-1,-2,2)
\\
&& \nonumber
          + 768 \* \Hhh(-1,-1,0) \* \zeta_2
          - 1024 \* \Hhh(-1,-1,3)
          - 1280 \* \Hhh(-1,0,0) \* \zeta_2
          + 512 \* \Hhh(-1,2,2)
          + 1024 \* \Hhh(-1,3,0)
\\
&& \nonumber
          + 1024 \* \Hhh(-1,3,1)
          - 256 \* \Hhh(1,0,0) \* \zeta_2
          - 256 \* \Hhh(1,1,0) \* \zeta_2
          + 512 \* \Hhhh(-2,-1,-1,0)
          - 512 \* \Hhhh(-2,-1,0,0)
\\
&& \nonumber
          + 256 \* \Hhhh(-2,0,0,0)
          + 512 \* \Hhhh(-1,-2,-1,0)
          - 1024 \* \Hhhh(-1,-2,0,0)
          + 512 \* \Hhhh(-1,-1,-2,0)
          - 512 \* \Hhhh(-1,-1,2,0)
\\
&& \nonumber
          + 512 \* \Hhhh(-1,2,0,0)
          + 512 \* \Hhhh(-1,2,1,0)
          + 1024 \* \Hhhhh(-1,-1,-1,0,0)
          - 1536 \* \Hhhhh(-1,-1,0,0,0)
          + 1024 \* \Hhhhh(-1,0,0,0,0)
          \biggr)
\\[-1mm]
&&
          - 1152 \* \Hhh(0,0,0) \* \zeta_3
          + 576 \* \Hhhhhh(0,0,0,0,0,0)
       \biggr\}
\:\: .
\eea

\renewcommand{\theequation}{\ref{sec:appD}.\arabic{equation}}
\setcounter{equation}{0}
\renewcommand{\thefigure}{\ref{sec:appD}.\arabic{figure}}
\setcounter{figure}{0}
\renewcommand{\thetable}{\ref{sec:appD}.\arabic{table}}
\setcounter{table}{0}
\section{The complete $\zeta_5$ contributions}
\label{sec:appD}
Here we finally present the exact expressions for the part of 
$\gamma_{\rm ns}^{\,(3)\pm}(N)$ which is proportional to $\,\zeta_5\,$:
\bea
\label{eq:gqq3+zeta5}
  \gamma_{\,\rm ns}^{\:(3)+}(N)\biggr|_{\zeta_5} &\!=\!&
\cff\,\, \* 320\,\*\zeta_5\* \Big(
       \frct{111}{12}
     +  6 \* \eta
     -  9 \* \eta^2
     + 14 \,\* \S(-2)
     \Big)
\nonumber \\ && \mbox{\hspn}
  +\, \cft\*\ca\,\, \* 320\,\*\zeta_5\* \Big(
     - \frct{59}{4}  
     - 12 \* \eta
     + 18 \* \eta^2
     - 22 \* \S(-2)
     \Big)
\nonumber \\ && \mbox{\hspn}
  +\, \cfs\,\*\cas\,\, \* 80\,\*\zeta_5\* \Big(
       \frct{67}{4} 
     + \frct{67}{3} \,\* \eta
     + \frct{58}{3} \,\* \S(1)
     - 58 \* \eta^2
     +  8 \* \S(1)\*\eta
     -  8 \* (\S(1))^2
     + 34 \,\* \S(-2)
     \Big)
\nonumber \\ && \mbox{\hspn}
  +\, \cf\*\cat\,\, \* \frct{80}{3}\,\*\zeta_5\* \Big(
       \frct{13}{4} 
     + \frct{40}{3} \,\* \eta
     - \frct{116}{3} \,\* \S(1)
     + 43 \* \eta^2
     - 16 \* \S(1)\*\eta
     + 16 \* (\S(1))^2
     -  6 \* \S(-2)
     \Big)
\nonumber \\ && \mbox{\hspn}
  +\, \dfFAnc\,\, \* 320\,\*\zeta_5\* \Big(
     - \frct{25}{2} 
     + \frct{25}{3} \,\* \eta
     + \frct{58}{3} \,\* \S(1)
     - 23 \* \eta^2
     +  8 \* \S(1)\*\eta 
     -  8 \* (\S(1))^2
     \Big)
\nonumber \\ && \mbox{\hspn}
  +\, \nf\,\*\cft\,\, \* 160\,\*\zeta_5\* \Big(
       \frct{3}{2}
     + \eta
     - 2 \* \S(1)
     \Big)
  \,+\, \nf\,\*\cfs\*\ca\,\, \* \frct{80}{3}\*\zeta_5 \* \Big(
     - \frct{3}{2}
     - \eta
     + 2 \* \S(1)
     \Big)
\nonumber \\ && \mbox{\hspn}
  +\, \nf\,\*\cf\*\cas\,\, \* \frct{80}{9}\,\*\zeta_5 \* \Big(
     - \frct{33}{2}
     - 25 \* \eta
     + 26 \* \S(1)
     + 12 \* \eta^2
     \Big)
\nonumber \\ && \mbox{\hspn}
  +\, \nf\,\*\dfFFnc\,\, \* \frct{1280}{3}\,\*\zeta_5\* \Big(
        3
     -  5 \* \eta
     -  2 \* \S(1) 
     +  6 \* \eta^2
     \Big)
\:\: , \\[3mm]
\label{eq:gqq3-zeta5}
  \gamma_{\,\rm ns}^{\:(3)-}(N)\biggr|_{\zeta_5} &\!=\!&
  \gamma_{\,\rm ns}^{\:(3)-}(N)\biggr|_{\zeta_5} 
\nonumber \\ && \mbox{\hspn}
+\, \cff\,\, \* 320\,\*\zeta_5\* \Big(
     - \frct{1}{6} 
     + \frct{29}{6} \* \eta
     +  7 \* \eta^2
     \Big)
  \,+\, \cft\*\ca\,\,\* 320\,\*\zeta_5\* \Big(
       \frct{1}{3} 
     - \frct{20}{3} \* \eta
     - 11 \* \eta^2
     \Big)
\nonumber \\ && \mbox{\hspn}
  +\, \cfs\*\cas\,\,\* 80\,\*\zeta_5\* \Big(
     - 1 
     + 4 \* \eta
     + 17 \* \eta^2
     \Big)
  \,+\, \cf\*\cat\,\,\* \frct{80}{3}\,\*\zeta_5\* \Big(
       \frct{5}{6} 
     + \frct{47}{6} \,\* \eta
     -  3 \* \eta^2
     \Big)
\nonumber \\ && \mbox{\hspn}
  +\, \dfFAnc\,\,\* 320\,\*\zeta_5\* \Big(
     - \frct{1}{6} 
     - \frct{13}{6} \,\* \eta
     \Big)
\:\: .
\eea
Since only sums with $w \leq2$ can occur with $\zeta_5$, the corresponding 
functional form is so restricted that a direct determination and verification 
is possible with eight even and odd moments.
  
\bigskip
{\footnotesize


\providecommand{\href}[2]{#2}\begingroup\raggedright\endgroup

}

\end{document}